\documentclass{cernrep}
\usepackage{graphicx,here} 
\usepackage{amsfonts,amsbsy,amssymb}
\font\tenimbf=cmmib10 at 10pt
\font\sevenimbf=cmmib10 at 6pt
\font\fiveimbf=cmmib10 at 4pt
\newfam\imbf
\textfont\imbf=\tenimbf
\scriptfont\imbf=\sevenimbf
\scriptscriptfont\imbf=\fiveimbf
\def\imb{\fam\imbf\tenimbf}
\def\ga{\mathrel{\mathpalette\fun >}}
\def\fun#1#2{\lower3.6pt\vbox{\baselineskip0pt\lineskip.9pt
\ialign{$\mathsurround=0pt#1\hfil##\hfil$\crcr#2\crcr\sim\crcr}}}  

\def\bea{\begin{eqnarray}}
\def\eea{\end{eqnarray}}
\def\beq{\begin{equation}}
\def\eeq{\end{equation}}
\def\Tdec{{T_{\rm dec}}}
\def\psat{{p_{\rm sat}}}
\def\LQCD{{\Lambda_{\rm QCD}}}
\def\lq{{\lambda_{\rm q}}}
\def\lg{{\lambda_{\rm g}}}
\def\pt{p_{_T}}
\def\kt{k_{_T}}

\def\mf{M_{_F}}
\long\def\comment#1{ }    
\def\be{\begin{equation}}
\def\ee{\end{equation}}
\def\lsim{\raise0.3ex\hbox{$<$\kern-0.75em\raise-1.1ex\hbox{$\sim$}}}
\def\gsim{\raise0.3ex\hbox{$>$\kern-0.75em\raise-1.1ex\hbox{$\sim$}}}
\newcommand{\sll}{\raise.15ex\hbox{$/$}\kern-.43em\hbox{$l$}}
\newcommand{\slepsilon}{\raise.15ex\hbox{$/$}\kern-.53em\hbox{$\epsilon$}}
\newcommand{\slvarepsilon}{\raise.15ex\hbox{$/$}\kern-.53em\hbox{$\varepsilon$}}
\newcommand{\slL}{\raise.15ex\hbox{$/$}\kern-.53em\hbox{$L$}}
\newcommand{\slP}{\raise.15ex\hbox{$/$}\kern-.53em\hbox{$P$}}
\newcommand{\slp}{\raise.1ex\hbox{$/$}\kern-.63em\hbox{$p$}}
\newcommand{\slq}{\raise.1ex\hbox{$/$}\kern-.63em\hbox{$q$}}
\newcommand{\slv}{\raise.1ex\hbox{$/$}\kern-.63em\hbox{$v$}}
\newcommand{\slR}{\raise.15ex\hbox{$/$}\kern-.53em\hbox{$R$}}
\newcommand{\slQ}{\raise.15ex\hbox{$/$}\kern-.53em\hbox{$Q$}}
\newcommand{\slK}{\raise.15ex\hbox{$/$}\kern-.53em\hbox{$K$}}
\newcommand{\slk}{\raise.15ex\hbox{$/$}\kern-.53em\hbox{$k$}}
\newcommand{\slSigma}{\raise.15ex\hbox{$/$}\kern-.53em\hbox{$\Sigma$}}
\newcommand{\slcalP}{\raise.15ex\hbox{$/$}\kern-.63em\hbox{$\cal P$}}
\newcommand{\slA}{\raise.15ex\hbox{$/$}\kern-.73em\hbox{$A$}}
\newcommand{\slbfA}{\raise.15ex\hbox{$/$}\kern-.73em\hbox{${\imb A}$}}
\newcommand{\slpartial}{\raise.15ex\hbox{$/$}\kern-.53em\hbox{$\partial$}}

\def\empile#1\over#2{\mathrel{\mathop{\kern 0pt#1}\limits_{#2}}}

\def\p{{\boldsymbol p}}
\def\q{{\boldsymbol q}}

\def\k{{\boldsymbol k}}

\def\x{{\boldsymbol x}}

\begin{document}

\title{PHOTON PHYSICS IN HEAVY ION COLLISIONS AT THE LHC}

\author{{\bf Conveners}: P.~Aurenche, O.~Kodolova, P.~Levai, I.~Lokhtin,
T.~Peitzmann,  K.~Redlich\\
{\bf Editor}: P.~Aurenche\\
{\bf Authors:}
F.~Arleo$^a$, P.~Aurenche$^b$, F.~Bopp$^c$, I.~Dadi\'c$^d$, G.~David$^e$,
H.~Delagrange$^f$, D.~d'Enterria$^{f,g}$, K.J. Eskola$^{h,i}$, F.~Gelis$^j$,
J.-Ph.~Guillet$^b$, S.~Jeon$^k$, Yu.~Kharlov$^l$, O.~Kodolova$^m$,
P.~Levai$^{n}$, J.H.~Liu$^{o}$, I.P.~Lokhtin$^m$, G.D.~Moore$^k$,
H.~Niemi$^{h,i}$,
A.~Nikitenko$^p$,T.~Peitzmann$^q$, P.~Petreczky$^e$,  J.~Ranft$^c$,
R.~Rapp$^r$, P.V.~Ruuskanen$^{h,i}$, K.~Redlich$^s$, S.S.~R\"as\"anen$^h$,
I.~Sarcevic$^{t}$, J.~Serreau$^u$, 
D.K.~Srivastava$^v$,
H.~Takai$^e$, S.~Tapprogge$^w$,
M.~Tokarev$^x$, I.N.~Vardanyan$^m$, M.~Werlen$^b$,  P.~Yepes$^o$}

\institute{
$^{~}$ ~\\
$^{~}$ ~\\
$^a$ ECT$^*$ and INFN, G.C di Trento, I-38050 Villazzano (Trento), Italy\\
$^b$ LAPTH, UMR5108 du CNRS associ\'ee \`a l'Universit\'e de Savoie,
F-74941 Annecy-le-Vieux, France\\
$^c$ Fachbereich Physik, Siegen Universit\"at, D-57068 Siegen, Germany\\
$^d$ Ruder Bo\v skovi\'c Institute, PO Box 180, HR-10002 Zagreb, Croatia\\
$^e$ Brookhaven National Laboratory, Upton, NY 11973, USA\\
$^f$ SUBATECH, CNRS/IN2P3, Ecole des Mines, F-44307 Nantes, France\\
$^{g}$ Nevis Laboratories, Columbia University, New York, NY 10027, USA\\
$^h$ Department of Physics, PB 35 (YFL), FIN-40014 University of Jyv\"askyl\"a,
Finland\\
$^i$ Helsinki Institute of Physics, PB 64, FIN-00014 University of Helsinki,
Finland\\
$^j$ Service de Physique Th\'eorique, CEA/DSM/Saclay, F-91191 Gif-sur-Yvette
Cedex, France\\
$^k$ Departement of Physics, McGill University, Montr\'eal, QC H3A-2T8, Canada\\
$^l$ Institute for High Energy Physics, Moscow Region, RU-142284 Protvino, Russia\\
$^m$ Institute of Nuclear Physics, Moscow State University, Moscow, Russia\\
$^{n}$ RMKI Research Institute for Particle and Nuclear Physics, P.O. Box 49,
Budapest 1525, Hungary\\
$^{o}$ T.W. Bonner Nuclear Lab., Rice University, Houston, Texas 77251-1892, USA\\
$^p$ Imperial College, Prince Consort Road, London SW7 2BZ, United Kingdom\\
$^q$ Utrecht University, NL-3584 TA Utrecht, The Netherlands\\
$^r$ NORDITA, Blegdamsvej 17, DK-2100 Copenhagen, Denmark\\
$^s$ Institute of Theoretical Physics, University of Wroclaw, PL-50204 Wroclaw,
Poland\\
$^{t}$ Department of Physics, University of Arizona, Tucson, Arizona 85721, USA\\
$^u$ Institute f\"ur Theoretische Physik, Universt\"at Heidelberg, D-69120,
Heidelberg, Germany\\
$^v$ Variable Energy Cyclotron Centre, 1/AF Bidhan Nagar, Kolkota 700 064,
India\\
$^w$ CERN, CH-1211 Gen\`eve, Switzerland\\
$^x$ JINR, Joliot-Curie 6, Moscow Region, RU-141980 Dubna, Russia\\
}
%
%
%

\maketitle 

\hfill

\begin{abstract}
 
Various pion and photon production mechanisms in high-energy nuclear collisions
at RHIC and LHC are discussed. Comparison with RHIC data is done whenever
possible. The prospect of using electromagnetic probes to characterize 
quark-gluon plasma formation is assessed.

\end{abstract}

\hfill
\newpage

\tableofcontents

\newpage
%
\section{INTRODUCTION}


The production of photons in heavy ion collisions is rather complex and, in the
standard approach, one roughly distinguishes three types of mechanisms: 

$-\ 1)$ the photon is produced in the hard interaction of two partons in the
incoming nuclei similarly to the well-known Chromodynamics (QCD) processes (QCD
Compton, an\-ni\-hi\-la\-tion, brems\-strahlung) in nucleon-nucleon collisions.
The rate is calculable in perturbative QCD and falls off at large transverse
momentum, $p_{_T}$, as a power law. Photons are also produced as decay 
products of hadrons, such as of $\pi^0, \eta, \cdots$, which are emitted in
hard QCD processes and at large  $p_{_T}$ the decay photon spectrum is also
power behaved;

$-\ 2)$ in the collision of two nuclei the density of secondary partons is so
high that the quarks and gluons rescatter and eventually thermalize to form a
bubble of hot quark-gluon plasma: at LHC the initial temperature of the plasma
is expected to be of the order of 1 GeV. It is assumed that the plasma evolves
hydrodynamically. Photons are emitted in the collisions of quarks and gluons
with an energy spectrum which is exponentially damped but which should extend
up to several GeV;  

$-\ 3)$ the QGP bubble expands and cools until a temperature of 150 to 200 MeV
is reached and a hadronic phase appears. As they collide the hot hadronic
resonances ($\pi^0,\ \rho,\ \omega$) emit photons until the freeze-out 
temperature is reached. The typical energy of such photons ranges from several
hundred MeV to several GeV. 

In this standard picture it is expected that the thermal production mechanisms
will produce an excess of photons with an energy of a few GeV. In the following
we critically discuss each stage of the production processes. One of the  main
problems is the fact that the theoretical predictions suffer from very large
uncertainties at each step of the above scenario. \\

The report is organized as follows:\\
chap. 2: nomenclature of photons according to their production mechanisms;\\
chap. 3: results of RHIC;\\
chap. 4: experimental aspects of photon detection at LHC;\\
chap. 5: inclusive photon and (background) $\pi^0$ production from
perturbative QCD;\\
chap. 6: inclusive thermal photon and $\pi^0$ production;\\
chap. 7: comparison of thermal and non-thermal photon and $\pi^0$ production
mechanisms;\\
chap. 8. preliminary studies on photon-jet and photon-particle correlations;\\
chap. 9: theoretical considerations on non equilibrium effects;\\
chap. 10: theoretical considerations on lattice calculations of dilepton
rates.\\
Useful information on nuclear cross sections on the one hand, and on
luminosities, acceptances, etc., on the other hand is collected in two
appendices.\\

Chapters 3 and 4 contain the experimental part of the report. The PHENIX results
on $\pi^0$ and $\gamma$ production are first reviewed before turning to a
discussion of the experimental capabilities of ALICE, CMS and ATLAS for
detecting pions and photons.\\

In Chaps. 5 to 8 quantitative studies of inclusive photon production are
presented. Both ``signal", {\em i.e.} direct photons, and ``background"  {\em
i.e.} photons from decay of resonances, are discussed. These studies are
carried out using presently available tools and models and, whenever possible,
uncertainties in the predictions are given. Proton-proton, proton-nucleus and
nucleus-nucleus collisions are treated in parallel. Predictions are made for
the ratio $\gamma_{\rm direct} / \gamma_{\rm all}$ or $\gamma_{\rm direct} /
\pi^0$ which determine if the extraction of a direct photon signal is feasible.
The production of low mass lepton pairs at large transverse momentum is
presented as a channel complementary to real photon production: it is based on
similar dynamics of production but suffers from different backgrounds. 

Whenever possible we use two alternative models. On the one hand, the
``standard" approach based on next-to-leading-order (NLO) QCD calculations to
describe the hard processes together with a hydrodynamic model to describe the
thermal evolution of the fireball produced in heavy ion collisions, and, on the
other hand, the Dual Parton Model (DPM) which combines soft and hard
leading-order (LO) dynamics and has been extremely successful in describing
hadron-hadron scattering as well as fixed target nucleus-nucleus scattering.
Surprisingly, predictions of the two models in nucleus-nucleus collisions turn
out to be very similar despite the fact that the treatment of final state
nuclear effects are quite different.  

Exploratory studies on photon-jet, photon-hadron and photon-photon correlations
are presented and the comparison between proton-proton and nucleus-nucleus
scattering is made. Only the LO approximation is used and, for the latter two
cases, only the signal is  considered. 

We conclude from these studies that:\\
-- thermal photon production manifests itself by an enhancement in the
inclusive photon spectrum at $\pt$ values below 10 to 15 GeV/c at the LHC.
The shape of the transverse momentum spectrum may also be indicative of the
production mechanisms;\\
-- the $\gamma_{\rm direct} / \pi^0$ ratio in nucleus-nucleus collisions should
be large enough to allow for the extraction of a direct photon signal, however
the uncertainties on the predictions are large mainly due to the poor knowledge
of the model parameters; \\
-- the lepton pair channel at large momentum transfer looks promising but
further detailed studies are necessary to determine if the large background
from uncorrelated pairs can be reliably subtracted;\\
-- correlation studies show characteristic changes of shapes in A+A collisions
compared to $p p$ collisions but, here again, further studies are necessary
concerning the background.

Many of the results presented are new, in particular, the comparison between
two alternative models as well as the extensive discussion on the uncertainties
at the LHC (role of the initial conditions of thermalization, chemical
equilibrium  {\em vs.} non equilibrium, $\cdots$), the studies of the lepton
pair spectrum as well as that of the correlations involving a photon.\\

Chapters 9 and 10 are of a different nature and address more fundamental issues.
Some of the basic hypotheses upon which the thermal production studies are
based are critically analysed. 

In Chap. 9, the relevance of the finite life-time of the thermal system is
adressed and it is shown that recent claims of a very large production rate of
photons due to this finite life-time are not tenable since they are based on a
defective modeling of the system. This is an original piece of work.
Furthermore, ways to deal with this problem in a realistic way are sketched. 

In Chap.10, the improved perturbative methods upon which the calculation of
thermal photon rates are based are compared with the non-perturbative lattice 
based method in a simple example, namely the production of a static lepton
pair. The two approaches seem to indicate a large discrepancy both in the
magnitude of the rate as well as in the functional dependence on the lepton
pair mass. The error bars in both predictions are very large however and it is
premature to conclude if a real contradiction between the two results exists.
This is a very interesting problem and, clearly, further studies are called
for. \\

This report is far from being the definitive work on photon,
dilepton and pion production in heavy ion collisions at the LHC. The results
are sufficiently interesting however to motivate more detailed theoretical and
phenomenological studies on these topics.

%
\section{PHOTONS AND PHOTONS}  
\label{sec:nomenclature}

We define the ``inclusive" photon spectrum in the usual sense: it is the
unbiased photon spectrum observed in a collision between two hadrons or a
hadron and a nucleus or two nuclei. This spectrum is built-up from a
``cocktail" of many components:

-- ``prompt" photons which are produced, early in the collision, in  hard QCD
processes. They are directly emerging from a hard process or produced by
bremsstrahlung in a hard QCD process. The associated spectrum is power behaved
and dominates at large transverse momentum;

-- ``thermal" photons which are emitted in  the collisions of quarks and gluons
in the QGP phase or in scattering of hadronic resonances in hot matter; their
spectrum is exponentially damped at large enough energy;

-- ``decay" photons are decay products of hadronic resonances (essentially
$\pi^0$ and $\eta$). These resonances can be either produced in hard QCD
processes (and the corresponding decay photon spectrum will be power behaved)
or at the end of the thermal evolution of the system;

-- ``direct" photons are the sum of ``prompt" and ``thermal" photons. They can
be obtained experimentally by subtracting from the inclusive spectrum the
contribution from the ``decay" photons which constitute a reducible
background.

In heavy ion collisions, the aim is the extraction of the ``thermal" signal: it
can only be done by subtracting from the ``direct" photon spectrum the
contribution of the ``prompt" photons (which are an irreducible background to
direct thermal photons) calculated from theory. Therefore it is of utmost
interest to correctly estimate the latter component. A prerequisite is a
precise control of the photon production rate in proton-proton and
proton-nucleus collisions.

One also defines in the context of thermal production ``soft" and ``hard"
photons. The ``soft" photons have an energy much less that the temperature of
the medium while the ``hard" ones have an energy of the order of the
temperature or larger. This terminology is somewhat different from that used in
the context of perturbative QCD. Only hard thermal photons are of interest for
phenomenological studied since soft ones are overwhelmed by the background.

%
%

\section{NEUTRAL PION AND PHOTON RESULTS FROM RHIC}

{\em G.~David}


The first three years of RHIC experiments brought spectacular results
at an impressive pace.  After producing the first Au+Au collisions
at $\sqrt{s}=130$~GeV June 12, 2000, the accelerator delivered
significant integrated luminosity (Table~\ref{tb:rhicruns}) and by
QM'01 (January 2001) many exciting analyses were
completed and presented.  Maybe the most intriguing observation
reported was the large suppression of high $\pt$ neutral pions
and charged hadrons in central collisions with respect to
peripheral or $p p$ 
collisions~\cite{ppg003}\footnote{$p p$ spectra were 
interpolated from measurements at lower and higher $\sqrt{s}$}
scaled with the
number of nucleon-nucleon collisions.  
In Run-2 RHIC operated at 
$\sqrt{s}=200$~GeV\footnote{Full design energy for Au+Au} 
producing both Au+Au~\cite{ppg014,starbacktoback} and polarized $p p$
collisions~\cite{ppg024}.  The suppression of high $\pt$  hadrons in central
Au+Au collisions has been confirmed and  the measurement extended to
$\sim$~10~GeV/c,  while the $p p$ data provided neutral pion spectra up to
13~GeV/c.  Therefore, the nuclear modification factor could be established with
$\pi^0$ measured in the same experiment.  However, it remained an open issue
whether the suppression is an initial state or final state effect. Proving its
versatility in Run-3 RHIC delivered D+Au collisions~\cite{ppg028,star2003} (and
once again polarized $p p$) which were analyzed extremely fast and the results
were published less than three  months after data taking. These results
essentially ruled out initial state effects as cause of the high $\pt$
suppression observed in Au+Au collisions at RHIC energies.  

Meanwhile, few and only preliminary results on inclusive and
direct photon production became public.  This is understandable
since the photon measurement is much more difficult than the
(correlation-type) $\pi^0$ measurement, and also because claiming
any {\it excess} photons over the abundant background from
hadron decays assumes that spectra of the contributing hadrons
themselves ($\pi^0,\eta,...${\it etc.}) are known to high
precision.  First published photon results from RHIC are expected
by the end of 2003, initially addressing the high $\pt$ region
where photon identification is least problematic.

One of the strengths of the RHIC program is a certain redundance
within and overlap between experiments.  In particular, photon
and $\pi^0$ measurements - while mostly done in PHENIX with the
electromagnetic calorimeters~\cite{nimpaper} - are also
possible both in STAR and PHENIX {\it via} photon conversion
which serves not only as a cross-check but helps
to extend the spectra to very low $\pt$.  In addition, even within
PHENIX there are two different electromagnetic
calorimeters using different technologies and
analyzed separately.  
The fact that one can make {\it independent measurements} of the
same observable within the {\it same experiment} greatly increases the
confidence in the ultimate results.

\begin{center}
 \begin{table}[ht]
  \begin{tabular}{|r||r|r|r|r|} \hline
   Run/date & Species  & $\sqrt{s}$ & Int. luminosity  &  
		Submitted publications   \\ \hline \hline 
   {\bf Run-1} &   &   & & \\
   June-Sep 2000 &   Au+Au &   130 GeV & 1$\mu$b$^{-1}$	& 
                                   {\bf PRL 88}, 022301 (2002) \\ \hline
   {\bf Run-2} &   &   & & \\
   Sep-Nov 2001 &   Au+Au &   200 GeV & 24$\mu$b$^{-1}$	&
                                   {\bf PRL 91},  072301 (2003)  \\
                       &   &   & & {\bf NPA 715} (2003) 683c \\
                       &   &   & & {\bf NPA 715} (2003) 691c \\
   Dec 2001 - Jan 2002 &   $p p$ &   200 GeV & 0.15pb$^{-1}$	& 
                                   hep-ex/0304038 v2 ($\rightarrow$PRL) \\ \hline
   {\bf Run-3} &   &   & & \\
   Nov 2002 - Mar 2003 &   D+Au &   200 GeV & 2.7nb$^{-1}$	& 
                                    {\bf PRL 91} 072303 (2003) \\
   Apr 2003 - Jun 2003 &   $p p$  &   200 GeV & 0.35pb$^{-1}$	& \\ \hline
  \end{tabular}
  \caption{ { \label{tb:rhicruns}}
	Overview of RHIC runs as of June 2003.  Integrated luminosity
  is given for PHENIX (which is typically the highest for the four
	RHIC experiments).
  In the last column we only list the {\it final} (submitted) publications 
  on $\pi^0$ and published QM'02 preliminary results on photons.
  }
 \end{table}
\end{center}

\subsection{$\pi^0$ spectra at RHIC}

\begin{figure}
\begin{minipage}[t]{75mm}
  \includegraphics[scale=0.32]{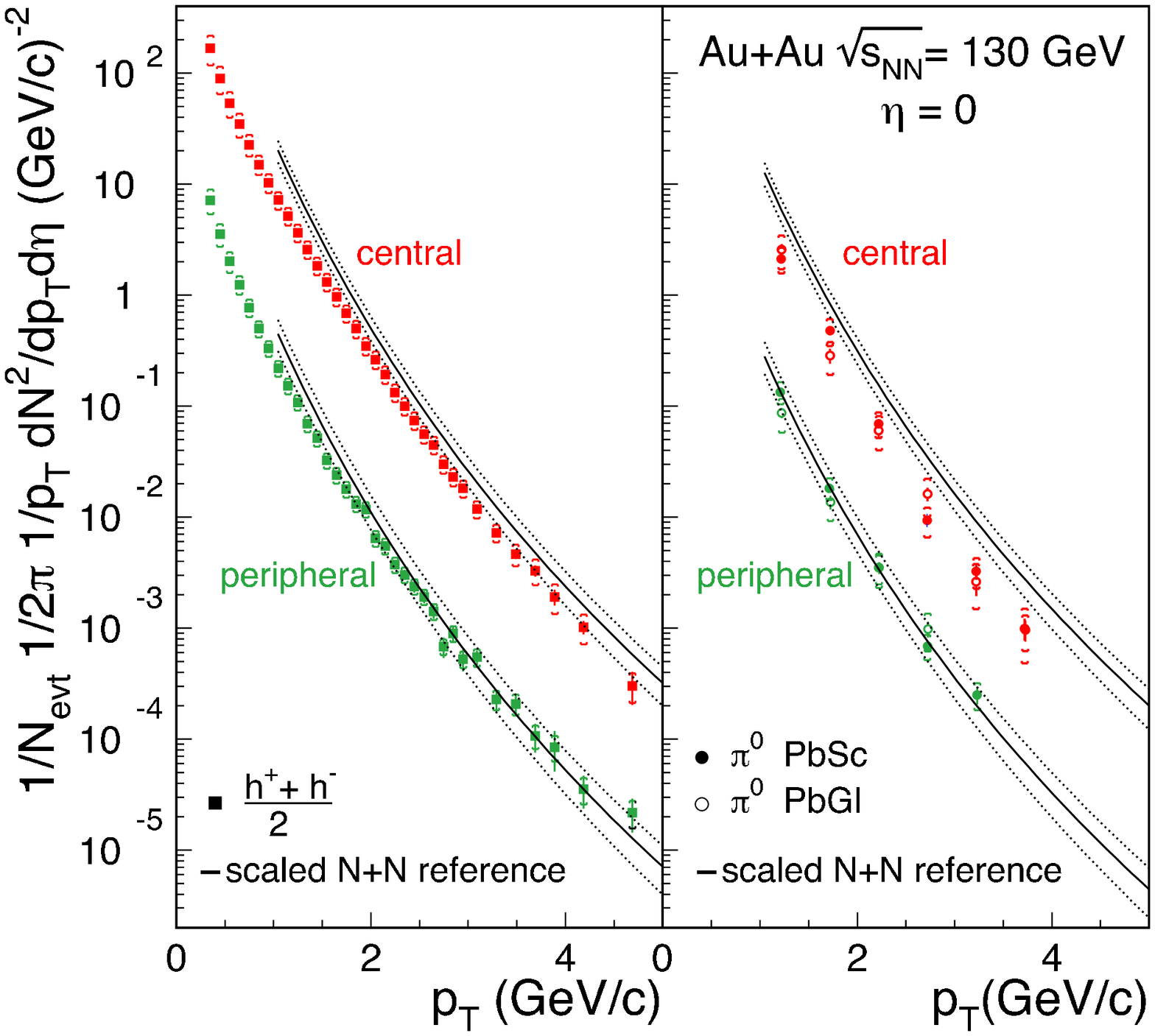}
\caption{
	PHENIX results from 130 GeV Au+Au collisions (Run-1).
	The yields per event at mid-rapidity for ({\em left}) 
	charged hadrons and ({\em right}) neutral pions
	are shown
	as a function of $\pt$ for 60--80\%~({\it lower}) and 
	0--10\%~({\it upper}) event samples, with
	the $\pi^{0}$ results from the PbSc and
	PbGl analyses plotted separately.
	The error bars indicate the statistical errors on
	the yield; the surrounding brackets indicate the
	systematic errors.  
	Shown for reference are the yields per collision in N+N 
	collisions, of charged hadrons and neutral pions
	respectively, 
	each scaled up by $\langle N_{coll} \rangle$ for the 
	class.  The bands indicate
	both the uncertainty in the N+N reference and in the 
	determination of $\langle N_{coll} \rangle$.
}
\label{fig:phenix_pi0_130}
\end{minipage}
\hspace{\fill}
\begin{minipage}[t]{75mm}
  \includegraphics[scale=0.3]{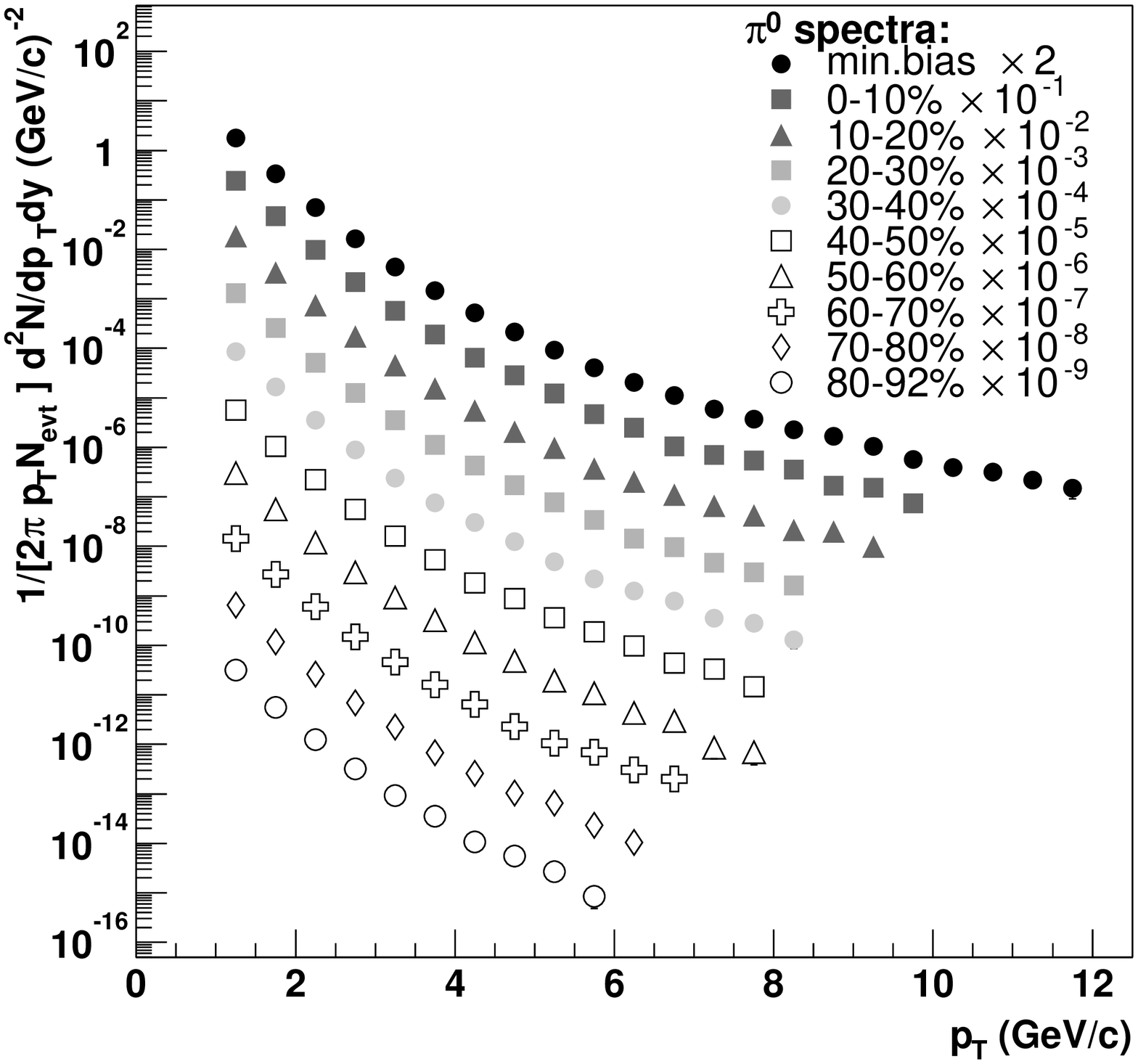} 
\caption{PHENIX results from 200~GeV Au+Au collisions (Run-2).
Invariant yields of $\pi^0$ at mid-rapidity are plotted
as a function of $\pt$ for minimum bias and 9 different centrality
selections (0-10\% is the most central, 80-92\% is the most
peripheral).  The yields are scaled for clarity.
}
\label{fig:phenix_pi0_200}
\end{minipage}
\end{figure}

One of the first and still most intriguing results from RHIC was the
observation in Run-1 ($\sqrt{s}=130$~GeV)
that in central Au+Au collisions the yield of
high $\pt$ $\pi^0$-s was strongly suppressed with respect to
expectations from $p p$ results at comparable 
energy\footnote{Since there are no $p p$ data at $\sqrt{s}=130$~GeV,
the reference spectrum was an interpolation of SPS and Tevatron
neutral and charged pion results; note that the uncertainty on
the reference spectrum is comparable to the total error of the
data on Fig.~\ref{fig:phenix_pi0_130}}
scaled by the calculated number of binary nucleon-nucleon 
collisions (Fig.~\ref{fig:phenix_pi0_130}), although the same collision 
scaling described the peripheral data adequately~\cite{ppg003}.  
Despite the large errors the effect was significant ($>2.5\sigma$),
but low integrated luminosity and an only partially instrumented 
calorimeter prevented PHENIX from exploring it past $\pt=4$~GeV/c.  
In Run-2 the combination of higher c.m. energy ($\sqrt{s}=200$~GeV),
much higher statistics, and a fully instrumented 
detector\footnote{In Run-1 only 3 of the 8 PHENIX electromagnetic
calorimeter sectors were instrumented and read out; in Run-2 all
8 sectors were operational and included in the analysis}
made it possible to extend the minimum bias $\pi^0$ $\pt$ spectra up to
12~GeV/c, and the semi-inclusive spectra up to 6-10~GeV/c,
depending on the centrality class (Fig.~\ref{fig:phenix_pi0_200}).

\begin{figure}
\begin{minipage}[t]{75mm}
  \includegraphics[scale=0.28]{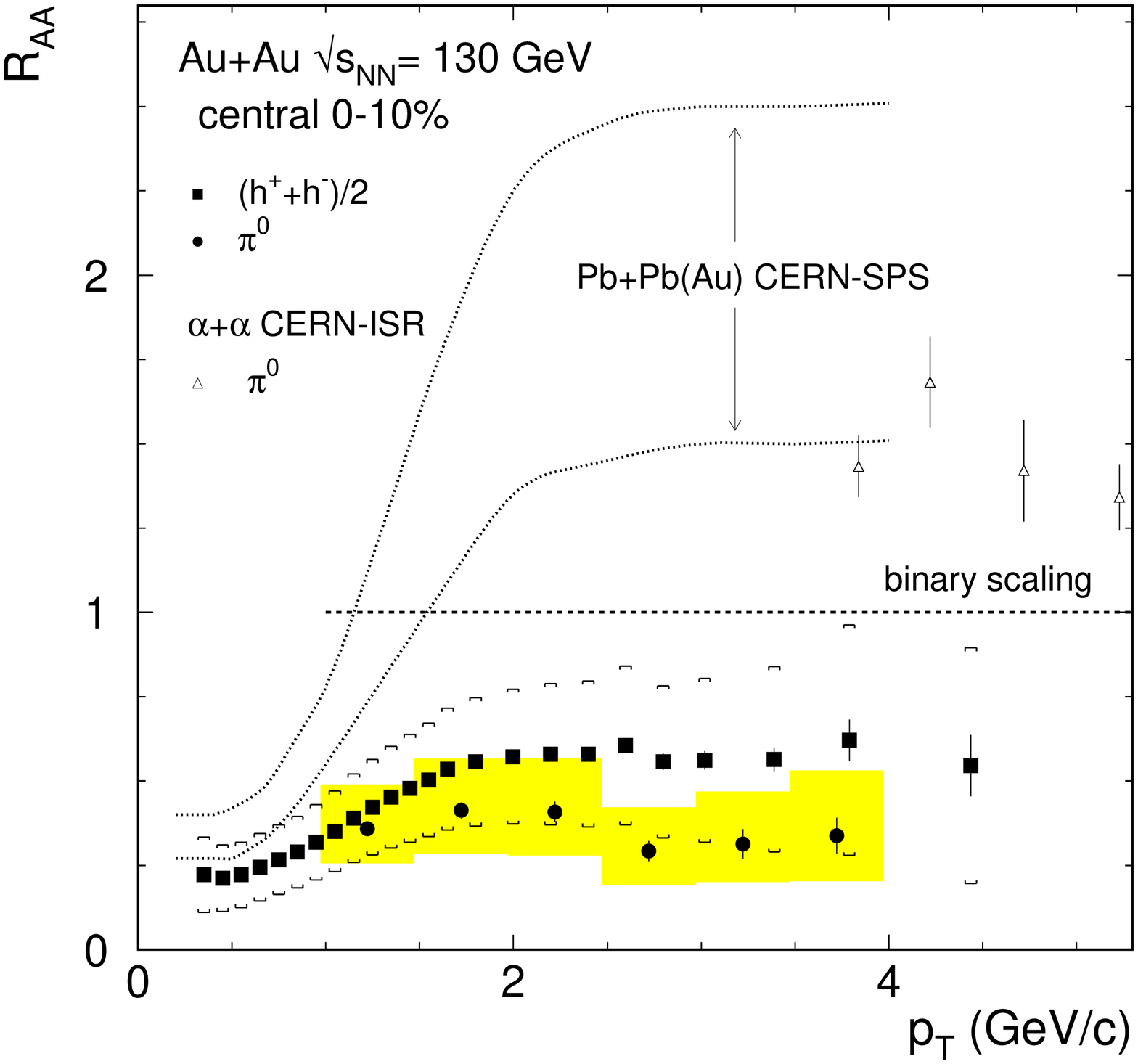}
\end{minipage}
\begin{minipage}[t]{75mm}
  \includegraphics[scale=0.35]{Sec03/phenix_Rcp_130_convert.eps.0}
\end{minipage}
\caption{
	PHENIX results from 130 GeV Au+Au collisions (Run-1).
	{\bf Left panel}:
	the ratio $R_{AA}$ for charged hadrons ($(h^{+}+h^{-})/2$)
	and neutral pions in central Au+Au collisions.
	The error bars indicate the statistical errors,
	the surrounding bands (shaded for
	$\pi^{0}$'s, brackets for $(h^{+}+h^{-})/2$)
	indicate the
	combined statistical and systematic errors on the ratio, 
	including the uncertainty in the $p p$ data and the 
	uncertainty in $\langle N_{coll} \rangle$.
	Also shown for reference are $R_{AA}$ for
	$\alpha+\alpha$ ($\sqrt{s}=31$~GeV) and for
	central Pb+Pb collisions ($\sqrt{s}=17.3$~GeV)
	measured at the CERN-SPS.
	{\bf Right panel}:
	Ratio of central to peripheral $\pt$ spectra 
	(both normalized with the calculated $N_{coll}$)
	for charged hadrons and $\pi^0$.
}
\label{fig:phenix_RAA_130}
\end{figure}

The deviation from simple scaling with the number of nucleon-nucleon
collisions $N_{coll}$
(or, more rigorously, with the nuclear overlap function
$T_{AB}$) is usually characterized by the {\it nuclear modification
factor} $R_{AA}$ defined as

\begin{equation} 
R_{AA}(\pt)\,=\,\frac{(1/N^{evt}_{AA})\,d^2N^{\pi^0}_{AA}/d\pt dy}
{\langle N_{coll}\rangle/\sigma_{pp}^{inel} \,\times\, 
d^2\sigma^{\pi^0}_{pp}/d\pt dy},
\label{eq:R_AA}
\end{equation}

\noindent
the expectation being that starting at some relatively low $\pt$ 
(1.5-2.0~GeV/c) which marks the approximate transition from
``soft'' to ``hard'' physics and where jets become dominant 
$R_{AA}$ reaches unity because
higher $\pt$ particles are produced in incoherent, large
momentum transfer parton-parton scatterings whose (small)
probability in turn is proportional to the number of
nucleon-nucleon collisions $N_{coll}$.  At lower 
$\sqrt{s}$ (SPS-energies) $R_{AA}$ even rises above unity
(Cronin-effect) due to multiple scattering of partons before 
the large $Q^2$ process initiates the jet.  
In stark contrast to that expectation of $N_{coll}$ scaling
or some enhancement $R_{AA}>1$,
PHENIX found already at $\sqrt{s}=130$~GeV (Run-1) that
for $\pi^0$ in central Au+Au collisions $R_{AA}$ never reaches
unity.  Instead, after reaching its maximum at $\pt\simeq1.5$~GeV/c
it {\it decreases} for higher transverse momenta
(Fig.~\ref{fig:phenix_RAA_130}) to about 0.3.
In other words, there is a factor of $\sim$3 suppression already 
around $\pt=3$~GeV/c.
On the left panel of Fig.~\ref{fig:phenix_RAA_130} $R_{AA}$ is
plotted for $\pi^0$ and charged hadrons ($(h^++h^-)/2$ in the
most central Au+Au collisions with $N_{coll}$ calculated from
the Glauber-model.  The suppression at higher $\pt$ is
even more dramatic when compared to the enhancement observed
in Pb+Pb at $\sqrt{s}=17.3$~GeV 
and $\alpha+\alpha$ at $\sqrt{s}=31$~GeV, also shown on the plot.
On the right panel of Fig.~\ref{fig:phenix_RAA_130} 
the ratio of central to 
peripheral data - both normalized with $N_{coll}$ - is shown.
The behavior is very similar to $R_{AA}$, but the $\pt$ coverage
is somewhat less since the peripheral spectrum has a smaller
range.  It should be pointed out that
$R_{AA}$ and the central/peripheral ratio 
(often referred to as $R_{CP}$) are dominated
by different systematic errors, so the combined ``message''
of the two results (significant suppression)
is even stronger than suggested by the
error-bars on any one of them.  Also, uncertainty on
$\sigma_{pp}^{inel}$ cancels to first order in the central 
to peripheral ratio (right panel).

The observed suppression was a very strong indication that an
extremely hot and dense medium has been created in central
Au+Au collisions which by some mechanism depleted the number
of high $\pt$ jets.  However,
it was unclear whether the very creation of the jets was
suppressed ({\it e.g.} due to initial state parton saturation) 
or the jet production itself was the same as in $p p$,
scaled by $N_{coll}$, but the jets lost a
significant fraction of their energy while traversing the
medium (as predicted by different models of jet quenching).  
Also, in case of energy loss the nature of the medium 
causing the loss (partonic or hadronic)
was unclear.  Finally, various theoretical scenarios predicted
similar suppression at 3-4~GeV/c (the upper limit of the
Run-1 results), but predicted different $\pt$ dependence of the 
suppression at transverse momenta beyond that range.

\begin{figure}
\begin{minipage}[t]{75mm}
  \includegraphics[scale=0.34]{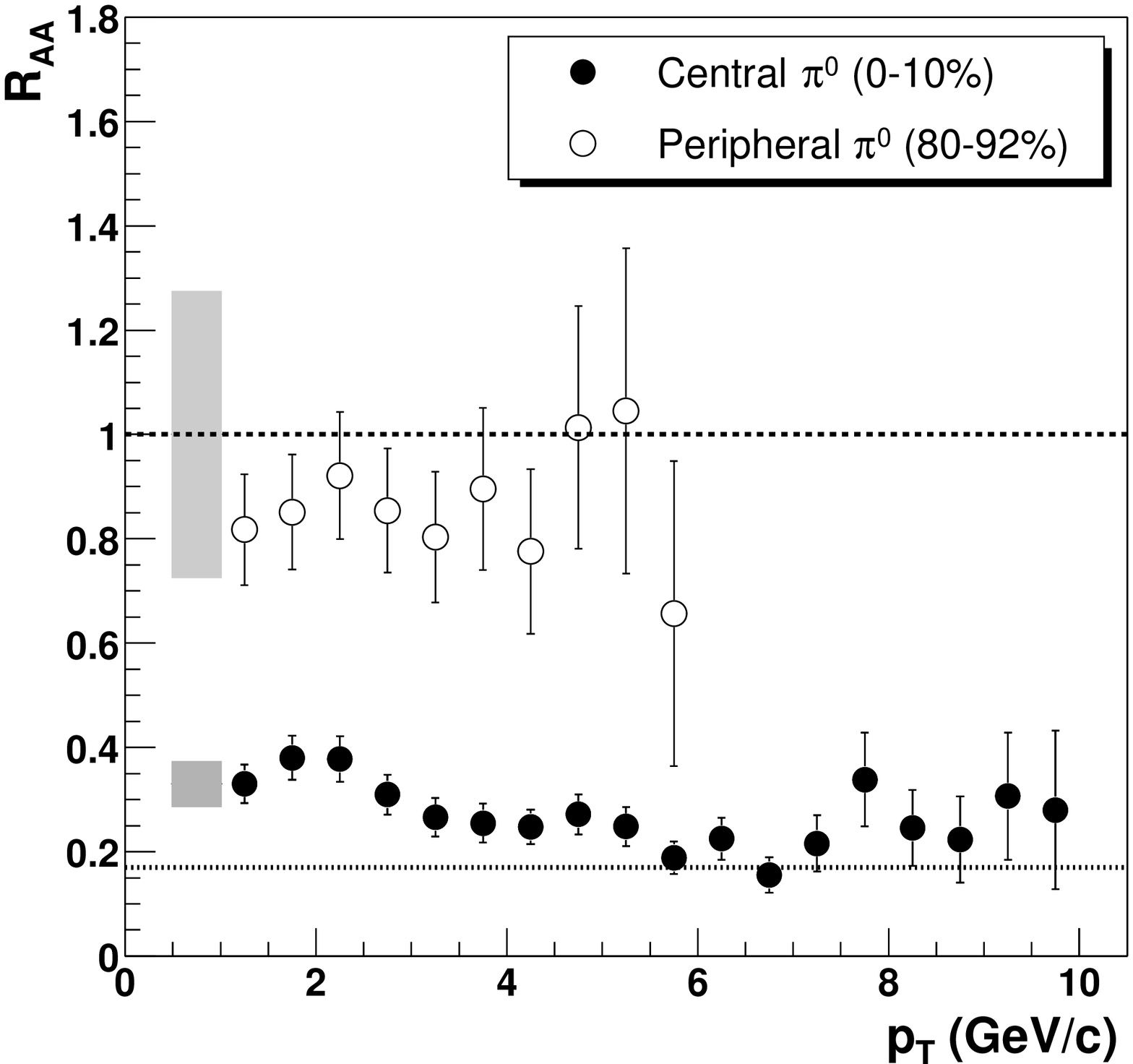}
\caption{PHENIX results from $\sqrt{s}$ = 200~GeV.
Nuclear modification factor $R_{AA}(\pt)$ for $\pi^0$ in central 
(closed circles) and peripheral (open circles) Au+Au collisions.
The error bars include all point-to-point experimental 
($p p$, Au+Au) errors. The shaded bands represent the 
fractional uncertainties in $\langle T_{AuAu} \rangle$ and in the $\pi^0$ yields 
normalization added in quadrature, which can move all the points up or down together
(in the central case the shaded band shown is the fractional error for the first point).
}
\label{fig:phenix_RAA_200}
\end{minipage}
\hspace{\fill}
\begin{minipage}[t]{70mm}
  \includegraphics[scale=0.35]{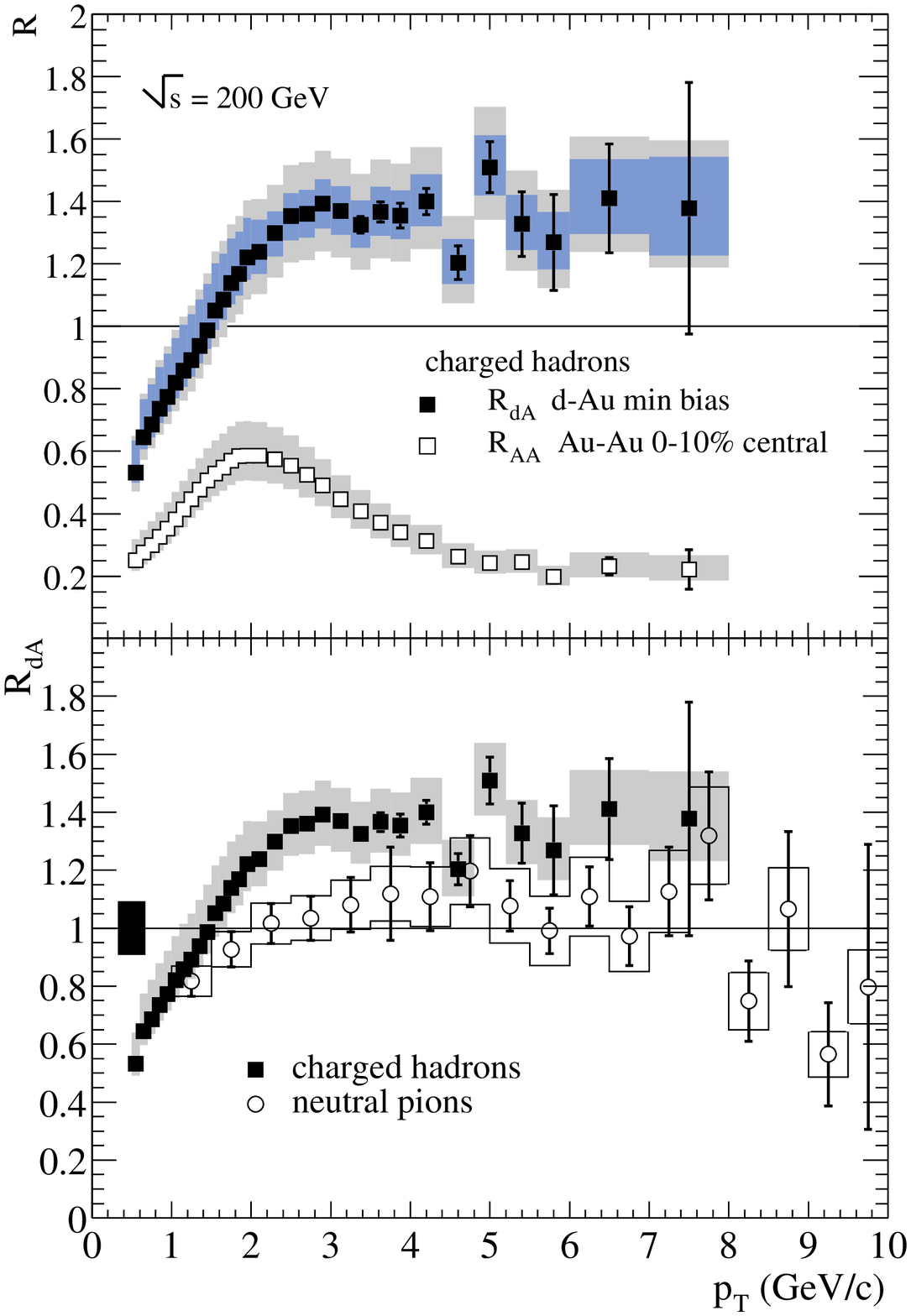}
\caption{
  Top: Nuclear modification factor $R_{\rm{dA}}$ for  $(h^+ + h^-)/2$ 
in minimum bias D+Au compared 
to $R_{\rm{AA}}$ in the 10\% most central Au+Au collisions. 
Inner bands show systematic errors which can vary with $\pt$,
and outer bands include also the normalization uncertainty.
Bottom: Comparison of $R_{\rm{DA}}$ for $(h^+ + h^-)/2$ and 
the average of the $\pi^0$ measurements in D+Au. 
The bar at the left 
indicates the systematic uncertainty in common for the charged 
and $\pi^0$ measurements.
}
\label{fig:phenix_RdA_200}
\end{minipage}
\end{figure}

In Run-2 RHIC delivered Au+Au {\it and} $p p$ collisions at
$\sqrt{s}=200$~GeV, both at sufficiently high integrated
luminosity such that the $\pi^0$ $\pt$ spectra could be
extended to 10-13~GeV/c.  The $p p$ results are described
elsewhere in this Report and in~\cite{ppg024}, but it should be
emphasized once again that measuring the reference $p p$ spectrum 
in the very same experiment ({\it i.e.} with similar systematics)
greatly reduces the systematic error on $R_{AA}$ proper.
The nuclear modification factor using $\pi^0$-s in the most 
peripheral and most central Au+Au collisions is shown on 
Fig.~\ref{fig:phenix_RAA_200}, where this time $R_{AA}$ was calculated
using the PHENIX $p p$ measurement.  Perfect scaling with $N_{coll}$
would mean $R_{AA}=1$.  Although averaging below one, the
peripheral data are certainly consistent with $N_{coll}$
scaling within 
errors\footnote{Errors are dominated by the fully correlated
normalization error shown as a grey band at the left side
of the plot.}.
However, in central collisions the suppression
is unambiguous in the entire $\pt$ 
range\footnote{Also, it should be pointed out that in
the region of $\pt$ overlap there is a very good agreement
between the $R_{AA}$ values from Run-1 and Run-2}.  
$R_{AA}$ reaches
its highest value (smallest suppression, $\sim$2.5) 
around 2~GeV/c, then decreases, and is constant within errors
above 4~GeV/c, giving a suppression factor of 4-5. 
While this result disfavored those quenching models that predict a
$\pt$ dependence of the 
suppression\footnote{For instance models including the LPM-effect
predict that $R_{AA}\propto\sqrt{\pt}$ asymptotically}, 
it did not differentiate
between initial state and final state 
effects\footnote{Although the presence of a hot, dense medium
that causes energy loss was further supported by the observation 
that back-to-back
jets virtually disappeared in central Au+Au 
collisions~\cite{starbacktoback}},
nor could it
distinguish between partonic and hadronic energy loss
scenarios.

Therefore, the major part of Run-3 at RHIC has been dedicated
to D+Au collisions at $\sqrt{s}=200$~GeV, under the
assumption that in D+Au collisions the gold nucleus remains
``cold'', and any effects due to a hot, dense medium in the 
final state would be absent.  On the other hand if an
initial state effect in the Au nucleus is responsible for the 
observed suppression, such suppression should manifest
itself in D+Au collisions, too.  Note, that the choice of
deuterium (D+Au) instead of proton ($p$Au) was motivated 
by purely technical 
reasons\footnote{Easier to collide because the 
magnetic rigidity of the two beams is closer to each other than
for $p$Au}, 
and it has been 
demonstrated that there is little if any difference in the
physics of +DAu and $p$Au collisions at these 
energies~\cite{ppg028} -- an interesting observation in its own right.  
The nuclear modification factor for D+Au
is shown on Fig.~\ref{fig:phenix_RdA_200} for charged
particles and neutral pions.  The suppression observed in
central Au+Au collisions is clearly absent 
here\footnote{These results have been published less than three
months after the data were taken and the analysis included
only a subset of all available data.  Therefore, within the quoted 
errors the $\pi^0$ data can not differentiate between $R_{dA}=1$ 
or some small Cronin-type enhancement.  A new analysis using the entire
dataset is underway, it is expected to have smaller errors and
it will investigate the centrality dependence of $R_{dA}$ as well}.
This result indicates
that the suppression in central Au+Au collisions is not
an initial state effect, nor does it arise from modification
of parton distribution functions in nuclei.
Further analyses including a detailed study of the centrality dependence
of $R_{dA}$ may shed further light on the mechanism of suppression.

There is a substantial effort in PHENIX to extract the $\eta$ yield 
from Run-2 Au+Au data and to establish the asymptotic $\eta/\pi^0$
ratio, but no results have been presented yet.  STAR published
$K^0_S$ spectra at $\sqrt{s}=130$~GeV using the 
$K^0_S \rightarrow \pi^+\pi^-$ channel~\cite{starK0S}, but
only at $\pt<1.5$~GeV/c.  Both $\eta$ and $K^0_S$ have $\pi^0$
decay channels and thus ``feed down'' into the $\pi^0$ spectrum
but due to the (two- and three-body) decay kinematics this
contamination (as well as
contributions from higher mass mesons) is negligible if compared
to current experimental errors~\cite{ppg014}.  

Using the published $\pi^0$ data at 130 GeV and the preliminary 200 GeV 
$\pi^0$ data a first study of $x_T$ scaling has been 
presented at the Fall 2002 DNP meeting~\cite{mjt}.  
At both energies suppression of high
$\pt$ $\pi^0$ with respect to point-like scaling from $p p$
collisions has been observed.  If the effect is due to 
shadowing of the structure functions rather than a final
state interaction with the hot medium, the yields
at a given $x_T$ and centrality should exhibit the same
suppression and the scaling exponent $n(x_T,\sqrt{s})$
should remain unchanged from $p p$ to Au+Au collisions.
The study found that, within systematic
errors, $x_T$ scaling with 
$n(x_T,\sqrt{s})=6.3 \pm 0.6$ applies for both peripheral and
central collisions in the range $0.025 \leq x_T \leq 0.06$.

\subsection{Inclusive and direct (non-hadronic) photons}

While $\pi^0$ spectra have already been published from all
three RHIC runs, so far only preliminary results were shown
on photons.  Though counterintuitive at first, measuring
photons - even the inclusive photon spectrum - is much more
difficult than measuring $\pi^0$-s.  
Neutral pions are usually reconstructed
from a 
correlation\footnote{Invariant mass $m_{\gamma\gamma}$ 
of {\it two} photons}
and - except for very high $\pt$ and/or very low multiplicities -
they are measured only {\it statistically}, calculating 
invariant mass from {\it all} photon candidates in an event rather 
than trying to identify both decay photons from a particular $\pi^0$.
Moreover, at low multiplicities ($p p$, D+Au, peripheral Au+Au) 
an accurate $\pi^0$ spectrum can be extracted without any
photon identification at all -- just by calculating the invariant
mass from pairs of {\it all} clusters in an event. 
Even in high multiplicity events good photon identification 
and effective rejection of other particles is not 
crucial\footnote{What's really important is to know the
{\it efficiency} of the photon identification (if any)
and the {\it smearing} of the photon energy due to overlaps
with other particles - which may push $m_{\gamma\gamma}$ out
of a reasonable invariant mass window causing a loss in the
raw $\pi^0$ count.  However, even large number of hadrons
mistakenly identified as photons ({\it contamination}) 
rarely cause any problems}.
Also, since the true mass of $\pi^0$ is known, the measurement is
``self-calibrating'' in the sense that any shift in the
observed centroid and any widening of the mass peak corresponds to 
a calculable shift in the (apparent) transverse momentum and
can be corrected for.  Finally, many of the pions created outside 
the collision vertex\footnote{In nuclear interactions with the detector
material, {\it i.e.} as ``classic'' background, or in decays
of long-lived hadrons, {\it i.e.} as irreducible ``physics'' background.} 
do not reconstruct with the proper
invariant mass due to the mismatch between the true and the
apparent opening angle of the two 
photons\footnote{Unless the direction of the photon is measured,
{\it e.g.} with a preshower detector or by an $e^+e^-$ conversion,
one has to calculate the ``apparent'' opening angle of all photon 
pairs under the assumption that they came from the vertex.}.  
Therefore,
many of those ``background'' $\pi^0$-s don't contribute to
the raw yields, and don't have to be corrected for.  

The situation with the inclusive photon spectrum is entirely
different.  Obviously particle identification and a very
precise knowledge of its efficiency is essential.  Usually
there is no straightforward way to check the energy (and $\pt$)
scale, although - depending on the slope - just 1\% error on the
energy can give up to 10\% error on the cross-section.
Unless the direction of the photon is measured, there is
little if any rejection/identification of photons not
coming from the collision vertex.

While inclusive photon spectra ($\gamma_{\rm inc}$) are important
in their own right, the exciting new physics is hidden in the
``direct'' or ``excess'' photon spectrum ($\gamma_{\rm dir}$), the 
difference of inclusive photons and photons from electromagnetic
decays of final state hadrons ($\gamma_{\rm decay}$).
Since $\gamma_{\rm decay}$ is simulated using fits to the measured
$\pi^0,\eta,...$ spectra,
reliable hadron spectra are a prerequisite to the direct
photon measurement and many (although not all) of the errors
on hadron spectra will propagate into the direct photon
spectrum.  

Finally, a good direct photon measurement 
will have to reveal not only the magnitude of photon excess 
over hadronic sources $\gamma_{\rm inc}-\gamma_{\rm decay}$ but also 
make the decomposition of
at least three overlapping spectra possible (hard scattering,
plasma phase, final state radiation).
Therefore,  the errors both on shape
and absolute normalization should be very 
small\footnote{Less than 10\%, preferably $\sim$5-6\%.
Results from WA98~\cite{WA98} provide a useful context:
their final publication quotes a $\pt$ dependent 5.7-8.9\%
systematic error on the photon excess.  To squeeze the errors
below 5\% is almost impossible in a real-life experiment.}.  
Once this ambitious goal is reached, one can work
backwards (starting with the highest available $\pt$) in unfolding the 
contributions of the different photon production
processes\footnote{Although
this will never be a purely experimental process: it will
always rely to some extent on theoretical calculations}.  
Once a high quality 
$\gamma_{\rm dir} = \gamma_{\rm inc}-\gamma_{\rm decay}$ ``direct'' 
(excess) photon spectrum is available,
starting at the highest transverse momenta one can try to 
fix the pQCD scale and improve upon fragmentation functions in $p p$
as well as establish the effects of the medium in Au+Au.  
Next, this pQCD contribution has to be subtracted from
$\gamma_{\rm dir}$. The result
at a few GeV/c is expected to be dominated by radiation from
the early plasma.  (Cross-checks with D+Au collisions will
decrease the uncertainties).  Then once again one can go lower
in $\pt$ ({\it i.e.} later in evolution time) to look for radiation
from the plasma - hadronic gas phase transition, and so on.  
Whereas the spectral shapes of different contributors are different, 
deconvolution of these spectra obviously requires unprecedented accuracy
on the $\gamma_{\rm dir}$ excess photon spectrum (which implies a comparable
accuracy of the $\pi^0$ spectra themselves).

\begin{figure}
 \begin{center}
    \includegraphics[width=15cm]{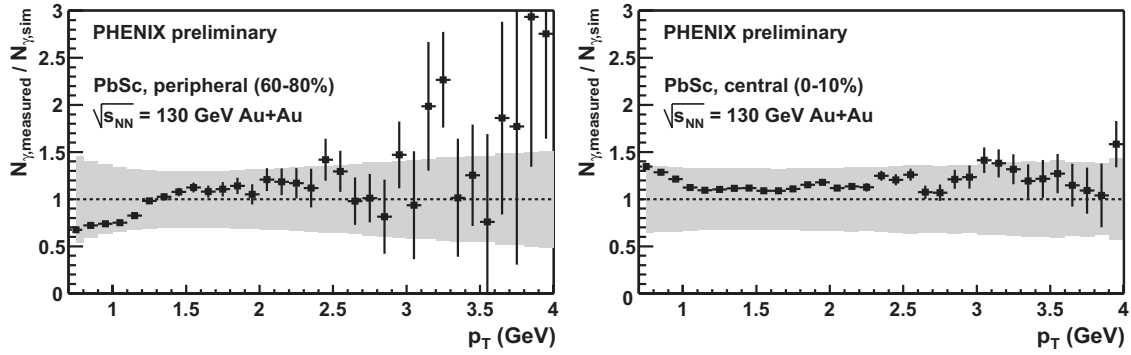} 
    \caption{Ratio of inclusive photon spectrum to the
      photon spectrum expected from hadron decays in
      $\sqrt{s} = 130$~GeV Au+Au collisions, in
      peripheral events (left panel) and central events
      (right panel).  The shaded bands are the systematic
      errors dominated by
      the uncertainties of the (measured) $\pi^0$ spectrum}
 \end{center}
\label{fig:phenix_photon_130}
\end{figure}

Preliminary results from PHENIX at $\sqrt{s}=130$~GeV 
(presented at QM'02) are
shown on Fig.~\ref{fig:phenix_photon_130} for peripheral
collisions (left panel) and central collisions (right panel).
The plots show the ratio $\gamma_{\rm inc}/\gamma_{\rm decay}$
of the inclusive photon spectrum and the photon spectrum
expected from final state hadron decays.  A fit to the
measured $\pi^0$ spectrum is used to calculate $\gamma_{\rm decay}$
with a fast Monte-Carlo program; 
$m_{_T}$ scaling is assumed for $\eta$ (higher mass mesons
are not included).  The band indicates the systematic errors,
completely dominated by the uncertainties of the fit itself.
Note that in Run-1 the $\pi^0$ measurement extended only
until $\pt=4$~GeV/c with large errors (Fig.~\ref{fig:phenix_pi0_130}).
Therefore, the fit was not 
very well constrained, both the absolute normalization
and the shape can vary significantly.  
Photon identification is based
solely on shower shape and timing - charged particle veto
is not applied.  Fluctuations of the points indicate that
the $1\sigma$ errors are {\it not} overestimated.  Within
errors the results are consistent with no photon excess
over known hadronic sources, but obviously the sensitivity
of the measurement is low.  

\begin{figure}
\begin{center}
\includegraphics[width=15cm]{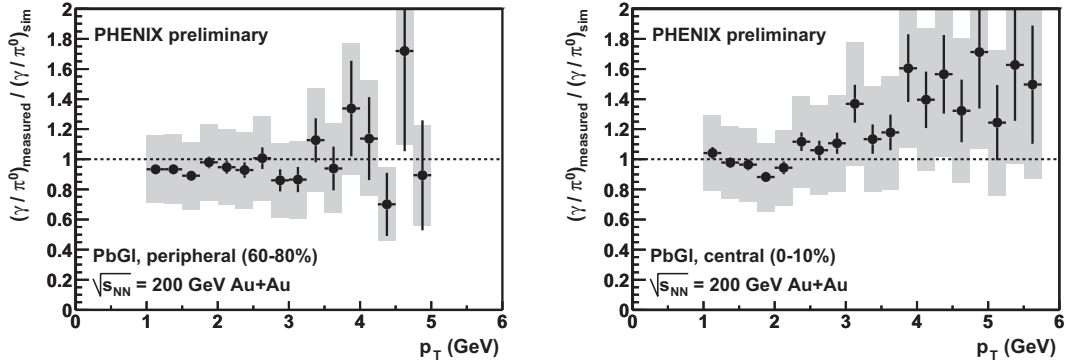} 
\caption{Measured over expected $\gamma / \pi^0$ ratio in Au+Au
  collisions at $\sqrt{s}$=200~GeV in peripheral (left panel)
  and central (right panel) collisions.  ``Expected'' means that all
  photons are assumed to come from hadron decays, and the double
  ratio eliminates certain systematic errors.  A ratio of unity means
  that all photons come from hadron decays.
  The shaded boxes represent
  the estimated $1\sigma$ systematic errors on the data points. }
\end{center}
\label{fig:phenix_photon_200}
\end{figure}

Preliminary results from PHENIX at $\sqrt{s}=200$~GeV 
(presented at QM'02) are
shown on Fig.~\ref{fig:phenix_photon_200} for peripheral
collisions (left panel) and central collisions (right panel).
Different from the 130~GeV data, here the double ratio
$(\gamma_{meas}/\pi^0_{meas})/(\gamma_{sim}/\pi^0_{sim})$ is
plotted, where the subscript $_{meas}$ refers to the
measured inclusive photon and $\pi^0$ spectra, 
$\pi^0_{sim}$ is a fit to the measured $\pi^0$ spectrum which 
is then used in a Monte-Carlo to generate the photon
spectrum $\gamma_{sim}$ expected from hadronic decays.
In the absence of any non-hadronic sources this double 
ratio would be exactly one, which is clearly the
case in peripheral collisions (left panel).  One should
appreciate the large, non-statistical fluctuations
of the points even at low $\pt$ - indicating that the
errors are once again not overestimated.  Data do not
extend as far in $\pt$ as for $\pi^0$-s, to avoid
large uncertainties of the $\pi^0$ fit in the tail and beyond
the measured range.  The double
ratio for central events (right panel) is still
consistent with no excess within errors, but obviously
exhibits a suggestive trend.

The STAR collaboration also presented preliminary results
on excess photons at QM'02~\cite{starconversion}.  In this
measurement photons converted to $e^+e^-$ pairs in the
silicon vertex tracker and the inner field cage of the
TPC were analyzed.  The experiment found that 
above $\pt=1.65$~GeV/c the contribution from $\pi^0$ to
the inclusive photon spectrum decreases in the most central
events (0-11\% centrality), indicating an increase in
contribution from other photon sources, possibly other 
electromagnetic decays or direct photons.  However,
there is a significant uncertainty in the normalization
of the $\pi^0$ spectra.

\subsection{Outlook}

Obviously, the carefully worded statements
from both experiments reflect the technical difficulties and
do not exclude a possibly substantial yield of direct photons.  
Results with reduced systematic errors can be expected within
a year and at larger $\pt$ will provide the first tests of
pQCD calculations.  However,
as emphasized in other sections of this Report, measuring the photon
excess over the yield expected from hadron decays is only the first
step: much of the interesting physics lies is hidden in the 
{\it composition} of this excess.
In order to disentangle the contributions from hard scattering,
the QGP, the mixed and the hadronic phases one has to measure this
excess both in a very wide $\pt$ range and with unprecedented
accuracy.  This is and continues to be the biggest challenge to
the experimentalists.

Therefore, it is instructive to briefly review the dominant sources of 
systematic errors on $\gamma_{\rm inc}/\gamma_{\rm decay}$ in the current 
analyses.  At RHIC the errors on 
$\pi^0$ are currently around 14-17\%, distributed about equally between
yield extraction, particle identification and effects of the
energy scale, and this determines the accuracy of the calculated
$\gamma_{\rm decay}$.  Contributions from other mesons are currently not
measured (although analyses of $\eta$ as well as $K^0_S$ above
1.5~GeV/c by charged pions are under way).

For the inclusive photon analysis photon efficiency has
about the same error as in the $\pi^0$ measurement, but
hadron contamination is added.  The energy
scale uncertainties have bigger weight (because of the lack
of ``self-calibration'', the possibility to cross-check the energy scale
using the invariant mass), and a major contributor to the
errors is the instrumental background.  Note that three of these
contributors (particle identification, contamination and background) 
become worse as one tries to move to lower transverse momenta
(thermal region).

On the other hand it is exactly at low $\pt$ where the
complementary measurements {\it via} conversion electrons in both
PHENIX and STAR offer advantages: since they provide directional
information, much of the instrumental background, and, in general,
photons not from the vertex can be eliminated.  Some upgrades
of current RHIC detectors point in this direction, too.
Also, new types of analyses are developed to reduce systematic
errors\footnote{One possibility is to compare results from
analyses where the origin of systematic errors is very different,
like comparing $R_{AA}$ and $R_{CP}$ in the $\pi^0$ analysis.}.
While trying to defeat all
sources of systematic errors is probably a futile excercise
one can make independent measurements of the same
quantity, within the same experiment, with very different
systematics, thus increasing the level of confidence in the
results.  Photon measurements are notoriously difficult:
the author thinks future LHC experiments are well advised
to include such ``redundancies'' from the very beginning.




%
\section{DETECTOR STUDIES, RESOLUTION}
\label{sec:detectors}
In this section one briefly describes the main features of ALICE, CMS and
ATLAS concerning photon and hadron detection. The acceptances for each detector
are summarized in appendix 2.


\subsection{Photon detection at ALICE}

{\em H.~Delagrange, D.~d'Enterria, T.~Peitzmann, Yu.~Kharlov}

The ALICE detector \cite{ALICE-TP} aims to study the physics of
strongly interacting matter at extreme energy densities, where the
formation of a new phase of matter, the quark-gluon plasma, is
expected. Among all the probes of the quark matter, photons could be
used to study the dynamics of strong interactions in hadronic
collisions. Owing to their small electromagnetic coupling, produced
photons do not interact with the surrounding matter and thus probe the
properties of the matter at the time of their production \cite{PPR-chap1}.

\subsubsection{Photon spectrometer PHOS in ALICE}

In ALICE, the photon spectrometer PHOS is designed \cite{PHOS-TDR} to
detect, identify and measure with high resolution the 4-momenta of
photons. Photon studies in heavy ion collisions require from the
detector a high discrimination power between photons and any other
kind of particles, charged and neutral hadrons or electrons. The best
possible resolutions in energy and position, as well as a reasonably
large acceptance provide a high neutral meson identification and,
thus, the high precision determination of the background for the
direct photon spectrum.

\begin{figure}[ht]
  \includegraphics[width=.48\textwidth]{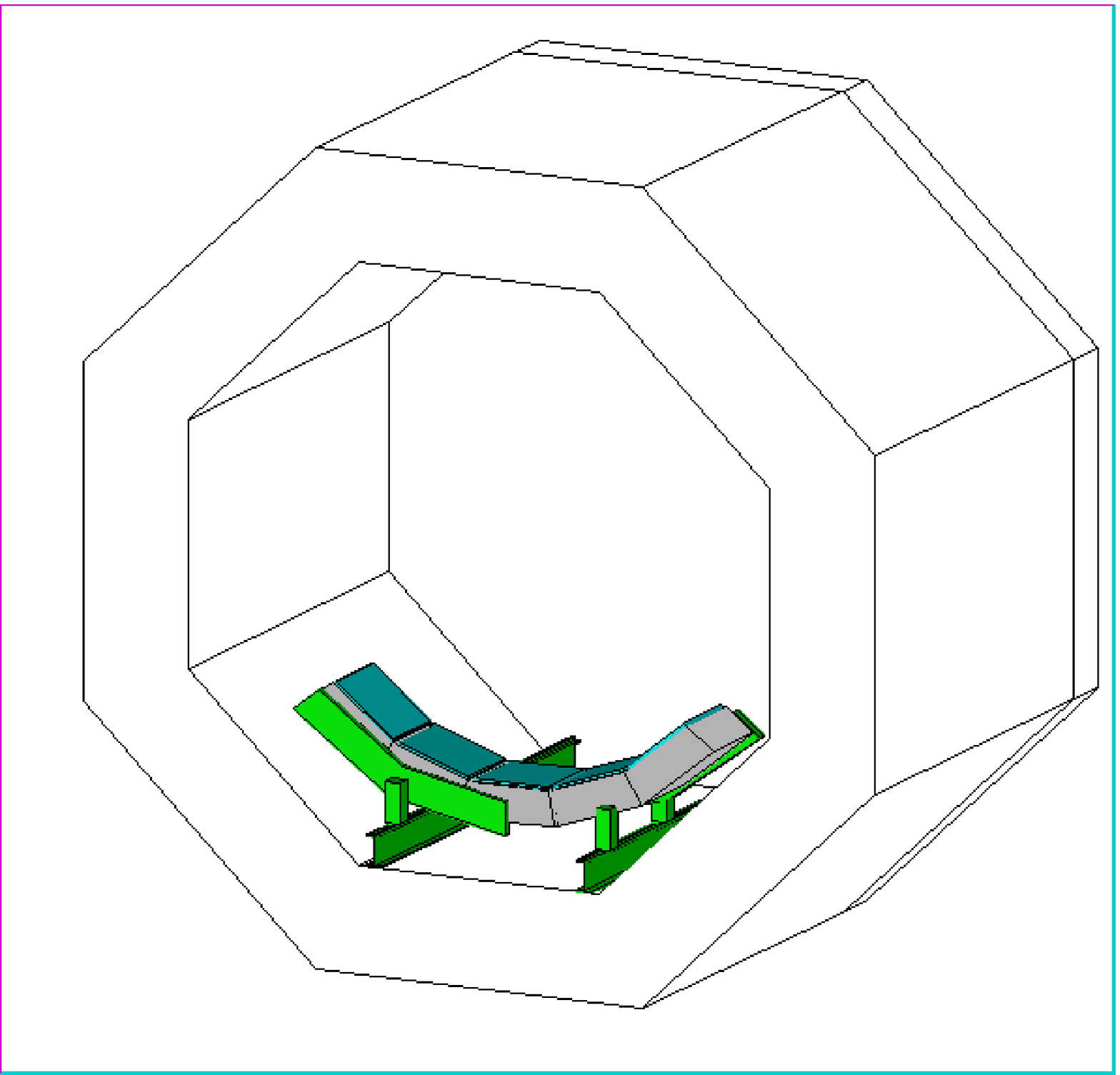}
  \hfill
  \includegraphics[width=.48\textwidth]{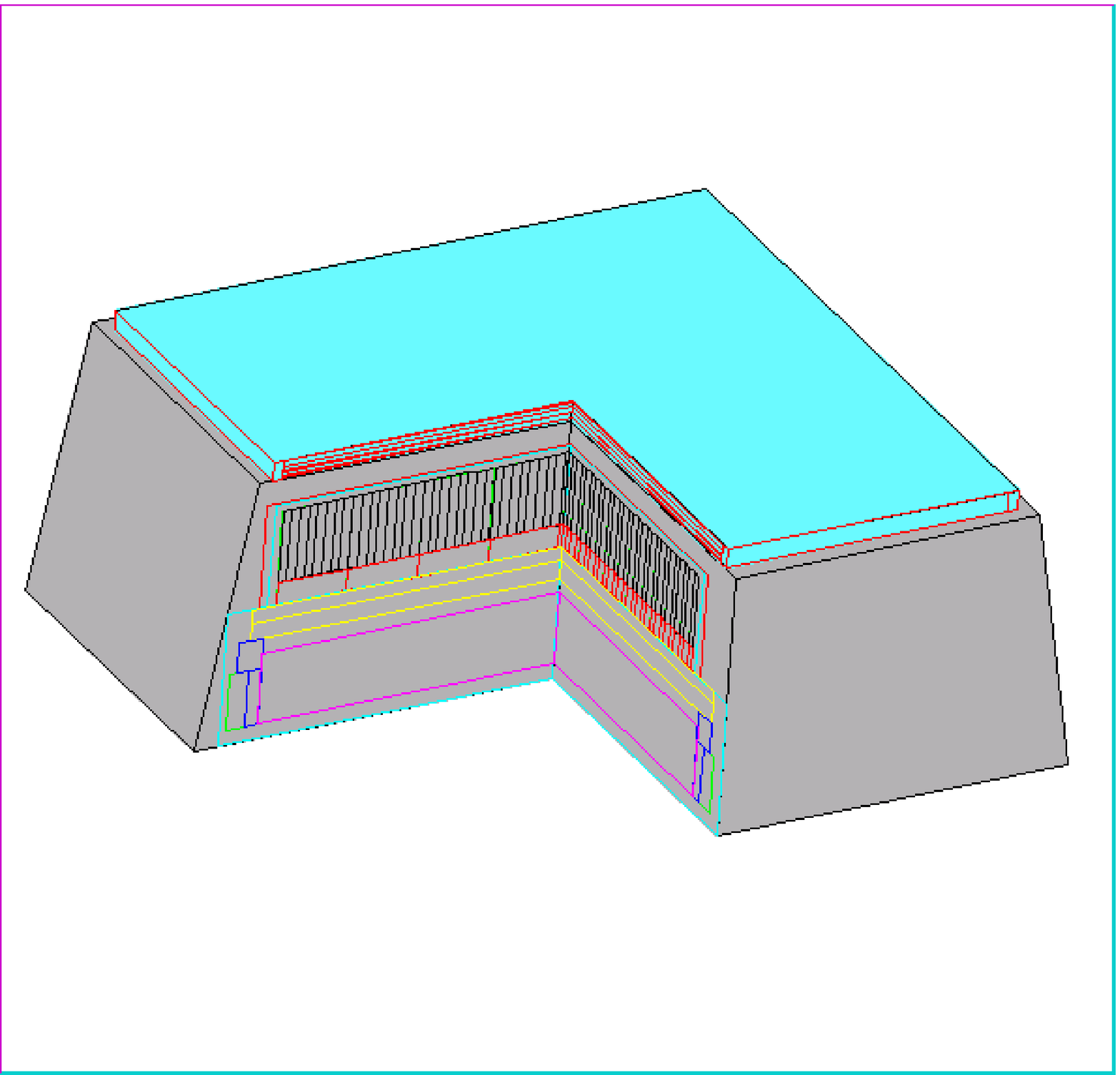}
\caption{The PHOS detector: PHOS inside the
ALICE solenoid magnet (left), and one PHOS module (right).} 
\label{fig:ALICE-PHOS}
\end{figure}

The final design of the photon spectrometer PHOS consists of five
identical modules positioned at the bottom of the ALICE detector
(Fig.\ref{fig:ALICE-PHOS}).
The PHOS modules are positioned at the distance 4.6~m from the beam
interaction point and installed at the azimuth angles $\pm 40.7^\circ,
\pm 20.3^\circ$, and $0^\circ$. Each module consists of the
electromagnetic calorimeter detector (EMC) and a charged particle veto
detector (CPV).

Each EMC module is constructed as a matrix of $64\times 56$ cells of
lead tungstate (PbWO$_4$) scintillator crystals. Each crystal,
elementary unit of the calorimeter, is an 18~cm long parallelepiped
providing 20~units of radiation length (X$_{\rm 0} = 0.89$~cm). It is
shaped with a squared cross-section of 22$\times$22~mm$^{2}$, to be
compared to the Moli\`ere radius of lead tung\-stanate, $r_{M} =
20$~mm. The scintillation light, in the visible near UV-wavelength
range, is read out by a 5$\times $5~mm$^{2}$ avalanche photo-diode
(APD) integrated with a low-noise pre-amplifier. The calorimeter is
operated at low temperature, $-25~^\circ$C, stabilized to $\pm
0.3^\circ$~C.  This operation mode on one hand enhances the
scintillation light output by a three fold factor and provides the
required high and constant energy resolution even for the less
energetic photons and, on the other hand, keeps the noise of the APD
low enough to provide a high signal to noise ratio. The electronic
chain associated to each crystal of the PHOS spectrometer delivers two
energy signals, one with low- and one with high gain, proportional to
the energy deposited in the crystal and a timing signal that measures
the time of the particle impact with respect to a trigger time
reference.

The CPV consists of multiwire proportional chambers with cathode pad
read-out. Each calorimeter module is covered by a CPV module with an
active area of 144.6$\times $144.6~cm$^{2}$, 1.3~cm deep and filled
with a gas mixture 70\% Ar and 30\% CO$_{\rm 2}$.  The total thickness
of the CPV module is 5.1~cm. Low-mass construction materials are used
for the CPV construction to minimize the material budget, radiation
length and detector mass.  The anode wires of the proportional chamber
are strung along the beam direction with a wire pitch of 5.65~mm and
placed at 5~mm above the cathode. The cathode is segmented in 64
(along beam direction)$\times $128 (across beam direction) rectangular
pads of 2.26$\times$1.13~cm$^{2}$, elongated along the anode wires. The
electron shower induced by the passage of a charged particle is
collected on the cathode and an induced charge
is retrieved from each pad of the CPV.

The PHOS acceptance in pseudorapidity is defined by $|\eta|<0.13$. Each
of five modules covers $17.8^\circ$ in azimuth angle.

\subsubsection{Intrinsic performance of PHOS}

The two parameters that describe the response of the EMC spectrometer
and play the most important role for photon identification are the
resolutions in energy and position. The energy resolution depends on
the spectrometer ability to collect most of the
electromagnetic shower energy, convert it into visible light and
transmit it to the APD, as well as on the APD photo-efficiency and
photon electron gain factor. The position resolution depends on the
segmentation of the spectrometer and energy resolution.

\begin{figure}[hbt]
  \centerline{
    \includegraphics[width=.80\textwidth]
		    {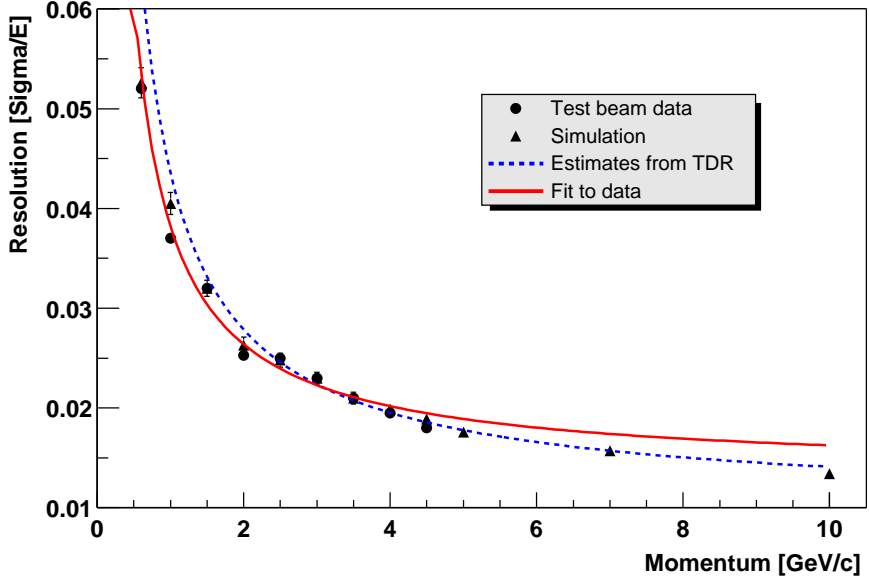}
  }
  \caption{Energy resolution of a $3\times 3$ EMC array measured in
    response to mono energetic electrons ($\bullet$) or calculated
    with simulations of mono energetic photons ($\blacktriangle$). The
    continuous line represents the result of the fit of
    Eq.~(\ref{PERF:Eresolution_equ}) to the data and extrapolated to
    10~GeV. The dashed line represents the values quoted in
    the TDR~\cite{PHOS-TDR}.}
  \label{fig:Eresolution}
\end{figure}

To determine experimentally these features an electron beam of energy
ranging from 0.6 to 4.5~GeV irradiated the central module of an array
of $3\times 3$ EMC modules. A Gaussian function was adjusted to the
distribution of total energy, \emph{E}, collected in the array. The
resulting resolution, $\sigma /E$, was compared
(Fig.~\ref{fig:Eresolution})
to the one obtained by the simulation performed in exactly the same
conditions as the experiment. 
The following parametrization was adjusted to the experimental resolution
dependence on electron energy:
\begin{equation}
  \label{PERF:Eresolution_equ}
  \frac{\sigma }{E}=\sqrt{\frac{a^{2}}{E^{2}}+\frac{b^{2}}{E}+c^{2}}
\end{equation}
where \emph{E} is in units of GeV, \emph{a} represents the
contribution of the electronic noise, \emph{b} the stochastic term,
and \emph{c} the constant term. 

The impact position on PHOS is reconstructed by calculating the
position of the center of gravity of the reconstructed cluster.  This
position is further corrected for the incidence direction of the
impinging photon. The previously discussed test beam measurements were
extended to verify the influence of the photon incidence on the
position resolution by tilting the array of EMC modules by 0, 3, 6 and
9$^\circ$. Fig.~\ref{fig:sigma_x(E,alpha)} shows the r.m.s. of the
Gaussian fit of this distribution versus the photon energies from 1 to
50~GeV for several incidence angles. 
\begin{figure}[ht]
  \parbox{0.47\hsize}{
    \centerline{
      \includegraphics[width=\hsize]{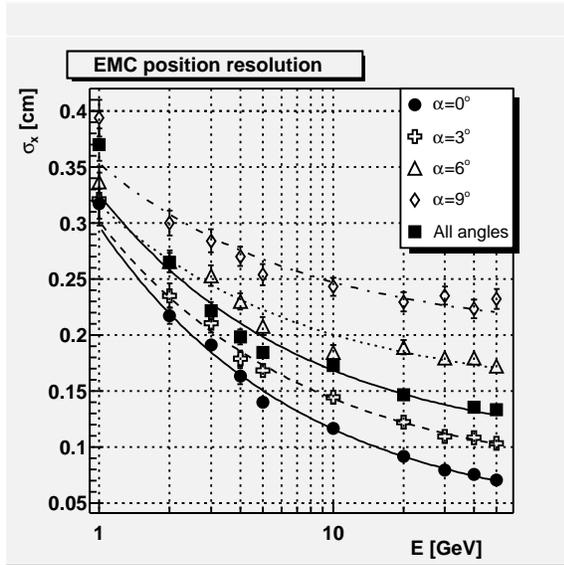}
    }
  }
  \hfill
  \parbox{0.47\hsize}{
    \caption{EMC position resolution versus photon energy for the
      incidence angles $\alpha=0,3,6$ and $9^\circ$ as well as for all
      possible incidence angles of photons emitted from the interaction
      point.}
    \label{fig:sigma_x(E,alpha)}
  }
\end{figure}

\begin{figure}[ht]
\parbox{0.48\hsize}{
  \centering
  \includegraphics[width=0.8\hsize]{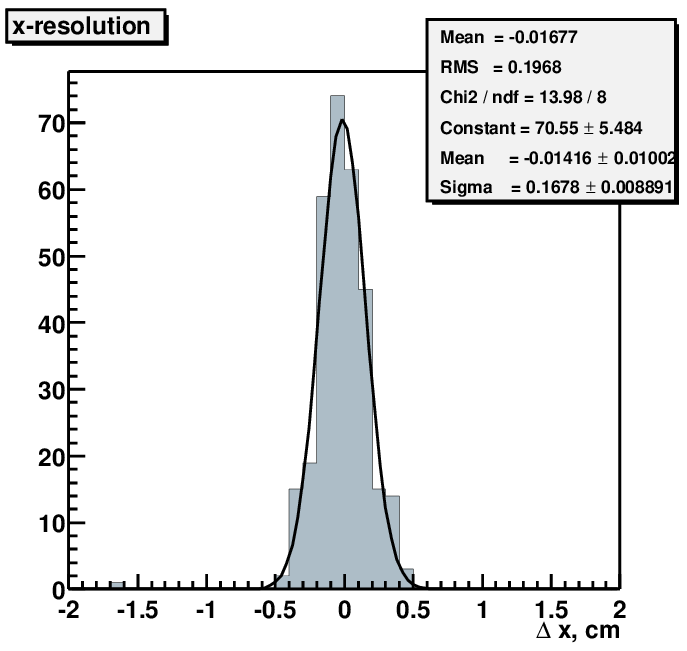}
  }
\hfil
\parbox{0.48\hsize}{
  \centering
  \includegraphics[width=0.8\hsize]{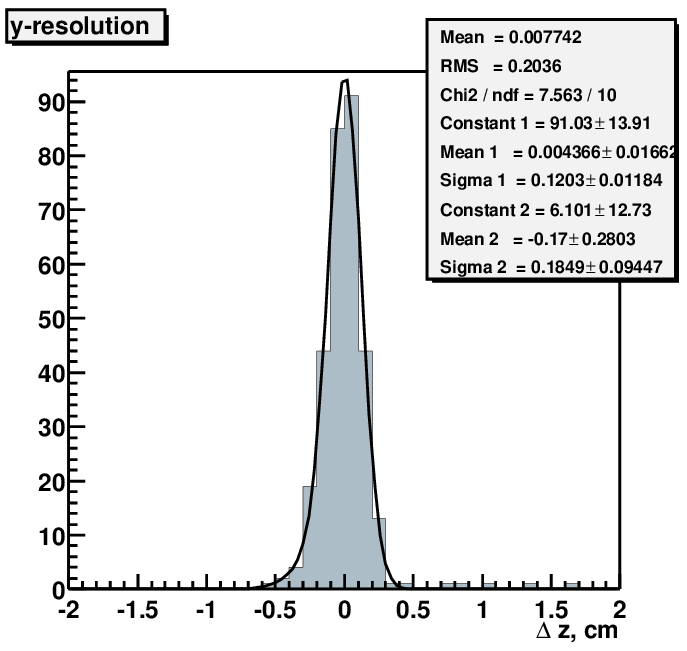}
  }
\caption{ Difference between the exact coordinate of the charged particle
  impact on the CPV, and the reconstructed coordinate. The plot for
  the $x$-axis (across the anode wires) is fitted by a single Gaussian,
  and for the $z$-axis (along the anode wires) is fitted by a sum of
  two Gaussians.}
\label{fig:cpv_resolution}
\end{figure}

The CPV detector is sensitive to any particle which initiates an
ionization process in the CPV gas volume. Therefore it will detect
charged particles with almost any momentum. The only parameter which
defines the response of the CPV, is the position resolution of the
charged track passing through the detector. The effective spatial
resolution of CPV was measured during beam-tests as $\sigma_x =
0.138$~mm (across the wires) and $\sigma_z = 0.154$~mm (along the
wires). Fig.~\ref{fig:cpv_resolution} illustrates the coordinate
resolution of the CPV.

\subsubsection{Particle identification}

Particle identification in PHOS is based on three methods:
\begin{itemize}
\item Time-of-flight (TOF) of the particles from the interaction point
  to EMC which allows to discriminate light particles (photons and
  high-energy hadrons) from slow heavy particles (low-energy
  nucleons);
\item Charge particle rejection which is based on matching of the
  reconstructed points in CPV and EMC;
\item Shower shape analysis which is based on the knowledge of the
  shower shape produced by different particles in the calorimeter.
\end{itemize}

\begin{figure}[ht]
  \vskip 5mm
  \centering
  \includegraphics[width=.9\textwidth]{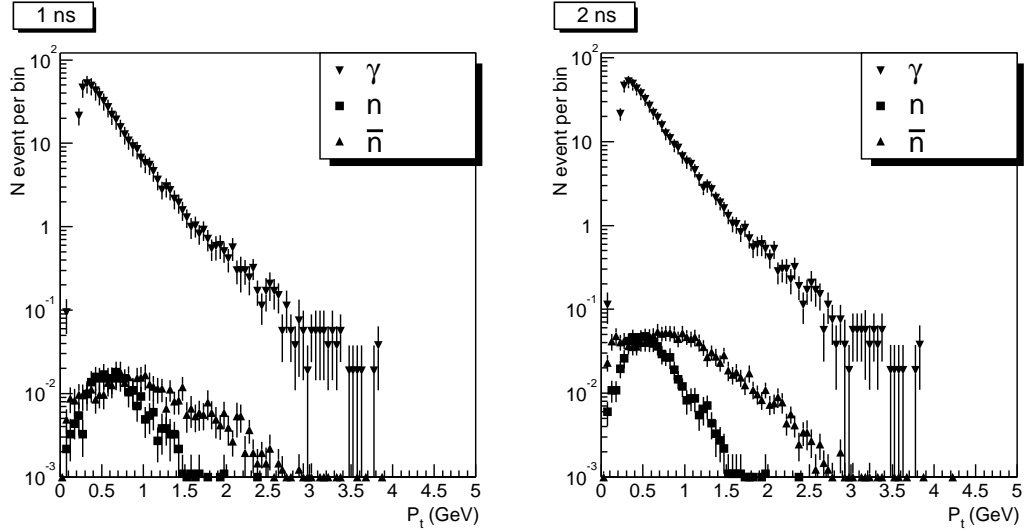}
  \caption{Spectrum of photons, compared to those of neutrons and
    antineutrons identified as photons by the TOF criterion with two
    TOF resolutions, 1 (left) and 2~ns (right) in Pb-Pb HIJING
    events.}
  \label{fig:TOF_spec}
\end{figure}

The performance of the time-of-flight depends on the time resolution
of EMC electronics. The time-of-flight of the photons from the
interaction point to PHOS is $15.3$~ns. Figure~\ref{fig:TOF_spec}
shows the spectrum of photons compared to those of neutrons and
antineutrons identified as photons by TOF with two TOF resolutions, 1
and 2~ns, in the most central Pb-Pb collisions, versus their
reconstructed energy in EMC.
The final time resolution for the EMC electronics is not selected yet,
so this figure provides the guideline for the design of the TOF
system.

\begin{figure}[ht]
  \centerline{
    \includegraphics[width=0.9\hsize]{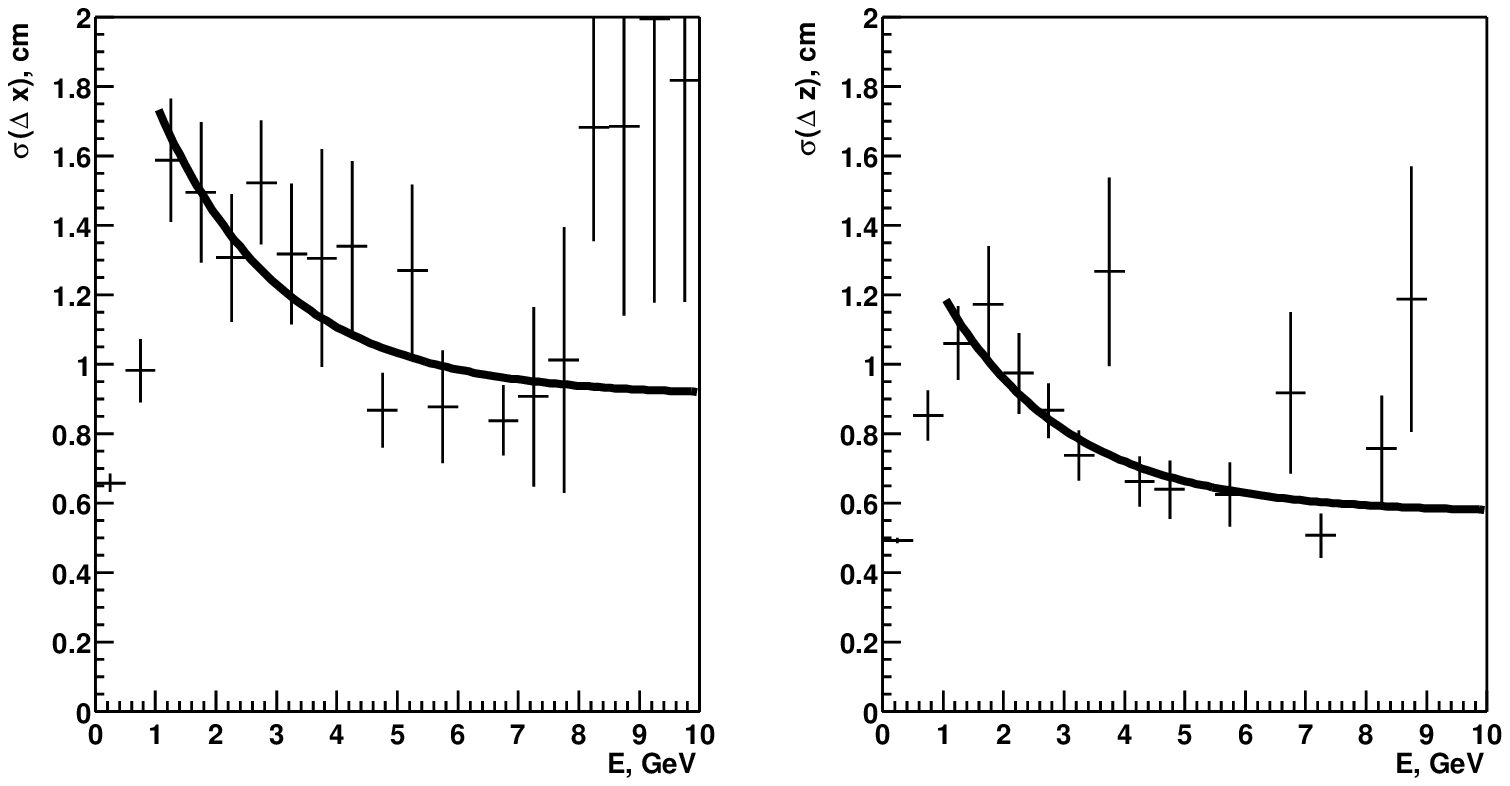}
  }
  \caption{RMS of the Gaussian fit of the EMC-CPV distance along the
    $x$-axis (i.e. across magnetic field) and $z$-axis (along the beam)
    for the charged pions produced with a uniform $p_T$ distribution
    versus the deposited energy.}
  \label{fig:CPV_match}
\end{figure}

Charged particles can be rejected in PHOS if a reconstructed point
in the EMC matches a CPV reconstructed point. Figure~\ref{fig:CPV_match}
shows the average deviation between the CPV and EMC reconstructed
points for charged pions versus their reconstructed energy, in the ALICE
magnetic field 0.5~T.

Shower shapes in EMC can be characterized by several
parameters. A set of 7 shower parameters has been chosen to identify
photons in EMC and discriminate them from hadrons and
$\pi^0$-mesons:
\begin{itemize}
\item lateral dispersion which is a mean squared deviation of the
  fired cells from the shower center;
\item shower main axes, $\lambda_0$ and $\lambda_1$ which are eigen
  values of the shower tensor in the $(x,z)$ plane;
\item shower sphericity which is a relative difference between
  $\lambda_0$ and $\lambda_1$.
\item the core energy which is a shower energy within a radius of 3~cm
  around the shower center;
\item the largest energy fraction in a single cell;
\item shower cell multiplicity.
\end{itemize}
These 7 parameters can be statistically correlated, and a set of 7
statistically independent parameters is found by diagonalizing the
covariance matrix of the shower shape in this 7-dimensional space.

\begin{figure}[ht]
  \centerline{
    \includegraphics[width=0.9\hsize]{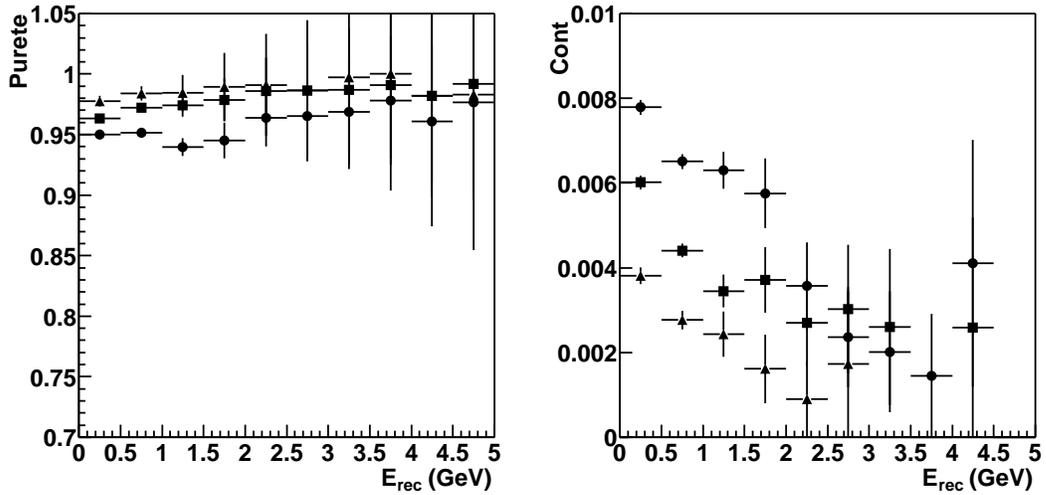}
  }
  \caption{Photon purity and antineutron contamination to the
  identified photon spectrum in HIJING Pb-Pb collisions, for three
  definitions of photon quality: low ($\blacktriangle$), medium
  ($\blacksquare$) and high ($\bullet$) purity photons.}
  \label{fig:Photon_pur_cont}
\end{figure}

Figure~\ref{fig:Photon_pur_cont} shows the purity of identified
photons by all three identification criteria, as well as the
contamination by antineutrons. Purity is a fraction of the
reconstructed particles identified as photons, which are really
photons, from the total number of reconstructed particles identified
as photons. Contamination in defined as a fraction of antineutrons
which are identified as photons, from the total number of
reconstructed particles and identified as photons particles. The
purity and contamination are shown for three definitions of photon
quality: low, medium and high purity photons.

\begin{figure}[ht]
  \centerline{
    \includegraphics[width=\hsize]{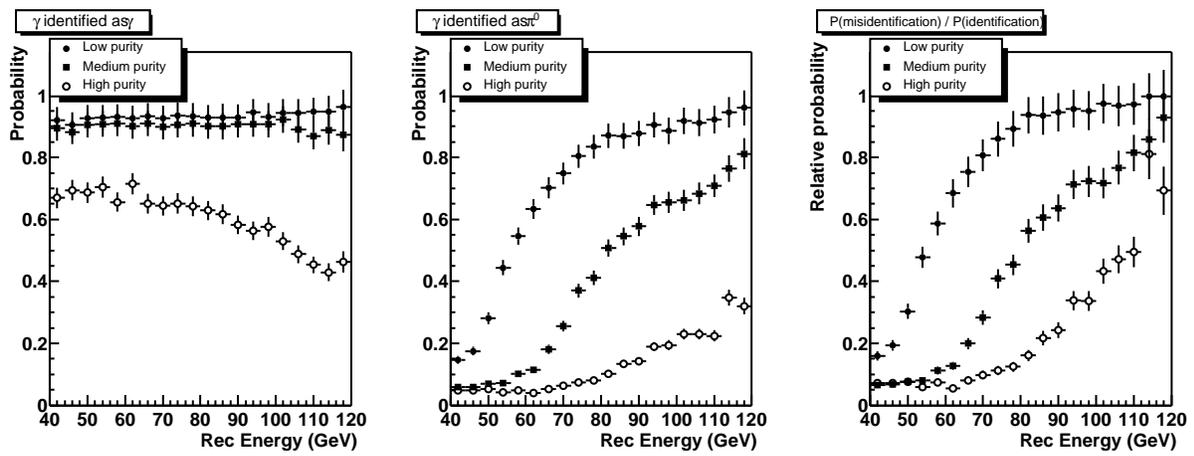}
  }
  \caption{Photon true identification probability, photon
  misidentification probability as $\pi^0$ and the ratio of the
  misidentification probability to the true one, for three values of
  photon purities.}
  \label{fig:photon_pi0}
\end{figure}

Shower shape analysis allows to distinguish photons and $\pi^0$-mesons
at high $p_T$, where both particles produce a single shower in
EMC. Figure~\ref{fig:photon_pi0} shows the true identification
probability of a photon at high $p_T$, a misidentification probability
of a photon as a $\pi^0$ and the ratio of the misidentification
probability to the true identification probability, for low, medium
and high purity photons.  This plot clearly illustrates that photons
can be distinguished from the $\pi^0$-mesons in a wide dynamical
range.




\subsection{Photon detection at CMS}

{\em O.L.~Kodolova, J.H.~Liu, I.P.~Lokhtin, A.N.~Nikitenko, I.N.~Vardanyan, P.~Yepes}

\subsubsection{CMS detector}

The Compact Muon Solenoid (CMS) is a general purpose detector designed primarily  
to search for the Higgs boson in proton-proton collisions at the 
LHC~\cite{CMS:1994}. The detector is optimized for 
accurate measurements of the characteristics of high-energy leptons and 
photons,  as well as hadronic jets in a large acceptance, providing unique
capabilities for ``hard probes'' in both $pp$ and $AA$ 
collisions~\cite{Baur:2000}.  

\vskip 0.7 cm 

\begin{figure}[htb]
\begin{center} 
\resizebox{120mm}{120mm} 
{\includegraphics{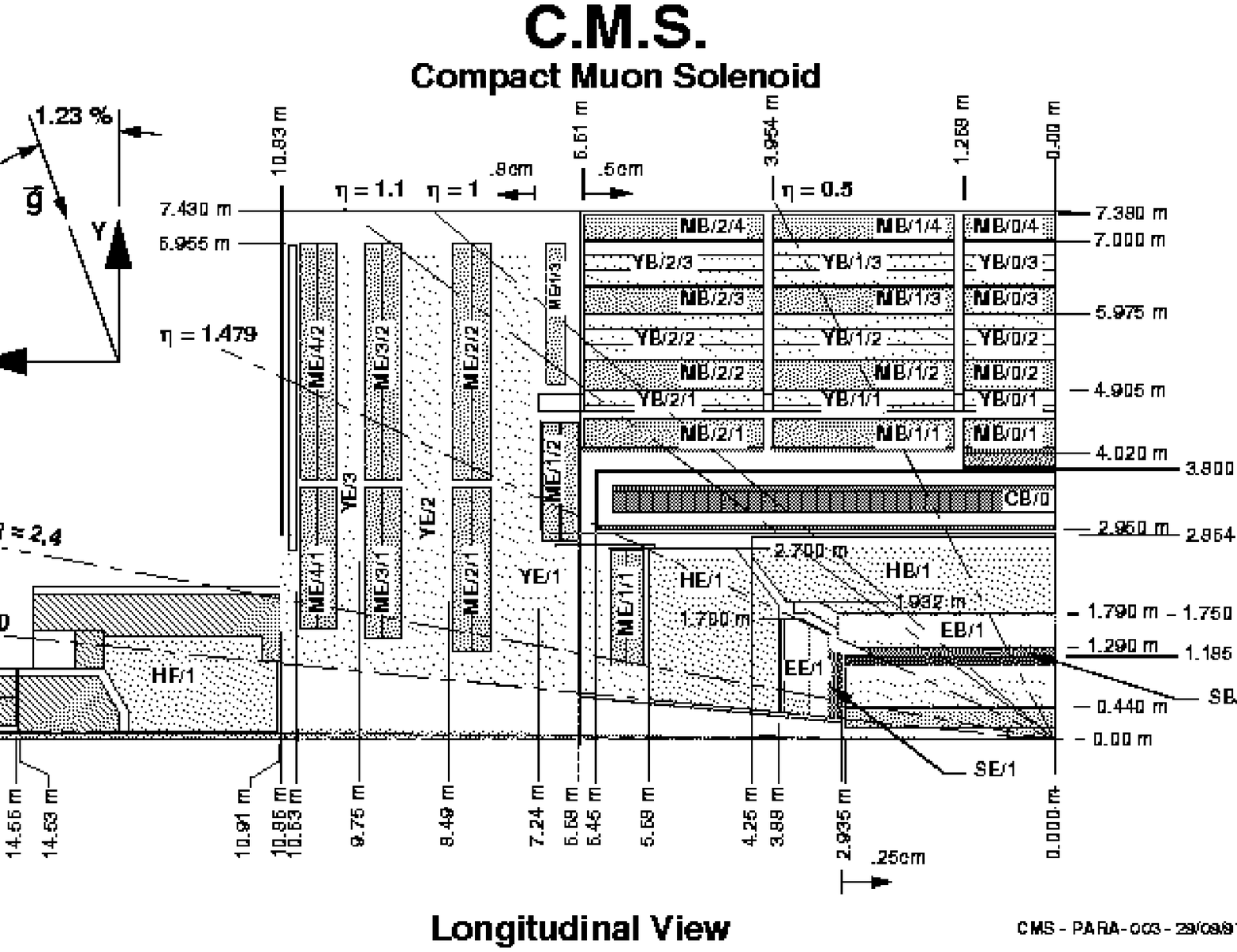}} 
\caption{\small  The longitudinal view of the CMS detector.}
\label{cmsph:fig1}
\end{center} 
\end{figure}

Detailed description of the detector elements can be found in the  
Technical Design Reports~\cite{CMSHCAL,CMSMUON,CMSECAL,CMSTRACK}. 
The longitudinal view of the CMS detector is presented in 
Fig.~\ref{cmsph:fig1}. The central element of CMS is the magnet, a $13$ m 
long solenoid with an internal radius $\approx 3$ m, 
which will provide 
a strong $4~T$ uniform magnetic field. The $4\pi$ detector consists of a $6$ m 
long and $1.3$ m radius central tracker, electromagnetic (ECAL) and hadronic
(HCAL) calorimeters inside the magnet and muon stations outside. 
The tracker and muon chambers cover the pseudorapidity region $|\eta|<2.4$, 
while the ECAL and HCAL calorimeters reach $|\eta|=3$. A pair of quartz-fiber 
very forward (HF) calorimeters, located $\pm 11$ m from the interaction point, 
cover the region $3<|\eta|<5$ and complement the energy measurement. The 
tracker is composed of pixel layers and silicon strip counters. The CMS muon 
stations consist of drift tubes in the barrel region (MB), cathode strip 
chambers in the endcap regions (ME), and resistive plate chambers in both 
MB and ME dedicated to triggering. The electromagnetic calorimeter is made of 
almost $76000$ scintillating ${\rm PbWO_4}$ crystals and the hadronic 
calorimeter consists of scintillator sandwiched between brass absorber plates. 
The main characteristics of the calorimeters are presented in
Table~\ref{calo:tab}. 
 
\begin{table}[htb]
\begin{center}
\caption{\small Energy resolution, $\sigma /E$, and granularity of the CMS 
calorimeters in the barrel (HB, EB), endcap (HE, EE) and very forward (HF) 
regions. The energy resolution is shown for the total energy of electrons 
and photons (ECAL) and transverse energy of hadronic jets (HCAL, HF).} 
\label{calo:tab} 

\medskip 

\begin{tabular}{|l|c|c|c|c|c|} \hline  
Rapidity coverage & \multicolumn{2} {c|} {$0<\mid\eta\mid<1.5$} 
& \multicolumn{2} {c|} 
 {$1.5<\mid\eta\mid<3.0$} 
  & {$3.0<\mid\eta\mid<5.0$} \\ \hline & & & & & \\ 
Subdetector & HCAL (HB) & ECAL (EB) & HCAL (HE) & ECAL (EE) & HF \\ 
& & & & & \\ \hline 
$\sigma /E=a/\sqrt{E} \bigoplus b$ & & & & & \\
$a$ & 1.16 & 0.027 & 0.91 & 0.057 & 0.77 \\
$b$ & 0.05 & 0.0055 & 0.05 & 0.0055 & 0.05 \\
\hline
granularity &  &  &  &  &
\\
$\Delta\eta \times \Delta\varphi$ & $0.087 \times 0.087$ & $0.0174 \times
0.0174$ & $0.087 \times 0.087$ & $0.0174 \times 0.0174$ 
& $0.175 \times 0.175$ \\
& & & & to $0.05 \times 0.05$ & \\ \hline 
\end{tabular}
\end{center}
\end{table}

\subsubsection{Photon triggering and identification} 

Photon identification, measurement and triggering in Pb$-$Pb collisions 
with the maximum estimated particle density, $dN^{\pm}/dy (y=0) = 8000$, 
have been studied~\cite{Baur:2000,Nikitenko:1998}  
with a full GEANT-based simulation of the CMS calorimetry 
and a parameterization 
of HIJING~\cite{hijing} data for the background.  

\begin{figure}[htb]
\begin{center} 
\resizebox{105mm}{105mm}
{\includegraphics{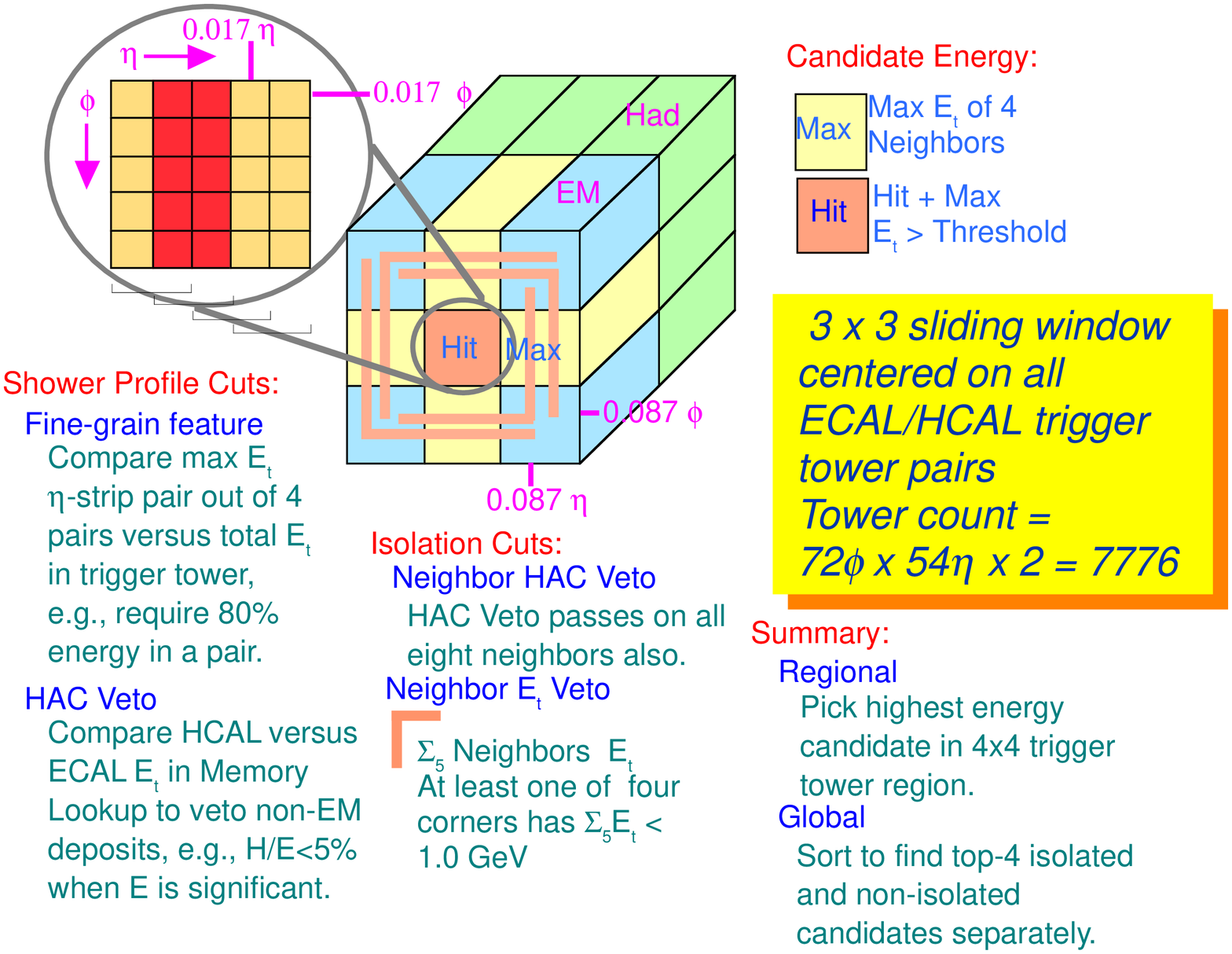}}

\vskip -3 mm

\caption {\small The CMS electron/photon trigger algorithm.}
\label{eg_tralgo}
\end{center} 
\end{figure}

The CMS electron/photon trigger algorithm developed for $pp$ collisions, see 
Fig.~\ref{eg_tralgo}, is suitable for 
triggering on energetic photons produced in the heavy ion collisions. 
Programmable thresholds on the cluster variables used in the algorithm have 
to be tuned to make it efficient. Estimates of the photon trigger rates have been made 
with two Algorithm Vetoes, see Fig.~\ref{eg_tralgo}, the Hadronic Veto 
($H/E<0.4$) and 
Neighbour $E_T$ Veto ($\sum_{5}$ Neighbours $E_T$ $< 25$ GeV). 
The rate of the single photon 
trigger is less than $1$ ($10$) Hz for a $50$ ($20$) GeV threshold.  With a 
such a threshold, the trigger 
efficiency is close to $100 \% $ for the $\gamma +$jet 
events useful for the off-line analysis~\cite{Nikitenko:1998}. 

\begin{figure}[htbp]
\begin{center} 
\resizebox{170mm}{170mm}
{\includegraphics{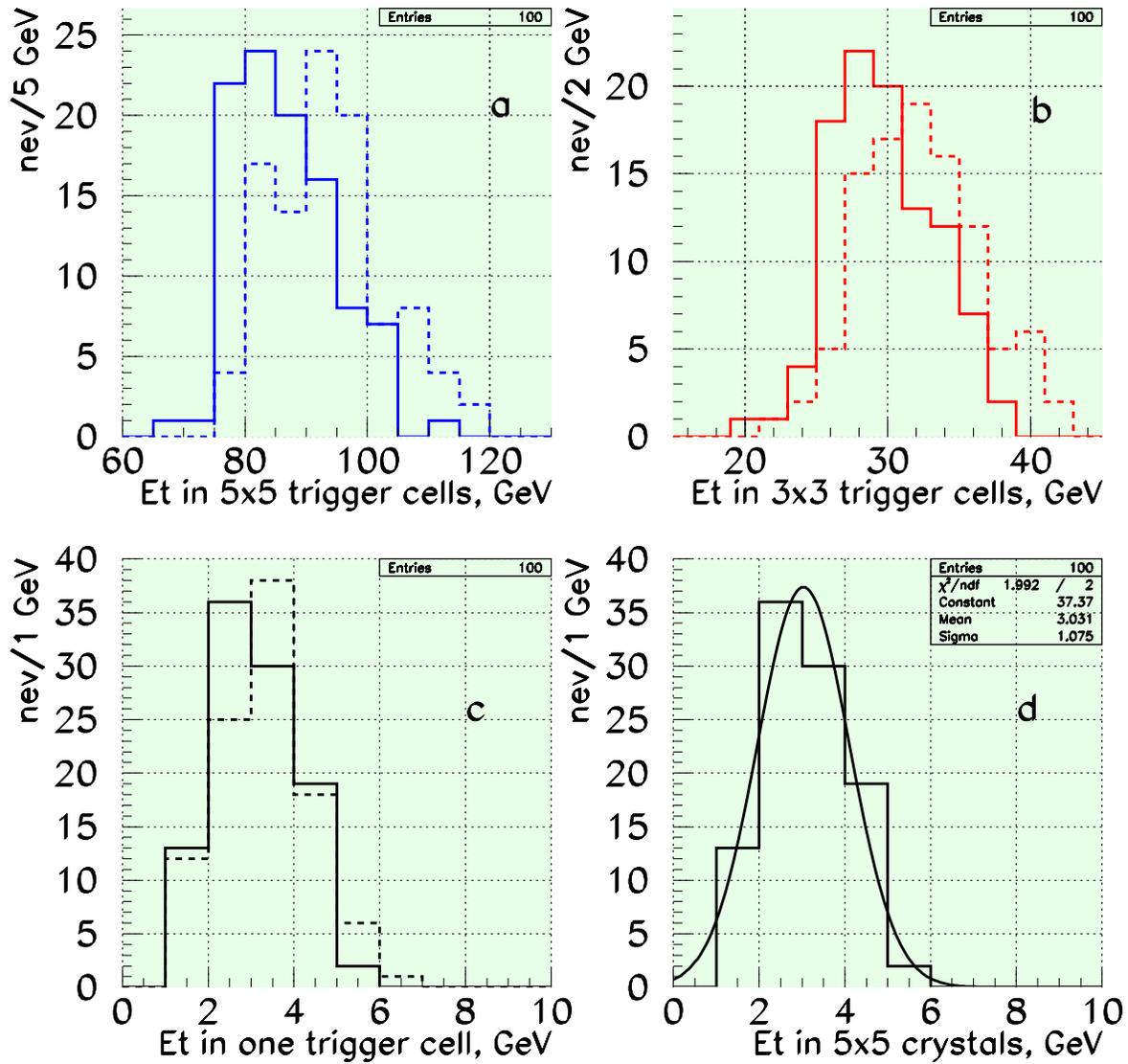}}
\caption {\small Distribution of the transverse energy of Pb$-$Pb event
deposited in trigger cells of $5 \times 5$ (a)
and $3 \times 3$ (b) crystals, not including the central one. 
(c) The transverse energy deposited in one trigger cell.  The solid 
histograms in (a)-(c) give the energy in the electromagnetic calorimeter while
the dashed histograms give the energy in the ECAL$+$HCAL
system. (d) A fit of the transverse energy deposited in
the electromagnetic part of one trigger cell.}
\label{isolation}
\end{center} 
\end{figure}

Apart from the trigger selections, single photon identification based on 
calorimeter isolation or on the use of calorimeter cells above a certain $E_t$ 
threshold (labeled ``cell $E_t$ cut'')  was considered.  The
photon energy may be measured 
in a $5 \times 5$ crystals cell (the trigger cell size) centred on the 
crystal with the highest response. 
Such a cell contains about $97\% $ of the photon energy. Identification may 
be based on a cut in the transverse energy, $E_{t}^{\rm isol}$, deposited in 
an area of $3 \times 3$ or $5 \times 5$ such cells, not including the central 
one. Distributions of  $E_{t}^{\rm isol}(5\times 5)$ and  
$E_{t}^{\rm isol}(3\times 3)$ are shown in Fig.~\ref{isolation}(a) and (b) 
for the energy from a Pb$-$Pb event deposited in the ECAL only and in the 
total ECAL$+$HCAL system. Only about $6\% $ of the transverse energy in the 
isolation area is measured by the 
hadron calorimeter, reflecting the softness of the charged particle spectra.  

The ``cell $E_t$ cut'' criterion is another method of photon identification 
which has been studied. It requires no energy above a given threshold 
deposited in every cell of the area around the central cell containing the 
photon. The transverse energy distribution in the cell 
is shown for Pb$-$Pb events in Fig.~\ref{isolation}(c).  A threshold of
$E_T=6.5$ GeV has been chosen for this distribution.  
This has been applied in an area of $7 \times 7$ 
cells, not including the central $3 \times 3$ trigger matrix since the 
trigger criteria must still be optimized and applied separately. 
``Cell $E_t$ cut'' gives us a rejection factor of $\sim 2.7$ for $\pi^0$ 
decays compared to a $14\% $ reduction of the single photon rate. 

The photon energy resolution is degraded due to the large ``pile up'' 
contribution in heavy ion collisions. In a $5 \times 5$ crystal matrix, we 
have extra energy deposited, with $\sim 1$ GeV, as seen in Fig.~\ref{isolation}(d), 
so that 
for a 120 GeV photon, the 0.64\% test beam resolution~\cite{Seez:1998} will 
be degraded to $0.80\%$ due to ``pile up''. This photon resolution is still
much better than the jet energy resolution. This can be improved by using 
$3 \times 3$ crystals matrix for energy measurement.  

\subsubsection{Photon reconstruction efficiency and resolutions} 

The capability of the CMS ECAL to reconstruct photons in heavy ion collisions 
was investigated in several $E_T$ intervals using a standard  
electromagnetic cluster reconstruction algorithm implemented in CMS object
oriented reconstruction package ORCA (version 6). The algorithm looks for 
crystals with energies above a certain threshold and creates a cluster in a 
$5 \times 5$ crystal matrix. The full GEANT-based simulation of the CMS 
calorimetry (CMSIM$\_$125 package) responses on single photons and 
HIJING central Pb$-$Pb event as a background were used in the analysis for the 
barrel and endcaps.  

\begin{figure}[htbp]
\begin{center} 
\resizebox{105mm}{105mm}
{\includegraphics{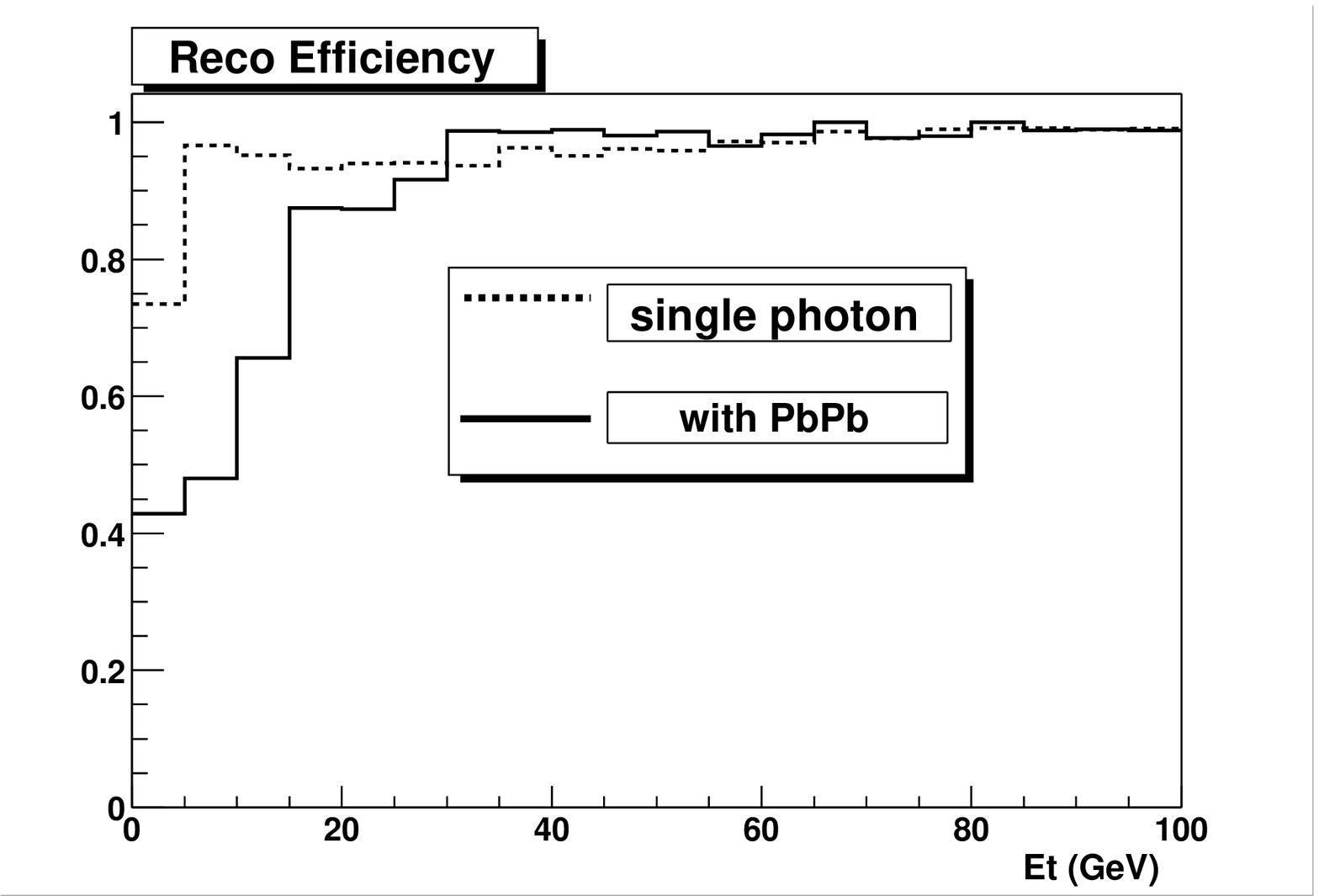}}
\caption {\small Photon reconstruction efficiency in Pb$-$Pb (solid) and $pp$ 
(dashed) events.} 
\label{eff-pt}

\vskip 0.7 cm 

\resizebox{105mm}{105mm}
{\includegraphics{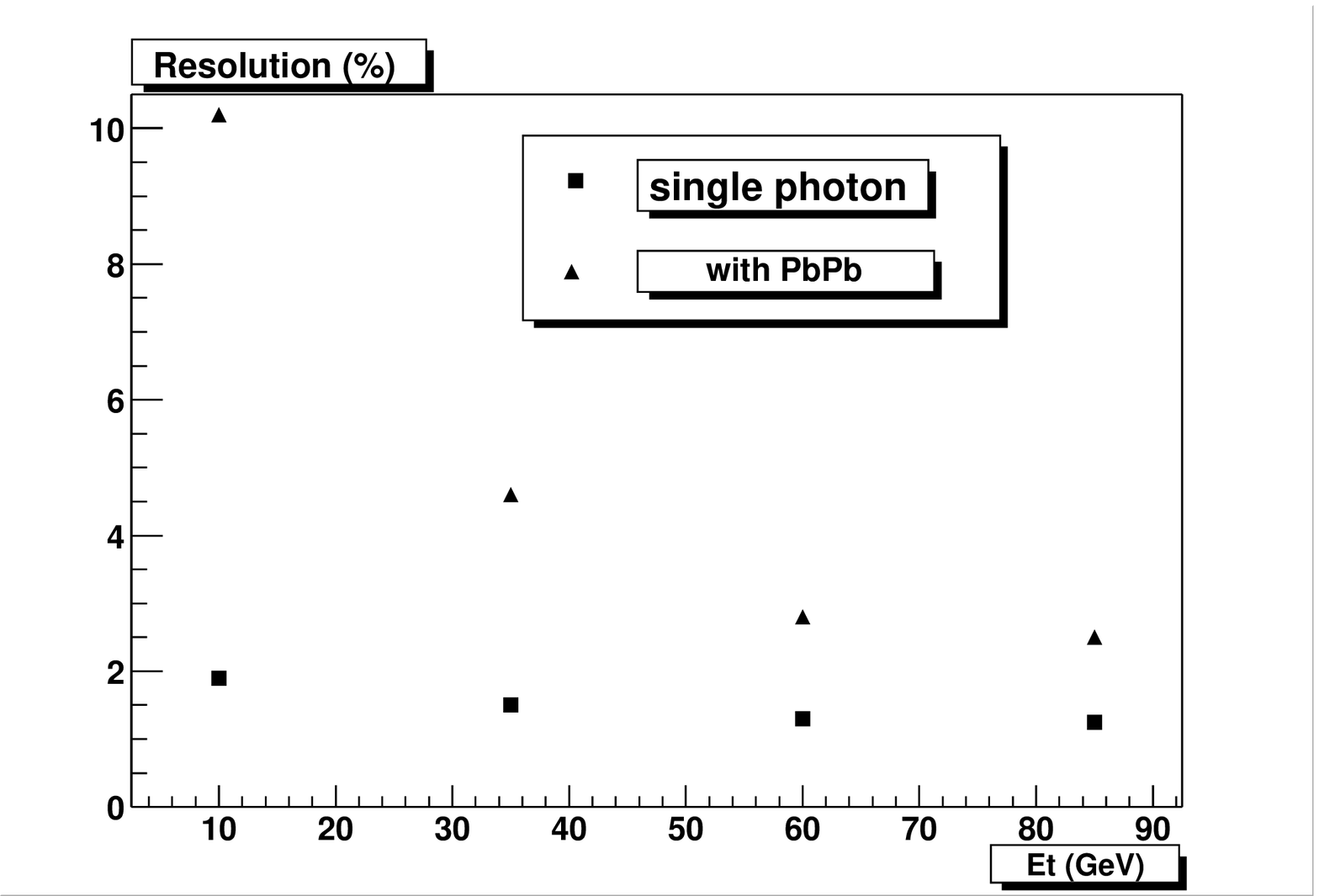}}
\caption {\small Photon transverse energy resolution in Pb$-$Pb (triangles) and $pp$ 
(squares) events.} 
\label{photres}
\end{center} 
\end{figure}

Photon reconstruction efficiency as a function of photon transverse energy in 
Pb$-$Pb and $pp$ events is shown in Fig.~\ref{eff-pt}. The estimated photon 
reconstruction efficiency for Pb$-$Pb collisions appears to be high enough,   
$\ga 80\% $, starting from $E_T \sim 15$ GeV. 
The efficiency dependence on the pseudorapidity is rather weak. The work on 
improvement of the photon reconstruction algorithm to increase reconstruction 
efficiency in high multiplicity environment is in progress. The photon 
spatial resolutions, $\sigma_{\varphi } \sim \sigma_{\eta }\sim 
0.005$, are practically the same for $pp$ and Pb$-$Pb collisions and do not 
depend significantly on the transverse energy. Such spatial 
resolution is better than $\eta - \varphi$ size of an ECAL cell, 
$0.0174\times 0.0174$. However, the influence of Pb$-$Pb background on energy 
resolution is more significant, as shown in Figure~\ref{photres}. At $E_T=10$ 
GeV, the transverse energy resolution degrades strongly in Pb$-$Pb events 
relative to $pp$, from $2\%$ to $10\%$. The difference decreases with increasing 
$E_T$ and becomes insignificant at $E_T \sim 100$ GeV. 

\subsubsection{Jet reconstruction} 

A detailed description of the jet reconstruction procedure in heavy ion 
collisions using the sliding window-type jet finding algorithm which 
subtracts the large background from the underlying event and a full 
GEANT-based simulation of the CMS calorimetry can be found in the chapter on
jets. The efficiencies and
background contamination levels are shown in Table~\ref{cmsjet:tab}, 
along with the transverse energy resolution for several values of jet
transverse energy in central Pb$-$Pb collisions assuming $dN^{\pm}/dy (y=0) = 
8000$.  The jet energy resolution is defined as $\sigma(E_T^{\rm 
reco}/E_T^{\rm gen})/ \langle E_T^{\rm reco}/E_T^{\rm gen} \rangle$ where
$E_T^{\rm reco}$ is the reconstructed transverse energy and $E_T^{\rm gen}$ 
is the transverse energy of all generated particles inside the given cone 
radius $R$.  Starting at $E_T=100$ GeV, jets can be reconstructed with $100\%$ 
efficiency and purity.  The purity is defined 
as number of events with true QCD jets divided by the number of events with 
all reconstructed jets.    

\begin{table}[hbtp]
\begin{center}
\caption{\small The jet purity, noise (contamination levels, false jets / generated 
jets) and transverse energy resolution ($|\eta|<0.3$, $R=0.6$) in central 
Pb$-$Pb collisions with $dN^{\pm}/dy (y=0) = 8000$.} 
\label{cmsjet:tab} 

\medskip 

\begin{tabular}{|c||c|c||c|} \hline  
$E_{T~{\rm min}}$ (GeV) & Purity & Noise  & $\sigma(E_T)/E_T(\%)$ \\ \hline 
 75 & 0.88 $\pm$ 0.03 & 0.083 $\pm$ 0.009 & 17.8 \\ \hline 
100 & 0.97 $\pm$ 0.03 & 0.011 $\pm$ 0.003 & 18.4 \\ \hline 
125 & 0.99 $\pm$ 0.03 & 0.004 $\pm$ 0.002 & 16.8 \\ \hline 
200 & 0.99 $\pm$ 0.03 & 0.001 $\pm$ 0.001 & 12.7 \\ \hline 
\end{tabular}
\end{center}
\end{table}

Although the jet transverse energy resolution is degraded by a factor 
$\sim 2$ in high multiplicity central Pb$-$Pb collisions compared to  
$pp$, the average measured jet energy in Pb$-$Pb collisions is 
the same as in $pp$.  Thus $pp$ interactions can be used as a baseline for 
heavy ion jet physics. 

It is important to note that the jet angular resolution, 
$\sigma_{\varphi }=0.045$ and $\sigma_{\eta }=0.05$ at $E_T^{\rm jet} = 100$ 
GeV, is still better than the $\eta - \varphi$ size of an HCAL 
tower $0.087\times 0.087$. Thus the angular position of a hard jet can be 
reconstructed in heavy ion collisions at CMS with high enough accuracy for 
analysis of jet production as a function of azimuthal angle and pseudorapidity.

\comment{




\subsection{Photon studies in the ATLAS detector}

{\em H.Takai, S. Tapprogge}

The ATLAS detector is designed to study high $p_T$ physics in
full luminosity proton-proton collisions at the LHC. Most of the detector
subsystems will be available for heavy ion collisions as well.
We are interested in the detection of photons from heavy ion collisions.
The highly segmented electromagnetic calorimeter will be heavily
used for this studies. We report on early assessment of
the detector capabilities in the heavy ion environment.


\subsubsection{The ATLAS Detector}

The ATLAS detector is designed to study proton-proton collisions at
the LHC design energy of 14~TeV in the center of mass.  The physics
pursued by the collaboration is vast and includes: Higgs boson search,
searches for SUSY, and other scenarios beyond the Standard Model, as
well as precision measurements of processes within (and possibly beyond)
the Standard Model. To achieve these goals at  full machine
luminosity of $10^{34} cm^{-2}s^{-1}$, ATLAS will have a precise
tracking system (Inner Detector) for charged particle measurements, an
as hermetic as possible calorimeter system, which has an extremely
fine grain segmentation, and a stand-alone muon system. An overview of
the detector is shown in Figure~\ref{fig:atlas}.

\begin{figure}[htbp]
\begin{center}
\includegraphics[width=12.5cm]{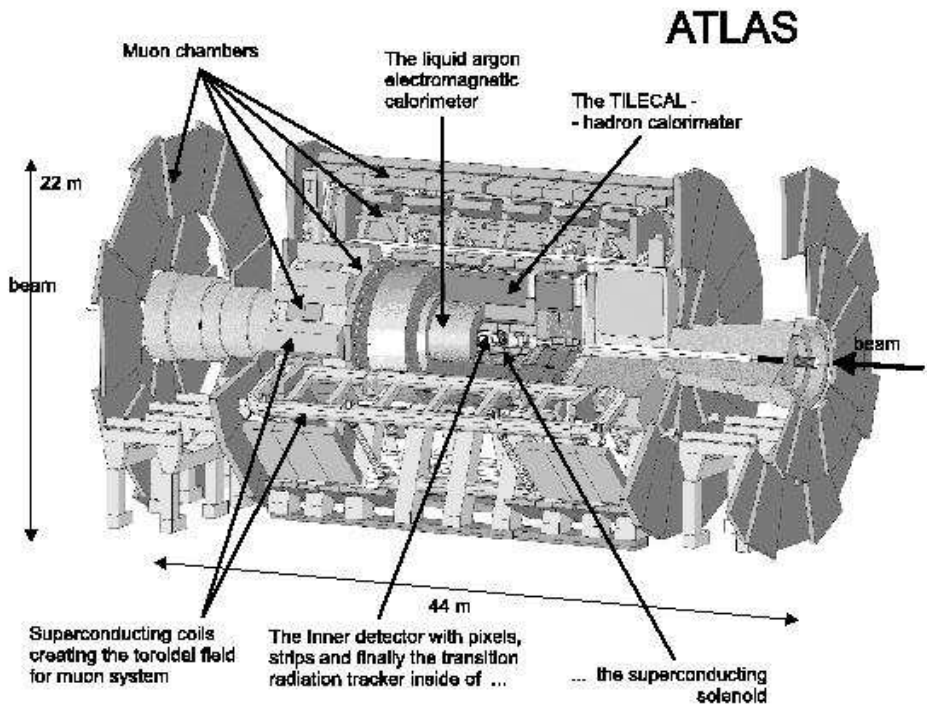}
\caption{The overall layout of the ATLAS detector}
\label{fig:atlas}
\end{center}
\end{figure}

The Inner Detector is composed of (1) a finely segmented silicon pixel
detector, (2) silicon strip detectors (Semiconductor Tracker (SCT))
and (3) the Transition Radiation Tracker (TRT).  The segmentation is
optimized for proton-proton collisions at design machine luminosity.
The inner detector system is designed to cover a pseudo-rapidity of
$\mid \eta \mid < 2.5$ and is located inside a 2~T solenoid magnet.

The calorimeter system in the ATLAS detector that surrounds the solenoid
magnet is divided into electromagnetic and hadronic sections and
covers pseudo-rapidity $\mid \eta \mid < 4.9$. The EM calorimeter is
an accordion liquid argon device and is finely segmented
longitudinally and transversely for $\mid \eta \mid \le 3.1$.  The first
longitudinal segmentation has a granularity of 0.003 x 0.1 $(\Delta
\eta \times \Delta \phi)$ in the barrel and slightly coarser in the
endcaps.  The second longitudinal segmentation is composed of $\Delta
\eta \times \Delta \phi = 0.025 \times 0.025$ cells and the last
segment $\Delta \eta \times \Delta \phi = 0.05 \times 0.05$ cells.  In
addition a finely segmented $(0.025 \times 0.1)$ pre-sampler system is
present in front of the electromagnetic (EM) calorimeter.  The overall
energy resolution of the EM calorimeter as determined in test beam measurements
 is $10\%/\sqrt{E} \oplus 0.5\%$. The calorimeter also has good pointing
resolution, $60 mrad/\sqrt{E}$ for photons and timing resolution
better than 200 ps for showers of energy larger than 20 GeV.

The hadronic calorimeter is also segmented longitudinally and
transversely.  Except for the endcaps and the forward calorimeters,
the technology utilized for the calorimeter is a lead-scintillator
tile structure with a granularity of $\Delta \eta \times \Delta \phi =
0.1 \times 0.1$.  In the endcaps the hadronic calorimeter is
implemented in liquid argon technology for radiation hardness with the
same granularity as the barrel hadronic calorimeter.  The energy
resolution for the hadronic calorimeters is $50\%/\sqrt{E} \oplus 2\%$
for pions.  The very forward region, up to $\eta = |4.9|$ is covered by
the Forward Calorimeter implemented as an axial drift liquid argon
calorimeter.  The overall performance of the calorimeter system is
described in~\cite{atlas1}.

The muon spectrometer in ATLAS is located behind the calorimeters,
thus shielded from most hadronic showers, and has a coverage of
$\mid \eta \mid < 2.7$. The spectrometer is implemented
using several technologies for tracking devices and a toroidal magnet
system, which provides a field of 4~T strength to have an independent
momentum measurement outside the calorimeter volume.  Most of the
volume is covered by MDTs (Monitored Drift Tubes). In the forward region,
where the rate is high, the Cathode Strip Chamber technology is chosen.
The stand-alone muon spectrometer momentum resolution is of the order
of $2\%$ for muons with $p_T$ in the range 10 - 100 GeV. 

The trigger and data acquisition system of ATLAS is a multi-level
system, which has to reduce the beam crossing rate of 40~MHz to an
output rate to mass storage of $\mathcal{O}(100)$~Hz. The first stage
(LVL1) is a hardware based trigger, which makes use of coarse 
granularity calorimeter data and dedicated muon trigger chambers only.
It has to reduce the output rate to about 75~kHz, within a maximum latency
of 2.5~$\mu$s. The High-Level Trigger (HLT) is composed of two stages,
the second level trigger (LVL2) and the event filter (EF), where further
reduction of the rate is achieved using algorithms implemented in 
software, making use of the full granularity and all sub-detectors. For
LVL2, the Region-of-Interest (RoI) concept is used to reduce the amount
of event data needed to only a few per cent. For
heavy ion physics where an interaction rate of 8 kHz is expected for
full luminosity Pb-Pb collisions, we expect to be able to record data with minimal
trigger requirements, e.g. centrality trigger.

The performance results mentioned have been obtained using a detailed
full simulation of the ATLAS detector response with GEANT and have 
been validated by an extensive program of testbeam measurements of 
all components.

\subsubsection{ATLAS and Photon physics}

Photons in ATLAS are detected in the EM calorimeter.  Early in the
detector design it was decided to include the capabilities of detecting the
Higgs boson through its decay into two photons.  Therefore the
calorimeter is highly optimized for high $p_T$ photon detection and good
rejection of $\pi^0$'s.  The EM calorimeter
sampling is divided in fine strips of $\Delta \eta \times \Delta
\phi = 0.003 \times 0.1$ to aid in the $\pi^0$ rejection, by identifying
overlapping photon showers.  Extensive
simulations of the performance in proton-proton environment have been
carried out. Figure~\ref{fig:gpi0} shows the results of this studies for 
photons with transverse energies up to 100 GeV.
\begin{figure}
\begin{center}
\includegraphics[width=8.cm]{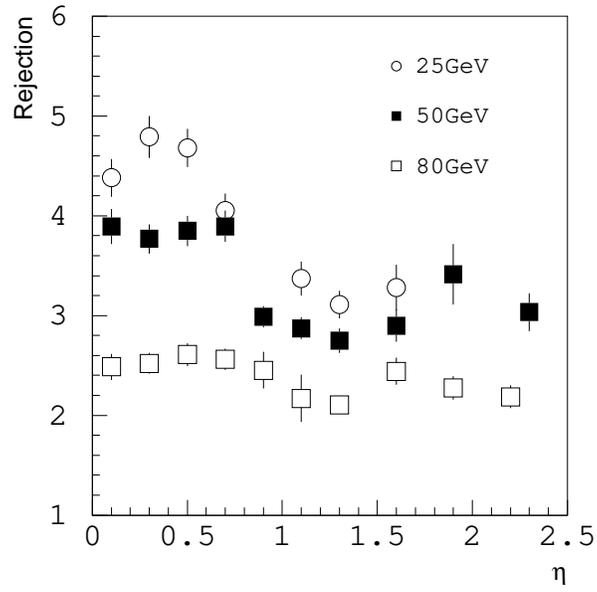}
\caption{Photon to $\pi^0$ rejection as function of $\eta$.}
\label{fig:gpi0}
\end{center}
\end{figure}
\begin{figure}
\begin{center}
\includegraphics[width=8.cm]{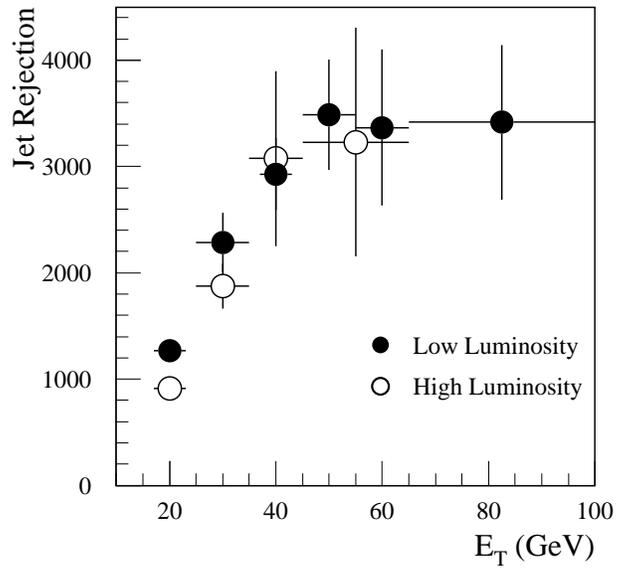}
\caption{Jet rejection after photon selection cuts as function of jet $E_T$ for low
and high luminosity proton-proton conditions.}
\label{fig:pi0jet}
\end{center}
\end{figure}
Heavy ion events are different in character when compared to high
luminosity proton-proton runs. Although the multiplicity is higher
there is no event pile-up. The underlying background to signals of interest,
for a typical central collision HIJING event,
comes from the soft particles produced in the collision. The majority
of the soft charged particle tracks ($p_T \leq 200 MeV/c$) are
curled up in the solenoid magnetic field. At the face
of the calorimeter the density of charged particles is approximately
$dN/dy \sim 2000$. The neutral particles
are in their vast majority low $p_T$ $\pi^0$s and they will tend
to deposit most of their energy in the first compartment, if not
in the calorimeter and solenoid cryostat walls ($\sim 1X_0$) 
However, the second
electromagnetic compartment could be relatively quiet and used
for photon studies. 

The major challenge in the study of single photons with the ATLAS detector will
be the mis-identification of $\pi^0$'s (or jets with a leading $\pi^0$) as a
photon.  Studies performed for high luminosity proton-proton runs indicate  good
performance for single photon identification. This is shown in
Figs.~\ref{fig:gpi0} and \ref{fig:pi0jet}. Because the performance is due to the
fine transverse segmentation coupled to a longitudinal segmentation, we do expect
similar performance for heavy ions.  Detailed simulation studies are under way.





%

\section{NON THERMAL PRODUCTION MECHANISMS}
\label{sec:pertQCD}

\noindent
{\em P.~Aurenche, F.~Bopp, H.~Delagrange, S.~Jeon, P.~Levai, J.~Ranft,
I.~Sarcevic, M.~Tokarev, M.~Werlen}

\vspace{.5cm}

For the discovery of the quark-gluon plasma using single photon production the
$p_{_T}$ range of interest is roughly { 1 GeV/c $<  p_{_T}  <$  10 GeV/c}:
indeed it is expected that thermal photon production in heavy ion collisions
will increase the production rate somewhere in this domain. It is therefore
important to also understand the non thermal production mechanism in the same
energy range. We shall consider both $\gamma$ and $\pi^0$ production since the
latter is a background to the former. To calculate the relevant rates the usual
theoretical tools at our disposal are the next-to-leading logarithm (NLO) QCD
calculations.  However, the considered $p_{_T}$ values are very small compared
to the center-of-mass energy of the collision and one is not far from the
(small $x$) kinematical boundary where perturbation theory may not be reliable.
We therefore supplement the QCD predictions with those from a model which
includes soft physics dynamics as well as semi-hard physics: such a
model (Dual Parton Model as implemented in \textsc{Phojet}/\textsc{Dpmjet}) has been successfully
confronted with data over a wide energy domain. We review each approach in
turn, before presenting phenomenological results.

\subsection{Theory : perturbative QCD approach at next-to-leading order (NLO)}
\label{sec:nloqcd}
{\em P.~Aurenche, H.~Delagrange, S.~Jeon, P.~Levai, I.~Sarcevic, M.~Werlen}

\subsubsection{Proton-proton collisions}

The  production cross section of a particle at large transverse momentum in
perturbative QCD is well known and has the usual factorisable form:%
\begin{eqnarray}
{d \sigma^{^{AB \rightarrow C}} \over d{\imb p_{_T}} dy} ={\sum}_{a,b,c}  \int
dx_a dx_b { dz \over z^2} F_{_{a/A}}(x_a,M) F_{_{b/B}}(x_b,M)  D_{C/c}(z,
M_{_F}) {d {\widehat \sigma}^{^{ab\rightarrow c}} \over d{\imb 
p_c{_{_T}}} dy_c}    (\mu, M, M_{_F}),
\label{eq:inclu-rate}
\end{eqnarray}
where the functions $F$ are the parton densities and $\ {\widehat
\sigma}^{^{ab\rightarrow c}}$ is the hard cross section between partons $a$ and
$b$, in the hadrons $A$ and $B$ respectively, to produce parton $c$. The
function $D_{_{C/c}}$ is the fragmentation function of parton $c$ into particle
$C$. In the case a photon is produced, an extra term has to be considered where
the  fragmentation function reduces to a $\delta(1-z)$ function:  in this case
the photon participates directly to the hard collision ($c = \gamma$) in
contrast to the bremsstrahlung process where the photon is produced in the
fragmentation of a quark or gluon $c$.\footnote{
	Both directly produced photons and bremsstrahlung photons are ``prompt"
	in the sense of Chap.~\ref{sec:nomenclature} and they contribute to the
	``direct" photon spectrum.}
The calculations have been carried out up to next-to-leading
order~\cite{pionlo,photonlo} in QCD (all functions $F,\ D,\
\sigma^{^{ab\rightarrow c}}$ are known in the NLO approximation) but there
remains a residual ambiguity related to the choice of the unphysical
renormalization scale $\mu$,  factorization scale $M$ and fragmentation scale
$M_{_F}$. 

For photon production extensive phenomenological studies have been carried out
in pro\-ton/(an\-ti-)pro\-ton scattering for $\sqrt s$ from 20 GeV to 1.8 TeV and the
situation was found to be rather
confused~\cite{{e706},{photonpheno},{pionpheno}}. Concerning results on fully
inclusive photon production in $p p$ and $\bar p p$, the theory is in
satisfactory agreement with all data from (fixed target) 20 GeV to (ISR) 63 GeV
with the same set of parameters (all scales set around $p_{_T}/2$ or slightly
smaller). There is one exception: the E706 data~\cite{e706} (at 31.6 and 38.8
GeV on Beryllium, but corrected by the experimental group to be compared to
proton-proton scattering) which are at least a factor 2 to 3 above the other
data and which do not have the same $p_{_T}$ dependence. For $\pi^0$ production
NLO theory and data are in agreement as far as the shape of the spectrum is
concerned but the data are systematically above theoretical
predictions~\cite{pionpheno} (with large fluctuations in the normalization of
experiments when compared to theory): one possible explanation is the poor
knowledge of fragmentation functions in the dominant large $z$ region which is
hardly constrained by $e^+ e^-$ data from which the fragmentation functions are
mostly derived. Indeed a large variation in the theoretical rates is observed
when using different fragmentation functions (BKK~\cite{bkk}, KKP~\cite{kkp},
Kretzer~\cite{kretzer}).  The recent RHIC data at 200 GeV~\cite{ppg024} will
certainly help clarify the phenomenology. 

In the present work, one will need the NLO calculations in a brand new
kinematical regime: $\sqrt{s} = 5.5$ and $14$ TeV and $\pt > 3$ GeV,
corresponding to  very small $x$ values. "Large $\pt$" NLO calculations have
never been tested before, in hadronic collisions, in this small $x$ kinematical
regime. For transverse momentum of 3 GeV, the typical  $x= 2 p_{_T}/ \sqrt{s}$
values  are of the order of $10^{-3}$ and much less if forward/backward
rapidities are explored. One may question the reliability of straightforward
NLO calculations in this domain. This is where ``recoil"
resummation~\cite{laili} is important but further studies are needed since the
present results are rather dependent on non-perturbative parameters and no 
convenient phenomenological calculations are available at present.  As
mentioned above the NLO predictions also suffer from the usual (factorization,
fragmentation, renormalization) scale uncertainties. In the following we follow
the usual (albeit arbitrary) practice to choose a common scale and let it
vary between $\pt/2$ and $2 \pt$.\footnote{
	Different choices were also tried such as varying $M=\mu$ and $M_{_F}$
	independently in the range specified above but the phenomenology for
	$\sqrt{s} = 200$ GeV and above is not affected.} 
The uncertainties on structure/fragmentation functions will be probed by using
different sets.

A specific feature of photon production at very high energy is related to the
fact that the brems\-strahlung component becomes large and dominant at small 
$x$. However this component is not really under control: in
particular the $G \rightarrow \gamma$ fragmentation channel is very important
but it is not constrained by previous data~\cite{bourhis}. This point was not
relevant for lower energies because of larger $x$ values or because it was
eliminated by isolation cuts in the collider data. However, at LHC, for small
$\pt$ this will introduce a large uncertainty on prompt photon production.

Among other possible uncertainties in the predictions, one should mention those
related to the "intrinsic" $\kt$. As will be seen below, the $\pi^0$ spectrum
obtained at RHIC turns out to be in very good agreement with NLO predictions
using standard scales (of the order of $\pt$) thus alleviating the need of
introducing "$\kt$ effects".\footnote{
	For an alternative approach see the recent papers of the Budapest
	group~\cite{levai}.}

To summarize: the main uncertainties in the predictions are, as usual, related
to the choice of scale values and the choice of fragmentation functions, the
latter being important even in the case of prompt photon production because of
the importance of the bremsstrahlung mechanism at small  $p_{_T}$ and large
energy. As for the uncertainties associated to the structure functions they
turn out to be relatively small (a few $\%$ only). These points will be
illustrated quantitatively in the phenomenological sections. At a more
fundamental level, we must admit that the NLO machinery is applied in a
kinematical region where it may not be justified but we have no possibility to
gauge the associated uncertainty until recoil resummation is understood, a
problem  which needs an urgent solution.

\subsubsection{Nuclear effects in $p$A collisions} 

One explains, in Appendix I, how to relate a $p$A hard cross section to the
corresponding proton-nucleon cross section
[Eqs.~(\ref{eq:binary_scaling_yiel1}),~(\ref{eq:binary_scaling_yiel11})].
The incoming proton undergoes multiple scattering on the nucleons, constituents
of the nucleus. The number of collisions $N_{coll}$ depends on the value of the
impact parameter $b$ or equivalently on the "centrality class", a high
centrality being obtained in collisions at small impact parameter. The Glauber
model used to describe the multiple collisions is based on the eikonal
approximation (independent scattering) and it is assumed that the parton
distributions of the nucleons, confined in the nucleus, are the same as those
of the free nucleon. It is known however that nuclear effects modify the
partons distributions. The nuclear structure functions are measured in Deep
Inelastic Scattering (DIS) of leptons on nuclei \cite{nmc}.   At small values
of $x$, for $x \le 0.07$, the nuclear structure function is found to be  less
than nucleon structure function  scaled by $A$, exhibiting the  so-called 
nuclear  shadowing.  As $x$ grows, the nuclear structure function gets bigger
than the nucleon structure function. This is known as anti-shadowing.  The 
kinematic region of interest at LHC energies is the region of  nuclear
shadowing~\cite{eskolaYR}. 

To calculate the rate of a hard process in proton-nucleus collisions one
therefore uses Eq.~(\ref{eq:inclu-rate}) to describe the proton nucleon
collision, where one of the partonic distribution $F_{a/{\rm A}} (x,M)$, say, is a
nuclear structure function properly normalized. Due to nuclear shadowing we
expect at LHC, a suppression of photon and pion production in proton-nucleus
collisions compared to the nucleon-nucleon case. The modification of the parton
distribution is written in a factorized form as 
\bea
F_{a/{\rm A}}(x,M) = S_{a/{\rm A}}(x,M)\,F_{a/N}(x,M) 
\label{nucl-struc}
\eea
\noindent
where $F_{a/N}(x,M)$ is the parton distribution function in a nucleon and
$S_{a/{\rm A}}(x,M)$ is the parton shadowing function. We assume here $F_{a/{\rm
A}}(x,M)$
to be normalized to one nucleon in the nucleus. Recent parametrizations of the
shadowing function of Eskola, Kolhinen and Salgado (EKS98), which are $M{~}^2$
dependent, distinguish between quarks and gluons
\cite{Eskola:1998df,Eskola:1998iy} and are shown to be in very good agreement with
the NMC data on $M^2$ dependence of $F_2^{\rm Sn}/F_2^{\rm C}$ \cite{arn}. Another
parametrization of nuclear parton distributions has been given in~\cite{hkm}. A
detailed comparison between these various sets is given in Eskola {\em et
al.}~\cite{eskolaYR}. 

In the infinite momentum frame where the nucleus is moving very fast, shadowing
is caused by high parton density effects at small $x$. The small $x$ partons
have a large longitudinal wavelength and can spatially overlap and recombine.
These recombination effects reduce the nuclear parton number densities and
hence the nuclear cross sections. Working in this frame enables one to treat
nuclear shadowing and parton saturation in nucleons on the same footing due to
the identical physical mechanism involved in both. Anti-shadowing is due to
longitudinal momentum conservation (momentum sum rule) in this frame.

Even though there has been considerable amount of theoretical work done on
nuclear shadowing and impressive progress made in understanding the physical
principles of nuclear shadowing \cite{arn}, we are far from having a precise
and quantitative description of nuclear shadowing. The scale dependence of the
nuclear structure functions is even less understood due to the limited range of
$Q{~}^2$ covered in fixed target experiments. Also, shadowing of gluons is not
well understood due to the fact that they cannot be directly measured in DIS
experiments. The working assumption is that high parton density effects are
negligible and DGLAP evolution equations are valid in which case the gluon
distribution function can be obtained from the scaling violation of the $F_2$
structure functions. This assumption, however, will break down at small values
of $x$ due to high parton density effects \cite{mnmob} and one will need to
measure the gluon distribution function differently.

Another nuclear effect that may be considered is that of the Fermi momentum in
the nucleus. In some approaches this contributes an extra nuclear "$\kt$" which
compounds with the intrinsic "$\kt$" in the nucleon, thus leading to an
appreciable increase of the cross sections~\cite{levai}.  As there is no need
of a nucleon "$\kt$" to describe proton-proton collisions, we likewise neglect
the small nuclear "$\kt$" one may introduce for proton-nucleus collisions.

\subsubsection{Nucleus-nucleus collisions} 
\label{sec:nuc-nuc}

The effect of multiple collisions is treated according to the Glauber model [see
Appendix I, in particular
Eqs.~(\ref{eq:binary_scaling_yiel1}), (\ref{eq:binary_scaling_yiel11})].
Besides nuclear shadowing discussed in the previous section, an important
effect in A+A collisions is the medium induced parton energy loss
effect~\cite{wiedemannYR}.  Fast partons produced in parton-parton collision
propagate through the hot and dense medium and through scatterings lose part of
their energy \cite{eloss} and then fragment into hadrons with a reduced energy.
While a dynamical study of the parton propagation in a hot and dense medium
created in a realistic heavy-ion collision and the modification of the
hadronization is most desirable, there is a phenomenological model \cite{hsw}
that takes this effect into account.   Given the inelastic scattering
mean-free-path, $\lambda_a$, the   parton type $a$ scatters $n$ times within a
distance $\Delta L$ before it escapes the system.   The modified fragmentation
function,  $zD_{\gamma/a}(z,\mf)$ is given in terms of the photon fragmentation
function $zD^0_{\gamma/a}(z,\mf)$ by
\cite{hsw} 
\begin{equation}
zD_{\gamma/a}(z,\Delta L,\mf)= \frac{1}{C^a_N}\sum_{n=0}^N P_a(n)
\bigg[z^a_nD^0_{\gamma/a}(z^a_n,\mf) + \sum_{j=1}^{n}
\bar{z}^j_aD^0_{\gamma/g}(\bar{z}^j_a,\mf)\bigg] 
\label{eq:frag} 
\end{equation}
where $z^a_n=z/(1-(\sum_{i=0}^{n}\epsilon^a_i)/E_{_T})$,
$\bar{z}_j^a=zE_{_T}/\epsilon_j^a$ and $ P_a(n)$ is the probability that a parton
of flavor $a$ traveling a distance $\Delta L$ in the nuclear medium will scatter
$n$ times. It is given by 
\begin{equation} 
P_a(n) = \frac{(\Delta L/\lambda_a)^n}{n!} e^{-\Delta L/\lambda_a}, 
\label{eq:probab} 
\end{equation}  and
$C^a_N=\sum_{n=0}^N P_a(n)$.   The first term in Eq. (\ref{eq:frag}) corresponds
to the  fragmentation of the leading parton $a$ with reduced  energy $E_{_T}-
\sum_{i=0}^{n}\epsilon^i_a$ after $n$ gluon emissions and the  second term comes
from the $j$-th emitted gluon having  energy $\epsilon^j_a$, where
$\epsilon^j_a$ is the energy loss of  parton $a$ after $j$ scattering.  One
should keep in mind, however, that Landau-Pomeranchuk-Migdal (LPM) effect in
QCD has been derived for static scatterers \cite{eloss}, which may not be
suitable approximation in case on hot QGP. One can study the effect of parton
energy loss on prompt photon and neutral pion  production at the LHC by
considering the following  cases of parton energy loss~\cite{jos,jji}:  1) constant parton
energy loss per parton scattering, $\epsilon^{~a}_{~n}=const$, 2) 
Landau-Pomeranchuk-Migdal energy-dependent energy loss, $\epsilon^{~a}_{~n}
\sim \sqrt {E^a_n}$ and 3) Bethe-Heitler energy-dependent energy loss, 
$\epsilon^{~a}_{~n} \sim E^a_n$. \footnote{ 
	It is shown that recently observed suppression of $\pi^0$ production in
	Au+Au collisions at RHIC, which is found to increase with $p_{_T}$
	increasing from $3$ GeV to $8$ GeV, is compatible with parton energy
	loss, when $\epsilon^a_n=0.06 E^a_n$~\cite{jji} (see
	sec.~\ref{sec05:pioprod}).}.
Alternative parametrizations of parton energy loss in a hot medium will be
considered in Chap.~\ref{correlations}~\cite{arleo2002}.

The energy loss mechanism affects the production of pions and the bremsstrahlung
production of photons, but not the production of photons directly emitted from
the hard scattering process. One therefore expects a stronger reduction of the
$\pi^0$ rate than the $\gamma$ rate when going from the nucleon-nucleon
collisions to A+A collisions and the ratio $\gamma_{\rm prompt} /\pi^0$ should
be enhanced in A+A collisions compared to $p p$ collisions.

\subsection{Theory: the combined Pomeron and (LO) perturbative QCD approach}
\label{sec:dpmjet}
{\em F.W.~Bopp, J.~Ranft}

Since, as mentioned previously, the relevant kinematical domain of interest is
at rather small $\pt$  we turn now to the discussion of a model which includes 
a "soft" physics component based on Regge theory as well as a "hard" component
based on lowest-order perturbative QCD. This model reproduces soft and
semi-hard data on production of hadrons from fixed target energies until 2 TeV.
The production of "prompt" photons is not implemented, while that of "decay"
photons is possible since the radiative decays of hadrons are included in the
model.

\subsubsection{Proton--proton collisions, the~Monte~Carlo~Event~Generator
PHOJET}

Hadronic collisions at high energies involve the production of particles with
low transverse momenta, the so-called \textit{soft} multiparticle production.
The theoretical tools available at present are not sufficient to understand
this feature from first QCD principles and phenomenological models are typically
applied in addition to pertubative QCD. The Dual Parton Model (DPM)
\cite{dpm-a} is such a model and its fundamental ideas are presently the basis
of many of the Monte Carlo (MC) implementations of soft interactions.

 \textsc{Phojet}-1.12 \cite{phojet-a,phojet-b} is a modern DPM and
 perturbative QCD based event generator
 describing   hadron-hadron interactions and also
hadronic interactions involving photons.
  \textsc{Phojet} replaces the original \textsc{Dtujet} 
  model \cite{dtujet}, which was the first implementation of this
  combination of  perturbative QCD and the DPM. 

The DPM combines predictions  of the large $N_c,N_f$ expansion of QCD
\cite{Veneziano74} and assumptions of duality \cite{Chew78} with Gribov's
reggeon field theory \cite{Gribov67a-e}. \textsc{Phojet}, being used for the
simulation of  elementary hadron-hadron, photon-hadron and  photon-photon
interactions with energies greater than 5 GeV, implements the DPM as a
two-component model using reggeon theory for soft interactions and (LO)
perturbative QCD for hard interactions. Each \textsc{Phojet} collision includes
multiple hard and soft pomeron exchanges, as well as initial and final state
radiation. In \textsc{Phojet} perturbative QCD  interactions are refered
to as hard pomeron exchange.  In addition to the model features as described in
detail in \cite{PhD-RE}, the version 1.12 incorporates a model for high-mass
diffraction dissociation including multiple jet production and recursive
insertions of enhanced pomeron graphs (triple-, loop- and double-pomeron
graphs). 

High-mass diffraction dissociation is simulated as pomeron-hadron or
pomeron-pomeron scattering, including multiple soft and hard
interactions \cite{Bopp98a}. To account for the nature of the pomeron being a
quasi-particle, the CKMT pomeron structure function \cite{Capella96a}
with a hard gluonic
component is used. These considerations refer to pomeron exchange
reactions with small pomeron-momentum transfer, $|t|$. For large
$|t|$ the rapidity gap production (e.g. jet-gap-jet events) is 
implemented on the basis the color evaporation model \cite{Eboli98a}.

For hard collisions \textsc{Phojet} uses the LO parton structure functions
GRV98(LO)\cite{Glueckreyavogt98}.  All color neutral strings in
\textsc{Phojet}  are hadronized  according to the Lund model as implemented in
\textsc{Pythia} \cite{pythia6-a,pythia6-b}. No parton fragmentation functions
are needed separately.

\textsc{Phojet} has been extensively tested against data in hadron--hadron 
collisions~\cite{PhD-RE}. In a  number of papers the four experimental  LEP
Collaborations compare many features of hadron production in $\gamma$
--$\gamma$ collisions to \textsc{Phojet}, a rather good agreement is usually
found. \textsc{Phojet} has been checked against practically all data on
transverse momentum distributions in $p p$ and  $\bar p p$ collisions  from
colliders~\cite{boppup}. In Fig.~\ref{fig:pp-pt} we plot this comparison.
Please note that the points in this Figure are from the \textsc{Phojet} Monte
Carlo while the data are represented by lines, fits to the data points. 
\begin{figure}[h]
\vskip-3.cm
\begin{center}
\includegraphics[height=20.0cm]{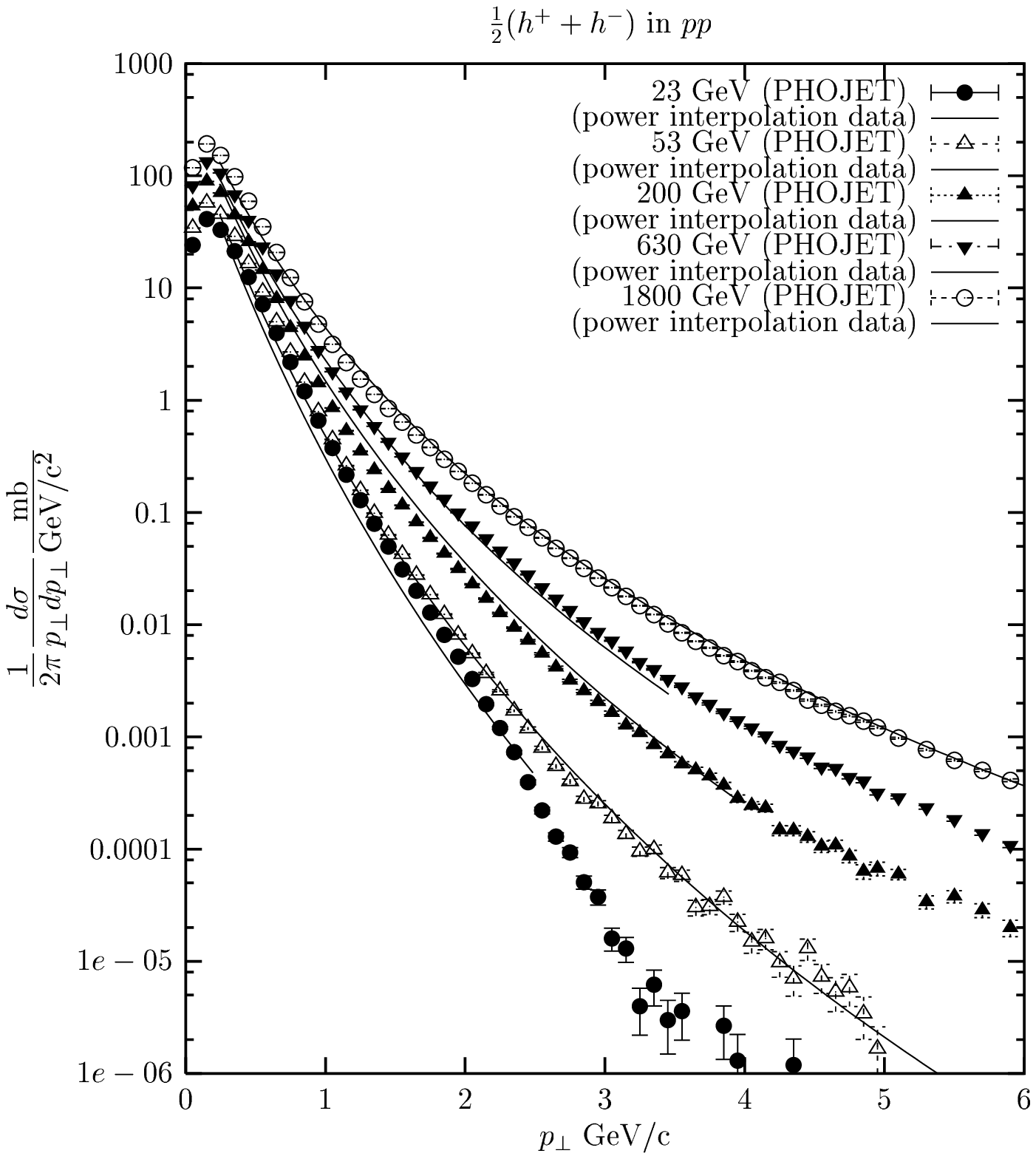}
\end{center}
\vskip-5cm
\caption{ Transverse momentum distributions of charged hadrons.  The
results of \textsc{Phojet} (points) are compared to experimental data 
represented by lines, fitted to the data points. The data at $\sqrt s$ = 23 
and 53 GeV are from the CERN--ISR~\cite{Alper75}, the data at 200 GeV are from
the  UA1 Collaboration~\cite{Arnison82} and the data at 630 and 1800 GeV are
from the CDF  Collaboration~\cite{Abe88}.
}
\label{fig:pp-pt}
\end{figure} 

\subsubsection{ 
Collisions involving nuclei, the~Monte~Carlo~Event~Generator ~\textsc{Dpmjet}-III}
The \textsc{Dpmjet}-III code system \cite{dpmjet1,dpmjet2}, is a MC event 
generator implementing Gribov--Glauber theory for collisions involving nuclei.
For all elementary nucleon-nucleon collisions it uses the  DPM  as implemented
in  \textsc{Phojet}.  \textsc{Dpmjet}-III is unique in its wide range of
applications simulating hadron-hadron, hadron-nucleus, nucleus-nucleus,
photon-hadron, photon-photon and  photon-nucleus interactions from a few GeV up
to  cosmic ray energies. 

Since its first implementations \cite{dpmjet2-a,dpmjet2-b,dpmjet2-c}
\textsc{Dpmjet} uses the Monte Carlo realization of the Gribov-Glauber multiple
scattering formalism according to  the algorithms of \cite{diagen} and allows
the  calculation of total, elastic, quasi-elastic and production cross sections
for any high-energy nuclear collision.  
This formulation of the Glauber model is somewhat more detailed than the model
described in Appendix~I.  In the model~\cite{diagen} the scattering amplitude 
is parametrized not only by the inelastic nucleon--nucleon cross--section, but
it is parametrized by using ${\sigma}_{tot}$, $\rho$ = ${\bf Re}
f(0)_{hN}$/${\bf Im} f(0)_{hN}$ and  the elastic slope $a$. ${\sigma}_{tot}$
and $a$ are taken as fitted by  {\sc Phojet} while for $\rho$ a
parametrization of experimental data is used. 
However, parameters needed for the collision scaling of the  NLO $\pi^0$ and
$\gamma$ cross sections ($\sigma_{inel}$, $N_{coll}$) are in  very close
agreement in \textsc{Dpmjet} with the ones determined in Appendix~I. To be
consistent we use in direct comparisons between \textsc{Dpmjet} and NLO results
always $\sigma_{inel}$ and $N_{coll}$ as determined by \textsc{Dpmjet}. No
collision scaling is used by \textsc{Dpmjet}, but of course it is easy to
calculate  $N_{coll}$ in \textsc{Dpmjet}. Realistic nuclear densities and radii
are used in  \textsc{Dpmjet}  for light nuclei and Woods-Saxon densities
otherwise.

During the simulation of an inelastic collision the above formalism samples the
num\-ber of ``woun\-ded'' nu\-cleons, the impact parameter of the collision and
the interaction configurations of the wounded nucleons. Individual 
hadron--nucleon interactions are then described by  \textsc{Phojet} including
multiple hard and soft pomeron exchanges, initial and final state radiation as
well as diffraction. 

As a new feature, \textsc{Dpmjet}-III allows the simulation of enhanced graph
cuts in non-diffractive inelastic hadron-nucleus and nucleus-nucleus
interactions. For example, in an event with two wounded nucleons, the first
nucleon might take part in a non-diffractive interaction whereas the second one
scatters diffractively producing only very few secondaries. Such graphs are
prediced by the Gribov-Glauber theory of nuclear scattering but are usually
neglected. Further features of  \textsc{Dpmjet}-III are a formation zone
intranuclear cascade \cite{fzic} and the implementation of certain baryon
stopping diagrams \cite{barystop-b}.

The \textsc{Dpmjet}-III code and further information are available from the
authors.
\textsc{Dpmjet}-III and earlier versions like \textsc{Dpmjet}-II have been 
extensively tested against data in hadron--nucleus and nucleus--nucleus
collisions~\cite{dpmjet1,dpmjet2,dpmjet2-a,dpmjet2-c}. The code is used for the
simulation of cosmic ray showers \cite{CORSIKA}.

The transverse momentum distributions according to \textsc{Dpmjet} are largely
determined by the properties of \textsc{Phojet}, which is called  for each
elementary interaction in \textsc{Dpmjet}. Note, all of these elementary
interactions are treated kinematically correctly.  All \textsc{Dpmjet} events
conserve  energy, momentum as well as additive quantum numbers like charge,
baryon number and strangeness strictly. Because of an appropriate treatment of
soft physics the transverse momentum distributions start at  $\pt$ = 0.

Applying \textsc{Dpmjet} to heavy ion collisions
at RHIC the first experience was, 
that the original hadron multiplicities and
pseudorapidity distributions were about one third too high compared to
the data. A new mechanism was needed to reduce $N_{\rm ch}$ and
$dN_{\rm ch}/d\eta |_{\eta = 0}$ in situations with a produced very dense
hadronic system. Such a mechanism, the fusion and percolation of soft
chains,  had indeed been conjectured since the
beginning of the Nineties. Introducing percolation and fusion of soft
chains into \textsc{Dpmjet}\cite{RER02}, it was indeed possible to get a
satisfactory agreement of  \textsc{Dpmjet} with the  $N_{\rm ch}$ and
$dN_{\rm ch}/d\eta$ measured at RHIC. The procedure desribed in
\cite{RER02} is quite time consuming and therefore not well suited for
high statistics \textsc{Dpmjet} runs as needed here. Therefore, we apply
here only an effective method, reducing  $N_{\rm ch}$ and
 $dN_{\rm ch}/d\eta$ in \textsc{Dpmjet} like in \cite{RER02} without
 spoiling energy--momentum conservation. This method
 would not distort the transverse momentum distributions, but it will
 reduce the cross sections at all $\pt$ by approximately the same
 30 \%  as  mentioned above.
 
 Calculating from \textsc{Dpmjet} transverse momentum 
distributions of decay photons (\textsc{Dpmjet} or \textsc{Phojet} 
do not calculate direct photon production) we use the possibility in the
code to declare certain resonances as stable. 
In this way we get separately 
(i) all
decay photons (having $\pi^0$ and $\eta$ decaying), 
(ii) all decay photons not coming fron $\pi^0$ decay (declaring $\pi^0$
as stable) and (iii) all
decay photons not coming from  $\pi^0$ or  $\eta$ decay (declaring
$\pi^0$ and $\eta$ as stable).  

\subsection{The $z-$scaling model}
{\em M.~Tokarev}

During the workshop, a phenomenological model to describe hadron and photon
production was also discussed: the $z-$scaling model.
This concept was suggested to analyze numerous experimental
data on high-$\pt$ hadron production in $pp, \bar pp$ and $p$A collisions
\cite{tok1}. The method of data analysis was developed for description of
$pp$, $\bar pp$ and $p$A interactions with  direct photon and jet production
in the high-$\pt$ region at the RHIC and LHC energies as well \cite{tok2}.
The scaling function is expressed via the invariant differential
cross section $Ed^{~3}\sigma/dp^3$ as follows
\begin{equation}
 \psi(z) = - \frac{\pi s}{\rho \sigma_{in}}J^{-1}
 E\frac{d^3\sigma}{dp^{3}}.
 \label{eq:r4}
 \end{equation}
Here $s$ is the collision center-of-mass energy squared,
$\sigma_{in}$ is the inelastic cross section, $J$ is the
corresponding Jacobian, and $\rho=dN/d\eta$ is the particle
multiplicity density.
The function $\psi(z)$ is normalized:
\begin{equation}
\int_{z_{min}}^{\infty} \psi(z) dz = 1.
\label{eq:b6}
\end{equation}
The relation allows us to give the physical meaning
of the scaling function $\psi(z)$ as a probability density to form
a particle  with the formation length.
The variable $z$  can be written in the form
\begin{equation}
z = z_0 \Omega^{-1}, \hspace*{0.5cm} {\rm{where}}  \hspace*{1cm}
\Omega(x_1,x_2)=(1-x_1)^{\delta_1}(1-x_2)^{\delta_2} .
\label{eq:r6}
\end{equation}
The  factor $z_0$ is
proportional to the transverse energy released in the
underlying collision of constituents. The expression
$\Omega^{-1}$ describes the resolution at which the collision of
the constituents can be singled out of this process.
The $\Omega(x_1,x_2)$ represents the relative number of all initial
configurations containing the constituents which carry fractions
$x_1$ and $x_2$ of the incoming momenta.
The $\delta_1$ and $\delta_2$ are the anomalous fractal
dimensions of the colliding objects (hadrons or nuclei).
The momentum fractions $x_1$ and $x_2$ are determined in a way to
minimize the resolution $\Omega^{-1}(x_1,x_2)$ of the fractal
measure $z$ with respect to all possible sub-processes
which satisfies 4-momentum conservation law.

Figure~\ref{fig:tokarev} shows  the data $\pt$  and $z$-presentation for $\pi^0$ meson
production in $pp$ collisions. The {f}irst one demonstrates the strong energy
dependence of the cross section as a function of the transverse momentum. The
$z$-presentation  reveals the energy  independence of $\psi(z)$ and, for large
enough $z$,  the power behavior, $\psi(z) \sim z^{-\beta}$, of the scaling
function. Based on the obtained results \cite{tok1,tok2} we can conclude that
verification of the asymptotic behavior of $\psi(z)$ for $\pi^0$ meson, direct
photon and jet production at LHC energies is of interest.
\begin{figure}[t]
\hspace*{-10cm}
\vskip -0.cm
\includegraphics[width=6.5cm]{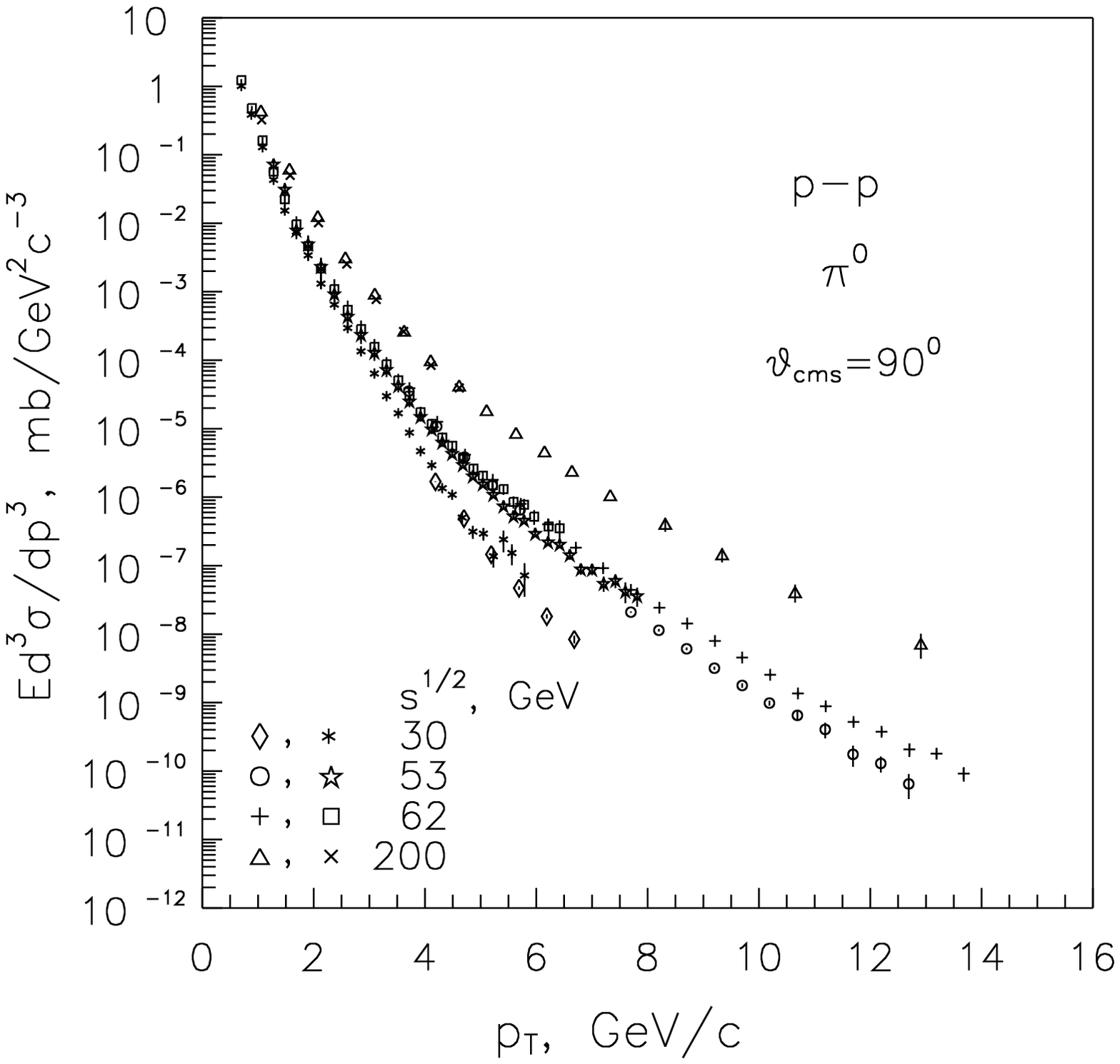}
\vskip -6.cm
\hspace*{8cm}
\includegraphics[width=6.5cm]{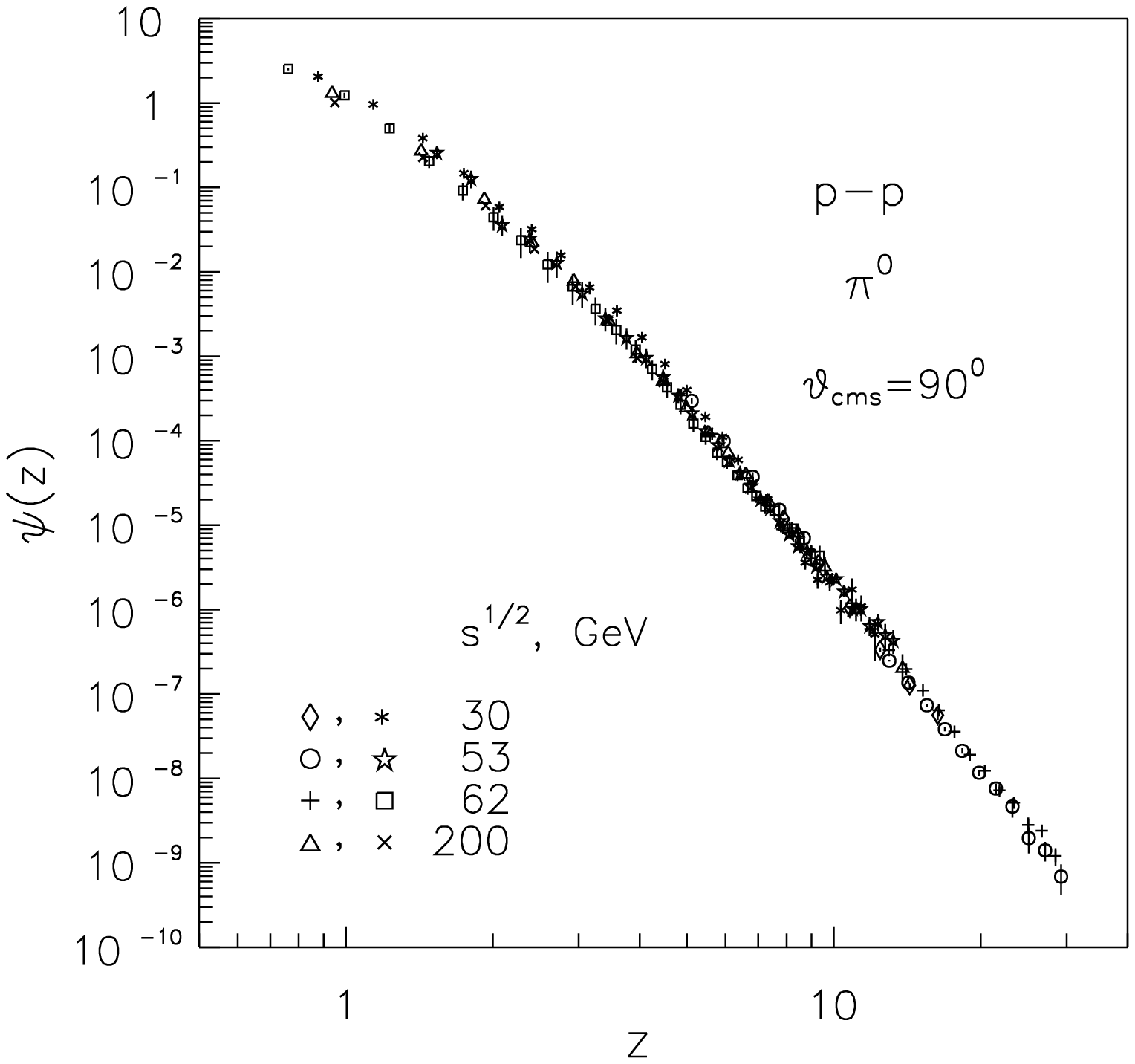}
\vskip 0.5cm
\hspace*{4cm} a) \hspace*{7cm} b)
\caption{
 (a) Dependence of the inclusive cross section of $\pi^0$ meson
production on transverse momentum in $pp$
collisions at $\sqrt s = 30-200~GeV$. Experimental data are
obtained at ISR and RHIC. (b) The corresponding scaling function.}
\label{fig:tokarev}
\end{figure}
Since no predictions are available for A+A collisions we do not pursue the
$z-$scaling approach further in this report. 

\vspace{1.cm}

{\em 
In the next sections of this chapter we compare the model predictions to the
$\pi^0$ transverse momentum distributions obtained at RHIC at $\sqrt s$ = 200
GeV, and then  make predictions for $\pi^0$ and $\gamma$ spectra at LHC
energies. It is found that both \textsc{Dpmjet} and NLO QCD agree with the data
and, furthermore, they give very similar results for the extrapolation to LHC
energies. A word of warning is necessary at this point concerning the
comparison between the two models in nucleus-nucleus collisions . We note that
the \textsc{Dpmjet} predictions represent the full picture of a heavy ion
collision while the NLO QCD predictions give a partial picture to which the
contribution of the quark-gluon plasma  should be added (see
Chap.~\ref{thermal}). In Chap.~\ref{thermal} it is
found that thermal contributions, in A+A collisions, are important in the
low $\pt$ range only. Comparing the predictions of the two models for A+A
collisions at LHC is therefore relevant only for $\pt >
10-15$~GeV/c. A detailed discussion of our results in A+A collisions
below this value of $\pt$ is given in Chap.~\ref{comparing}.
} \\

%
%

\subsection{Phenomenology of $\pi^0$ production} 
\label{sec05:pioprod}

Our standard NLO predictions are obtained using the CTEQ5M~\cite{cteq5} or,
equivalently, CTEQ6M ~\cite{cteq6} parton distributions and the KKP~\cite{kkp}
fragmentation functions.  In Fig.~\ref{fig:pp200pio}, the experimental $\pi^0$
transverse momentum spectrum in $p p$ collisions at 200~GeV~\cite{ppg024} is
compared to theoretical predictions. One notes the remarkable agreement of the
data with the NLO QCD estimates as well as with the \textsc{Dpmjet}
predictions. The \textsc{Dpmjet} model parameters had been tuned to fit charged
particle spectra (see Fig.~\ref{fig:pp-pt}) and no new adjustement was
necessary to obtain the $\pi^0$ spectrum. 

The spread in the NLO QCD estimates (grey band in the figure) reflects the
uncertainties associated to the choice of scales (all scales equal varying
from  .5 $p_{_T}$ to 2 $p_{_T}$): if the uncertainty is roughly [$+ 100\%,\ -
60\%$] at $p_{_T}=3$ GeV/c it  reduces to $\pm$ 40\% at $p_{_T}=12$ GeV/c. The
upper limit of the NLO predictions is obtained with the smallest scales and the
lower limit with the largest scales. Varying the scales independently (for
example $\mu = M \ne M_{_F}$) one still obtains results in the grey band of the
figure with only slight changes in the slope in $p_{_T}$ (for $p_{_T}$ below 6
GeV/c). Using the BKK~\cite{bkk} fragmentation functions rather than the KKP
lowers the theoretical predictions by about 25\% while the
Kretzer~\cite{kretzer} functions reduce the predictions by about 70\% (lowest
$p_{_T}$) to 45\% (highest $p_{_T}$).  Concerning CTEQ6M, the 40 different sets
predict cross sections within $[-5\%, +3\%]$ for $\pt < 10$~GeV/c and within
$[-6\%, + 10\%]$ at $\pt = 14$~GeV/c.
Turning now to MRS99~\cite{mrs99} rather than CTEQ hardly makes any changes:
decrease of the cross section by about 10\% at low $\pt$ and stability for $\pt
> 6$ GeV/c. \\
From this discussion one concludes that the largest uncertainties in the
theoretical predictions are related to the choice of scales and to the
fragmentation functions, a remark which also holds true at LHC energies.
Concerning the fragmentation functions, it should be recalled that KKP give the
best phenomenology for $\pi^0$ production at lower 
energies~\cite{bgpw2002,e706-2003}.

\begin{figure}[htbp]
\begin{center}
{\includegraphics[width=14cm]{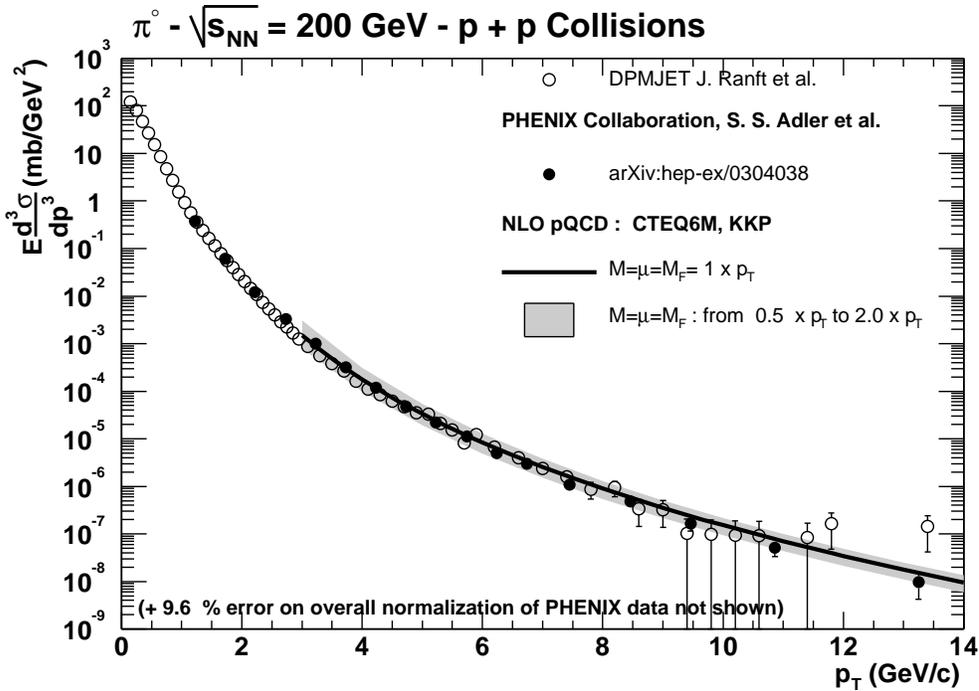}}
\end{center}
\vskip-0.5cm
\caption{ 
PHENIX data on $\pi^0$ production in $p p$ collisions at 200 GeV, compared to
various theoretical predictions. The grey band indicates the range the NLO QCD
predictions for scales ranging from 0.5 $\pt$ to 2 $\pt$ and the open points are
the \textsc{Dpmjet} predictions. CTEQ6~\cite{cteq6} structure and
KKP~\cite{kkp} fragmentation functions are used. The theoretical predictions
are evaluated by averaging the invariant cross sections in an interval $\pm
0.35$ unit of rapidity around rapidity 0. 
}
\label{fig:pp200pio}
\end{figure} 
\begin{figure}[htbp]
\begin{center}
{\includegraphics[width=15cm]{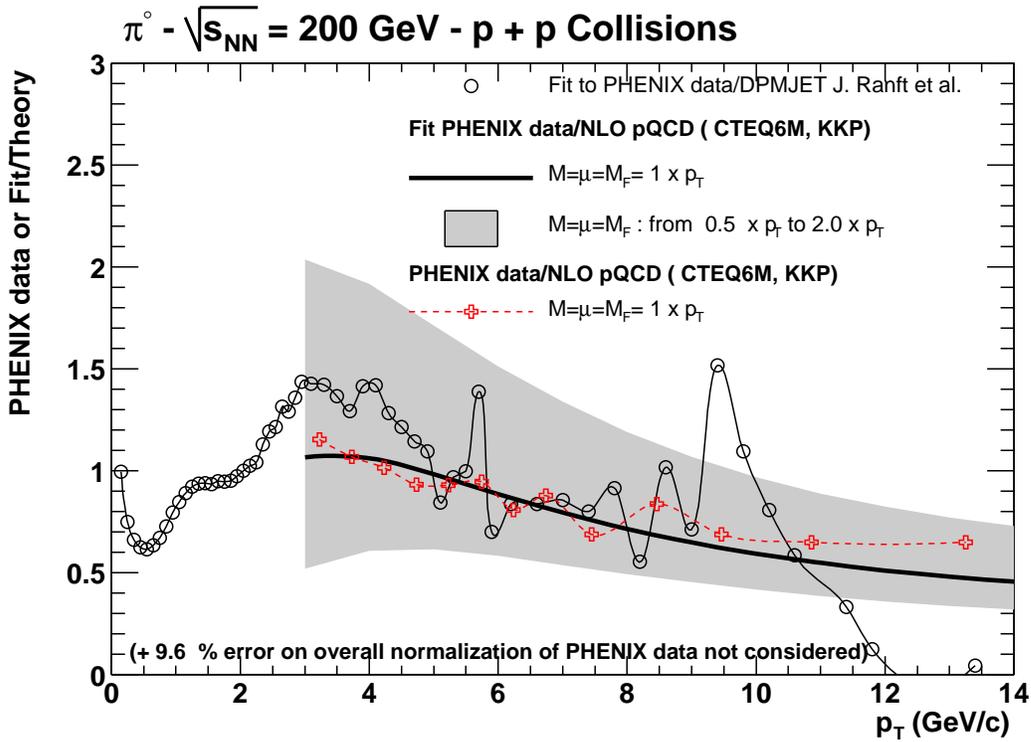}}
\end{center}
\vskip-0.5cm
\caption{ PHENIX data on $\pi^0$ production in $p p$ collisions at 200
GeV normalized to various theoretical predictions shown in
Fig.~\ref{fig:pp200pio}. The experimental error bars on the data points
(crosses) are not shown.}
\label{fig:rat-pp200pio}
\end{figure} 

We also show on a linear plot, in Fig.~\ref{fig:rat-pp200pio}, the data
normalized to the theoretical predictions. The solid line shows the ratio of a
fit of the PHENIX data to the standard NLO predictions while the dashed line,
with open crosses, shows the data normalised to the NLO estimates: the
agreement between theory and experiment is remarkable given the fact that no 
parameters has been adjusted to fit the data. In view of the high quality, and
large $p_{_T}$ coverage, of the RHIC present and forthcoming data it would be
interesting to perform a joint fit of $e^+ e^-$ data and RHIC data to better
constrain the fragmentation functions specially in the high $z$ region where
$z$ is the fragmentation variable. Another advantage would be the possibility
to obtain a better control on the gluon fragmentation function which appears as
a next-to-leading effect in $e^+ e^-$ while it is a leading term in $p p$
collisions: this would help un-correlate the gluon fragmentation parameters
from the other parameters in the fit. The agreement between data and \textsc{Dpmjet} is
also excellent even at the lowest $\pt$ values. The large fluctuations at
high $p_{_T}$ are related to the low statistics of the model.

We turn now to LHC energies. The theory predictions are shown, for $p p$
collisions, at  5.5~TeV (Fig.~\ref {fig:etrap55}) and 14~TeV (Fig.~\ref
{fig:etrap14}).  In each case we display two plots, the first one with log-log
scales to emphasis the low $\pt$ region and the other one with semi-log scales
up to $\pt = 100$~GeV/c.
\begin{figure}[htbp]
\begin{center}
\vskip-1.cm
\includegraphics[height=10.cm]{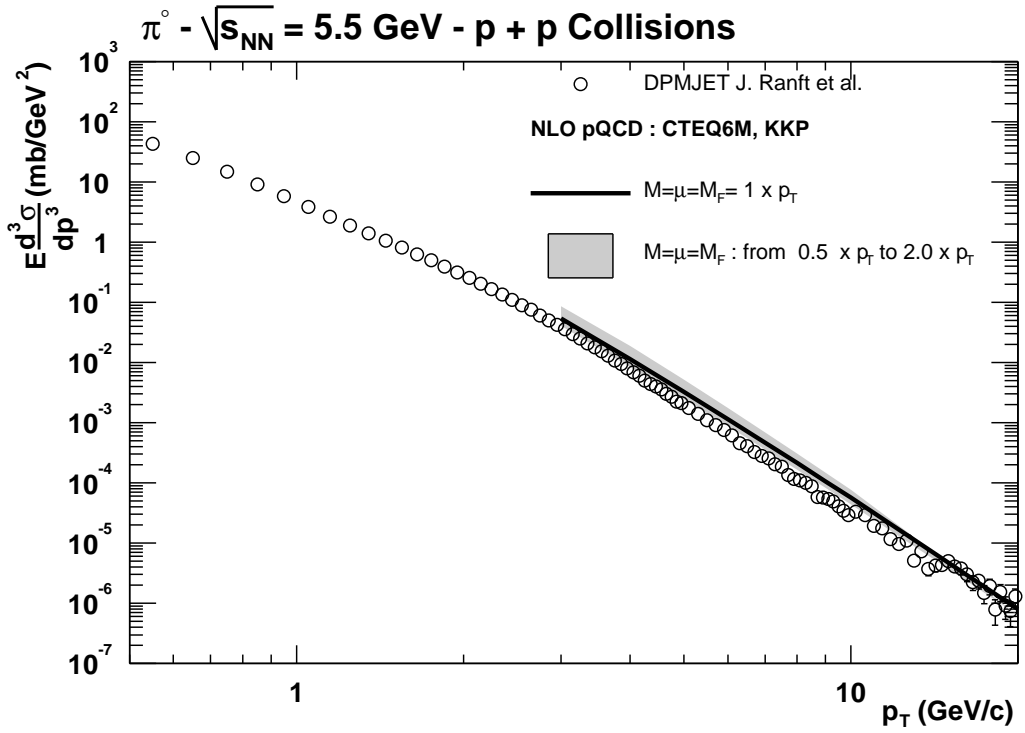} 
\includegraphics[height=10.0cm]{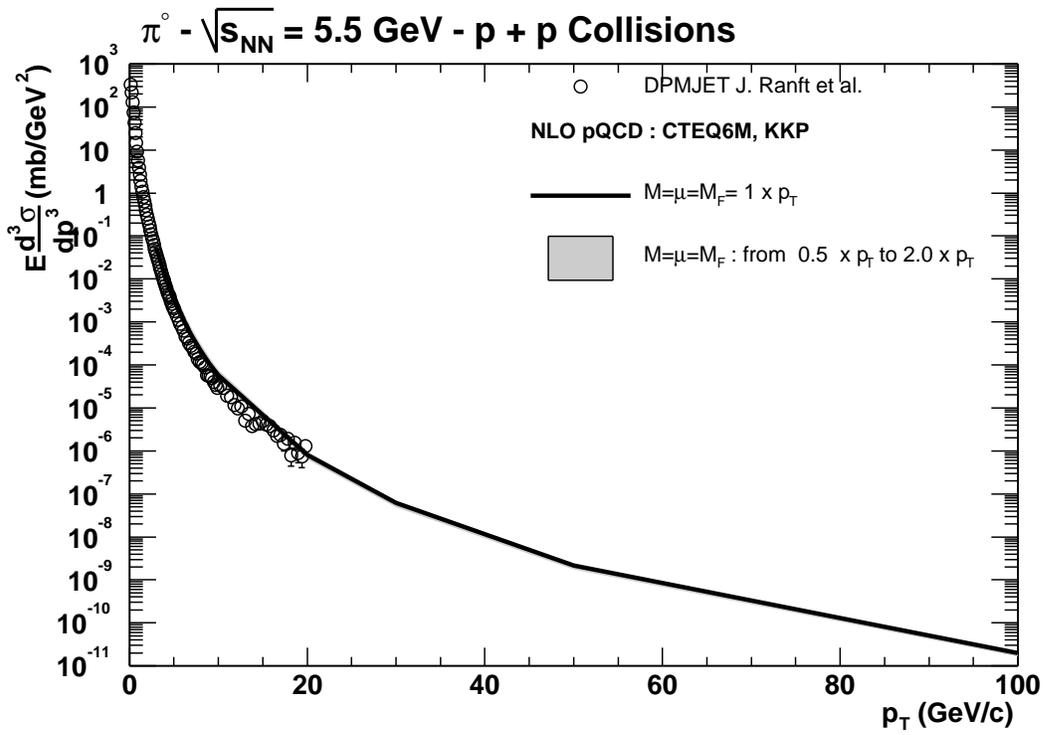} 
\end{center}
\vskip-0.5cm 
\caption{ Theoretical predictions for the $\pi^0$ transverse momentum
spectrum at 5.5 TeV. The conventions used are as in Fig.~\ref{fig:pp200pio}.
The invariant cross section is calculated at 0 rapidity.
} 
\label{fig:etrap55} 
\end{figure}  
\begin{figure}[htbp]
\begin{center}
\vskip-1.cm
\includegraphics[height=10.cm]{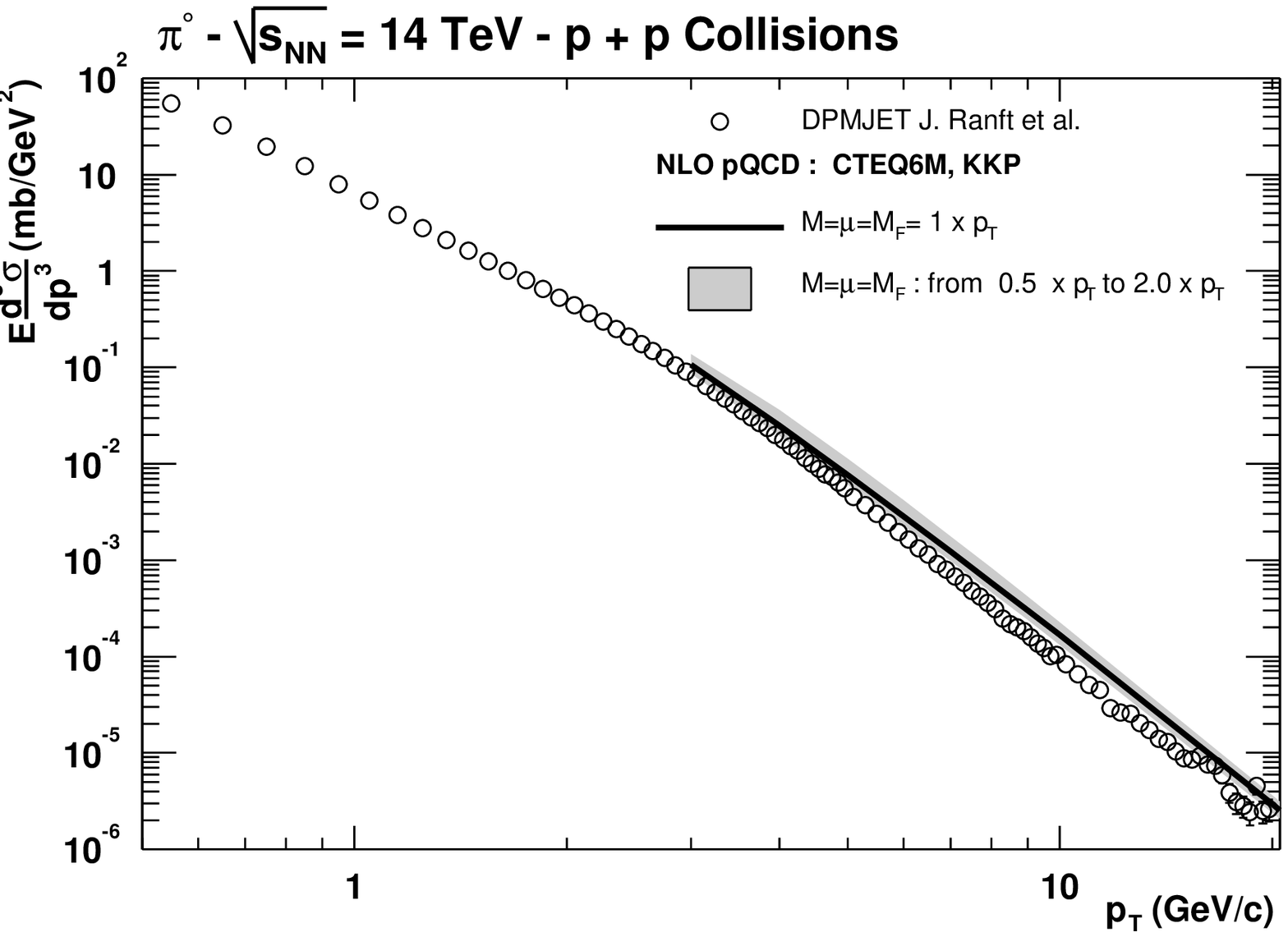}
\includegraphics[height=10.0cm]{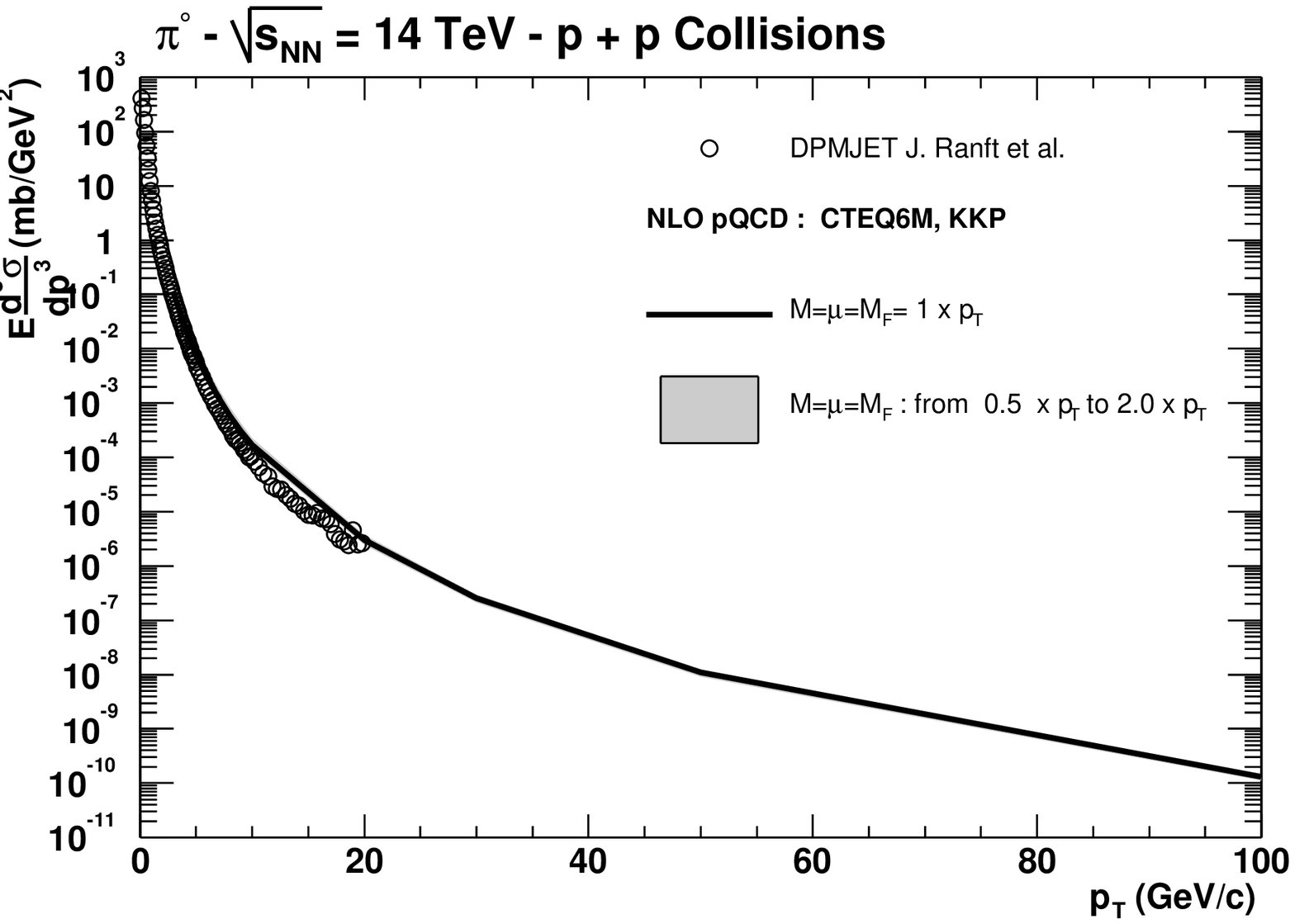}
\end{center}
\vskip-0.5cm
\caption{  
Same as Fig.~\ref{fig:etrap55} but at $\sqrt s = 14$ TeV.}
\label{fig:etrap14}
\end{figure} 
One notes again the good overall agreement between \textsc{Dpmjet} and NLO QCD
results in the overlapping region. Furthermore the use of NLO calculations
appears reasonable down to $\pt = 3$~GeV/c below which one does not dare make
perturbative QCD predictions!  In fact, at $\pt = 3$~GeV/c, one notes a change
of slope in the \textsc{Dpmjet} predictions, illustrating the effects of soft
physics (pomeron exchanges) in the model: the slope of the distribution
decreases leading to a finite prediction at $\pt = 0$.  

The scale variations in the NLO QCD predictions are still rather large at low
transverse momentum (for $\pt = 3$~GeV/c it is [$+60\%,\ -25\%$] at $\sqrt s =
5.5$~TeV and [$+25\%,\ -20\%$] at $\sqrt s = 14$~TeV) but reduces to roughly
$\pm 10\%$ at $\pt = 100$~GeV/c for both energies. These NLO results  tend to
be slightly less steep than the \textsc{Dpmjet} prediction and larger even when
choosing the largest scales. This is due to using different input parameters
for the perturbative sector in the two calculations. Keeping with the CTEQ
family of 40 parametrizations the NLO predictions agree within $\pm 10\%$ at
$\pt=3$~GeV/c and within $\pm 2.5\%$ at $\pt=100$~GeV/c. 
As mentioned above using MRS99 and BKK fragmention functions would lower the
NLO predictions which would then overlap the \textsc{Dpmjet} ones. A power
behavior of the predictions holds for $\pt$ between 3~GeV/c and 20~GeV/c: one
finds for \textsc{Dpmjet} a behavior of type 
\begin{eqnarray}
{d \sigma^{\pi^0} \over d{\imb p_{_T}} dy}
&\sim& {1 \over \pt^{5.91}},\ {\mbox{\rm at 5.5~TeV}}  \\
&\sim& {1 \over \pt^{5.63}},\ {\mbox{\rm at 14~TeV}},
\label{eq:secfit-dpm}
\end{eqnarray}
while for the standard (CTEQ6M, KPP, all scales equal to $\pt$) NLO predictions
the behavior is slightly harder:
\begin{eqnarray}
{d \sigma^{\pi^0} \over d{\imb p_{_T}} dy}
&\sim& {1 \over \pt^{5.45}},\ {\mbox{\rm at 5.5~TeV}},  \\
&\sim& {1 \over \pt^{5.09}},\ {\mbox{\rm at 14~TeV}}\ .
\label{eq:secfit-nlo}
\end{eqnarray}
\begin{figure}[htbp]
\begin{center}
\includegraphics[height=10.0cm]{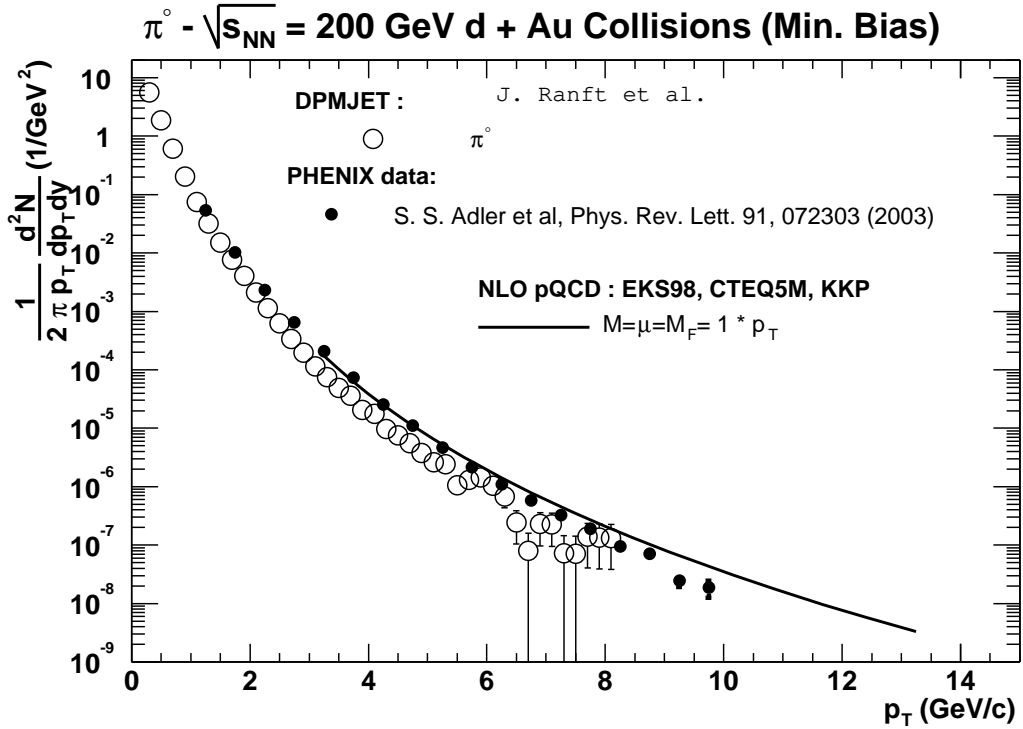}
\end{center}
\vskip-0.5cm
\caption{  
The pion transverse momentum spectrum for deuterium-gold collisions at
$\sqrt s = 200$ GeV. The solid line represents the standard NLO predictions using
the EKS shadowing parametrization. The theoretical predictions
are evaluated by averaging the invariant cross sections in an interval $\pm
0.35$ unit of rapidity around rapidity 0.
}
\label{fig:dau200etrap}
\end{figure} 
%
\begin{figure}[t]
\begin{center}
\includegraphics[height=10.0cm]{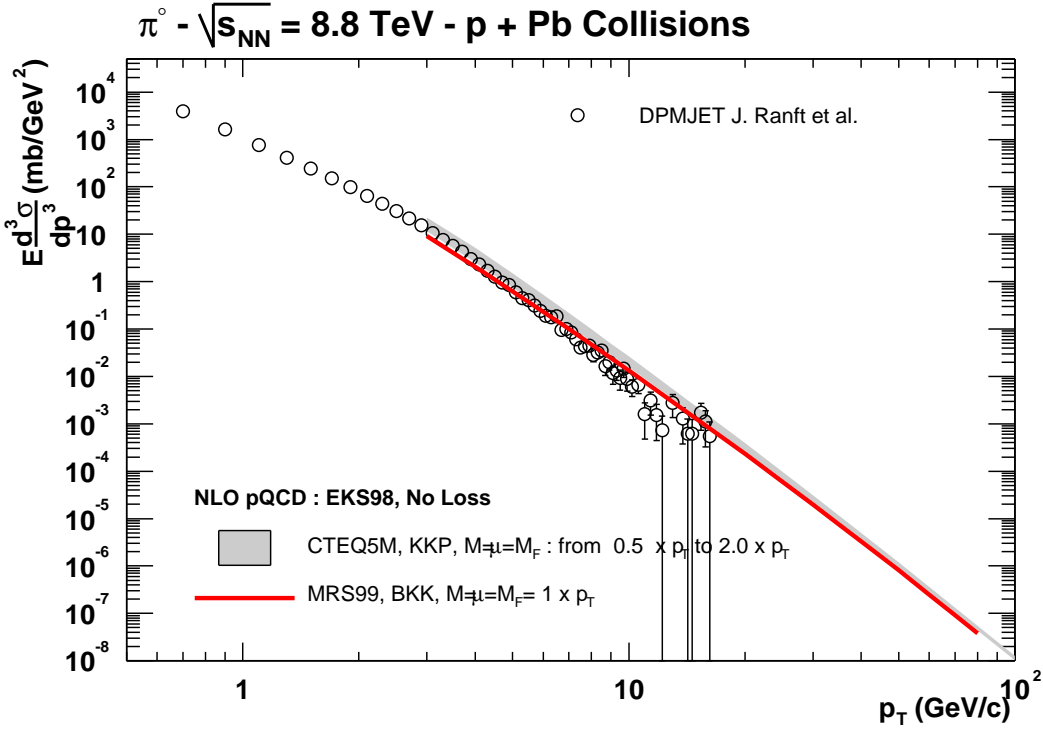}
\end{center}
\vskip-0.5cm
\caption{  
The pion transverse momentum spectrum at $\eta=0$ for proton-lead collisions at
$\sqrt s = 8.8$ TeV. The grey band represents the standard NLO predictions as in
previous figures. The solid line is the NLO QCD result using MRS99~\cite{mrs99}
and BKK~\cite{bkk} parametrizations with all scales equal to $\pt$. All NLO
predictions include EKS~\cite{Eskola:1998df,Eskola:1998iy} shadowing.
}
\label{fig:ppb88etrap}
\end{figure} 
\begin{figure}[h]
\begin{center}
\vskip -1.cm
\includegraphics[height=10.0cm]{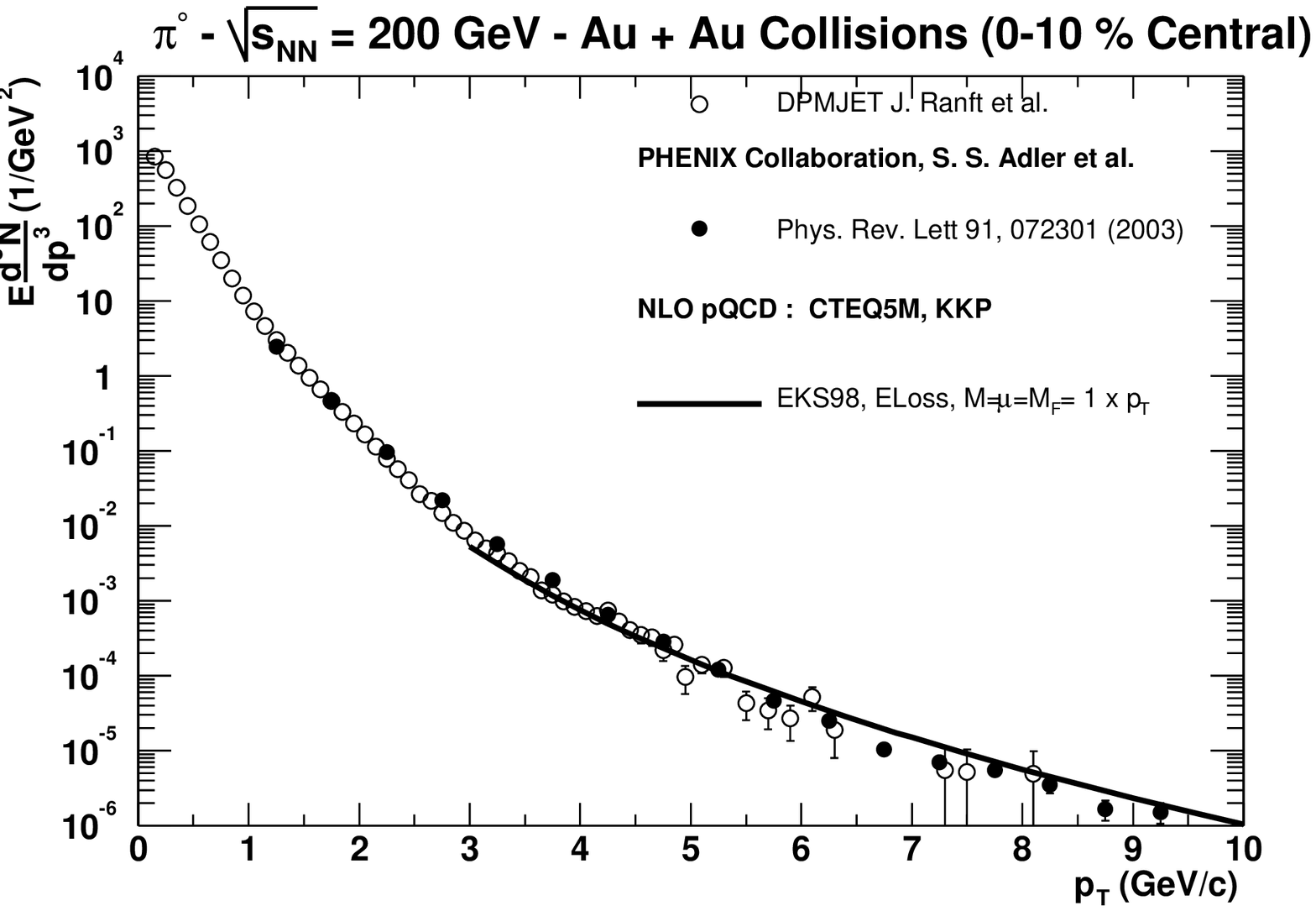}
\end{center}
\vskip-.5cm
\caption{
Comparison between PHENIX data on gold-gold collisions at $\sqrt s =$ 200~GeV
and theoretical predictions. The theoretical predictions are evaluated by
averaging the invariant cross sections in an interval $\pm 0.35$ unit of
rapidity around rapidity 0. 
}
\label{fig:auau200etrap}
\end{figure} 

We consider now deuterium-gold scattering at 200~GeV and proton-lead scattering
at $\sqrt s = 8.8$~TeV and we display, in figs.~\ref{fig:dau200etrap} and
\ref{fig:ppb88etrap}, the \textsc{Dpmjet} results together
with the NLO QCD predictions, using the shadowing model
of~\cite{Eskola:1998df,Eskola:1998iy}. At RHIC the agreement between data and
the standard NLO predictions is good although the latter may not be 
steep enough and tend to fall above the data for $\pt > 7$~Gev/c. The
magnitude of uncertainties of the NLO calculations is the same as for $p p$
collisions. The \textsc{Dpmjet} predictions have the right order of magnitude at
low $\pt$ but they tend to decrease somewhat too fast as $\pt$ increases. 
Work is in progress to understand this point~\cite{ranft-dAu}.

For LHC in $p$Pb collisions the two models are compatible within the error bars
but the slope of the cross section is slightly steeper for \textsc{Dpmjet}. As
before the grey band shows the range of predictions for our standard inputs
while the solid line is obtained using MRS99 and BKK functions with all scales
equal to $\pt$. The latter result is equivalent to the standard one with all
scales equal to $2 \pt$.

Turning now to A+A collisions, the comparison of the $\pi^0$ spectrum in
gold-gold collisions at  $\sqrt s = 200$ GeV together with the theoretical
estimates is shown in Fig.~\ref{fig:auau200etrap}. The agreement between theory
and experiment is surprisingly good over the whole $\pt$ range. In the DPM
approach no jet quenching and no shadowing are introduced but a string fusion
mechanism is at work which is calibrated to reproduce the particle
multiplicity. It also reproduces the minimum bias rapidity distributions. In
the NLO QCD approach the theory is consistent with the data at large $\pt$
provided a jet energy loss mechanism is introduced. In the figure we display
the result with  model 3) of sec.~\ref{sec:nuc-nuc}, {\em i.e.} Bethe-Heitler
energy loss with $\epsilon^{~a}_{~n} = 0.05 E^a_n$ for gluons and
$\epsilon^{~a}_{~n} = 0.025 E^a_n$ for quarks and with
$\lambda_a=0.5$ for gluons and 1 for quarks~\cite{jos,jji} (see
Eq.~(\ref{eq:probab}) and the discussion following it). Below $\pt = 5$~GeV/c
the NLO QCD predictions fall below the data: this is the region where thermal
production of $\pi^{~0}$ may become relevant as will be discussed in
Chapters~\ref{thermal} and \ref{comparing}.
\begin{figure}[htbp]
\begin{center}
\includegraphics[height=10.0cm]{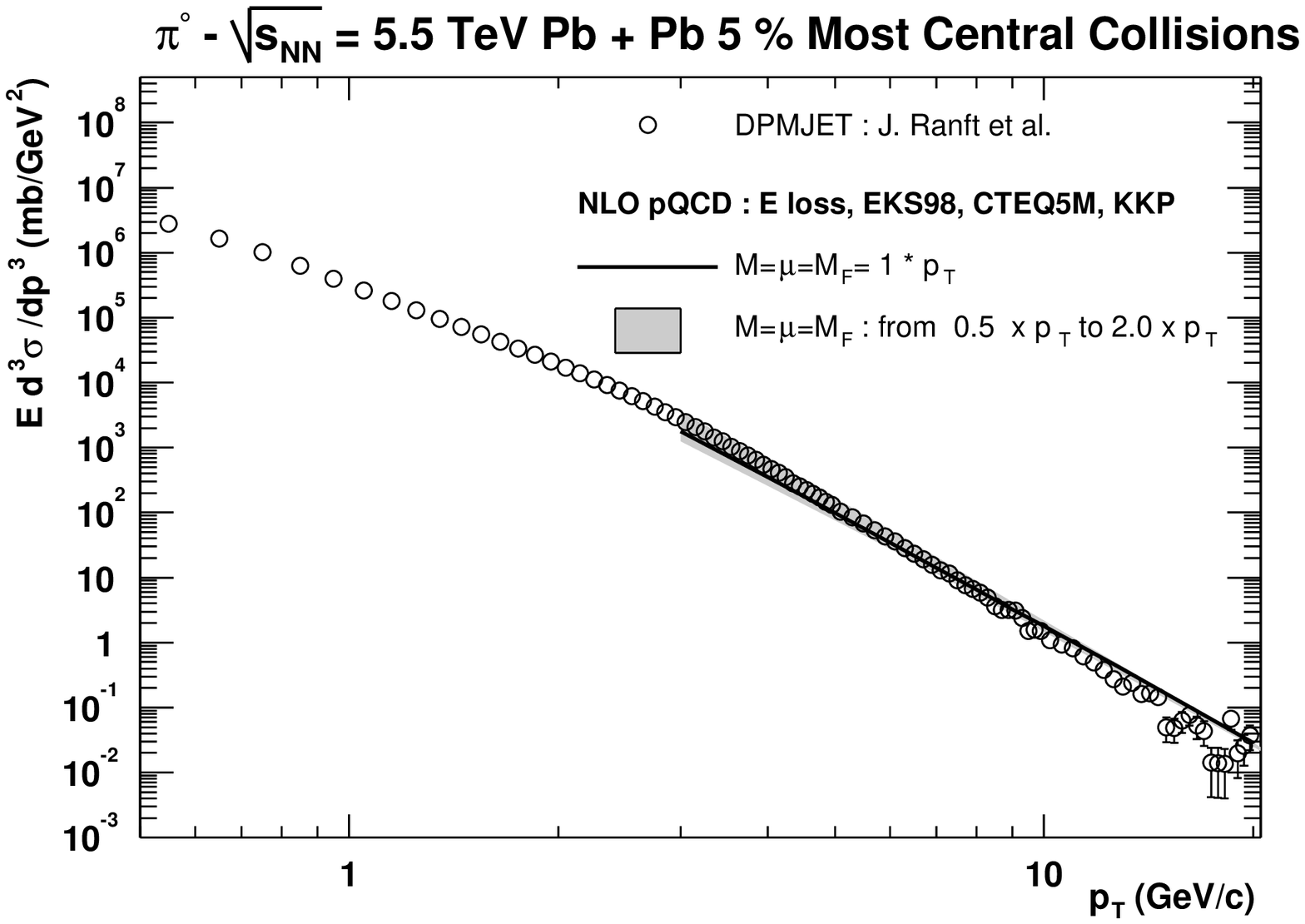}
\includegraphics[height=10.0cm]{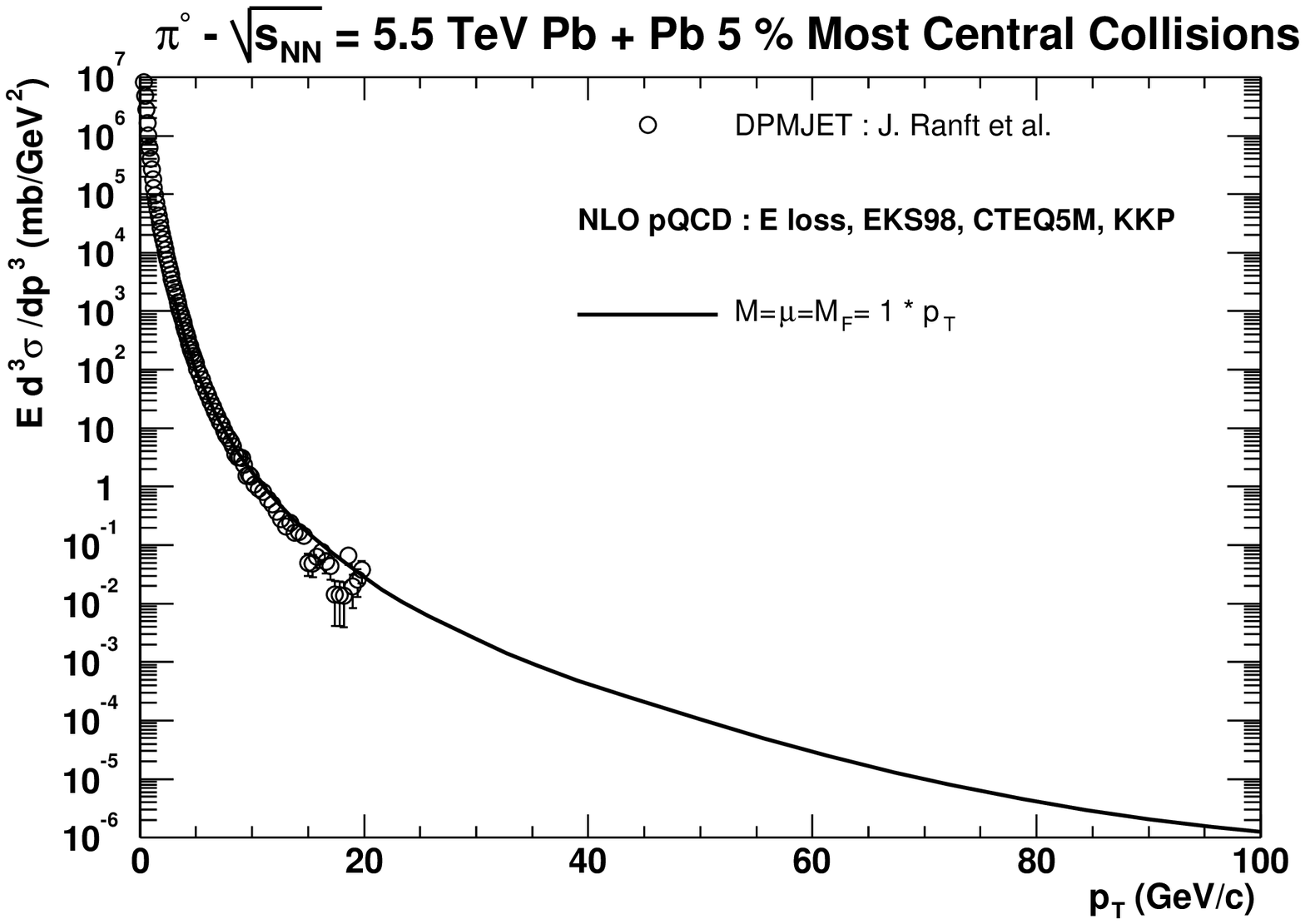}
\end{center}
\vskip-.5cm
\caption{  
Same as Fig.~\ref{fig:etrap55} for lead-lead collisions at $\sqrt s = 5.5$
TeV.}
\label{fig:pbpb55etrap}
\end{figure} 

The extrapolation to lead-lead collisions at $\sqrt s = 5.5$ TeV  are
compatible (see Fig.~\ref{fig:pbpb55etrap}) even though the treatment of final
state effects in both models is completely different. The \textsc{Dpmjet}
results are slightly steeper and one finds, in the range 3 $< \pt$~[GeV/c] $<$
20,
\begin{eqnarray}
{d \sigma^{\pi^0} \over d{\imb p_{_T}} dy}
&\sim& {1 \over \pt^{5.66}},\ {\mbox{\rm for NLO QCD}}  \\
&\sim& {1 \over \pt^{6.24}},\ {\mbox{\rm for \textsc{Dpmjet}}}\ ,
\label{eq:AAsecfit}
\end{eqnarray}
At $\pt=$ 5 GeV/c one expects, integrated over azimuthal angle and per unit of
rapidity, 160 events/${\mbox{\rm GeV}}^{~2}$/sec.

\comment{
{\bf To obtain the number 160 i use the cross section read from figure
~\ref{fig:etrap55}, i multiply by 6.28 for the azimuthal integration  and by 5
$10^{26}$ which is the Pb+Pb luminosity given in Appendix II. Is this correct?}
}

%
%
%
%

\subsection{Phenomenology of prompt and decay photon production}
\begin{figure}[htbp]
\begin{center}
\vskip-1.cm
\includegraphics[height=10.0cm]{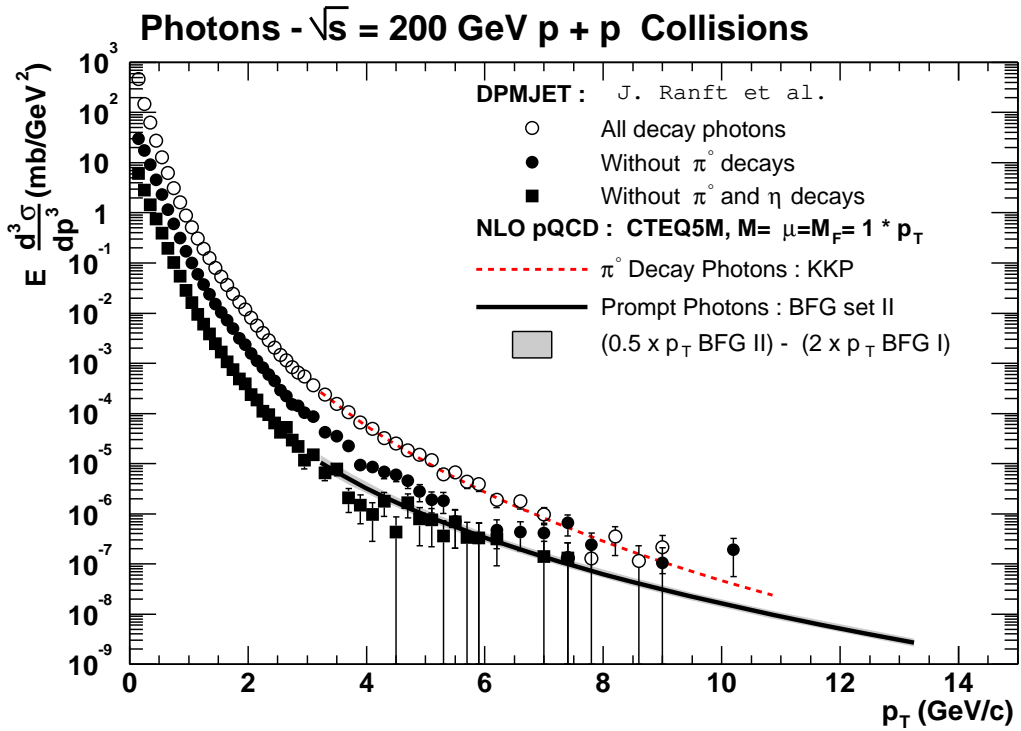}
\end{center} 
\vskip-.5cm
\caption{ 
Comparison of ``prompt" and ``decay" spectra in $p p$  collisions at 200 GeV.
CTEQ5~\cite{cteq5} structure functions are used for the NLO QCD predictions.
For the prompt photon spectrum the BFG~\cite{bourhis} parametrization of
fragmentation functions is used while the KKP~\cite{kkp} fragmentation
functions are used for the NLO QCD estimates of decay photons. The theoretical
predictions are evaluated by averaging the invariant cross sections in an
interval $\pm 0.35$ unit of rapidity around rapidity 0.
} 
\label{fig:pp200phoetrap} 
\end{figure} 
\begin{figure}[htbp]
\begin{center}
\vskip-2.cm
\rotatebox{-90}{\includegraphics[height=15.0cm]{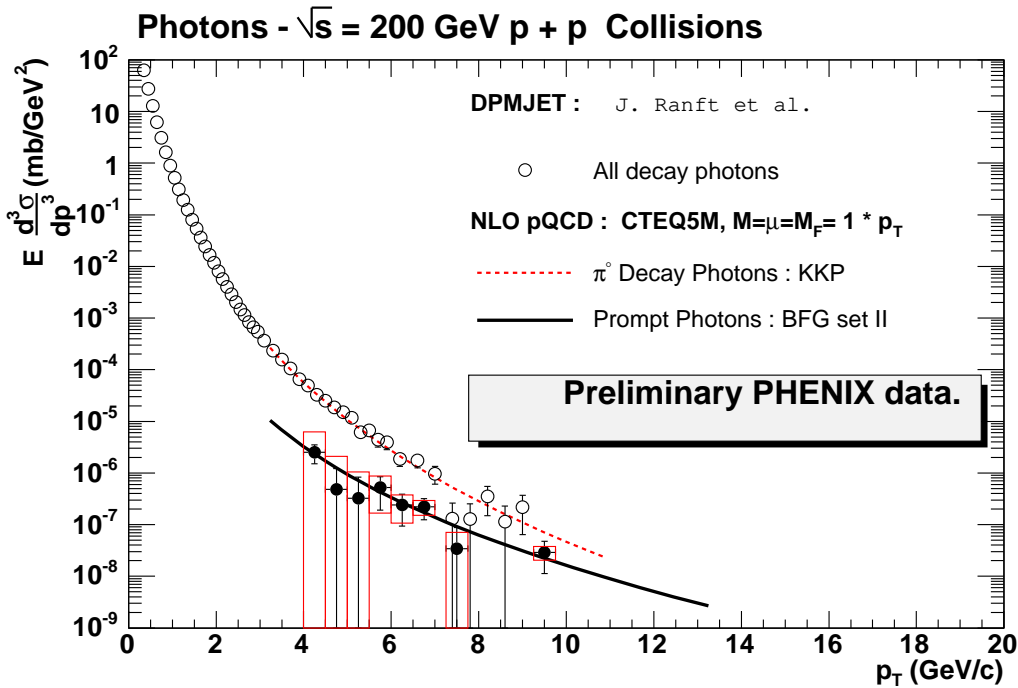}}
\end{center} 
\vskip-.5cm
\caption{ 
Comparison of the theoretical prompt photon spectrum in $p p$ collisions at 200
GeV, shown in the above figure, with the preliminary PHENIX data (full dots).
The data points are collected from a figure shown in ref.~\cite{{Frantz:04}}.
The theoretical predictions of decay photons are those of
Fig.~\ref{fig:pp200phoetrap}.
} 
\label{fig:pp200pho-direct-data} 
\end{figure} 
\begin{figure}[htbp]
\begin{center}
\includegraphics[height=10.0cm]{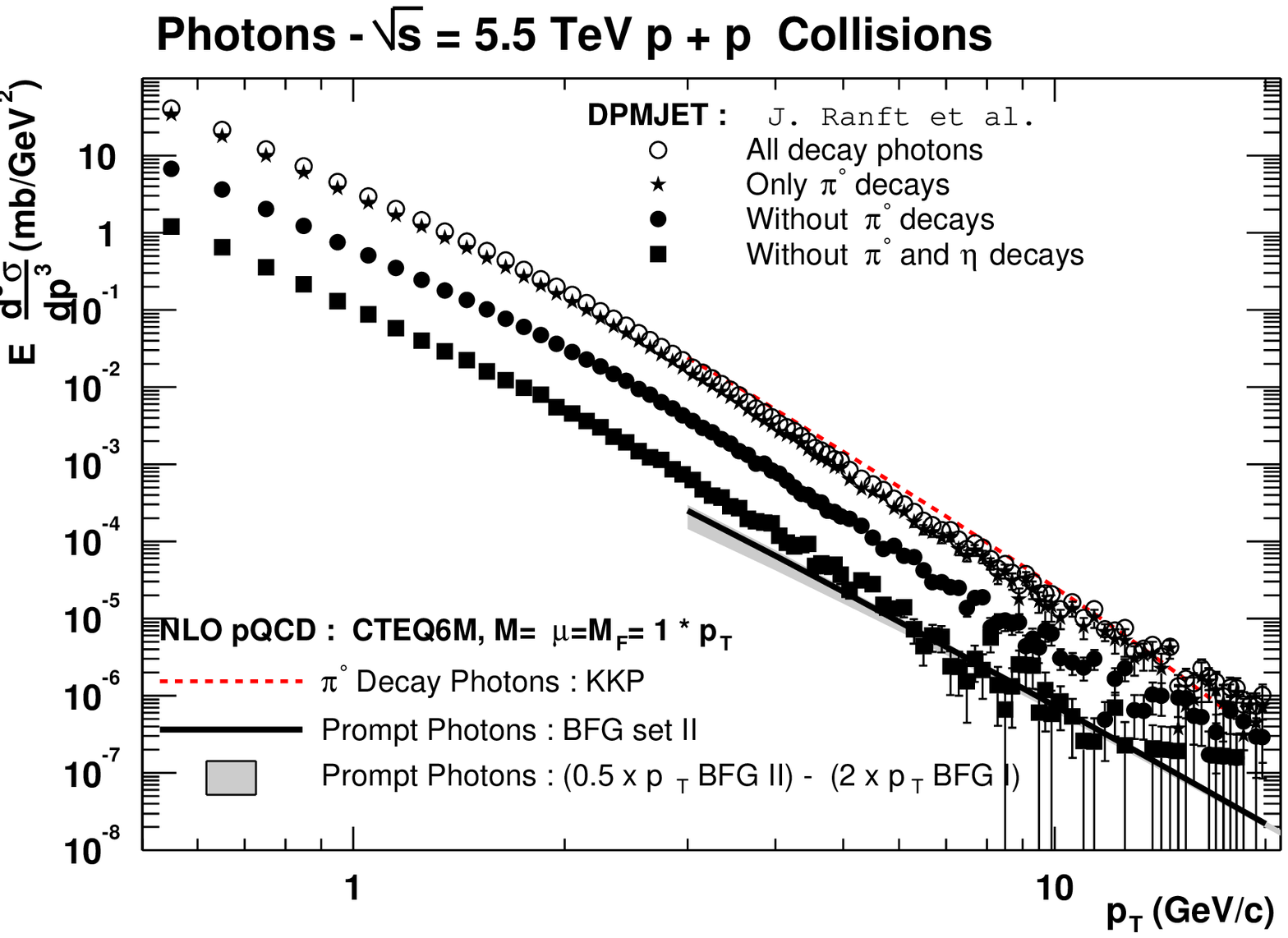}
\includegraphics[height=10.0cm]{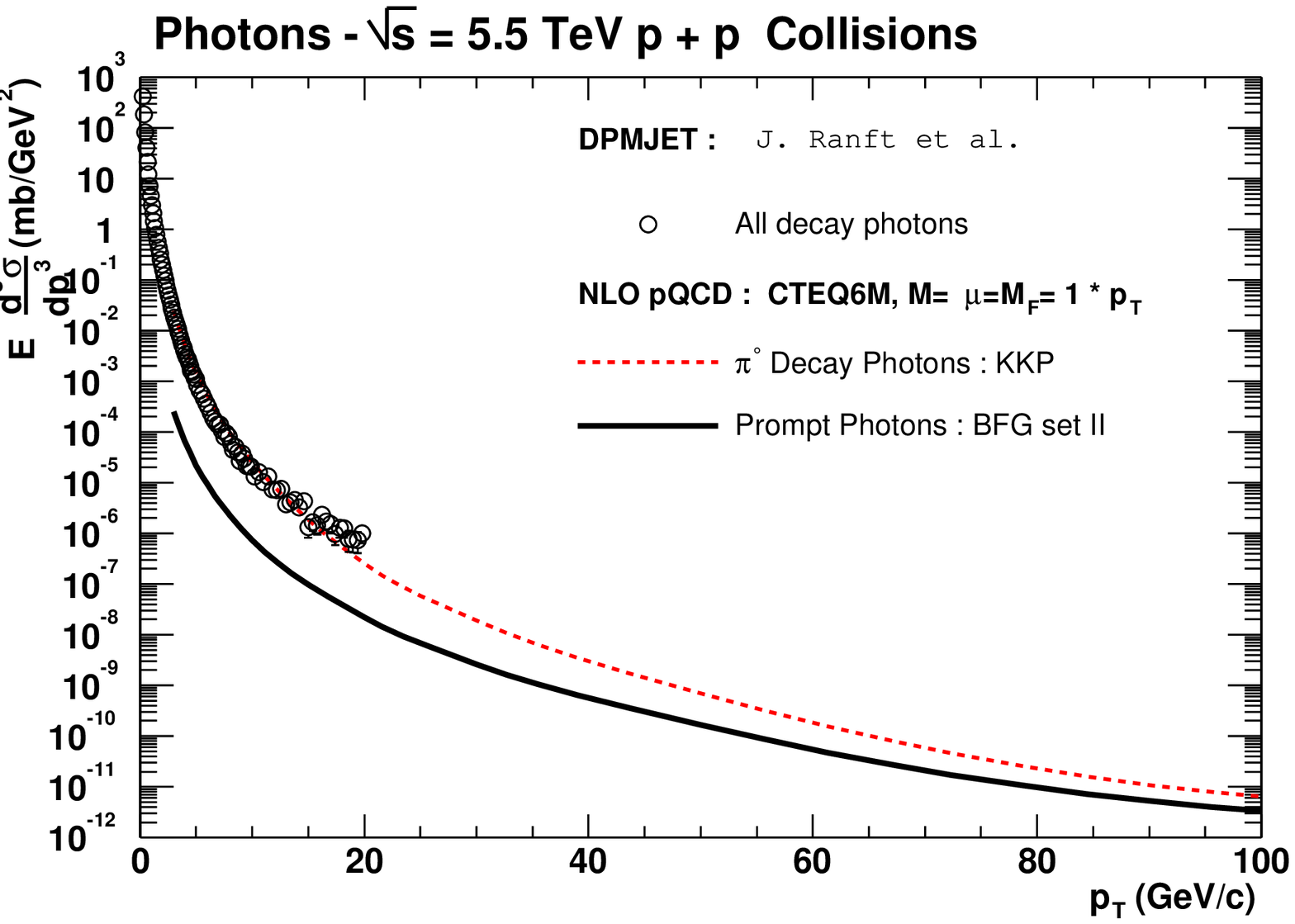}
\end{center} 
\vskip-.5cm
\caption{ 
Comparison of ``prompt" and ``decay" spectra in $pp$  collisions at $\sqrt{s}
=$5.5 TeV. CTEQ6~\cite{cteq6} structure functions are used for the NLO QCD
predictions. For the prompt photon spectrum the BFG~\cite{bourhis}
parametrization of fragmentation functions is used while the KKP~\cite{kkp}
fragmentation functions are used for the NLO QCD estimates of decay photons.
The cross section is evaluated at 0 rapidity.
} 
\label{fig:pp55phoetrap} 
\end{figure} 
\begin{figure}[htbp]
\begin{center}
\includegraphics[height=10.0cm]{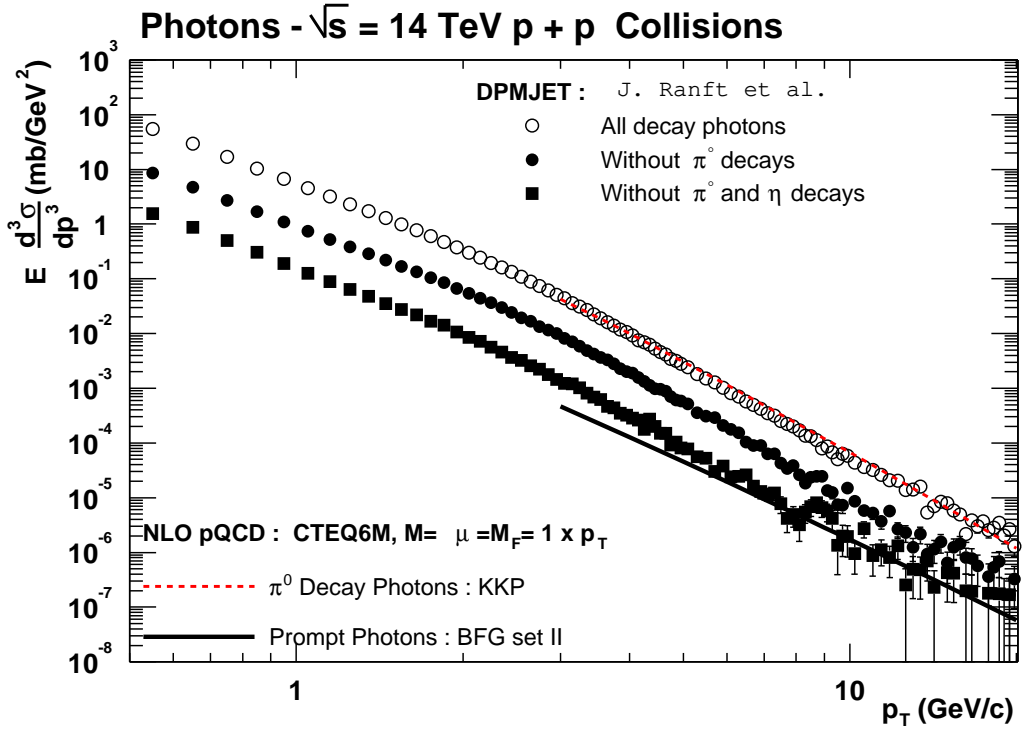}
\includegraphics[height=10.0cm]{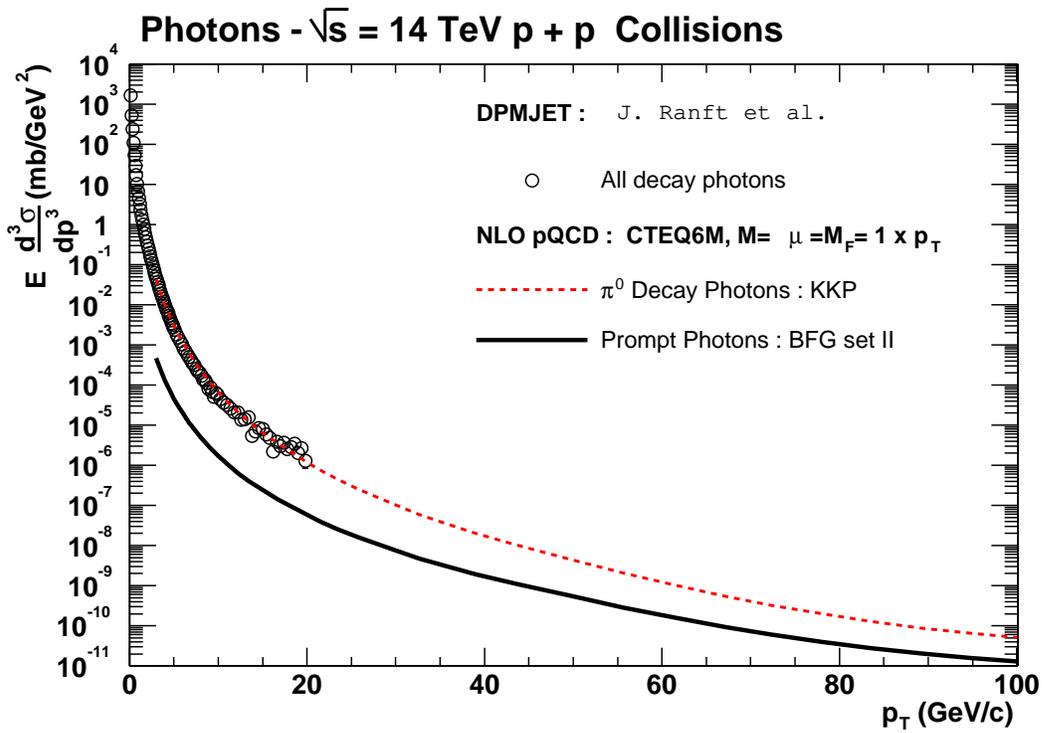}
\end{center} 
\vskip-.5cm
\caption{Same as Fig.~\ref{fig:pp55phoetrap}  at $\sqrt{s} =$~14~TeV.}
\label{fig:pp14phoetrap} 
\end{figure} 
\begin{figure}[htbp]
\begin{center}
\includegraphics[height=18.0cm]{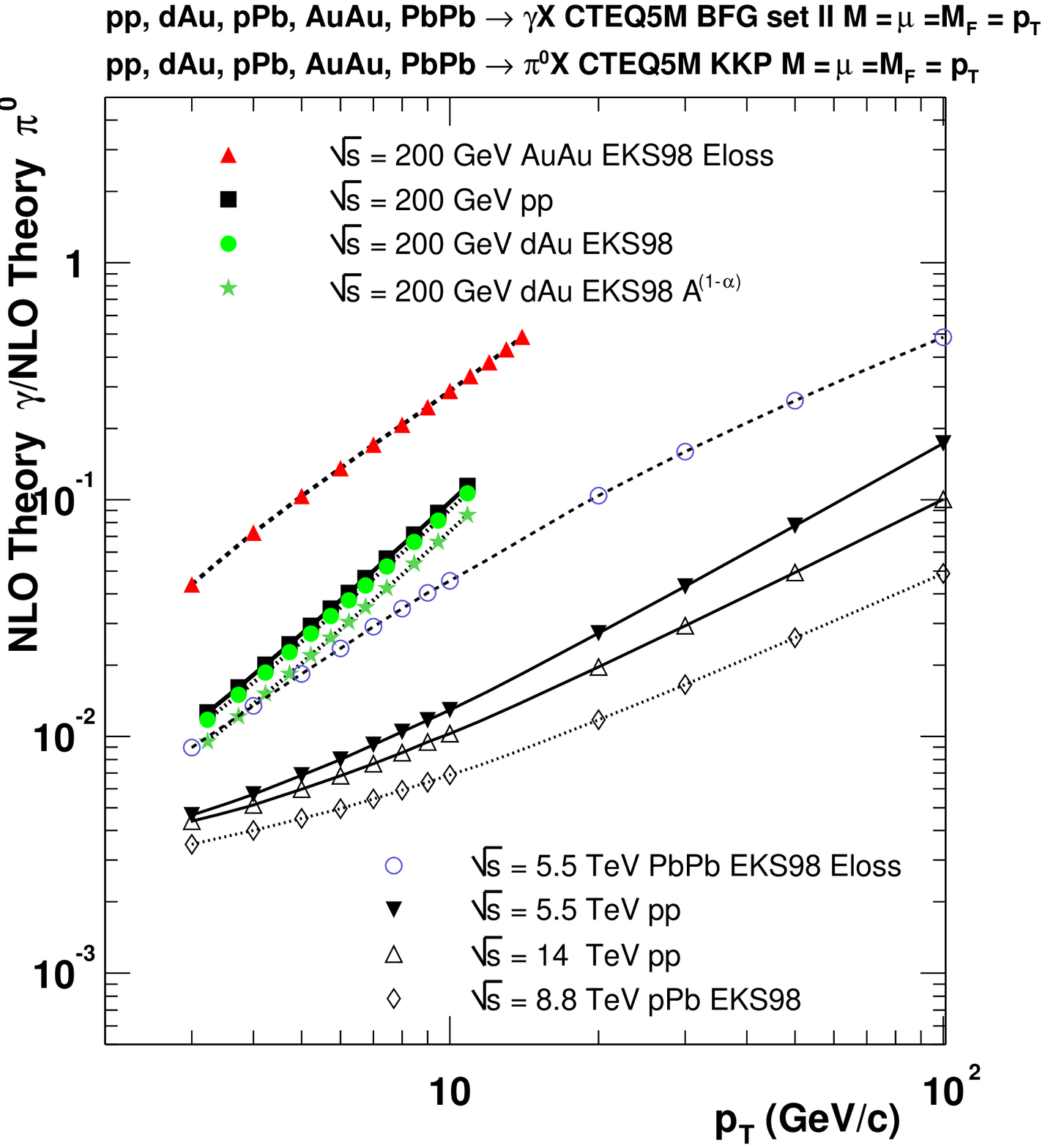}
\end{center}
\vskip-.5cm
\caption{
NLO QCD predictions for the ratio  $\gamma_{\rm prompt}/ {\pi^0}$ in
proton-proton, proton-nucleus and nucleus-nucleus collisions at 200~GeV and
5.5~TeV. The Au+Au ratio at $\sqrt s =$ 200~GeV and the Pb+Pb ratio at $\sqrt s
=$ 5.5~TeV are modified by thermal effects at the lower $\pt$ end of the
spectrum and are predicted to be higher than shown (see
sec.~\ref{sec:compare-photon}).
}
\label{fig:pp200-55-ratios}
\end{figure} 
In this section we discuss the photon spectrum and compare the rates of
production of ``prompt'' and ``decay'' photons. While for $\pi^0$ production we
have two independent models at our disposal, for the prompt photon spectrum one
can, at the moment, only obtain NLO QCD predictions . For decay photons
the \textsc{Dpmjet} code explicitely includes the radiative decay of resonances
($\pi^{~0}$, $\eta$ and other hadronic resonances of the low mass meson and
baryons multiplets) so that one will be able to assess the importance of
various channels to the decay photon spectrum. One can also obtain a 
prediction for the full decay photon spectrum from NLO QCD via standard
convolution formulae based on the $\pi^{~0}$ spectrum. 
Assuming a power behavior for the $\pi^0$ spectrum, 
${d \sigma^{\pi^0} / d p_{_T}} \sim  \pt^{-n}$, one can derive the following
formula for the ratio
of the decay photon spectrum over the $\pi^0$ spectrum\footnote{
	Slightly more elaborate fits to the $\pi^0$ cross sections are used
	to derive the decay photon spectra in the NLO QCD estimates.
}~\cite{fm1984}:
\begin{equation}
R_{\gamma_{\rm decay}/\pi^0} = {2 \over n-1}.
\label{eq:phot-to-pi}
\end{equation}
No isolation cuts have been applied when calculating the photon rates.

The results are displayed in figs.~\ref{fig:pp200phoetrap} to
\ref{fig:pbpb55phoetrap}. In all figures, the solid line indicates the
``standard NLO QCD" prompt photon predictions based on CTEQ5M or CTEQ6M for the
structure functions and BFG~set~II~\cite{bourhis} for the fragmentation
functions of partons into a photon. Using (the less favored but still
compatible with present data) BFG~set~I~\cite{bourhis} would lower the
predictions. The grey band, when shown, indicates the range of uncertainties in
the NLO calculations: the lowest prediction is obtained with BFG~set~I and
all scales set to 2 $\pt$ while the highest one is obtained with 
BFG~set~II and all scales set to 0.5 $\pt$.
For the decay
photon spectra, we show as a dashed line the estimates of NLO QCD using a
generalization of Eq.~(\ref{eq:phot-to-pi})~\cite{fm1984}. Concerning the \textsc{Dpmjet}
predictions, the open circles indicate the full photon decay spectrum, the full
circles the spectrum with the $\pi^0$ decay contribution removed and the full
squares the spectrum with both $\pi^0$ and $\eta$ decays removed.

At $\sqrt{s} = $~200~GeV, one sees that the decay photon spectrum dominates
the prompt photon one and, furthermore, one notices the excellent agreement 
between the \textsc{Dpmjet} and the NLO QCD estimates which is not surprising
since both models agreed very well in their $\pi^0$ spectrum predictions.
The range of uncertainties, due to changes of scales and of photon
fragmentation functions, in the NLO QCD rate of prompt photons is less than
$\pm$~50\% at low $\pt$ and much smaller at high $\pt$. Displayed in 
Fig.~\ref{fig:pp200pho-direct-data} is the comparison between the theoretical
predictions of prompt photon production and the experimental results recently
obtained by the PHENIX collaboration~\cite{Frantz:04}: the agreement is
very satisfactory.

At LHC, Figs.~\ref{fig:pp55phoetrap} and \ref{fig:pp14phoetrap}, the same
features are found except that the prompt photon rate is relatively smaller, at
the level of the decay photon spectrum with the $\pi^0$ and $\eta$
contributions removed. At low $\pt$, the uncertainty in the prompt photon
spectrum is largely due to the choice of photon fragmentation functions: for 
$\sqrt{s}=$ 5.5~TeV or 14~TeV, using BFG~set~I~\cite{bourhis} decreases the
predictions by a factor of up to  $2.5$ at $\pt \sim$~3 GeV/c, where the
brems\-strahlung component is important, but by $10\%$ or less for $\pt$
above $20$~GeV/c: this large variation at low $\pt$ is associated to
uncertainties in the gluon fragmentation into a photon which is hardly
contrained by present data. 
%
At LHC the uncertainty associated to the CTEQ structure
functions is much smaller: $\pm 10\%$ at $\pt=3$~GeV/c and about $\pm 2.5\%$ at
$\pt=100$~GeV/c similarly to the results for $\pi^0$ production. As for
predictions based on the MRS99 parametrization, they are reduced by less than
$15\%$  compared to our standard results, with the largest reduction occuring
at the small values of transverse momentum. 

In more details, the results for $p p$ scattering at LHC can be summarized as
follows: 1) at the lower end of the spectrum, there is almost 10 times more
photons from  $\pi^0$ decays than from all other sources taken together; 2)
decays from $\eta$'s are again about 10 ten times larger than those from
``other" resonances; 3) the level of production of prompt photon is similar to
that of decay photons from ``other" hadronic resonances. 

At RHIC as well as at LHC, the ratio $\gamma_{\rm prompt}/ \gamma_{\rm decay}$
slowly increases with $\pt$.  To see this it is customary to display the ratio
$\gamma_{\rm prompt}/ {\pi^0}$: this is done in Fig.~\ref{fig:pp200-55-ratios}
where the predictions for RHIC and LHC are compared. The ratios are rather
small, specially at LHC, and isolation cuts will certainly be necessary to
increase them.
\begin{figure}[htbp]
\begin{center}
\includegraphics[height=10.0cm]{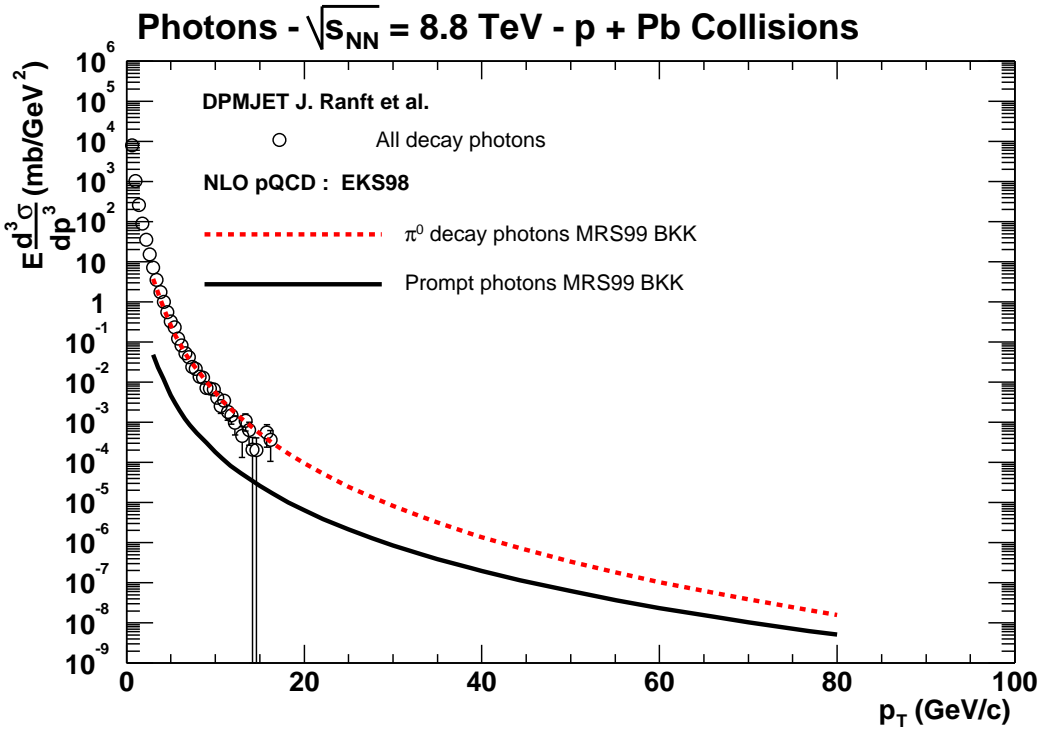}
\end{center}
\vskip-.5cm
\caption{ Comparison of ``prompt" and ``decay" spectra in proton-lead
collisions at 8.8 TeV.   
}
\label{fig:ppb88phoetrap}
\end{figure} 
\begin{figure}[htbp]
\begin{center}
\includegraphics[height=10.0cm]{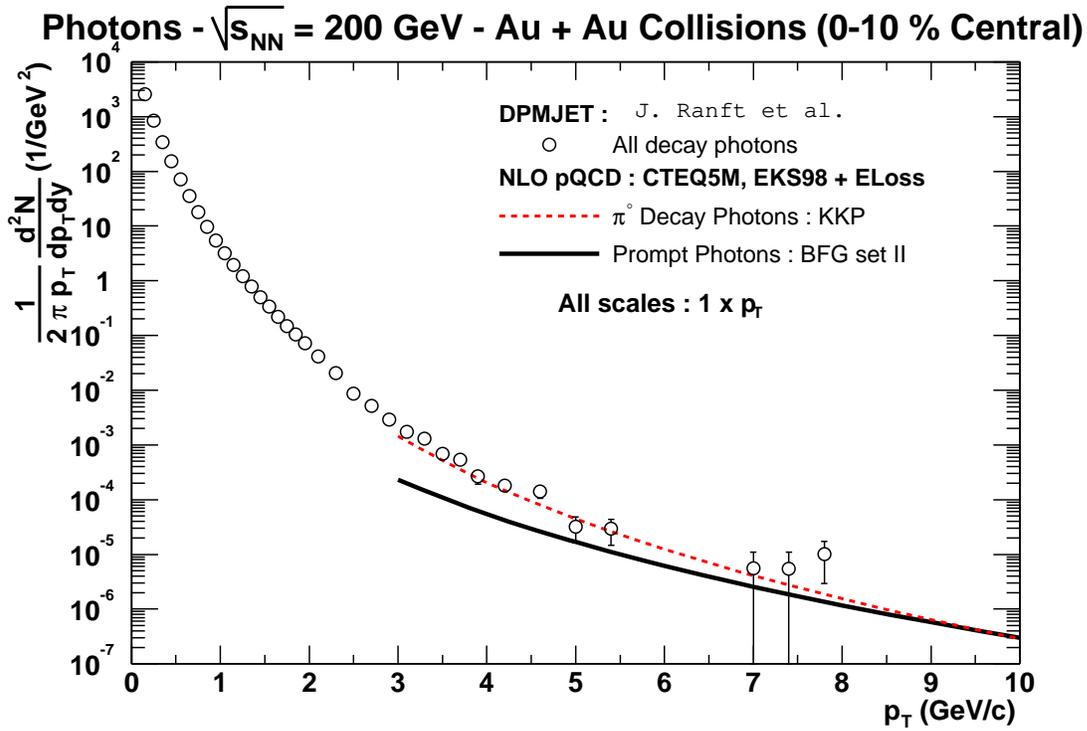}
\end{center}
\vskip-.5cm
\caption{  
Comparison of the prompt and the decay photon spectra in Au+Au collisions at
$\sqrt s = 200$~GeV in NLO QCD. The \textsc{Dpmjet} results for decay photons
are also shown. The theoretical predictions are evaluated by averaging the
invariant cross sections in an interval $\pm 0.35$ unit of rapidity around
rapidity 0.
}
\label{fig:auau200phoetrap}
\end{figure} 
\begin{figure}[htbp]
\begin{center}
\includegraphics[height=10.0cm]{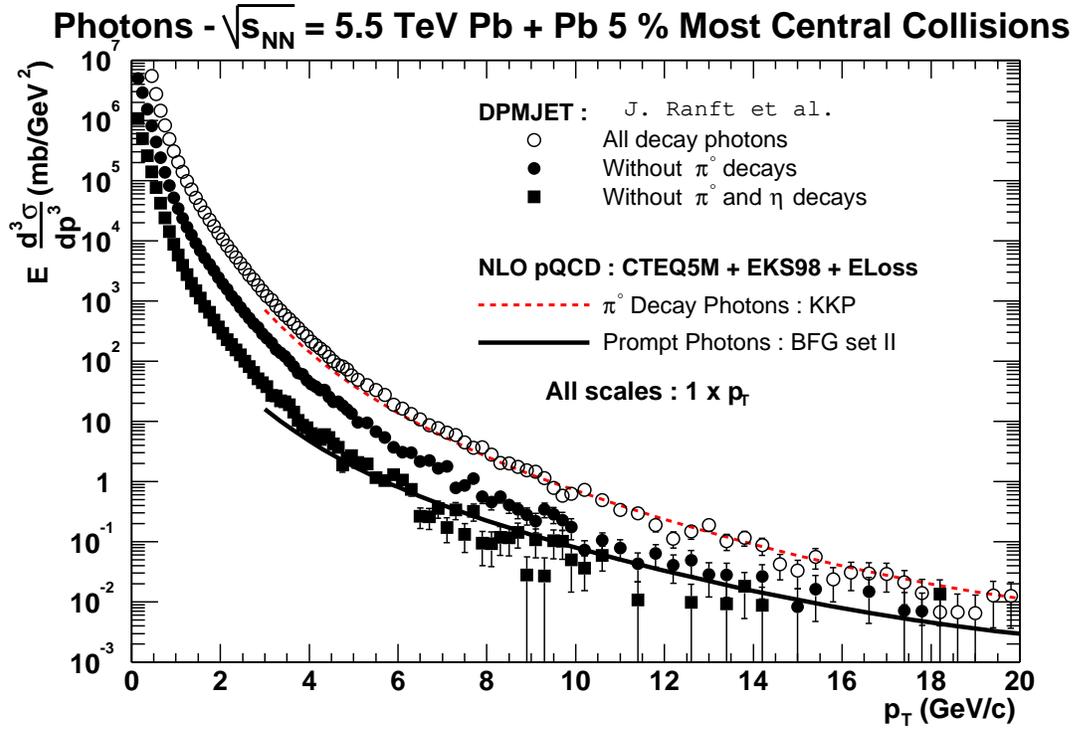}
\includegraphics[height=10.0cm]{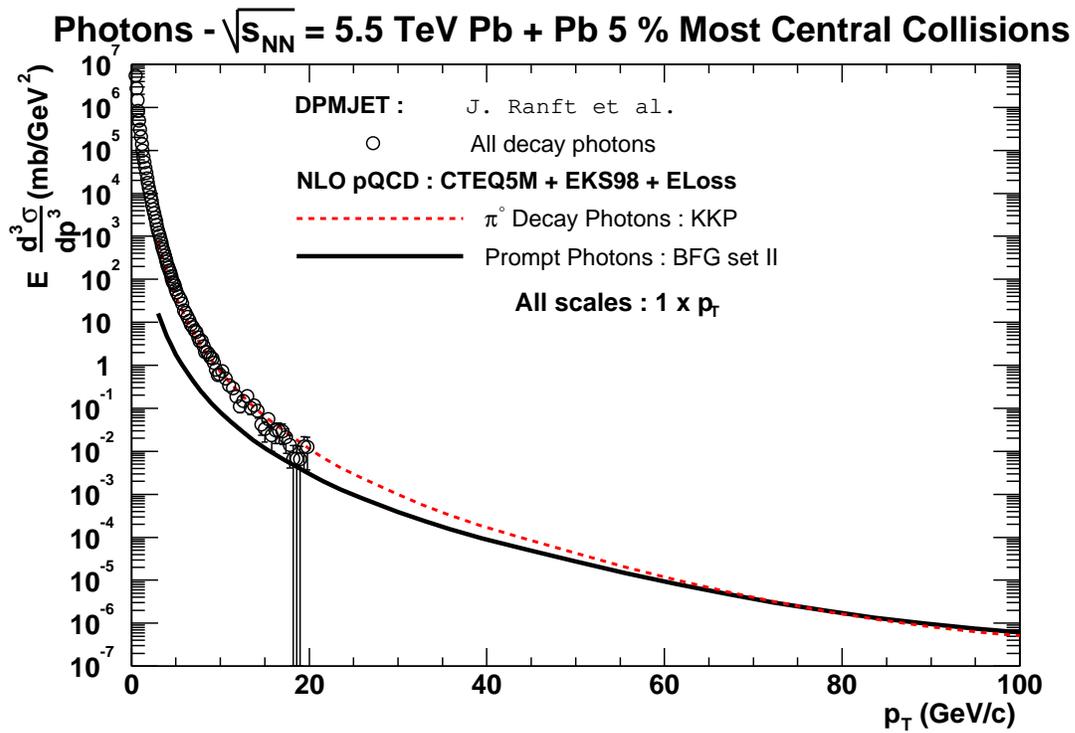}
\end{center}
\vskip-.5cm
\caption{ 
Comparison of ``prompt" and ``decay" spectra in lead-lead collisions at 5.5 TeV.
}
\label{fig:pbpb55phoetrap}
\end{figure} 

In Fig.~\ref{fig:ppb88phoetrap} the results for $p$Pb scattering at LHC are
summarized: the \textsc{Dpmjet} predictions are in good agreement with the NLO
QCD estimates based on the MRS99 and BKK parametrizations, although they tend to
have a slightly steeper slope as already noticed in Fig.~\ref{fig:ppb88etrap}
for pion production. As expected the prompt photon signal is much below the
decay photon spectrum.

The ratio $\gamma_{\rm prompt}/ {\pi^0}$ is shown in
Fig.~\ref{fig:pp200-55-ratios} for $d$Au collisions at 200~GeV and $p$Pb
collisions at 8.8~TeV. At RHIC very little change is seen compared to $p p$
collisions at the same energy. We also looked at the effect of assuming an
$A^{1-\alpha}$ dependence of the $d$Au cross sections, with $\alpha=1.08$ for 
$\pi^0$ production and $\alpha=1.04$ for $\gamma$ production. It is seen to
have a moderate effect on the estimate of the ratio. For $p$Pb scattering at
LHC a small energy dependence is observed compared to $p p$ scattering at
5.5~TeV with a larger decrease of the ratio at  the larger $\pt$ values.

We turn finally to the case of nucleus-nucleus collisions. The results for RHIC
are shown in Fig.~\ref{fig:auau200phoetrap}. One sees that in Au+Au collisions
the ratio  $\gamma_{\rm prompt}/ \gamma_{\rm decay}$ increases with $\pt$ and
approaches 1 at $\pt = 10$~GeV/c. The value of this ratio is much higher than
for $p p$ scattering at the same energy. In the NLO QCD model this is due to
the ``jet quenching" mechanism which decreases the rate of decay photons more
than that of prompt photons: indeed only the bremsstrahlung component is
affected in the latter case. An estimate of the relative increase of prompt
photons in A+A collisions is also seen by plotting the ratio $\gamma_{\rm
prompt}/ {\pi^0}$ as done in Fig.~\ref{fig:pp200-55-ratios}: this ratio is
approximately 3 times higher in Au+Au collisions than in $p p$ collisions.

The same pattern is seen at 5.5 TeV (Fig.~\ref{fig:pbpb55phoetrap}) where prompt
and decay photons become comparable above $\pt = 60$~GeV/c, in contrast with
the $p p$ case as seen in Fig.~\ref{fig:pp55phoetrap}. Looking at the medium
$\pt$ range one should be able to extract a prompt photon spectrum from the data
for $\pt > 10$~GeV/c since the ratio $\gamma_{\rm prompt}/ \pi^0$ becomes of the
order of 5\% or larger as shown in Fig.~\ref{fig:pp200-55-ratios}. As for RHIC
this ratio is two or three times larger than in $p p$ collisions.
In this figure only the non thermal contribution to heavy ion collisions is
taken into account. Thermal production of pions and photons occurs at low $\pt$
values in heavy ion collisions. At LHC it is important below $\pt =$ 10-15
GeV/c and the ratios turn out to be larger than shown as will be discussed in
sec.~\ref{sec:compare-photon}.


\subsection{Small mass lepton pairs at large transverse momentum}
\label{dilepton-non-thermal}

As seen in the previous section the prompt photon rate is much below the rate
of decay photons at low $\pt$ values. The dominant backgrounds are the decays $\pi^0 \rightarrow
\gamma \gamma$ and $\eta \rightarrow \gamma \gamma$. If one considers instead
of real photon production, the emission of virtual photons (lepton pairs) part
of this background can be eliminated. For example, considering the production
of an electron pair in the mass range $M_{e^+e^-} = $ [0.2,0.6]~GeV/c$^2$ one
gets rid of the $\pi^0$ background (the lepton pairs from the Dalitz decays of
the $\pi^0$ are below this mass range) and one stays below the $\rho, \omega,
\phi \rightarrow e^+ e^-$ decays. The $\eta$ background is also reduced because
of  the mass constraints on the Dalitz pairs.

From the theoretical point of view, the production rate of small mass lepton
pairs at relatively large transverse momentum is very similar to that of real
photon. To get a rough estimate of the rate of production of a Drell-Yan pair,
integrated over a given mass range, we follow the procedure of \cite{abf} where
the virtual photon mass is neglected, compared to the other scales $\pt$ and
$\sqrt s$, everywhere except in the bremsstrahlung component: indeed, in the
fragmentation process, the virtual photon mass, rather than $\Lambda_{QCD}$, acts
as a cut-off of the final state collinear singularity. Following this
procedure, we find that the rate of production of an electron pair in a finite
mass range is given, with a good accuracy, by the simple relation
\begin{eqnarray}
{d \sigma^{e^+ e^-} \over d{\imb p_{_T}} dy} \simeq C_ {e^+ e^-}\ \alpha \ 
{d \sigma^{\gamma} \over d{\imb p_{_T}} d y}
\label{eq:virtphotrate}
\end{eqnarray}
where $C_ {e^+ e^-} \sim 0.3$ for 0.2~GeV/c$^2$  $< M_{~e^+e^-}<$ 0.6~GeV/c$^2$  valid in the
range 2~GeV/c $< \pt <$ 100~GeV/c, and $C_ {e^+ e^-} \sim 0.2$ for 1~GeV/c$^2$ $<
M_{e^+e^-} <$ 3~GeV/c$^2$ in the range 4~GeV/c $< \pt <$ 100~GeV/c.  These numbers
are independent of the fragmentation functions used. These estimates refer only
to the "prompt" production mechanism.
\begin{figure}[htbp]
\begin{center}
\includegraphics[height=10.0cm]{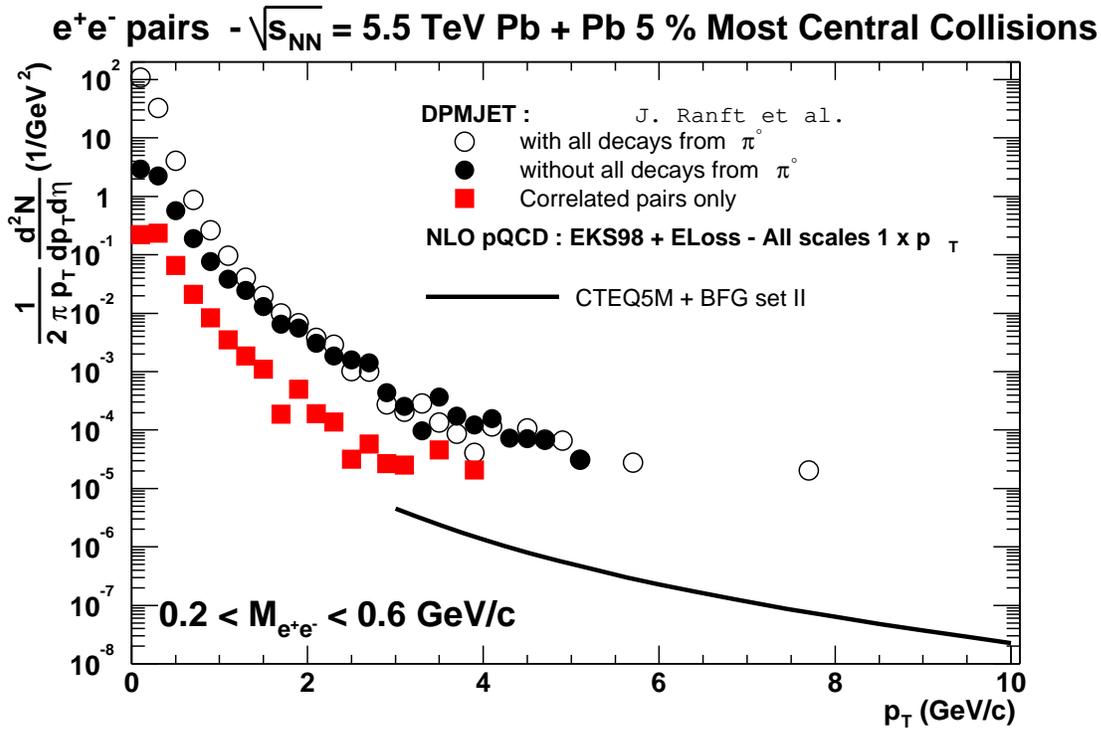}
\end{center}
\vskip-.5cm
\caption{  
Lepton pair spectrum at $\eta = 0$ in Pb+Pb collisions as a function of $\pt$
for $\sqrt{s} =  5.5$~TeV. The solid line indicates the prompt (signal) NLO QCD
predictions while the squares and the points represent backgrounds of different
origins calculated using \textsc{Dpmjet}. Thermal production will increase the
signal (see sec.~\ref{sec:comp-leptons}).
}
\label{fig:pbpb55e+e-etrap}
\end{figure} 
Using, from Appendix II, the luminosity factors for ALICE at $\sqrt s =$ 5.5
TeV, the rate of production of a prompt lepton pair in the mass range
0.2~GeV/c$^2$ $< M_{~e^+e^-} <$ 0.6~GeV/c$^2$ is estimated to be 6 pairs per
second with $\pt = 3$~GeV/c and 0.1 pair per second with $\pt = 8$~GeV/c. The
corresponding numbers in lead-lead collisions are 0.4 and 0.02 respectively.
We should say at this point, that thermal production of dileptons (see next
chapter) will increase the direct production of lepton pairs.

The production mechanism we just discussed suffers from a background with can be
estimated using \textsc{Dpmjet}: indeed this code contains the Dalitz decays of
resonances as well as the semi-leptonic decay processes of charmed and heavy
flavor resonances.  We show in Fig.~\ref{fig:pbpb55e+e-etrap} the prompt lepton
pair spectrum for $0.2$~Gev/c$^2 < M_{~e^+ e^-} < 0.6$~Gev/c$^2$
(Eq.~(\ref{eq:virtphotrate})) together with the background from lepton pairs
originating  from the same resonances (correlated pairs). One sees that, as
expected, at high enough $\pt$ the signal becomes comparable to the background.

Unfortunately, this estimate of the background is incomplete as one expects a
huge contribution from uncorrelated pairs  due to the large combinatoric factor
when collecting pairs of leptons: in a typical \textsc{Dpmjet} event one
expects about 180 $e^-$ and $e^+$ including $\pi^{~0}$ decays but only 17 $e^-$
and $e^+$ excluding $\pi^{~0}$'s. This is shown in the figure by
the open dots (including $\pi^{~0}$ decays) and full dots (exluding $\pi^{~0}$
decays): obviously the situation is not as favorable since the background is
now two orders of magnitude above the estimated signal! Note however the nice
feature that the combinatoric background from $\pi^{~0}$'s disappears for $\pt
> 2$~GeV/c. One way to tame the huge background may be to subtract from the
data the spectrum of like charge pairs in the same kinematic range or pairs
constructed from different events. Using low statistics runs of \textsc{Dpmjet}
one finds that subtracting the like-charge pair spectrum from the unlike-charge
pair spectrum one gets numbers which are very close to the correlated pair
background whenever the transverse momentum of the pair is larger than
1.5~GeV/c~\cite{ranft2003}. The statistical fluctuations are large and clearly
further studies are necessary to remove the uncorrelated background.

\section{PHOTONS FROM THERMAL, EXPANDING FIREBALL}
\label{thermal}

\noindent
{\em  P.~Aurenche, F.~Gelis, G.~Moore, H. Niemi, R.~Rapp, K.~Redlich,
P.V.~Ruuskanen, S.S.~R\"as\"anen, D.K.~Srivastava}

\vspace{.5cm}
For the calculation of thermal photons --- photons from secondary
interactions among particles in the expanding fireball of matter
produced in a heavy ion collision --- the emission rates from hot matter
and the space-time evolution of the matter are needed.  The equilibrium
emission rates in both QGP and the Hadron Resonance Gas (HRG) are
reviewed below. The space-time evolution of expanding matter is
described in terms of relativistic hydrodynamics with the assumption of
longitudinal boost invariance.

There are uncertainties both in the emission rates and the description of the
nuclear fireball.  One such uncertainty affecting both the emission and the
evolution of produced matter arises from the fact that the initially produced
quark-gluon matter is not expected to be in chemical equilibrium.  E.g., in the
pQCD + saturation model \cite{Eskola:2002qz} described below, most of produced
partons are gluons.  Assuming kinetic equilibrium and describing the deviation of
number densities from equilibrium values in terms of fugacities, $\lambda_i$, we
typically obtain $\lambda_g \sim 0.6...1$ and $\lambda_q \sim 0.2$ in the
initially produced parton system.  The quark fugacity is assumed to be the same
at LHC energies as the antiquark fugacity.  The effect of non-equilibrium values
of fugacities on the photon emission rates is studied below.  Using rate
equations for reactions which change the parton numbers
\cite{Biro:1993qt,Elliott:1999uz}, we will study in detail also the time
evolution of fugacities.

The description of the space--time evolution of a fireball produced in a
nucleus--nucleus collision must begin with characterization of the
initial state of produced matter.  Information on initial state is
obtained from calculations of particle production from primary
interactions in nuclear collisions.  There are several model
calculations, some more detailed than the others, of hadron spectra in
heavy ion collisions at RHIC energies.  We use these calculations or,
when necessary, their extensions at the LHC energy as the first step in
obtaining the initial state for the space-time evolution of produced
matter.  The most important quantities in determining the initial
conditions are the total hadron multiplicity, $dN/dy$, which fixes the
the total initial entropy and the production time scale, $\tau_0$.  The
initial densities are proportional to $dN/dy$ and $1/\tau_0$.  The
models for particle production differ also in the transverse dependence
of nuclear densities.

After reviewing the emission rates both in the quark-gluon plasma and in the
hot hadronic gas, we consider in Sec.~\ref{sec:part-prod} different models for
particle production in order to establish a reasonable range of uncertainties
in the initial state. The hydrodynamic evolution is dealt with in the following
section, while the predictions for the rates of thermal pions and photons are
discussed in Secs.~\ref{sec:spectra_hadron} to \ref{sec:pion_photon}. The last
section is devoted to estimating the yield of small mass lepton pairs in the
quark-gluon plasma.

\subsection{Thermal photon emission rates}
\label{sec:thermalrates}

The rate of production, per unit time and volume, of a real photon of momentum
($E,\imb p$) in a system in thermal equilibrium is calculated via the
formula~\cite{Weldo3,GaleK1}:
\begin{equation}
  E {{dN}\over{dt d^3{\imb x} d^3{\imb p}}}=-
  {1\over{(2\pi)^3}}\; {1 \over \exp(E/T) -1}\,
  {\rm Im}\,\Pi^{^{R}}{}_\mu{}^\mu(E,{\imb p})\; ,
  \label{eq:realphot}
\end{equation}
where $\Pi^{^{R}}{}_\mu{}^\mu(E,{\imb p})$ is the retarded photon
polarization tensor. The pre-factor $1 / (\exp(E/T) -1)$ provides the expected
exponential damping when $E \gg T$.

A loop expansion of $\Pi^{^{R}}$ is constructed with
effective propagators and
vertices. For production of photons in the quark-gluon plasma the basic degrees
of freedom are quarks and gluons and the corresponding physics will be
discussed in the next two sections, while for production in hot hadronic matter
the relevant degrees of freedom are mesons ($\pi$, $\rho$, $a_1$, $\cdots$) and
baryons, and this will be the subject of Sec.~\ref{sec:hotgas}.

\subsection{Photon emission rates from the chemically equilibrated quark-gluon
plasma}
\label{sec:qgp}
{\em P. Aurenche, F. Gelis, G. Moore}

The most complete calculations are those based on the improved perturbation
theory of Braaten and Pisarski, {\em i.e.} on the Hard Thermal Loop  (HTL)
resummed effective theory \cite{BraatPis1,FrenkT1}. In the HTL approach, the
quarks and the gluons acquire an effective mass of the order of
$\sqrt{\alpha_s} T$, where $\sqrt{\alpha_s}$ is the strong coupling, due to
their interactions in the plasma. Quark and gluon exchange mechanisms on a long
distance (equivalently small momentum transfers) are also modified and they
become effectively screened by mass effects of ${\cal O}(\sqrt{\alpha_s} T)$.
Thus, thermal effects tend to regularize the infra-red behavior of the theory.
Estimates of Eq.~(\ref{eq:realphot}), in the HTL effective theory, based on
1-loop, 2-loop and multi-loop contributions have been performed: each case
corresponds to different physical processes. We review them in turn.

\subsubsection{Real photons}

At one-loop order in the HTL theory, the production of a hard ($ E \gg T$) real
photon is given by the diagram in Fig.~\ref{fig:processes-1}. One of the
fermion leg in the loop may have a soft momentum flowing though it, the other
one beeing necessarily hard. Following the HTL approach, thermal corrections
are resummed on the soft line and an effective fermion propagator should be
used. When taking the discontinuity of the diagram one of the cut fermions is
necessarily space-like (denoted $Q$ in the figure) in the case of real photon
production. Cutting through the effective quark propagator of momentum $Q$
exhibits Landau damping {\em i.e.}, the exchange of a virtual quark in the
scattering of quarks and gluons in the medium. In this way one obtains the
processes contributing to photon production, namely Compton scattering and $q
\bar q$ annihilation into a photon and a gluon.  As a result of resummation, the
potential singularity at 0 momentum transfer in the  fermion propagator $Q$ is
screened by an effective mass $m_{\rm q}$. Taking into account this thermal
correction to the quark propagator, the imaginary part of the photon
polarization tensor can be calculated~\cite{Kapusta:qp,BaierNNR1}. For hard
photons, it reads:
\begin{equation}
{\rm Im}\,\Pi^{^{R}}{}_\mu{}^\mu(E,{\imb p}) = 4\pi
{ \frac{5\alpha\alpha_s}{9}}T^2
\left[\ln\left(\frac{E T}{m_{\rm q}^2}\right)
-\frac{1}{2}-\gamma_{_{E}}+\frac{7}{3}\ln(2)+\frac{\zeta^\prime(2)}{\zeta(2)}
+ {\cal O}({T \over E}) \right]\;
\label{eq:rate-1-loop}
\end{equation}
\begin{figure}[htbp]
\begin{center}
\resizebox*{12cm}{!}{\includegraphics{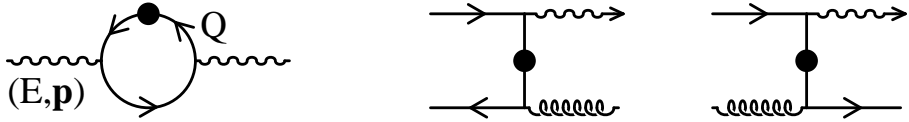}}
\end{center}
\caption{\label{fig:processes-1}
Left diagram: one-loop contribution to hard real photon production; right
diagrams: examples of diagrams contributing to  photon production,
annihilation and Compton scattering. The symbol $\bullet$ on the quark
propagator indicates that the effective propagator is used.}
\end{figure}
Note that the mass $m_{\rm q}$ is given by $m_{\rm q}^2=\pi\alpha_s C_f
T^2$ with $C_f\equiv(N_c^2-1)/2N_c$.  The numerical factor $5/9$ is
the sum of the quark electric charges squared for $2$ flavors (u and
d); for $3$ flavors (u, d and s), this factor should be replaced by
$6/9$. It is clear from the expression above that the quark thermal mass
acts as the cut-off of a logarithmic collinear singularity.

For some time this was thought to be the final answer for the photon rates at
${\cal O}(\alpha_s)$. It then became clear that some formally higher order
processes (see Fig.~\ref{fig:processes-2}) are in fact strongly enhanced by
collinear singularities. The corresponding processes are bremsstrahlung
emission and off-shell annihilation. A common property
of these two diagrams is that they have an off-shell quark next to the vertex
where the photon is emitted, and the virtuality of this quark becomes very
\begin{figure}[htbp]
\begin{center}
\resizebox*{12cm}{!}{\includegraphics{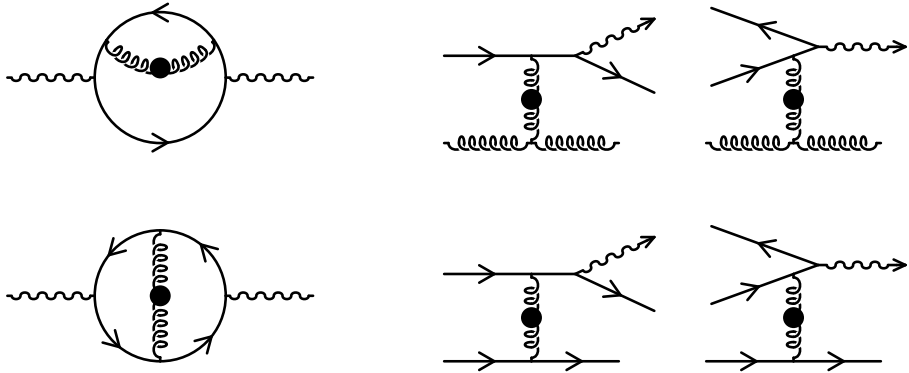}}
\end{center}
\caption{\label{fig:processes-2}
Left most diagrams: two-loop contributions to hard real photon production;
right diagrams: bremsstrahlung and off-shell annihilation processes. The symbol
$\bullet$ on the gluon propagator indicates that the effective propagator is
used.}
\end{figure}
small if the photon is emitted forward. Again,  the quark thermal mass $m_{\rm
q}$ prevents these diagrams from being truly singular. However, contrary to the
one loop diagrams, the singularity is linear instead of logarithmic, and it
brings a factor $T~^2/m_{\rm q}^2$. Combined with $\alpha_s^2$ from the
vertices, these diagrams turn out to be also of order ${\cal
O}(\alpha_s)$~\cite{AurenGKP2,AurenGKZ1,Steffen:2001pv,AurenGZ4}.
For 3 colors and 2 light quark flavors, the ${\cal
O}(\alpha_s)$ contribution of these two diagrams is exactly:
\begin{equation}
{\rm Im}\,\Pi^{^{R}}{}_\mu{}^\mu(E,{\imb p}) = \frac{32}{3\pi} \,
\frac{5\alpha\alpha_s}{9}\left[\pi^2 \frac{T^3}{E} + E T\right]\; .
\end{equation}
In this formula, the term in $1/E$ comes from the brem\-sstra\-hlung diagram
and dominates for soft photons, while the term linear in $E$ comes from
off-shell annihilation which therefore dominates for very hard photons ($E \gg
T$).\footnote{
        For 3 flavors, the same formula holds but the numerical prefactor is
        replaced by a very complicated expression~\cite{AurenGZ4}.}
%
%

Given the enhancement in the diagrams of Fig.~\ref{fig:processes-2}, one may
wonder if  higher order diagrams also contribute to the same ${\cal
O}(\alpha_s)$ order. To discuss the issue in physical terms, it is
convenient to define the concept of {\sl photon formation time}. Consider a
virtual quark of momentum $R\equiv P+Q$ which splits into an on-shell quark of
momentum $Q$ and a photon of momentum $P$ (see the processes in
Fig.~\ref{fig:processes-2}). The photon formation time can be identified with
the lifetime of the virtual quark, which is itself related to its virtuality by
the uncertainty principle. A simple calculation gives:
\begin{equation}
t_{_{F}}^{-1}\sim { \delta E}=r_0-\sqrt{{\boldsymbol
r}^2+m_{\rm q}^2}  \approx \frac{E}{2 q_0 r_0} \left[{\boldsymbol q}_\perp^2+{
m_{\rm q}^2} \right]\; ,
\label{eq:tF}
\end{equation}
where the 3-momentum of the photon defines the longitudinal axis. The collinear
enhancement in the diagrams of Fig.~\ref{fig:processes-2}, due to the small
virtuality of the quark that emits the photon, can be rephrased by saying that
it is due to a large photon formation time, of ${\cal O}(1/\alpha_s T)$.
If the formation time is large, the quark can rescatter in the medium, while
emitting the photon, as it is shown in Fig.~\ref{fig:LPM}. This can be
partially taken into account by introducing a collisional width $\Gamma\sim
\alpha_s T\ln(1/\alpha_s)$ on the quarks in the calculation of the diagrams of
Fig.~\ref{fig:processes-2}~\cite{AurenGZ2}. A large sensitivity to this
parameter was found at leading order, thereby indicating that an infinite
series of diagrams must be resummed in order to fully determine the ${\cal
O}(\alpha_s)$ photon rate. This phenomenon is nothing but a manifestation of
the Landau Pomeranchuk Migdal (LPM) effect
\cite{LandaP1,LandaP2,Migda1}.
\begin{figure}[h]
\begin{center}
\resizebox*{6cm}{!}{\includegraphics{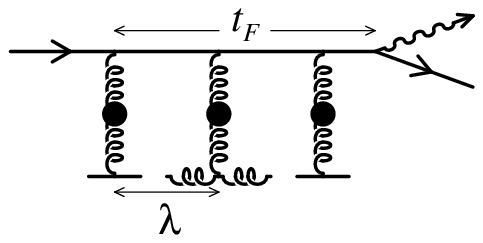}}
\end{center}
\caption{\label{fig:LPM} A ladder correction to bremsstrahlung.}
\end{figure}

A considerable progress was made recently in \cite{ArnolMY1}: it was shown that
there are infrared cancellations between diagrams of different topologies, and
that these cancellations remove any sensitivity to the magnetic scale.
Physically, this cancellation can be interpreted as the fact that ultrasoft
scatterings are not efficient in order to induce the production of a photon. As
a consequence, only the ladder family of diagrams needs to be resummed in order
to obtain the complete leading ${\cal O}(\alpha{}_s)$ photon rate. The
resummation of this series of diagram can then be performed in two steps
summarized in Fig.~\ref{fig:equations}.
\begin{figure}[h]
\begin{center}
\resizebox*{15cm}{!}{\includegraphics{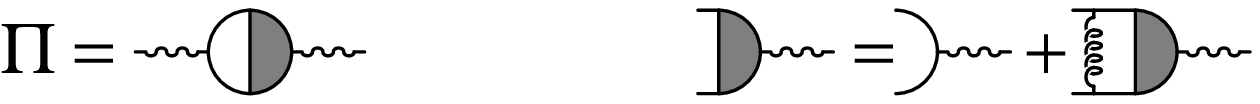}}
\end{center}
\caption{\label{fig:equations}  Resummation of ladder diagrams.}
\end{figure}
The first one is a Dyson equation for the photon polarization tensor,
whose explicit form is \cite{ArnolMY1,ArnolMY2,ArnolMY3}:
\begin{eqnarray}
{\rm Im}\,\Pi^{^{R}}{}_\mu{}^\mu(E,{\imb p}) \approx
{\alpha N_c}
\int_{-\infty}^{+\infty}\!\!dq_0
\,[n_q(r_0)-n_q(q_0)]\;
\frac{q_0^2+r_0^2}{(q_0r_0)^2}
\,{\rm Re}\int
\frac{d^2{\boldsymbol q}_\perp}{(2\pi)^2}\;
{\boldsymbol q}_\perp\cdot{\boldsymbol f}({\boldsymbol q}_\perp)\; ,
\label{eq:AMY}
\end{eqnarray} with $r_0\equiv E+q_0$, $n_{q}(r_0)\equiv 1/(\exp(r_0/T)+1)$ the
Fermi-Dirac statistical weight, and where the dimensionless function
${\boldsymbol f}({\boldsymbol q}_\perp)$ denotes the resummed vertex between
the quark line and the transverse modes of the photon (this is represented by
the shaded vertex in the above pictures). In the Dyson equation, this function
is dotted into a bare vertex, which is proportional to ${\boldsymbol q}_\perp$
(the photon has a transverse polarization and therefore the coupling vanishes
for a collinear emission by a quark).  The second equation, that determines the
value of ${\boldsymbol f}({\boldsymbol q}_\perp)$, is a Bethe-Salpeter equation
that resums all the ladder corrections \cite{ArnolMY1,ArnolMY2,ArnolMY3}:
\begin{equation}
\frac{i}{t_{_{F}}}{\boldsymbol f}({\boldsymbol q}_\perp)
=
2{\boldsymbol q}_\perp
+4\pi \alpha_s C_f T \!\! \int
\frac{d^2{\boldsymbol l}_\perp}{(2\pi)^2} \,
{\cal C}({\boldsymbol l}_\perp) \,
[{\boldsymbol f}({\boldsymbol q}_\perp+{\boldsymbol l}_\perp)-{\boldsymbol
f}({\boldsymbol q}_\perp)]\; ,
\label{eq:integ-f}
\end{equation}
where $t_{_{F}}$ is the time defined in Eq.~(\ref{eq:tF}) and where
the collision kernel has the following expression: ${\cal
  C}({\boldsymbol l}_\perp)={m^2_{\rm debye}}/{{\boldsymbol
    l}_\perp^{\ 2}} ({{\boldsymbol l}_\perp^{\ 2}+{ m^2_{\rm
      debye}}})$ \cite{AurenGZ4}.
Note that in the Dyson equation, the quark propagators are dressed (by
resumming self-energy corrections  which is akin to introducing the collisional
width or damping rate) to match  the resummation performed for the vertex, so
that gauge invariance is preserved. It is this dressing on the quark
propagators which is responsible for the term $-{\boldsymbol f}({\boldsymbol
q}_\perp)$ under the integral in Eq.~(\ref{eq:integ-f}).  From this integral
equation, it is easy to see that each extra rung in the ladder contributes a
correction of order $\alpha_sT q_0 r_0/ E m_{\rm q}^2$, in which the
$\alpha_s$ drops out.  Therefore, all these corrections contribute to ${\cal
O}(\alpha_s)$ to the photon rate. Note again that the only parameters of the
QGP that enter this equation are the quark thermal mass $m_{\rm q}$ and the
Debye screening mass, $m_{\rm debye}$. This integral equation was solved
numerically in \cite{ArnolMY2}, and the results are displayed in
Fig.~\ref{fig:LPM-photon}.
\begin{figure}[htbp]
\begin{center}
\resizebox*{!}{7cm}{\rotatebox{-90}{\includegraphics{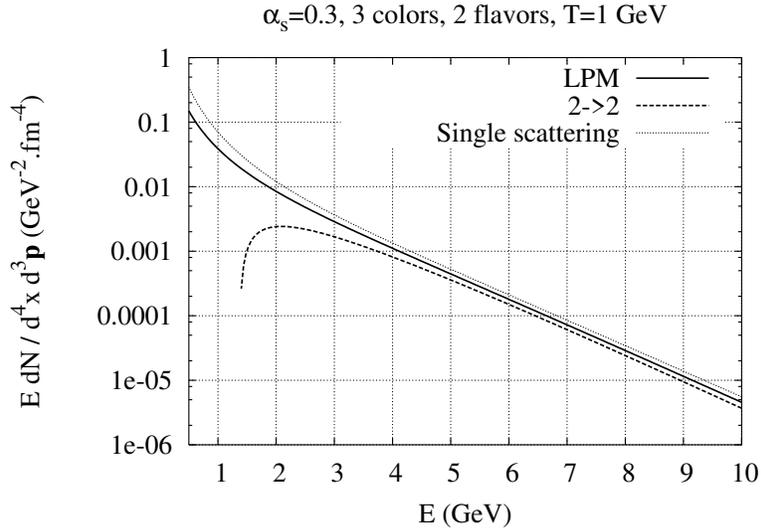}}}
\end{center}
\caption{\label{fig:LPM-photon} ${\cal O}(\alpha_s)$ contributions
  to the photon production rate in a QGP. The parameters used in this
  plot are $\alpha_s=0.3$, 3 colors, 2 flavors and $T=1$~GeV.}
\end{figure}
In this plot, `LPM' denotes the contribution of all the multiple scattering
diagrams, while `$2\to 2$' denotes the processes of figure
\ref{fig:processes-1}. The single scattering diagrams
(Fig.~\ref{fig:processes-2}) are also given so that one can appreciate the
suppression due to the LPM effect.

\subsubsection{Small mass lepton pairs}
\label{thermaldileptonrates}

For the reasons given in Sec.~\ref{dilepton-non-thermal}, it is also
interesting to consider the production of a small mass leptons pair at hard
momentum.  The calculation parallels that described above with added technical
complications due to the mass of the virtual photon (lepton pair). 
The relation between the rate of production of a pair of mass $M$ at 4-momentum
$(E, {\imb p})$ and the virtual photon polarization tensor is
\begin{equation}
  {{dN}\over{dt d^3{\imb x} dE d^3{\imb p}}}=-
  {1\over{12\pi^4}}\; {\alpha \over M^2} \; {1 \over \exp(E/T) -1}\,
  {\rm Im}\,\Pi^{^{R}}{}_\mu{}^\mu(E,{\imb p})\; .
  \label{eq:virtphot}
\end{equation}
Neglecting the mass of the pair compared to its energy, the imaginary part of
the photon polarization tensor at one-loop is still given by
Eqs.~(\ref{eq:rate-1-loop}) but the collinear cut-off (quark thermal mass) in
the logarithmic term now involves the photon mass~\cite{AltheR1,ThomaT1}. The 
two-loop~\cite{AurenGZ3} and the multi-loop~\cite{AurenGMZ1} calculations have
been carried out. In the latter case, it is convenient to separate the
contributions of the  transverse and longitudinal polarizations. For the
transverse case, one obtains  again Eqs.~(\ref{eq:integ-f}), (\ref{eq:AMY})
while  a related integral equation has been derived and solved for the case of
the longitudinal polarization~\cite{AurenGMZ1}.

\vspace{0.5cm}


\subsection{Hard photon production  from  hot out-of-chemical equilibrium QGP}
\label{sec:out-of-chem}
{\em F. Gelis, K. Redlich}

As a first step we consider only the basic reactions for photon production due
to Compton scattering and annihilation of quarks (see
Fig.~\ref{fig:processes-1}). We assume that off-equilibrium effects on the
momentum of QGP constituents can be parameterized by a modification of their
distributions. A HTL-resummation is also supposed to be valid to screen soft
and collinear (mass) singularities.


To account for non-equilibrium effects in partonic medium we follow the
approximations described in detail in \cite{chou,Eijck,LeBellac2}. It
amounts simply to replace the thermal particle energy distribution
$n(k{~}_0)$ by their non-equilibrium counterparts, i.e. in general by
Wigner distributions $n(k{~}_0, X)$. In terms of the cumulant expansion
of the initial density operator, this amounts to approximate the
initial correlations by the second cumulant only, i.e. to neglect all
non-gaussian correlations. As a further approximation, we ignore for
the time being the possible dependence on the center-of-mass
coordinate $X$, essentially assuming a homogeneous and isotropic
medium\footnote{This is a valid assumption if there is a scale
  separation between the {\it macroscopic} scale (i.e. the scale of
  hydrodynamical inhomogeneities, encoded in the variable $X$) and the
  microscopic scale corresponding to the process under study
  \cite{GelisSS1}. The $X$ dependence is reintroduced later
  by making the fugacities $X$ dependent.}. Practically, we take the
following distributions for quarks and gluons:
\begin{equation}
 n_q(k_0) \equiv \left\{
 \begin{array}{c}
  n_q(|k_0|) \\
  1 - n_q(|k_0|))
 \end{array} \right.
 \quad\quad
 n_g(k_0) \equiv \left\{
 \begin{array}{c}
  n_g(|k_0|) \\
  -(1+n_g(|k_0|))
 \end{array} \right.
 \quad\mbox{for}\quad
 \begin{array}{c}
  k_0 > 0 \\
  k_0 < 0
 \end{array}
 \label{25}
\end{equation}
with the J\"uttner parameterizations
\begin{equation}
 n_{g(q)} (|k_0|) \equiv \frac{\lambda}{e^{|k_0|/T} \mp\lambda}\; ,
\label{eq:juttner}
\end{equation}
expressed by introducing the fugacity parameter $\lambda$
\cite{groot}, which is assumed to be energy independent. Obviously
$\lambda \not= 1$ in the case of chemical non-equilibrium.

Neglecting the slow macroscopic scale $X$ compared to the fast
microscopic scale, the structure of the resulting perturbation theory
is very similar to the equilibrium Closed Time Path formalism. For
instance, the $21$ components of the non-equilibrium bare propagators
for bosonic and fermionic fields,
\begin{eqnarray}
 &&iD_{21}(P) = 2\pi \varepsilon(p_0) \delta(P^2) (1+n_g(p_0))\; ,\quad
 iS_{21}(P) =  2\pi\delta(P^2) \slP \varepsilon(p_0)
 (1- n_q(p_0))\; , \label{eq:propfree}
\end{eqnarray}
have formally the same structure as the corresponding ones in
equilibrium, but depend now on the modified distribution functions
(Eq.~\ref{25}).  However, for a dressed propagator the above is not in
general valid, as the appearance of some additional out-of-equilibrium
terms is to be expected \cite{AltheR1,Altherr2,Bedaque}. Therefore, in order to
calculate a physical rate with this formalism, one needs to account not only
for the modifications of the particle momentum distributions but also for
modifications of their propagators.

The rate of real photon emission, already given in
Eq.~(\ref{eq:realphot}) in terms of the retarded photon polarization
tensor, can also be written in terms of its component $\Pi{~}_{12}$ in the Closed
Time Path formalism:
\begin{equation}
 E\frac{dN}{dtd^3\x d^3\p} = \frac{i}{2(2\pi)^3} {\Pi_{12}}{}_\mu^\mu (E,\p),
 \label{62}\; .
\end{equation}
As in the equilibrium case the simplest diagrams, in the loop expansion of
$\Pi{~}_{12}$, correspond to photon production by annihilation and Compton
processes
$
 q + \bar q \to g + \gamma, \;
 q (\bar q) + g \to q (\bar q) + \gamma\; ,
$
as shown in Fig.~\ref{fig:processes-1}. In the off-equilibrium situation one
needs first to determine the appropriate generalization of the effective quark
propagator. In the un-resummed theory it is known that ill-defined terms of the
form $\delta(Q^2) / Q^2$ are generated. However, in the resummed
approach one can show that quark propagator can be approximated by a form
similar to the equilibrium one~\cite{our3,our4}:
\begin{equation}
 S_{12}^{^{HTL}}(Q) = -n_q(q_0)\; {\rm Re}\,\left[
  \frac{i}{\slQ - \Sigma^{^{HTL}}_{_{R}}(q) + i\varepsilon q_0} \right]\; ,
 \label{412}
\end{equation}
where $\Sigma^{^{HTL}}$ is the quark HTL self-energy, evaluated with
non-equilibrium distribution functions.  The only non-equilibrium effect on
this HTL self-energy amounts to a modification of the quark thermal mass $m_q$
(the prefactor in $\Sigma^{^{HTL}}$):
\begin{eqnarray}
  m_{\rm q}^2 &=& \frac{g^2}{\pi^2} C_f \int_0^\infty E dE [n_g(E)
   + n_q(E)]= \pi\alpha_s C_f T^2 {2 \over3} \left(\lambda_g +
\frac{\lambda_q}{2}\right)\; .
 \label{eq:newmq}
\end{eqnarray}
which reduces to the expression given in the previous section in the equilibrium
case ($\lambda{}_g =\lambda{}_q=1$).

The hard photon production rate from a system away from equilibrium can then be
calculated using well established methods in equilibrium field theory. The
final result in the leading order in $\alpha{}_s$ is~\cite{our4} ({\em cf}
Eq.~\ref{eq:rate-1-loop} for the equilibrium result):
\begin{equation}
  E \frac{dN}{dtd^3\x d^3\p}
= e_q^2 \frac{\alpha\alpha_s}{2\pi^2} \lambda_q T^2
   e^{-E/T} \left[ \frac{2}{3}(\lambda_g + \frac{\lambda_q}{2})
  \ln \left(\frac{E T}{m_{\rm q}^2(\lambda_q,\lambda_g)} \right) + \frac{4}{\pi^2} C(E ,
  T, \lambda_q, \lambda_g) \right]\; ,
 \label{eq:resc}
\end{equation}
with
\begin{eqnarray}
  C(E, T, \lambda_q, \lambda_g) &\equiv&  \lambda_q
  \left[ -1 + (1-\frac{\pi^2}{6}) \gamma +
  (1-\frac{\pi^2}{12}) \ln\frac{E}{T}  + \zeta_- \right]
  \nonumber\\
  & & + \lambda_q\lambda_g  \left[ \frac{1}{2} - \frac{\pi^2}{8} +
   (\frac{\pi^2}{4} -2) (\gamma+\ln\frac{E}{T}) + \frac{3}{2} \zeta'(2)
  + \frac{\pi^2}{12}\ln 2 + (\zeta_+ - \zeta_-) \right]
  \nonumber \\ \label{416}
  & & + \lambda_g \left[ \frac{1}{2} + (1-\frac{\pi^2}{3})\gamma +
   (1-\frac{\pi^2}{6} )\ln\frac{E}{T} - \zeta_+ \right]\; ,
\end{eqnarray}
where $\gamma$ is Euler's constant, $\zeta'$ the
derivative of Riemann's function and where we have defined:
\begin{equation}
 \zeta_+ \equiv \sum_{n=2}^\infty \frac{1}{n^2} \ln(n-1) \simeq 0.67, \qquad
 \zeta_- \equiv \sum_{n=2}^\infty \frac{(-)^{n}}{n^2} \ln (n-1) \simeq -0.04\; .
 \label{417}
\end{equation}
The above result shows that, as in the equilibrium medium, the
generalized thermal mass provides a self consistent cut-off for
the logarithmic singularity. The dynamical screening of the mass
singularity seen in Eq.~(\ref{eq:resc}) (here given for the distributions
Eq.~(\ref{eq:juttner}))
actually does not depend on the explicit form of the non-equilibrium
quark and gluon distribution functions. Changing the parameterization
of these functions enters only through the redefinition of the mass
parameter in Eq.~(\ref{eq:resc}) and the constant $C$ in Eq.~(\ref{416}), keeping
at the same time the functional form of the rate unchanged. For
$\lambda{}_q=\lambda{}_g=1$ Eqs.~(\ref{eq:resc}), (\ref{416}) reproduce
the well known one-loop results of the previous section.

The out-of-equilibrium approach discussed above for the case of
Compton and $q\bar{q}$ annihilation processes can also be extended to
the more complicated processes of Fig.~\ref{fig:processes-2}. The main
effect of the fugacities comes from the fact that the number of
emitters is reduced. To take this into account, one just need to
rewrite the statistical functions in Eqs.~(\ref{eq:realphot}) and
(\ref{eq:AMY}) as:
\begin{eqnarray}
\frac{1}{\exp(E/T)-1}[n_q(p_0+E)-n_q(p_0)]=-n_q(p_0+E)n_q(-p_0)\; .
\label{eq:stat-dist}
\end{eqnarray}
Since the equilibrium quark distribution function satisfies
$n_q(-x)=1-n_q(x)$, one sees readily that the right hand side of the
previous formula gives the combination of distribution functions one
expects from kinetic theory in each of the kinematical domains: for
$p_0>0$ (bremsstrahlung of a quark) we get $n_q(p_0+E)(1-n_q(p_0))$,
for $-E<p_0<0$ (quark-antiquark annihilation with a scattering) we get
$n_q(p_0+E)n_q(|p_0|)$, and for $p_0<-E$ (bremsstrahlung of an
antiquark) we get $n_q(|p_0|)(1-n_q(|p_0+E|))$.  Given its
interpretation in terms of kinetic theory, the right hand side of
Eq.~(\ref{eq:stat-dist}) can be generalized\footnote{But it would not
  be correct to use the out-of-equilibrium $n_q$ in the left hand side
  $[n_q(p_0+E)-n_q(p_0)]/(\exp(E/T)-1)$. In fact, the equality of
  Eq.~(\ref{eq:stat-dist}) is only valid with equilibrium
  distributions.} to out of chemical equilibrium situations by simply
replacing the Fermi distribution by the form of
Eq.~(\ref{eq:juttner}). For this to make sense, the out-of-equilibrium
distribution at negative arguments must be defined in such a way that
$n_q(-x)=1-n_q(x)$ holds, which is indeed the case (see
Eqs.~(\ref{25})).

The other places where the fugacities enter in the calculation of the
multiple scattering diagrams is via the quark thermal mass $m_{\rm
q}^2$ and via the Debye mass $m_{\rm debye}^2$. We recall here the expression
for these masses:
\begin{eqnarray}
&&m_{\rm q}^2=\frac{g^2 C_{_{F}}T^2}{6}(\lambda_g+\frac{\lambda_q}{2})\; ,\\
&&m_{\rm debye}^2=\frac{g^2T^2}{3}(N_c\lambda_g+N_f \frac{\lambda_q}{2})\; .
\end{eqnarray}

It is interesting to remark that using fugacities $\lambda_q,\ \lambda_g < 1$
will obviously reduce the photon rate, because the number of emitters is
reduced, however, since the thermal quark mass is also decreased, the
collinear enhancement at work in bremsstrahlung and off-shell annihilation
processes is increased. The two effects partly compensate~\cite{DuttaSMKC}


\subsection{Thermal photon production from hadronic matter}
\label{sec:hotgas}
{\em R. Rapp}


The objective in the following is to assess the emission rate of photons in a
hot hadron gas as expected to be formed under conditions characteristic for LHC
heavy-ion collisions. The rate is calculated acording to
Eq.~(\ref{eq:realphot}).

\subsubsection{$\pi\rho a_1$ gas and form factors}
\label{sec_piroa1}

Early calculations~\cite{Kapusta:qp} of photon radiation from hadronic
matter were guided by the analogue to lowest-order pQCD processes in a QGP,
{\it i.e.}, Compton and annihilation graphs involving the most abundant
constituents in a meson gas. These were $\pi\pi \to \rho\gamma$,
 $\pi\rho\to \pi\gamma$  as well as $\rho\to\pi\pi\gamma$ (and
$\omega\to\pi\gamma$), which are  based on the $\pi\pi\rho$-vertex with
a coupling constant determined from the free $\rho$ ($\omega$) decay.
At photon energies in excess
of $E\simeq$~1~GeV, $t$-channel pion exchange in the second reaction
completely dominates the rate. Shortly thereafter~\cite{Xiong:1992ui},
the $a_1$ $s$-channel pole graph was recognized as a potentially
relevant contribution, due its large coupling to $\pi\rho$ states.
At sufficiently high energies, $s$-channel resonance graphs are,
however, suppressed due to $1/(s-m_R^2)$ powers
in the intermediate propagators ($m_R$: resonance mass),
rendering $a_1$ $t$-channel graphs increasingly important.
A systematic evaluation of photon production from a $\pi\rho a_1$
gas, based on the Massive Yang-Mills (MYM) Lagrangian
approach, has been performed in ref.~\cite{Song:1993ae}.
Here, chiral symmetry provides relations between
couplings, leaving  4 free parameters (including non-minimal
coupling terms) allowing for a reasonable phenomenology for the
free $\pi\rho a_1$ system with electromagnetic decays.
Two different sets implying different
off-shell behaviors lead to total rates that differ by no more than
50\%. The dominant contribution to the rate at photon energies
beyond $\sim$~0.5~GeV was confirmed to be due to $t$-channel ($\pi$ and $a_1$)
exchanges in the $\pi\rho\to\pi\gamma$ reaction~\cite{Song:1993ae}.
A convenient parametrization of the pertinent results, which will be
used below, has been provided in ref.~\cite{Song:1998um}.
\begin{figure}[!ht]
\begin{center}
\includegraphics[width=11cm]{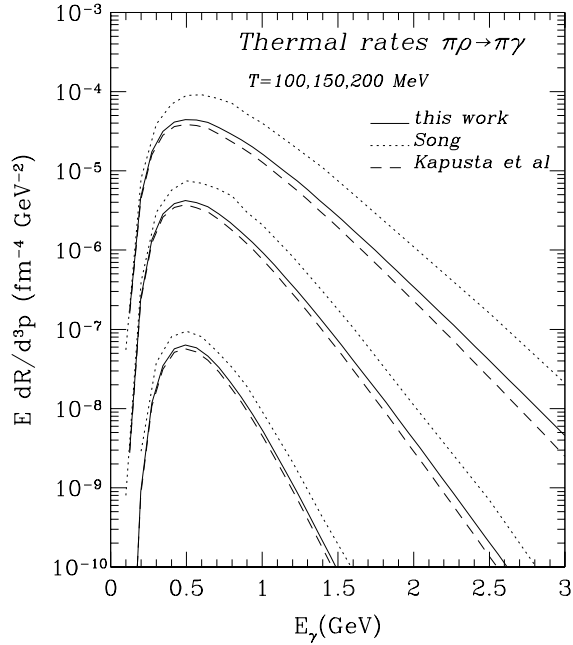}
\caption{Thermal photon production rates from the  $\pi\rho \to \pi\gamma$
reaction within $SU(2)$ chiral lagrangians including vector mesons using
the Massive Yang-Mills~\protect\cite{Song:1993ae} (dotted line) and
Hidden-Local Symmetry~\protect\cite{Halasz:1997xc} (full line).
The dashed line is from the earlier work of
ref.~\protect\cite{Kapusta:qp} which does not include $a_1$ degrees of freedom.}
\end{center}
\label{fig_rate-su2}
\end{figure}

An important element in applying effective hadronic models
at moderate and high momentum transfers is the use of vertex
form factors to simulate finite hadronic-size effects, which are not
accounted for in the $\pi\rho a_1$ calcuations of Ref.~\cite{Song:1993ae}.
For similar reactions their effect has been
studied in Ref.~\cite{Kapusta:qp} and found to give a typical net
suppression over the bare graphs by a significant factor $\sim$3 at
photon energies $E\simeq 2.5$~GeV. We perform a rough estimate of
their effect in the present context as follows. Consider pion-exchange
assuming a standard dipole form factor,
\begin{equation}
F(t)=\left(\frac{2\Lambda^2}{2\Lambda^2-t}\right)^2 \ ;
\end{equation}
the average four-momentum transfer in  $t$-channel
pion exchange can be approximated according to
\begin{equation}
\frac{1}{(m_\pi^2-\bar t)^2} = \frac{1}{4E^2} \int\limits_0^{4E^2}
  dt \ \frac{1}{(m_\pi^2-t)^2} \simeq  \frac{1}{m_\pi^2 4E^2} \ ,
\end{equation}
that is
\begin{equation}
-\bar t \simeq 2 \ E \ m_\pi
\end{equation}
at sufficiently large $E>m_\pi$. Upon multiplying the rate parametrizations
of the $\pi\rho a_1$ gas~\cite{Song:1998um} (dotted line in Fig.~\ref{fig_HGff})
by $F(\bar t)^2$ (with $\Lambda=1$~GeV~\cite{Rapp:1999qu})
we obtain the dashed curve in Fig.~\ref{fig_HGff}.
\begin{figure}[!ht]
\begin{center}
\includegraphics[width=9cm,angle=-90]{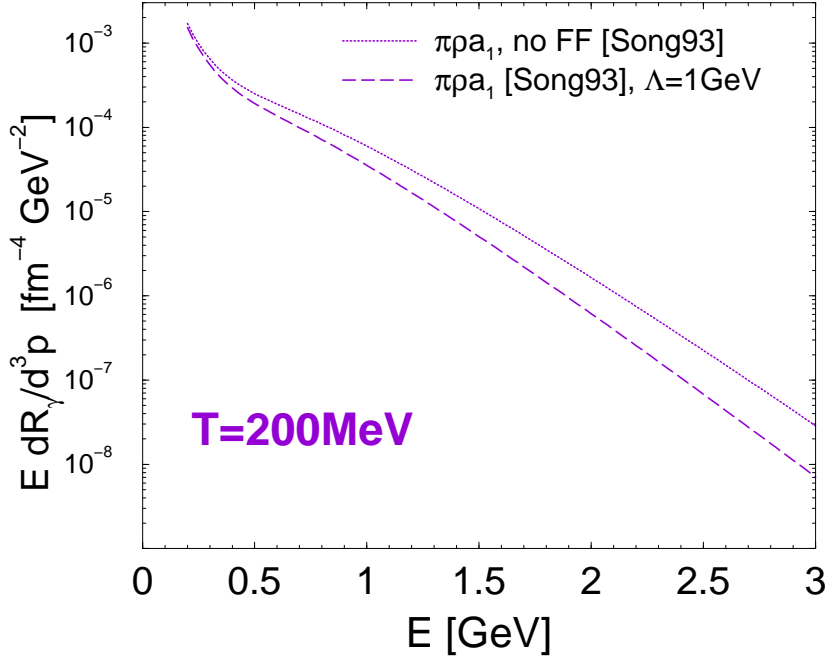}
\end{center}
\caption{Effect of hadronic form factors on the photon production
rates from a $\pi\rho a_1$ gas in the MYM approach of
ref.~\protect\cite{Song:1993ae}.}
\label{fig_HGff}
\end{figure}
The reduction of the rate in the 2-3~GeV region amounts
to a factor of 3-4, quite in line with the exemplary microscopic
calculation of Ref.~\cite{Kapusta:qp}.

We believe that the calculation including formfactors should provide
a reasonable baseline for the $SU(2)$ case.

\subsubsection{Additional mesonic sources}
\label{sec_addmes}

An obvious contribution to photon production not accounted for
in the previous section stems from radiative decays of heavier
mesonic resonances which can not be easily treated in a chiral
framework. In the many-body approach of Refs.~\cite{Rapp:1999qu,Rapp:1999us}
the electromagnetic correlator has been evaluated including all mesonic
resonances up to about 1.4~GeV in mass which exhibit significant
coupling to final states including rho-mesons (and thus,
employing VDM, photons), These are
$\omega(782)$, $h_1(1170)$, $f_1(1285)$, $\pi(1300)$, $a_(1320)$,
$K^*(892)$ and $K_1(1270)$. Pertinent photon rates are displayed
in Fig.~\ref{fig_mesres}.
In particular, the $\omega\to\pi\gamma$ decay generates a large
low-energy strength, consistent with the early results
of Ref.~\cite{Kapusta:qp}. Note that all hadronic vertices carry
(dipole) formfactors with typical cutoff parameters of around
1~GeV, as extracted from an optimal fit to measured hadronic
and radiative branching ratios within VDM (see Ref.~\cite{Rapp:1999qu}
for details). Contributions from still higher-mass mesonic resonances
(such as $\omega(1420)$ and $\omega(1650)$) have recently been
computed~\cite{RG02} and turned out to be negligible.
\begin{figure}[!ht]
\begin{center}
\includegraphics[width=9cm,angle=-90]{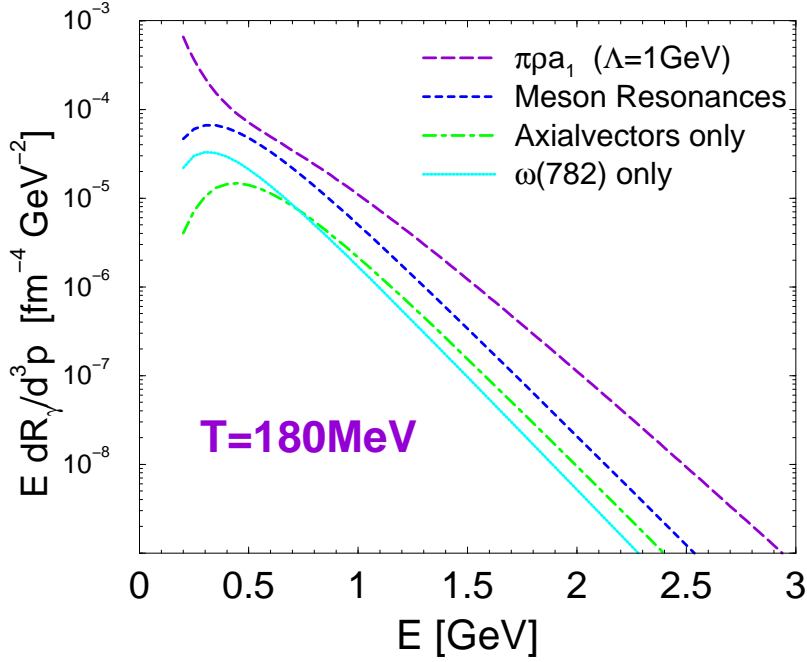}
\end{center}
\caption{Various sources of thermal photon production from a hot
meson gas.}
\label{fig_mesres}
\end{figure}

Another significant photon source can be expected from additional
$t$-channel exchanges, in particular $\omega(782)$ exchange
in $\pi\rho\to\pi\gamma$, and pion-exchange in $\pi K\to \gamma K^*(892)$,
$\pi K^*(892) \to K\gamma$. Results for the former, which do not seem
to be available in the literature yet, have been obtained recently~\cite{RG02},
indicating an emission strength that is small at low energies but exceeds the
in-medium many-body results of ref.~\cite{Rapp:1999qu} beyond
$E\simeq 1.5-2$~GeV.
For the kaon-induced processes, one can use $SU(3)$ symmetry
to estimate the relevant couplings from the analogue $SU(2)$
processes, {\it i.e.}, $\pi \pi\to \gamma \rho$, $\pi \rho \to \rho\gamma$,
and account for the smaller kaon ($K^*$) densities ($n_{K+\bar K}/n_\pi \sim$~40\% 
at maximal SPS energies and beyond).
One finds that the strange analogues make up less than 10\% in the
former case, and about 20\% in the latter.

\subsubsection{Baryonic sources}
\label{sec_bar}
Finally we have to address the role of baryons.
Naively, one might not expect significant contributions at
collider energies where the {\em net} baryon densities are small.
However, since the photon is a $CP$ eigenstate, the relevant
quantity for its production is the {\it sum} of baryon and
anti-baryon densities, which is not small at temperatures
close to $T_c$ even in a {\em net} baryon-free
environment~\cite{Rapp:2000pe}. Moreover, since the corresponding
anti-/baryon content survives the subsequent hadronic evolution in
a heavy-ion collision~\cite{Tserruya:2002yg,Rapp:2002fc}, their
effects on photon and dilepton observables remain appreciable also
in the later stages.

Much like in the mesonic case, baryon-resonance decays ought to be
rather suppressed at energies beyond 1 GeV. No explicit
computation of $t$-channel exchanges has been reported so far. However,
within the chiral reduction formalism, contributions from
nucleons have been assessed in ref.~\cite{Steele:1997tv}.
Within a hadronic many-body framework~\cite{Rapp:1997fs,Rapp:1999us,Rapp:1999qu},
primarily constructed for low-mass dilepton applications,
the e.m.~correlator has been evaluated incorporating a rather
extensive set of baryonic processes.  These results can be
straightforwardly extrapolated to the photon point, and the
corresponding rate is shown by the solid line in Fig.~\ref{fig_bar}.
\begin{figure}[!ht]
\begin{center}
\includegraphics[width=9cm,angle=-90]{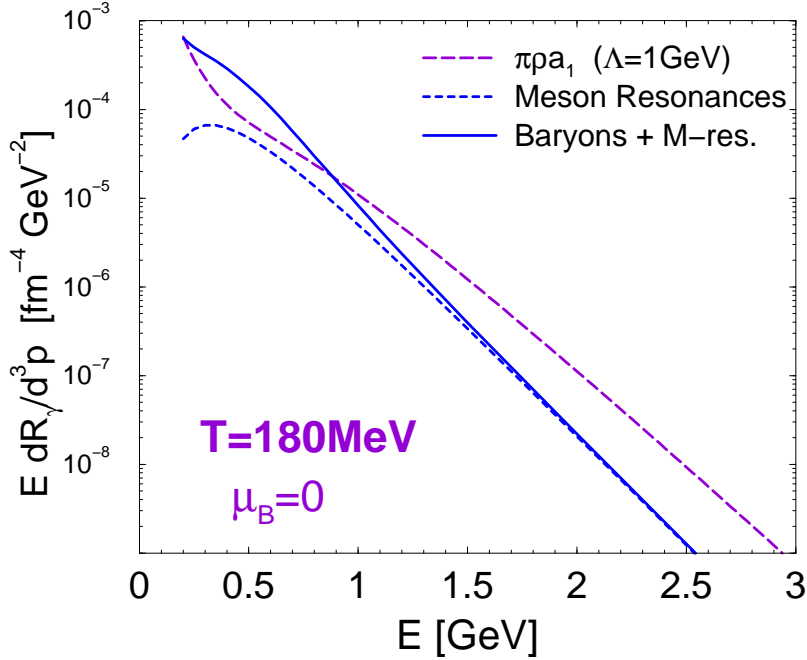}
\end{center}
\caption{Thermal photon production rate (under conditions resembling
LHC energies) for a $\pi\rho a_1$ gas~\protect\cite{Song:1993ae} including
formfactors (long-dashed lines), and within the many-body spectral function
approach of ref.~\protect\cite{Rapp:1999us} (full line). The total emissivity
corresponds to the sum of the two rates; $\omega$ $t$-channel and strange-meson
processes, not included in the figure, are expected to induce an additional increase
of about 30-40\%  at energies above 1.5~GeV.}
\label{fig_bar}
\end{figure}
More specifically,
both the mesonic resonance contributions as discussed above
(short-dashed line), as well as baryonic
resonance decays (most notably $\Delta(1232)$, $N(1520)$)
and pion-exchange processes involving nucleons and deltas,
are included coherently. The baryon-induced emission can be considered as
rather reliable for baryon densities up to at least normal nuclear
matter density, $\rho_0=0.16$~fm$^{-3}$, being  constrained by photoabsorption
spectra on nucleons and nuclei~\cite{Rapp:1997ei} (at zero temperature).
At comparable baryonic densities this approach yields about a factor
of two more photons than what has been found in the chiral
reduction approach~\cite{Steele:1997tv} when coupling on-shell pions and
nucleons (see Ref.~\cite{Steele:1999hf} for an update
including $\Delta(1232)$ and $N(1520)$ resonances).

Under conditions relevant for LHC energies (as chosen in the figure),
processes involving (anti-) baryons are the prevailing source of thermal
photons  at energies  $E< 1$~GeV
(similar to what has been found for low-mass dileptons~\cite{Rapp:2000pe}).
Beyond, the $t$-channel processes within the light meson gas sector
take over, dominating the rate at higher energies.

It is thus suggestive to regard the combination of the $\pi\rho a_1$
emission rate (including form factor effects), supplemented with $\omega$
$t$-channel and kaon-induced proceses, with the hadronic many-body results
($a_1$ contribution removed) as a semi-realistic
approximation to the full hadron gas emissivity.


\subsection{Particle production from primary interactions in nuclear
collisions}
\label{sec:part-prod}
{\it K.J. Eskola, J.~Ranft, P.V. Ruuskanen}

Results from model calculations of initial particle production are the
average numbers, transverse energies and momentum spectra of produced
gluons and quarks.  How to obtain from the momentum space densities the
space-time densities needed for the hydrodynamic calculation is
explained in detail in Sec.~\ref{sec:inicond} using the pQCD +
saturation calculation of (mini)jet production \cite{Eskola:2002qz} as
an example.

The essential quantities in determining the initial state for the
hydrodynamic expansion are (i) the total multiplicity $dN/dy$, (ii) the
initial time scale $\tau_0$ and (iii) the shape of the transverse
distribution.  Production models differ on all these and our aim in
discussing model calculations is to estimate what are the present
uncertainties in these quantities.  We can then vary the initial
conditions accordingly to explore the range of uncertainty in predicting
the thermal photon spectrum in the nucleus--nucleus collision at LHC.

In this subsection, we consider different approaches to primary
production:  (i) pQCD + final state saturation
(minijet) \cite{Eskola:2002qz} and
(ii) initial state saturation, the colour glass condensate model
\cite{Gribov:tu}, which both describe the produced matter as a
parton system.  We also discuss the string based model \textsc{Dpmjet}, of Ranft and
collaborators, already presented in Sec.~\ref{sec:dpmjet} (see
refs.~\cite{dpmjet1,dpmjet2,fzic,barystop-b}).

\subsubsection{Perturbative QCD + saturation model}
\label{pQCDsat}

In the pQCD + saturation model \cite{Eskola:2002qz} one assumes that at
collider energies the production of final state particles is dominated
by collinearly factorized parton (minijet) production above a momentum
scale, $\psat \gg \LQCD$, determined from an assumption that saturation
reduces parton production below this scale and gives only a minor
contribution which can be effectively included in the contribution from
partons above $\psat$.

The numerical minijet calculation \cite{Eskola:1988yh,Eskola:1996ce}
gives the cross sections $d\sigma_i(\Delta y)/dp_{_T}$ in the rapidity
interval $\Delta y$ for parton species $i=g,q$ and $\bar q$.  For
$\Delta y\lsim 1$ (around $y=0$), the cross section is proportional to
$\Delta y$ and the differential cross section can be defined as
\be
\frac{d\sigma}{dydp_{_T}^2}=\frac{1}{\Delta y}
\frac{d\sigma(\Delta y)}{dp_{_T}^2}
= \frac{1}{\Delta y}\int_{\Delta y} dy_1 \int dy_2 \sum_{ijkl}
f_{i/A}(x_1,Q^2) f_{j/A}(x_2,Q^2)
\frac{d\hat\sigma}{d\hat t}^{ij\rightarrow kl}
\label{eq:sigmaDeltay}
\ee
with nuclear parton distributions $f_{i/A}$, momentum fractions
$x_{1,2}$, scale $Q=p_{_T}$ and the partonic cross sections $d\hat\sigma$.
The basic quantities for the calculation of initial densities in
nucleus--nucleus collisions (cf. Sec.~\ref{sec:inicond}) are the
integrated and $p_{_T}$-weighted cross sections
\bea
\int_{p_0}^\infty dp_{_T}\frac{d\sigma_i}{dydp_{_T}} &=&
\frac{1}{\Delta y}\int_{p_0}^\infty dp_{_T}\frac{d\sigma_i(\Delta y)}
{dp_{_T}}=\frac{1}{\Delta y}\sigma_i(p_0,\Delta y) =
\frac{d\sigma_i(p_0)}{dy} \,,\\
\label{eq:sigmajet}
\int_{p_0}^\infty dp_{_T} p_{_T}\frac{d\sigma_i}{dydp_{_T}} &=&
\frac{1}{\Delta y}\int_{p_0}^\infty dp_{_T} p_{_T}
             \frac{d\sigma_i(\Delta y)}{dp_{_T}}=
\frac{1}{\Delta y}\sigma_i\langle E_{_T}\rangle(p_0,\Delta y) =
\frac{d\sigma_i\langle E_{_T}\rangle(p_0)}{dy} \,.
\label{eq:sigmaET}
\eea
When multiplied with the nuclear overlap function $T_{AA}$ (see
Eq.~\ref{eq:nuc_overlap} and appendix 1 for further details),
$\sigma_i(p_0,\Delta y)$ gives the average number and $\sigma_i\langle
E_{_T}\rangle(p_0,\Delta y)$ the average transverse energy of partons $i$
in a rapidity interval $\Delta y$ with transverse momenta above $p_0$.
The number of partons is well-defined only in the leading order, at
higher orders it is not an infrared safe quantity.  The transverse
energy, however, is well-defined, and it has been computed to
next-to-leading order in \cite{Eskola:2000ji,Eskola:2000my}.  The factor
$K=1.6$ from such a computation for the LHC energy\footnote{$K$ depends
on $\sqrt{s}$, scale choice and PDFs.} is used to estimate the
effect
of the NLO contributions both to the number and the transverse energy of
produced partons.  In what follows, the number of partons is used in
estimating the chemical composition of produced partons but the initial
energy is determined from the produced total transverse energy alone.
Nuclear shadowing, as given by the EKS parametrization
\cite{Eskola:1998df,Eskola:1998iy}, is included in the calculation of
these cross sections.

To close the calculation, the cut-off momentum $p_0$ for central $AA$
collisions is fixed as the saturation scale, $\psat$, from a geometric
condition
\be
T_{AA}(0)\frac{\sigma(\psat,\Delta y)}{\Delta y}\frac{\pi}{\psat^2} =
\pi R_A^2\,,
\label{eq:sat}
\ee which is solved numerically. We fix $\Delta y=1$.  The condition
above can be interpreted as giving the parton density at which the
produced partons begin to overlap leading to fusions which suppress
the further production of softer partons.  The condition can also be
interpreted as the beginning of overlap of primary interaction volumes
that can also be expected to suppress the further yield of partons.

Once the saturation momentum $\psat$ has been obtained from the
condition above, the initial production cross sections
$\sigma_i(\psat,\Delta y)$ and $\sigma_i\langle E_{_T}\rangle(\psat,\Delta
y)$ can be calculated.  Although the pQCD+saturation model is an
effective (non-microscopic) approach to describe the bulk of the
initial parton production at central rapidities of nearly central $AA$
collisions, it provides enough information to fully determine the
initial state for the hydrodynamic evolution if the initial time is
given.  With the hydrodynamic evolution \cite{Eskola:2001bf} it has
correctly predicted the multiplicities in central Au+Au collisions at
RHIC energies $\sqrt{s}=56$, 130 and 200 GeV
\cite{Eskola:2001bf,Eskola:2001dd}, and also given a successful
description of the transverse momentum spectra of pions, kaons and
(anti)protons \cite{Eskola:2002wx}.

Using the notation $\sigma(1)=d\sigma/dy$, the integrated partonic
cross sections at $\sqrt {s}=5.5$ TeV, the energy of lead--lead
collisions at LHC, are $\sigma_g(1)=135$ mb, $\sigma_q(1)=6.46$ mb,
$\sigma_{\bar q}(1)=6.14$ mb, while the $p_{_T}$-weighted cross section
becomes $\sigma\langle E_{_T}\rangle(1)=468$ mb\,GeV. The saturation
scale is $\psat=2.03$~GeV.  Below in Sec.~\ref{sec:inicond},
initial energy density and initial values of gluon and (anti)quark
fugacities are constructed from these cross sections.

\subsubsection{ Initial state parton saturation and color glass
condensate
}  \label{sec:CGC}

A much discussed theoretical approach to particle production in
heavy-ion collisions at collider energies has been based on the
assumption that the initial state parton densities saturate and non-linear
dynamics becomes dominant \cite{Gribov:tu}.  This means that the parton
(gluon) densities in a nucleus (or in a proton) at small $x$ are so
high, that gluons interact coherently.  From the point of view of the
color fields, the high density or large occupation numbers of the field
quanta can be described as the formation of a "color glass condensate"
which can be treated in terms of a classical effective field theory
\cite{McLerran:1993ni,McLerran:1999hj}.  Quantum corrections (to the
classical parton distributions of a nucleus) have also been considered
\cite{Kovchegov:1996ty,Jalilian-Marian:1996xn}; see also
\cite{Kovchegov:2001sc,Kovchegov_YR} and references therein.

Applicability of saturation models can be tested, and the saturation
scales extracted, in deep inelastic $lA$ scattering. For the free
proton a geometric scaling of the structure function $F_2$ at small
values of $x$ has been found
\cite{Golec-Biernat:1998js,Golec-Biernat:1999qd,Stasto:2000er}.
Similar analyses have also been performed for the nuclei
\cite{Freund:2002ux,Armesto:2002ny} but due to the limited amount of
nuclear DIS data available, it is not yet fully clear how well the
assumption of high (saturated) parton density is met in heavy-ion
collisions with the nuclei available in nature and at the highest
collider energy available currently at RHIC or in the near future at
the LHC (see also \cite{Gyulassy:1997vt}).

In the effective field theory approach to gluon production in $AA$
collisions, when boost invariance is assumed, it becomes possible to
choose the gauge in such a way that the problem can be formulated as a
dimensionally reduced 2+1--dimensional theory.  After this the problem
is amenable to numerical approach \cite{Krasnitz:1998ns} utilizing
lattice regularization.  For analytical studies of primary gluon
production in $AA$, see \cite{Kovchegov:2000hz} and references therein.

The earlier numerical calculations with lattice regularization were
performed using SU(2) symmetry \cite{Krasnitz:1999wc} and cylindrical
nuclei and the treatment of colour neutrality was not yet adequate
\cite{Krasnitz:2000gz}.  Recently Krasnitz {\it et al.}\ have formulated
the calculation using SU(3) \cite{Krasnitz:2001qu,Krasnitz:2002mn}, more
appropriate nuclear geometry for spherical nuclei, and also imposing
local colour neutrality in the transverse overlap region of collision
\cite{Krasnitz:2002mn}.  Local colour neutrality leads to a rapid
decrease of colour field strength outside the nucleus and to more
a realistic treatment
of multiplicity and transverse energy production.  However, the average
transverse momentum, obtained in this model, is too large and has
remained a puzzle for several years.  Quite recently, Lappi has solved
this problem, and a factor $\sim 2$ reduction in the ratio $E_{_T}/N$ has
been discovered \cite{Lappi:2003bi}.

The lattice approach does not, however, give a value for the saturation
scale (or the colour source density $\mu$) itself; the overall
normalization must be obtained from elsewhere.  For RHIC phenomenology,
the authors of \cite{Krasnitz:2002mn} suggest two sets of results which
now in light of the latest results of \cite{Lappi:2003bi} lead to
following, qualitatively different descriptions of the final state:
\begin{itemize}
\item
In the case of a smaller scale $\mu$, the total transverse energy
$E_{_T}\sim g_s^4R_A^2\mu^3$ produced from the classical fields
approximately equals the measured result, whereas the number of
initially produced partons ($N\sim g_s^2R_A^2\mu^2$) is only $\sim$half
of the multiplicity of hadrons measured in the experiment.  In this case
the only evolution in the final state would be the fragmentation of
partons to a couple of hadrons on the average.  In this picture, which
corresponds to the scenario suggested by Kharzeev, Levin and Nardi
\cite{Kharzeev:2001gp}, one would expect the photon and lepton pair
emission after the primary interactions to be very rare.
\item For a larger saturation scale, the number of partons is close to
the measured number of hadrons but the initially produced transverse
energy is about 2.5  times bigger than the measured one \cite{Lappi:2003bi}.
In this case, production must be followed by strong collective expansion,
initially dominantly in the longitudinal direction, which allows for a
transfer of energy into the longitudinal motion. This case corresponds to
the evolution suggested by pQCD+saturation+hydrodynamics
\cite{Eskola:2001bf}.
\end{itemize}

The two cases described above represent the extreme ends of the
possibilities for the final state and real collisions can be somewhere
between these two.  In general, the production of partons would be
followed by some degree of fragmentation, then by thermalization stage
and finally by the hydrodynamic expansion.  How thermalization may occur
is studied e.g. in
\cite{Eskola:1988hp,Shuryak:wc,Mueller:1999pi,Mueller:1999fp,Bjoraker:2000cf,Baier:2000sb,Baier:2002bt}.
In Sec.~\ref{sec:inicond}, when discussing the range of initial
conditions for the hydrodynamic expansion, we take, at fixed
multiplicity, these alternative scenarios to indicate different
choices of the initial time for the hydrodynamic expansion.

In the model of Kharzeev, Levin and Nardi \cite{Kharzeev:2001gp} the
partonic final state at RHIC energies corresponds to the first
alternative of Krasnitz, Nara and Venugopalan described above.  Like in
the original form \cite{Gribov:tu}, parton saturation is formulated in
the initial state in terms of unintegrated gluon distributions.  The
rapidity distributions of produced gluons are computed as $2\rightarrow
1$ gluon collisions and they are found to be proportional to the
integrated gluon distribution and the number of participant nucleons.
In this scenario, the produced gluons just fragment into a couple of
hadrons and there is no strongly interacting medium, no collectivity nor
any energy transfer from transverse to longitudinal motion:  the
pseudorapidity distribution of final hadrons is directly that of
produced gluons.  With negligible amount of secondary collisions, photon
and lepton pair emission after primary interactions should be
negligible.  (This, of course does not concern the decays of final
hadrons which produce plenty of photons and lepton pairs.)

The energy dependence in the model \cite{Kharzeev:2001gp} enters through
the dependence of the saturation scale on the center-of-mass energy,
obtained in
\cite{Golec-Biernat:1998js,Golec-Biernat:1999qd,Stasto:2000er} from DIS
of the free proton.  Also the number of participants is $\sqrt s$
dependent.  The overall normalization in \cite{Kharzeev:2001gp} is,
however, left free.  If one assumes that the ratio of the multiplicity
of final hadrons to that of the initially produced gluons is energy
independent, the normalization can be fixed at, say, the RHIC energy
$\sqrt{s}=130$ GeV, and predictions can be made for $dN_{AA}/d\eta$
for final hadrons at other energies.  In this way, the model
\cite{Kharzeev:2001gp} predicts the charged particle multiplicity for
central Pb+Pb at $\sqrt{s}=5.5$~TeV to be $dN_{\rm ch}/d\eta\sim 2200$.
This number is somewhat smaller than in the pQCD+saturation model
\cite{Eskola:2002qz,Eskola:2001bf} and it is included in the
multiplicity range considered in Sec.~\ref{sec:inicond}.

\subsubsection{Particle production in \textsc{Dpmjet} dual parton model.}

In the \textsc{Dpmjet} Monte Carlo model all hadrons are produced from soft and
hard chains, soft chains resulting from soft pomeron exchange and hard chains
(jets and minijets) from (LO) perturbative QCD.  The Dual Parton Model
(\textsc{Dpmjet}) is treated in detail in Sec.~\ref{sec:dpmjet}. The assumption
is that the initial energy density is to be defined by the soft and hard chain
end partons before they decay into hadrons and not by the  final state hadrons as
in the original Bjorken formula\cite{Bjorkenf}. The information on the
transverse location and the energy-momentum 4--vectors of the chain-end partons
are stored for each event.  Since \textsc{Dpmjet} describes the evolution of
the final state in the momentum space, a connection between the rapidity $y$
and the longitudinal space-time rapidity $\eta$, as well as the production
timescale, must be specified.  Below, when discussing the initial conditions
for the hydrodynamical evolution in detail, we compare the energy densities
from \textsc{Dpmjet} and the pQCD + saturation calculation.

\subsection{Hydrodynamic expansion of thermal matter}
\label{sec:hydrodynamics}
{\it H. Niemi, P.V. Ruuskanen, S.S. R\"as\"anen, K. Redlich, D.K.~Srivastava}

The dynamic evolution of the expanding matter in local thermal
equilibrium is governed by the hydrodynamic equations
\be
\partial_\mu T^{\mu\nu} = 0
\ee
which express the energy and momentum conservation as they are
transferred
by pressure gradients and flow from one region to another during the
expansion.  The energy--momentum tensor
\be
T^{\mu\nu} = (\epsilon+p)u^\mu u^\nu-pg^{\mu\nu}
\ee
is expressed in terms of
$\epsilon$, the energy density, $p$, the pressure, and $u^\mu$, the
4--velocity of flow.  At LHC energy the net baryon number in the central
rapidity region is very small and will be neglected in the following.

At LHC the total rapidity interval is large and, over a few units, the
central rapidity region is anticipated to be relatively flat and it
should be reasonable to impose boost invariance in the central rapidity
region.  The rapidity of flow is then equal to the space-time rapidity,
$\eta_{\rm flow}=\eta= (1/2)\log[(t+z)/(t-z)]$, not only for the initial
conditions but throughout the expansion.  We will also restrict the
calculations to zero-impact-parameter collisions with azimuthal
symmetry.  The densities and the transverse flow velocity depend then
only on $\tau=\sqrt{t^2-z^2}$ and $r$ and can be expressed as
$\epsilon=\epsilon(\tau,r)$ and $v_r(\tau,r)$.  This simplifies the
equations greatly and numerical solutions are easily obtained once the
distributions $\epsilon (\tau{}_0,r)$ and $v_r(\tau{}_0,r)$ are fixed at
initial time $\tau_0$ and the Equation of State (EoS) is specified.  In
all calculations, the initial transverse velocity is taken to be zero,
$v_r(\tau{}_0,r)=0$.

\vspace{0.5cm}

\subsubsection{The Equation of state}
\label{sec:EoS}
\def\aa{\pi^2/90}
\def\aaf{\frac{\pi^2}{90}}

To solve the hydrodynamic equations an Equation of State (EoS) has to
be specified. We study the expansion using different EoSs. We
investigate the expansion and the photon emission also when
quarks and gluons are not in chemical equilibrium. The deviation from
chemical equilibrium is described using a multiplicative fugacity for
gluons and (anti)quarks. For the collisions at LHC we assume that the
net baryon number is zero.


We mainly use an EoS which describes the high temperature phase as an
ideal gas of massless quarks and gluons both in kinetic and chemical
equilibrium.  The low temperature phase is a gas of all hadrons and
hadron resonances with masses below 1.4 GeV interacting with a repulsive
mean field.  This, as well as a bag constant, is needed for consistent
treatment of transition between the two phases as a first order phase
transition.  With $N_f=3$ the bag constant $B$ and the mean field
constant $K$ are chosen to be $B^{1/4}=243$ MeV and $K=450$ MeV fm$^3$ giving
$T_c=167$ MeV for the transition temperature.  In
Ref.~\cite{Sollfrank:1997} this is called the EoS$\,$A.

\def\lg{\lambda_g}
\def\lq{\lambda_q}
\def\lqbar{\lambda_{\bar q}}

The expected gluon dominance in parton-based approaches of primary
production means that even if the produced particles are close to
kinetic equilibrium, they will be far from chemical equilibrium.  We
describe this by using multiplicative fugacity factors in the EoS.  The
equilibrium energy density
\be
\epsilon = \epsilon_q+\epsilon_{\bar q}+\epsilon_g+B
\ee
with $\epsilon_i(T)=3a_iT^4$, $a_g=16\aa$ and
$a_q=a_{\bar q}=(21/4)N_f\aa$ is then replaced by
\be
\epsilon(T,\lambda_i)=\lambda_q\epsilon_q+
\lambda_{\bar q}\epsilon_{\bar q}+\lambda_g\epsilon_g+
B(T_c,\lambda_i)\,.
\label{eq:eps_fugacities}
\ee
Since the net baryon number will be small at central rapidities as indicated
also by the result $\sigma_q\simeq\sigma_{\bar q}$ from the minijet calculation
at the LHC energy, we will take $\lq=\lqbar$.

A similar expression can be written for the pressure showing explicitly
that the relation $\epsilon^{\rm th}=3p^{\rm th}$ remains unchanged for
the thermal parts of pressure and energy density.  At large
temperatures, say $T>2T_c$, the thermal parts dominate, indicating that
the evolution of the flow at high temperatures is unaffected by
fugacities becoming different from~1.  However, the relation between the
temperature and the energy density depends strongly on fugacities:
small fugacities indicate a decrease in the number of effective degrees
of freedom and an increase in the temperature at fixed energy density.
This has a large effect on the photon emission rates.


As discussed above, the value of the bag constant $B$ is chosen so that $T_c=167$
MeV in chemical equilibrium.  Obviously this can also be done with fugacities
different from 1. In this case $B$ depends not only on $T_c$ but also on the
fugacities.  As there are no counterparts of fugacities in the hadron gas we
take the phase transition temperature $T_c$ to be a constant, independent of
fugacities.  This avoids the ambiguity of having a phase transition to hadron
gas at different temperatures.  Once $T_c$ is fixed, the bag constant will
depend through the Gibbs criterion on fugacities only:
\be
p_{\rm HG}(T_c) = p_{\rm QGP}^{\rm th}(T_c, \lambda_i) -
B(\lambda_q,\lambda_g)\,.
\label{eq:Gibbs_criterion}
\ee
It turns out that the evolution of fugacities is fast enough
so that, in the central region of the transverse plane, they approach values
close to 1 before the temperature reaches $T_c$. This supports the
choice of constant $T_c$.

This kind of description of phase transition cannot be made with arbitrarily
small fugacities.  As will be discussed in the next subsection, a
straightforward application of the minijet results for the parton cross sections
would lead to very small values of fugacities at the edge of the nuclei in the
transverse plane.  To avoid a situation where the effective number of degrees of
freedom in the parton matter becomes so small that the phase transition from the
hadron gas to parton matter would not occur at all, we have to limit the values
of fugacities from below.  The numerical values of these limits depend on the
details of the hadron gas phase.  If the hadron gas consists of all hadrons and
hadron resonances with masses below $1.4$ GeV, we take limits to be $\lg\ge0.60$
and $\lq\ge0.17$.

\begin{figure}
\begin{minipage}[c]{6.6cm}
\centerline{\includegraphics[width=6.0cm]{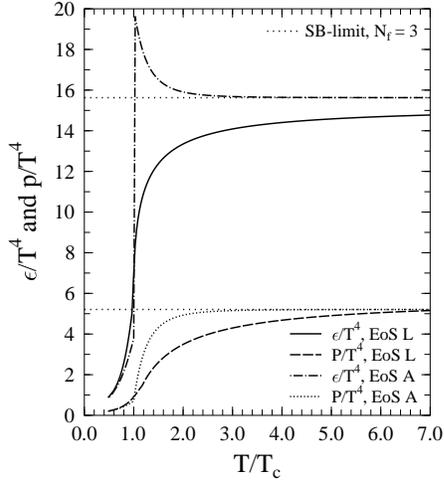}}
\end{minipage}
\hfill
\begin{minipage}[c]{8.6cm}
      \caption{Energy density and pressure as function of temperature
               for EoS$\,$A and a parametrization of lattice data.}
\label{fig:EoS}
\end{minipage}
\end{figure}

\def\EoSL{EoS$\,$L}
\def\EoSA{EoS$\,$A}

We also consider the expansion using results for the EoS from lattice
calculations.  As a starting point we use a parametrization of lattice
results on the pressure and energy density as a function of temperature
at $T\ge T_c$~\cite{Schneider:2001nf}.
We join this parametrization with the EoS of Hadron
Resonance Gas (HRG) at $T_c$.  We constrain the parametrizations to be
consistent with the thermodynamic relations $s=(\epsilon+p)/T=\partial
p/\partial T$.  The resulting parametrization, \EoSL\, is shown in
Fig.~\ref{fig:EoS} for the energy density and the pressure together with
$\epsilon$ and $p$ in the \EoSA.\ Since for the energy density the
ideal-gas Stefan--Boltzmann limit is reached from below on the lattice,
the temperature at given energy density is higher than in the case of
\EoSA\ when the limit is reached from above.  The ratio of energy
density to pressure in plasma is quite similar in both cases when $T$ is
not in the vicinity of $T_c$.

It turns out that the high temperature region completely dominates the
total photon emission from the plasma.  For this reason the difference
in the emission for \EoSA\ and \EoSL\ comes from the difference in the
temperature when the calculation is done starting from initial
conditions with the same energy density.  As will be seen below, the
difference in the spectrum is visible but not very significant.

\vspace{0.5cm}

\subsubsection{Initial conditions for hydrodynamics}
\label{sec:inicond}
\def\d2r{{{\rm d}^2{\bf r}}}
\def\dNch{{dN_{\rm ch}/dy}}

Models of particle production are formulated in momentum space.  To
obtain the initial conditions needed to initialize the
hydrodynamical calculation, we must correlate the formation time
$\tau{}_0$ and the position with the momenta of produced quanta.

At high collision energy the incoming nuclei are strongly Lorentz
contracted and the time it takes for them to pass through one another is
short.  The collision region can then be approximated as a plane
perpendicular to the beam direction, $z=0$, at time $t=0$.  Particles
produced in the collision will move with a longitudinal velocity
$v_z=z/t$ or rapidity equal to the space-time rapidity,
$y=\eta=(1/2)\log[(t+z)/(t-z)]$.  Combining this with the formation time
and incorporating the nuclear geometry allows us to calculate the
produced energy in a volume element $dz\d2r=\tau_0d\eta\d2r$ from the
momentum spectrum of produced particles.

Depending on the approach, nuclear geometry enters the calculation
somewhat differently.  In the pQCD + saturation calculation, based on
factorization of hard processes, the momentum distribution of produced
partons per unit transverse area is
\be
\frac{1}{\d2r}\frac{dN(r)}{dydp_{_T}}=
\frac{d\sigma}{dydp_{_T}} [T_A(r)]^2\,,
\ee
where the nuclear thickness function $T_A(r)$ is given in terms of
$\rho_A(z,r)$, the nucleon density in the nucleus $A$, as the integral
$T_A(r)=\int dz\rho_A(z,r)$.  Nuclear geometry enters here explicitly
as the factor $[T_A(r)]^2$ expressing the nucleon-nucleon luminosity per
unit transverse area.  Nuclear size appears also in the calculation of
the cross section both through the saturation condition~(\ref{eq:sat})
and nuclear shadowing in the structure functions.  Since $E_{_T}=p_{_T}$ for
massless partons, the energy density of produced partons in a volume
element with the longitudinal size $\tau_0d\eta=\tau_0dy$ is obtained as
the $p_{_T}$-weighted integral of the parton distribution:
\be
\epsilon(\tau_0,r) = \frac{1}{\tau_0\d2r}
\int dp_{_T}\,p_{_T}\frac{dN(r)}{dydp_{_T}}\,.
\ee
Using Eq.~(\ref{eq:sigmaET}),
we obtain from the pQCD + saturation model
\be
\epsilon(\tau_0,r) = \left(\int_{\psat}^\infty
dp_{_T}\,p_{_T}\frac{d\sigma}{dydp_{_T}}\right)\frac{[T_A(r)]^2}{\tau_0}
= \frac{1}{\Delta y}\sigma\langle E_{_T}\rangle(\psat,\Delta y)
            \frac{[T_A(r)]^2}{\tau_0}\,.
\label{eq:iniepsilon}
\ee
This formula shows that in the minijet calculation the production is
proportional to the number of binary collisions per unit transverse
area.


As mentioned above, in the \textsc{Dpmjet} Monte Carlo model the initial energy
density is defined in terms of the soft and hard chain-end partons before
they decay into hadrons.  A grid is introduced in the transverse plane of
the collision and the Bjorken formula is used to calculate the transverse
energy of chain-end partons separately in each of these transverse space
bins.  This is done separately in bins of different transverse momentum
using formation time in the Bjorken formula which depends on the transverse
momentum, $\tau\approx 1/p{}_T$. The formation time is normalized to $\tau=1$
fm/c in the lowest $p_{_T}$ bin ($0\le p{}_T\le 0.8$ GeV/c), which contains
essentially all the soft chain ends.

The total energy is obtained by summing the contributions from different
transverse energy bins.  In Fig.~\ref{fig:eps_comp} we compare the
energy densities from \textsc{Dpmjet} and the pQCD + saturation calculation.  At
the maximum the \textsc{Dpmjet} result is 15 \% below that from the minijet
calculation and within the uncertainties of the models they agree quite
well.  It should be mentioned that the \textsc{Dpmjet} calculation gives also the
rapidity distribution over the whole rapidity range, whereas the minijet
calculation cannot be extended over more than a few units at central
rapidities.  Since our hydrodynamical calculation is based on boost
invariant flow in the longitudinal direction, we cannot utilize the
rapidity dependence of the \textsc{Dpmjet} result.

\begin{figure}
\begin{minipage}[c]{7.6cm}
\centerline{\includegraphics[width=7.0cm]{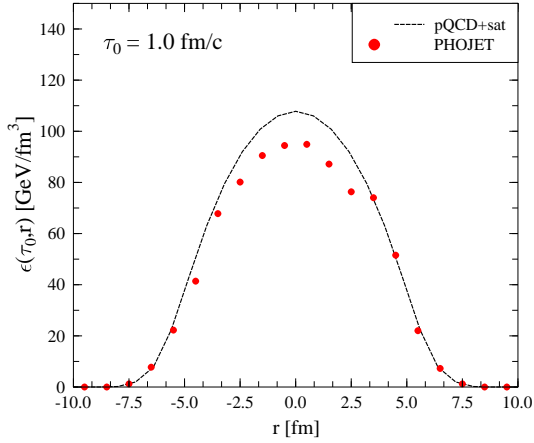}}
\end{minipage}
\hfill
\begin{minipage}[c]{7.6cm}
      \caption{
      Transverse dependence of the energy density $\epsilon(\tau,r)$ at time
      $\tau=1.0$ fm/c from the pQCD+saturation model compared
      with results from the PHOJET calculation (circles)  based on dual
      parton string model. }
\label{fig:eps_comp}
\end{minipage}
\end{figure}

In the models based on initial state saturation, production is related
to the number of participants which for central, zero-impact-parameter
collisions is proportional to $T_A(r)$. Details in implementing the
production dynamics may modify the transverse dependence but we will
take the binary collision density $[T_A(r)]^2$ and the participant
density $T_A(r)$ to represent the range of variation in the {\it shape}
of transverse distribution of initial energy density.

We next consider the choice of initial time $\tau_0$ for the
hydrodynamic evolution.  In models based on parton production, the
formation time scale of primary partons is the inverse of the saturation
momentum, $\tau_0\sim 1/\psat$.  However, as was discussed earlier, the
time needed for the system to thermalize, $\tau_{\rm th}$, can be longer
than the production time of partons
\cite{Mueller:1999fp,Bjoraker:2000cf,Baier:2000sb}, especially if the
primary production is followed by fragmentation.  Usually it is argued
to indicate that thermal and hydrodynamical description of expansion
cannot be started earlier than at $\tau_{\rm th}$.  On the other hand,
collisions drive the system towards equilibrium and they cannot be
ignored since they contribute both to the buildup of collective motion
and the emission of photons and lepton pairs even before full
thermalization has been achieved.  Applying hydrodynamics with initial
time $\tau_0\sim\tau_{\rm prod}<\tau_{\rm th}$ means assuming that it
provides a reasonable approximation for treating the effects due to
collisions which drive the matter to thermal equilibrium.  Especially,
the high transverse momentum photons might otherwise be underestimated
since the collision energies of partons are largest at the earliest
stage.

In the pQCD + saturation model the assumption of early thermalization
can be supported with two arguments:  First, for the kinetic pressure
and energy density of massless particles the relation $p=\epsilon/3$
holds for any isotropic momentum distribution.  Second, in the pQCD +
saturation model the number density and the energy density of produced
partons is approximately consistent with the thermalization assumption.
This would indicate that the collision rate can be similar to that in
thermal system with the same energy density.

In the pQCD minijet calculation the saturation momentum at
$\sqrt{s}=5.5$ TeV is 2.03 GeV giving $\tau_0=0.1$ fm/c. This is
the earliest initial time to be used in exploring the photon emission.

Results for the multiplicity are very similar from the minijet model and
the \textsc{Dpmjet} calculations as implied by the similarity of initial
conditions, Fig.~\ref{fig:eps_comp}.  They both give values $dN_{\rm
ch}/dy\sim3000$ for the total multiplicity and agree within the
uncertainties of the models, like the choice of the saturation scale in
the minijet calculation or the possible entropy generation after the
primary production.

In the numerical evaluation of classical field equations the saturation
scale is a parameter and the multiplicity changes with different choices
of the scale.

Calculations based on the initial state saturation formulated in terms
of the initial gluon densities \cite{Kharzeev:2001gp} predict the energy
dependence of the multiplicity and once the normalization is fixed from
the RHIC data, the prediction for the charged multiplicity is around
$dN_{\rm ch}/dy\sim2200$ at LHC.  Even though the authors indicate that,
at least at RHIC, the evolution in final state, other than
fragmentation, is not important, we exploit their result by considering
an initial distribution $\propto T_A(r)$ normalized to fit the predicted
multiplicity from their model.

Results from classical effective field theory on lattice in the
transverse plane can change, depending on the choice of the saturation
scale, by a large factor (order of 2 at RHIC energy)
\cite{Krasnitz:2002mn} for the parton multiplicity.  The larger
multiplicities are comparable with those from the minijet calculation
and the lower ones lead to a final state similar to the one from the
saturation calculation in terms of initial gluon distribution
of incoming nuclei \cite{Kharzeev:2001gp}.  To get an idea how much the
multiplicity may affect the photon emission we scale the initial
conditions also upwards to cover the interval $2000\ \lsim~\ dN_{\rm
ch}/dy\ \lsim\ 4000$ or from $\sim3000$ to $\sim6000$ in the total
multiplicity.

Even though the final state energy density at $\tau=1$ fm/c from the
\textsc{Dpmjet} model is quite similar to other calculations as shown in
Fig.~\ref{fig:eps_comp}, it is not clear that the photon emission should
be the same.  In the \textsc{Dpmjet} calculation also strings are present in the
final state and the space-time evolution might be very different from
that of parton gas, predicted from the other calculations.  Neither is
it clear how well the use of the photon emission rates based on pQCD
calculation with parton degrees of freedom apply if strings play an
important role among the degrees of freedom in the final state.

{\it To summarize:} The two most important parameters for the
determination of the initial state, the multiplicity $dN_{\rm ch}/dy$ and
the initial timescale $\tau_0$, have a considerable range of
uncertainty.  Also the maximum values of the temperature in the final
state are affected by the transverse shape of initial distributions and
since the photon emission has a very strong temperature dependence,
these effects are seen in the calculated photon spectrum.  To estimate the
uncertainty in the calculation of thermal photon emission we explore the
following values of the input parameters for the initial conditions:
\begin{itemize}
\item
$2000\ \lsim \ dN_{\rm ch}/dy\ \lsim \ 4000$,
\item
$0.1 {\rm fm/c}\le \tau_0\le 1.0\ {\rm fm/c}$ and
\item
$\epsilon(\tau_0,r)\propto [T_A(r)]^\kappa$ for $\kappa=1$ and 2.
\end{itemize}

\subsubsection{Initial conditions and the evolution of fugacities}
\label{sec:fugacities}
\def\sigmaET{\sigma\langle E_{_T}\rangle}

Gluon dominance in the initial production will lead to small (anti)quark
fugacities.  This turns out to have an important effect on the emission
rates.  At fixed temperature the decrease of fugacities leads to smaller
densities which decreases the emission rate.  However, at fixed energy
density the decrease of fugacities increases the temperature which in
turn increases the emission rate.  Thus the effects from the lack of
chemical equilibrium are complicated and the question if the emission
rate increases or decreases can only be answered with a detailed
calculation~\cite{Aurenche:work}.  This means that we have to determine
not only the initial values and the evolution of the temperature $T$ but
also of the fugacities $\lambda_g$, and $\lambda_q$.

We determine the initial values of fugacities in the transverse plane at
the initial time $\tau_0$ from the information on jet cross sections
$\sigma_g$ and $\sigma_q\approx\sigma_{\bar q}$ and the cross section
for transverse energy production, $\sigmaET$.  In doing so we also have
to impose limits for the values of fugacities to assure thermodynamic
consistency.

In Sec.~\ref{sec:inicond} we constructed the initial energy density from
$\sigmaET$, Eq.~(\ref{eq:iniepsilon}).  We can define the number
densities of gluons and (anti)quarks, $n_g$ and $n_q$ respectively, in
the same way by using $\sigma_g$ and $\sigma_q$ instead of $\sigmaET$.
If the temperature is known, the fugacities can be determined from the
relation between the equilibrium thermal density $\hat n_i(T)$ and the
real density $n_i$:
\be
n_i=\lambda_i\hat n_i(T),\ i=g,q\,.
\label{eq:fugacity}
\ee
The equilibrium density $\hat n_i(T)$ is given as $\hat n_i=b_iT^3$ with
$b_g=16\zeta(3)/\pi^2,\ \zeta(3)=1.20206$ for gluons and $b_q=b_{\bar
q}=9 N_f\zeta(3)/ 2 \pi^2$ for (anti)quarks.  In fact, to determine the
fugacities and the temperature the equations both for the number
densities and the energy density, Eq.~(\ref{eq:eps_fugacities}) must be
solved simultaneously.  An extra complication arises from the fugacity
dependence of the bag constant $B$.

We will not go to all the details of solving the equations for the
temperature and the fugacities here; it obviously can be done.  However
we must discuss the range of values for fugacities for which a
consistent equation of state can be constructed.  In Sec.~\ref{sec:EoS}
we stated a lower limit for the fugacities to ensure that the hadron gas
will be the stable low temperature phase.  It turns out that if we solve
the fugacities and the temperature from the initial conditions using
Eqs.~(\ref{eq:fugacity}) and (\ref{eq:eps_fugacities}) we will obtain,
at the edges of the transverse nuclear overlap, fugacities which violate
this condition.  Also, in the central region of the transverse plane,
gluon fugacities with values $\lg>1$ are obtained.  If the fugacity is
written in terms of chemical potential, values exceeding one will lead
to a singular unphysical phase space distribution function.  To avoid
this, we limit the gluon fugacity to be $\lg\leq 1$.  If solving the
Eqs.~(\ref{eq:fugacity}) and (\ref{eq:eps_fugacities}) gives $\lg>1$ we
take $\lg=1$ or equivalently $n_g=\hat n_g(T)$ and keep the quark
fugacity unchanged.  New temperature and number densities can then be
recalculated using Eqs.~(\ref{eq:eps_fugacities}) and
(\ref{eq:fugacity}).  The lower limits at the edges of the nuclear
overlap are imposed in the same way.  It turns out that when these
physically motivated limits are applied at the initial time, they are
not violated by evolution of fugacities during the expansion on the
parton matter.

\begin{figure}
\begin{minipage}[c]{7.6cm}
\centerline{\includegraphics[width=7.0cm]{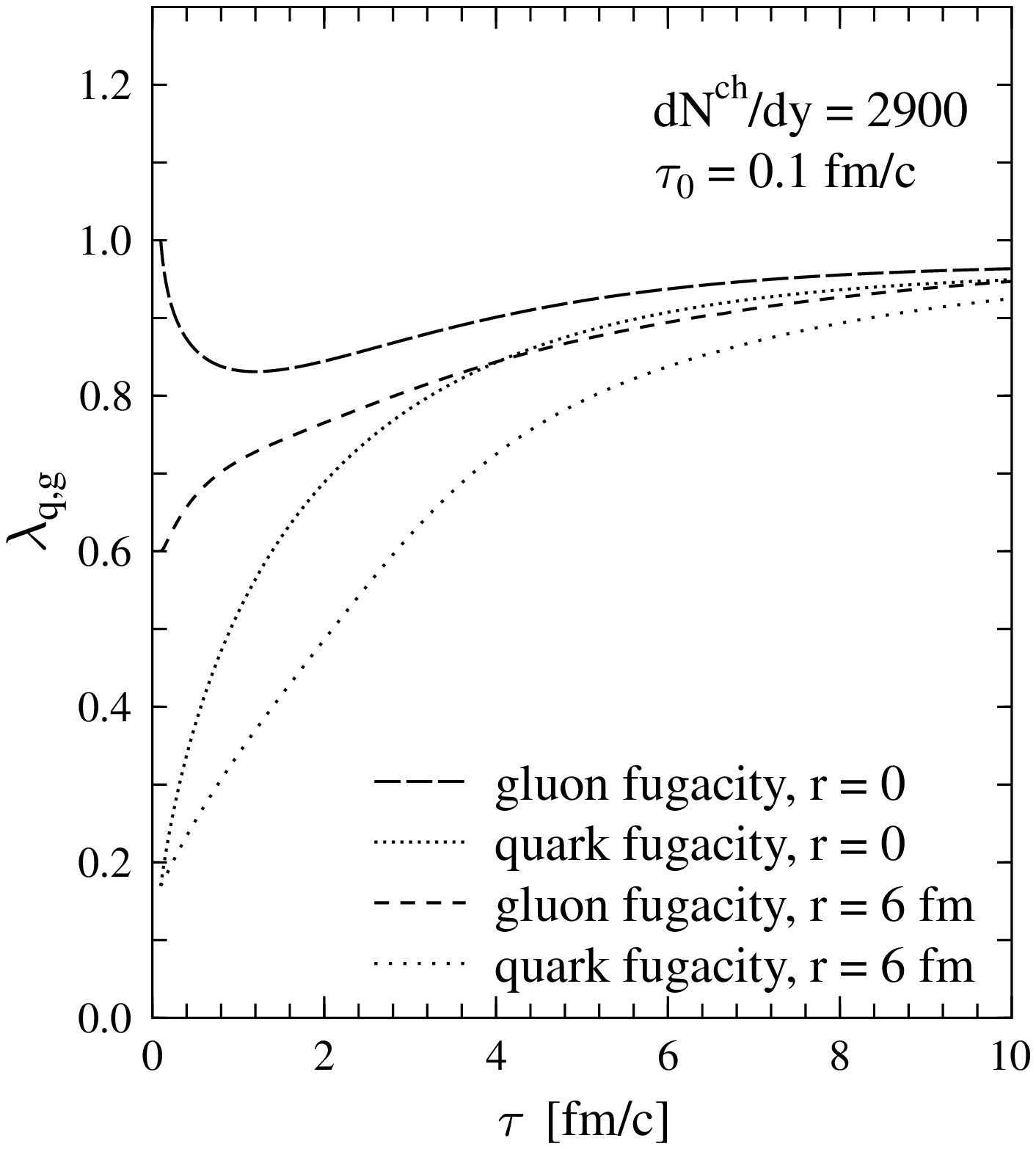} }
\vspace{-0.1in}
   \caption{
    Evolution of fugacities starting from the pQCD + saturation initial
    state. $\lambda_g(\tau,r)$ and $\lambda_q(\tau,r)$ are shown as
    function of time for $r=0$ and $r=6$ fm.}
\label{fig:evolvefug}
\end{minipage}
\hfill
\begin{minipage}[c]{7.6cm}
\vspace{-0.2in}
\centerline{\includegraphics[width=7.0cm]{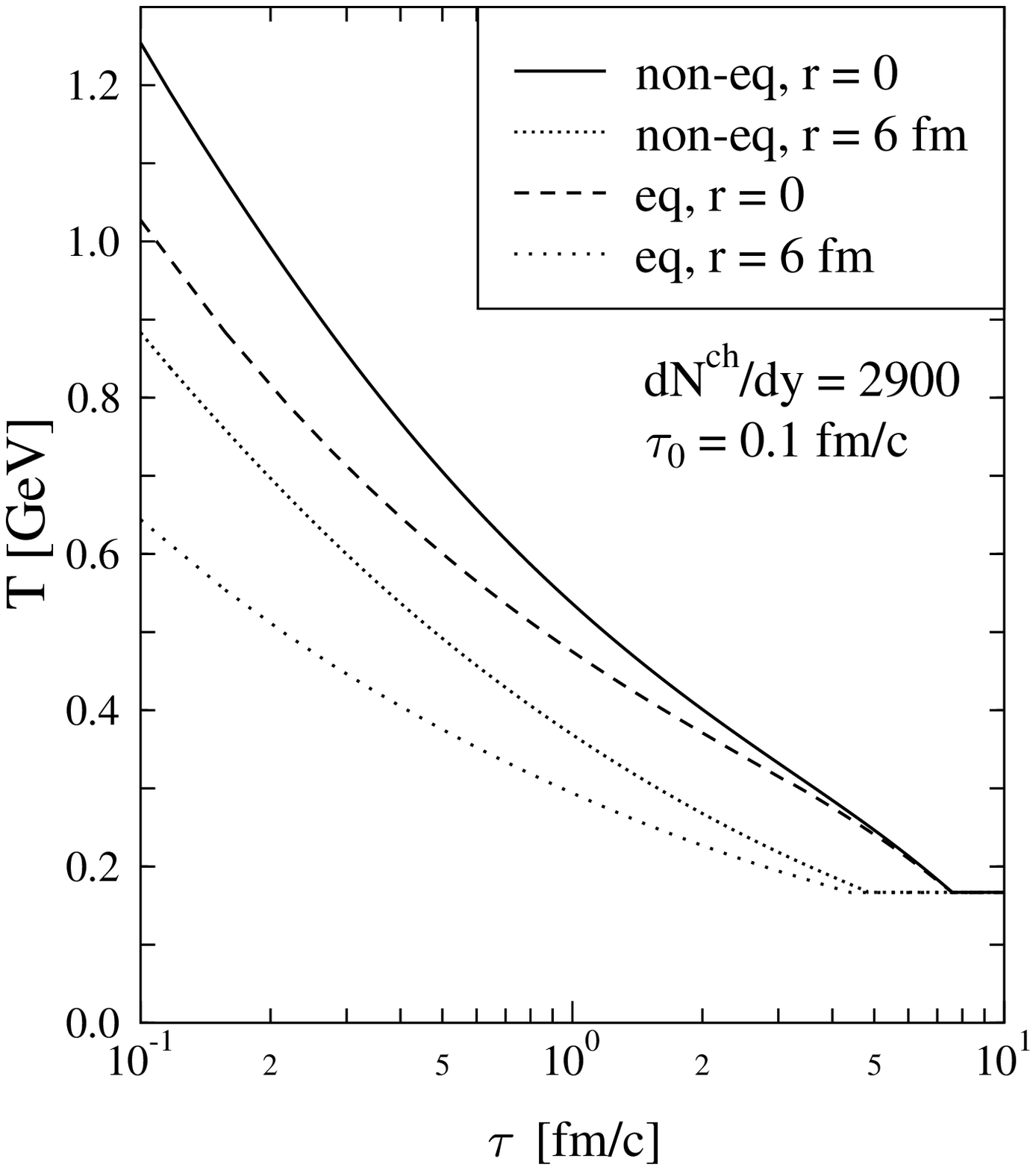} }
\vspace{-0.1in}
   \caption{
     The evolution of temperature for the same situations as for the
     fugacities in Fig.~\ref{fig:evolvefug}.}
\label{fig:evolvetemp}
\end{minipage}
\end{figure}

The evolution of fugacities is determined by the rates of number changing
reactions for gluons and quarks.  There is a considerable uncertainty in the
parton cross sections which are needed in the rate calculations.  We
follow the studies of Bir\'o et al.~\cite{Biro:1993qt} and Elliott and
Rischke~\cite{Elliott:1999uz})
and include the processes $gg\to q\bar q$ and
$gg\to ggg$ as well as their inverse reactions into the rate equations.
The rate equations for the gluon and quark fugacities can be written as (see
Eqs.~(23) and (24) in Ref.~\cite{Biro:1993qt})
\be
\partial_\mu(\lambda_i\hat n_i(T) u^\mu)=
C_i(T,\lambda_g,\lambda_q)\,,\ \ i=g,q\,.
\label{eq:hydro}
\ee
with the collision terms
\bea
C_g&=&\hat n_gR_3\lg(1-\lg)-
2\hat n_gR_2\lg\left(1-\frac{\lq\lambda_{\bar q}}{\lg^2}\right)\,,\\
C_q&=&\hat n_gR_2\lg\left(1-\frac{\lq\lambda_{\bar q}}{\lg^2}\right)\,,
\label{eq:colli}
\eea
where, even though we have taken $\lq=\lambda_{\bar q}$, we have written
the quark and antiquark fugacities explicitly to show the origin of
different terms. For the rate parameters $R_i$ we use the results
derived in \cite{Biro:1993qt}:
\bea
R_2&\approx& 0.24 N_f\alpha_s^2\lambda_gT\ln(1.65/\alpha_s\lambda_g)\,,\\
R_3&=& 1.2\alpha_s^2T(2\lambda_g-\lambda_g^2)^{1/2}\,.
\label{eq:rate}
\eea
Equations (\ref{eq:hydro}-\ref{eq:rate}) are solved numerically
simultaneously with the hydrodynamic equations and the change in the EoS
due to the change in fugacities is included in the calculation.

In the beginning of the expansion with large parton densities the
evolution towards thermal equilibrium is fast but it slows down as the
matter expands and rarefies.  In Fig.~\ref{fig:evolvefug} we show the
time dependence of the fugacities in the parton matter at $r=0$ and
$r=6$ fm starting from the pQCD + saturation initial state at
$\tau_0=0.1$ fm/c.  At $r=0$ gluon fugacity $\lambda_g=1$, but since the
number of quarks and antiquarks is small, the rate for $gg\to q\bar q$
reduces the number of gluons faster than the the inverse reaction and
reaction $gg\to ggg$ can balance it.

The evolution of fugacities affects also the evolution of the
temperature.  In Fig.~\ref{fig:evolvetemp} the temperature is shown as
function of time for the same conditions as the fugacities in
Fig.~\ref{fig:evolvefug}. To see the significance of the lack of
chemical equilibrium, we show the evolution of the temperature for the
same initial conditions also when full equilibrium is assumed. Initially
the temperature is higher when the matter is out of chemical
equilibrium. This happens because the effective number of degrees of
freedom is $16\lambda{}_g + 21 N_f \lambda_q / 2 $ instead of
$16 + 21 N_f / 2$ of
the equilibrium case.
As the number of effective degrees of freedom increases with time
the temperature drops faster in non-equilibrium matter and approaches that in
equilibrium matter. However, as will be seen below, from the point of view of
photon emission, the higher initial temperature turns out to 
approximately compensate for the lower parton densities.

\subsection{Hadron spectra from a hydrodynamical calculation}
\label{sec:spectra_hadron}
{\it H.Niemi, P.V. Ruuskanen, S.S. R\"as\"anen }

The calculations of hadron and photon spectra proceed somewhat
differently.  Since the mean free paths of photons are larger than the
size
of the fireball, they are emitted throughout the expansion of matter
from the whole thermal volume.  Hadrons escape only from the surface of
the fireball or after the density in the whole volume is so small that
no further interactions take place.  We first consider the hadrons and
then the photons.

As the matter expands, distances between particles become large,
collisions cease and momentum distributions freeze out.  The condition
for the freeze-out to happen is usually expressed locally in terms of
the energy density or temperature reaching a given value.  This
determines a three-dimensional freeze-out surface $\sigma^\mu(x)$ in
space--time.  The prescription of Cooper and Frye \cite{Cooper:1974mv}
to convolute the flow and the thermal motion along the freeze-out
surface
\bea
 \pi{dN\over d^3{\bf p}\slash E}
&=& {dN\over dydp_{_T}^2} = \pi\int_\sigma
d\sigma_{\mu}(x)p^{\mu}f(x,p;T(x)) \\
&=& {g\over 2\pi}
\sum_{n=1}^\infty (\pm 1)^{n+1} \int_\sigma r\tau \big[-p_{_T} I_1(n
\gamma_rv_r{p_{_T}\over T}) K_0(n\gamma_r {m_T\over T})\,d\tau \nonumber\\
& & +m_T I_0(n \gamma_rv_r{p_{_T}\over T})K_1(n\gamma_r{m_T\over T})\,dr
\big]\,,
\label{eq:CooperFrye}
\eea
is used to calculate first the spectra of all hadrons and hadron
resonances. The second expression above is valid for cylindrically
symmetric, boost invariant flow with $v_r$ the radial flow velocity,
$\gamma_r=1/\sqrt{1-v_r^2}$, and $K_0$ and $I_0$ Bessel functions. We
then follow the chains of all possible two and three
body decays and collect the spectra of final stable hadrons.  The stable
hadrons can be interpreted as the absolutely stable hadrons or also
those with weak decays or both weak and electromagnetic decays.  The
treatment of hadrons with weak decays can be important, e.g.\ in the case
of $\Lambda$'s, depending if the measurement can separate the feed-down
from heavy hyperons or not.  Electromagnetic decays, as a source of
photons, are of special interest but we display later also
the spectrum of $\pi^0$'s before their decay.

\begin{figure}[htb]
\begin{minipage}[c]{8.6cm}
\centerline{\includegraphics[width=8.0cm]{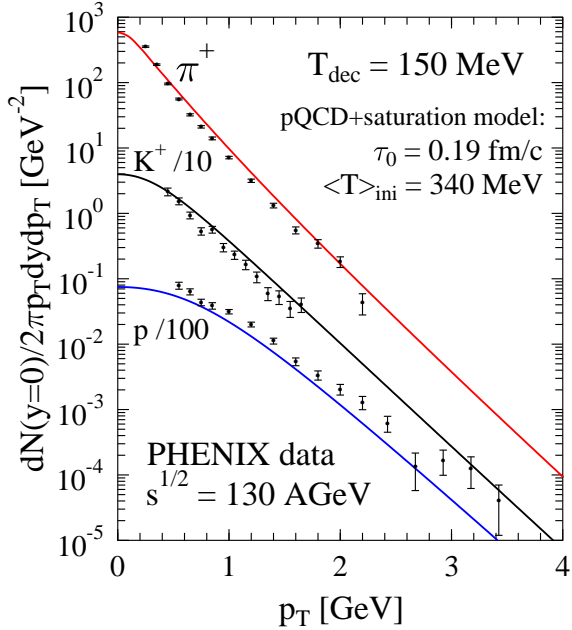} }
\end{minipage}
\hfill
\begin{minipage}[c]{6.6cm}
   \caption{
    Transverse momentum spectra of positive pions, kaons and protons
    measured by PHENIX Collaboration at $\sqrt{s} =130$ GeV shown with
    results from a hydrodynamical calculation with initial conditions
    from the pQCD + saturation model.}
\label{fig:phenix_pos}
\end{minipage}
\end{figure}

Since the transverse momentum distribution of the decay photons depends
strongly on the transverse spectra of hadrons, we show in
Fig.~\ref{fig:phenix_pos} the measured spectra of positive pions,
kaons and protons at RHIC \cite{Adcox:2001mf} with results from a
hydrodynamical calculation using the initial conditions from the pQCD +
saturation model~\cite{Eskola:2002wx}. The results describe the data quite well at
small momenta, but it is clear that a hydrodynamic calculation cannot
describe the hadron spectra at transverse momenta larger than $\sim
4-5$ GeV.  The fraction of hadrons with $p_{_T}\gsim5$ GeV from all hadrons
is small and in thermal models the dominant fraction of low transverse
momentum hadrons is assumed to come from the thermalized matter with
collective motion.  From this point of view the calculation of decay
photons using the hadron spectra from a hydrodynamical calculation gives
a reasonable estimate in the range of low momenta, say up to
$p_{_T}\sim3-4$ GeV.

Similarly, the spectra of hadrons from the thermal fireball at LHC
energy can be calculated for any given choice of initial conditions.  As
an example of hadron spectra at LHC we show here the spectra of positive
pions and kaons and protons in Fig.~\ref{fig:posit_5500} obtained
with initial conditions from the pQCD + saturation
model~\cite{Eskola:work} which,
as already pointed out above, reproduce well the data at RHIC.  To show
the uncertainty in the shape due to the decoupling procedure, results
with decoupling temperatures $\Tdec=120$ and 150 MeV are displayed.  All
spectra become flatter at later decoupling and the role of the flow is
seen from the change getting larger with increasing mass. We will show
later how this uncertainty is transferred to the spectrum of
photons from hadron decays.

\begin{figure}[htb]
\begin{minipage}[c]{8.6cm}
\centerline{\includegraphics[width=8.0cm]{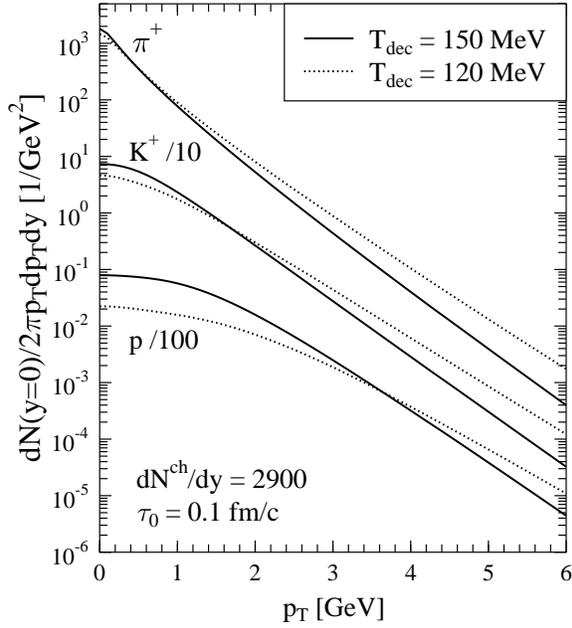} }
\end{minipage}
\hfill
\begin{minipage}[c]{6.6cm}
   \caption{
    Transverse momentum spectra of positive pions, kaons and protons
    calculated at $\sqrt{s} =5.5$ TeV from the pQCD + saturation
    initial conditions~\cite{Eskola:work}. The two sets of curves
    correspond to two different choices of decoupling temperature,
    $\Tdec=120$ and 150 MeV.}
\label{fig:posit_5500}
\end{minipage}
\end{figure}

Note that even though the range of parametrization of initial conditions
have been motivated by different model calculations {\it it does not
mean that such spectra are predictions by the authors of these models}
in some of which hydrodynamic expansion may not be needed at all.  When
using a thermal model for describing the hadron spectra and the
electromagnetic emission we have used these calculations to argue what
could be the reasonable range of input parameters and we always make the
assumption that there is a hydrodynamical expansion stage between the
primary production and the formation of final hadrons.  As reviewed in
Sec.~\ref{sec:CGC}, e.g.\ in \cite{Kharzeev:2001gp} no thermal stage is
assumed to occur.

\subsection{Spectra of thermal photons}
\label{sec:spectra_thermal}
{\it H. Niemi, P.V.~Ruuskanen, S.S. R\"as\"anen, D.K.~Srivastava}
\def\Cdot{\!\cdot\!}

The photon spectra are obtained by integrating the emission rate
over the whole space--time volume of the fireball.
We use the emission rates which are discussed in detail
in Secs.~\ref{sec:qgp} for the equilibrium QGP,
in Sec.~\ref{sec:out-of-chem} for the QGP in kinetic equilibrium but
out-of-chemical equilibrium and in \ref{sec:hotgas} for the hadron
resonance gas.

The local emission rate is a function of the local temperature and the energy
of the photon in the rest system of the matter, 
$d N (\omega^*,T) /d^4 x$. In the
case of flowing matter this is the co-moving frame of the fluid element.
If $p^\mu = (E, \imb p)$ is the four--momentum of the photon in the fixed frame where
the photon is observed and $u^\mu$ the four--velocity of the emitting
fluid element in that frame, the energy of the photon in the co-moving
frame is $\omega^*=p\Cdot u=p_\mu u^\mu$.  Since $d^3{\imb p}/E$ is
Lorentz invariant, we obtain
\be
E \frac{dN_\gamma}{d^3{\imb p}} =
\int d^4x \frac{dN(\omega^*,T)} {d^4x \ {d^3{\imb p}/E}}\,.
\ee
At the transition temperature $T_c$ the matter may enter into a mixed
phase which is characterized by the volume fractions of each phase.
In the mixed phase the emission rate is taken to be the sum of rates in
QGP and hadron gas at $T=T_c$, weighted with the volume fraction of the
phase.

\begin{figure}[htb]
\begin{minipage}[c]{7.6cm}
\centerline{\includegraphics[width=7.0cm]{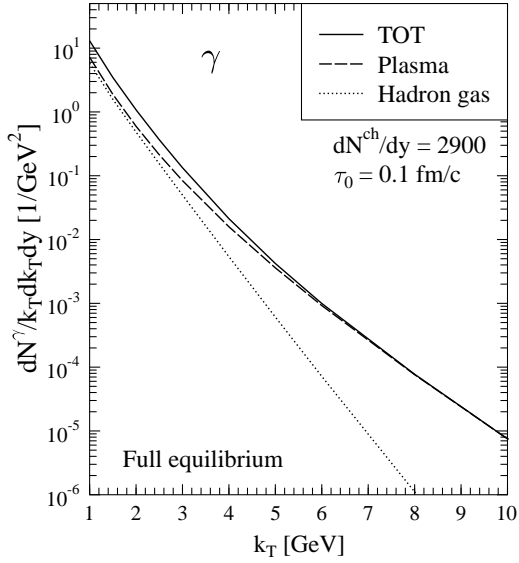} }
\end{minipage}
\hfill
\begin{minipage}[c]{7.6cm}
   \caption{
     Our baseline spectrum of thermal photons: full equilibrium with
     \EoSA, $dN_{\rm ch}/dy=2900$ and $\tau_0=0.1$ fm/c.
     Contributions from plasma and hadron gas are shown separately.
     }
\label{fig:baseline}
\end{minipage}
\end{figure}

We next consider the photon spectra which are obtained when
$u^\mu(r,\tau)$ and $T(r,\tau)$ are calculated using different initial
conditions and equations of state.  We start with the spectrum, shown in
Fig.~\ref{fig:baseline}, from the minijet calculation with the pQCD +
saturation model, used as a baseline and plotted in most of the figures
to facilitate the comparison of different results.  In this baseline
case we assume full equilibrium in the plasma phase, both kinetic and
chemical, the charged multiplicity is $dN_{~\rm ch}/dy=2900$ and we have
taken $\tau_0=1/\psat\approx 0.1$ fm/c. A similar calculation at RHIC energy,
$\sqrt{s}=$ 200~GeV has been presented in Ref.~\cite{Rasanen:2002qe}.

The large-$p_{_T}$ region of the spectrum is completely dominated by the
emission from plasma during a short period at earliest times.  The
transverse flow plays no role at that time and the slope is determined by
temperatures close to the maximum temperature.  With these initial
conditions plasma gives the main contribution also at smaller values of
transverse momentum.  Photons from the hadron gas phase have a much
steeper slope than those from QGP.  This slope is not given by the
temperature of the hadron phase alone but by the combination of
temperature and the flow.  Actually the situation is quite similar with
that of the hadron spectra.  This is shown by the fact that the slope of
the (almost massless) pions in Fig.~\ref{fig:posit_5500} is very close
to that of photons from hadron phase in Fig.~\ref{fig:baseline}.

\begin{figure}
\begin{minipage}[c]{7.6cm}
\centerline{\includegraphics[width=7.0cm]{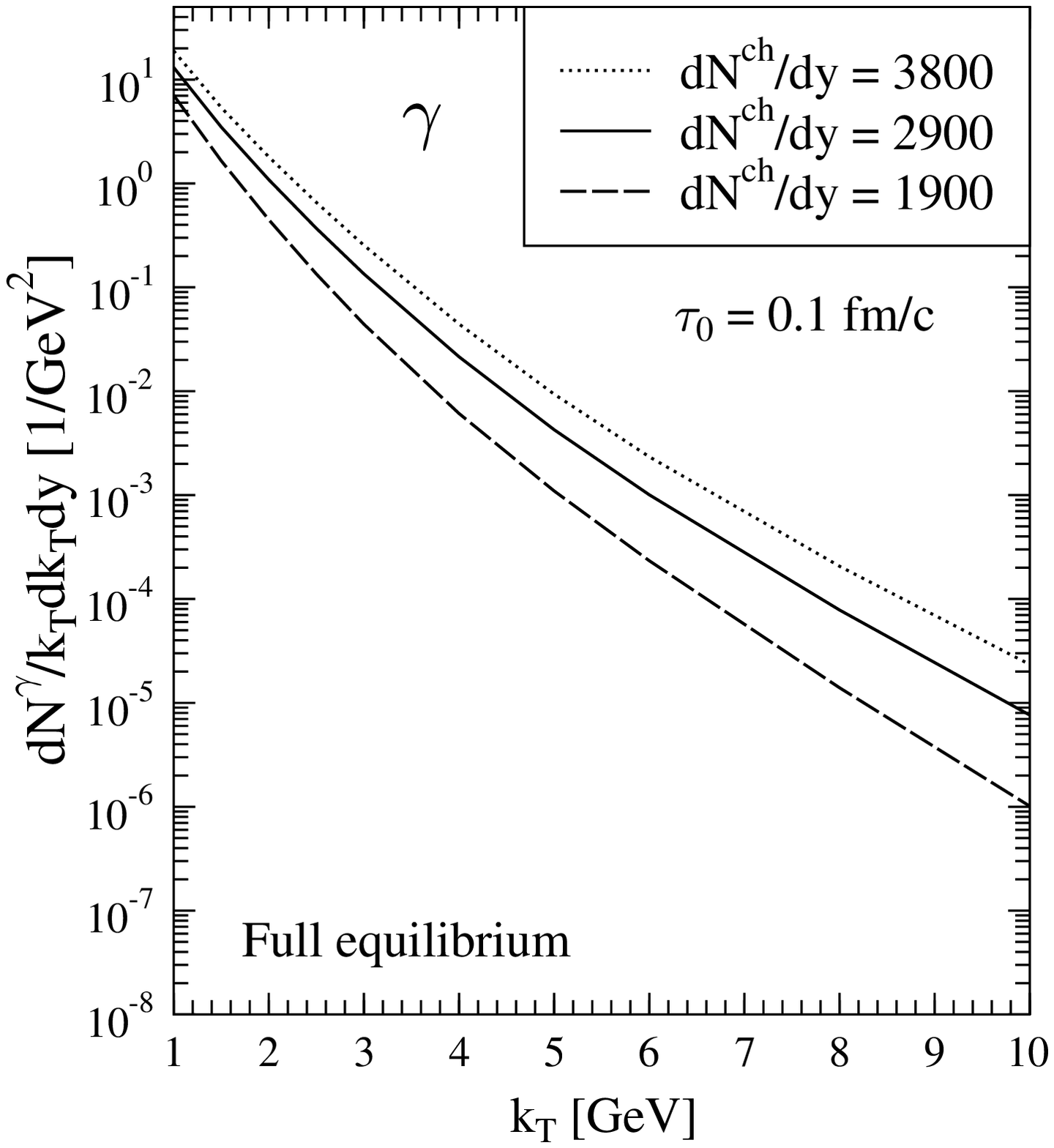}}
\vspace{-0.3in}
    \caption{
     Thermal photon spectra in Pb+Pb collisions at LHC energy
     for different charged hadron multiplicity assuming full thermal
     equilibrium throughout the expansion.
     }
\label{fig:gamma_dNchdy}
\end{minipage}
\hfill
\begin{minipage}[c]{7.6cm}
\vspace{-0.2in}
\centerline{\includegraphics[width=7.0cm]{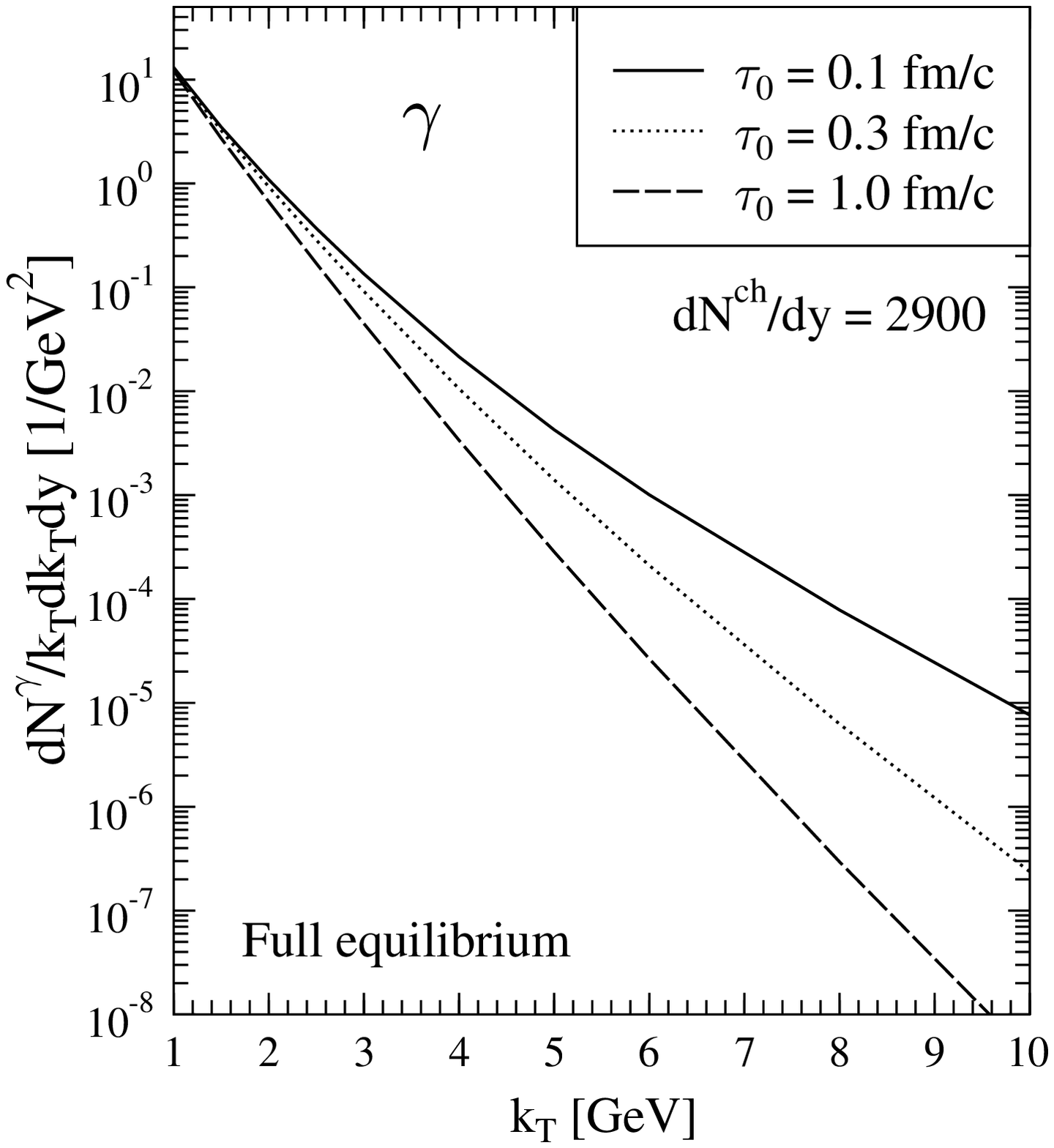} }
\vspace{-0.3in}
   \caption{
   As in Fig.~\ref{fig:gamma_dNchdy} but as a function of initial time
   $\tau_0$ for fixed charged multiplicity $dN_{\rm ch}/dy=2900$.
   }
\label{fig:tau_0}
\end{minipage}
\end{figure}

In Fig.~\ref{fig:gamma_dNchdy} spectra of thermal photons are shown
for charged multiplicities $dN_{~\rm ch}/dy=1900,\ 2900$ and 3800.  To
see the effect from the change of the multiplicity alone, same value
$\tau_{~0}=0.1$ fm/c for the initial time has been used in each case.  As will
be seen later, the transverse momentum window of interest for observing thermal
photons may extend from $p_{_T} \sim$ 2 GeV up to $p_{_T} \sim 10$ GeV in the
favourable case. A change in charged particle multiplicity from 1900 to 3800
increases the photon rate by more than a factor of five at the lower end and
almost a factor of 10 at the upper end of the interval.  Since the dependence
of decay photons is close to linear in hadron multiplicity, a strong dependence
on the multiplicity in the relative strength of thermal photons to those from
hadronic sources can be expected.

A large uncertainty in the photon emission comes from the uncertainty on
the production and thermalization times of the initial partons.  In all
production calculations based on QCD the initial state is dominated by
gluons due to the SU(3) properties of the couplings among gluons and
(anti)quarks.  Thus the system is far from chemical equilibrium after
the production stage.  It may also be far from kinetic equilibrium and
there are arguments that equilibrium is reached only on a timescale of
2\ldots3 fm/c \cite{Mueller:1999fp,Bjoraker:2000cf,Baier:2000sb}.
We will consider the chemical
equilibration separately below.  Here we just consider the effect of the
initial time $\tau_{~0}$ in the hydrodynamical calculation.

We already argued in favour of a shorter timescale for the initial time
of hydrodynamic expansion, $\tau_0$, than obtained from the
thermalization arguments.  In a parton-based approach the production
timescale is provided by the momentum scale at which the production is
expected to saturate, $\tau_{\rm prod} \sim 1/\psat$.  The momentum
distribution will not be thermal initially but the partons can scatter
from the time $\tau_{\rm prod}$ on and some of these interactions
produce photons.  A kinetic calculation of photon emission starting
from different initial distributions would be better justified but we
would like to argue that by taking $\tau_0 = \tau_{\rm prod}$ we
effectively simulate the photon emission also from those interactions
which bring the matter towards kinetic equilibrium.  In
Fig.~\ref{fig:tau_0} we show the variation when the initial time is
changed.  It is seen to be very strong at large $p_{_T}$ as is expected
since the magnitude of local relative momenta or, in the thermodynamic
language, the temperature decreases rapidly due to the longitudinal
expansion.

When we compare these spectra with photons from other sources, we will see that
the observation of thermal photons, or more generally photons from secondary
collisions among initially produced partons, is possible only if the produced
partons can begin to scatter at early times immediately after production when
they still have large relative momenta.

\begin{figure}[htb]
\begin{minipage}[c]{7.6cm}
\centerline{\includegraphics[width=8.0cm]{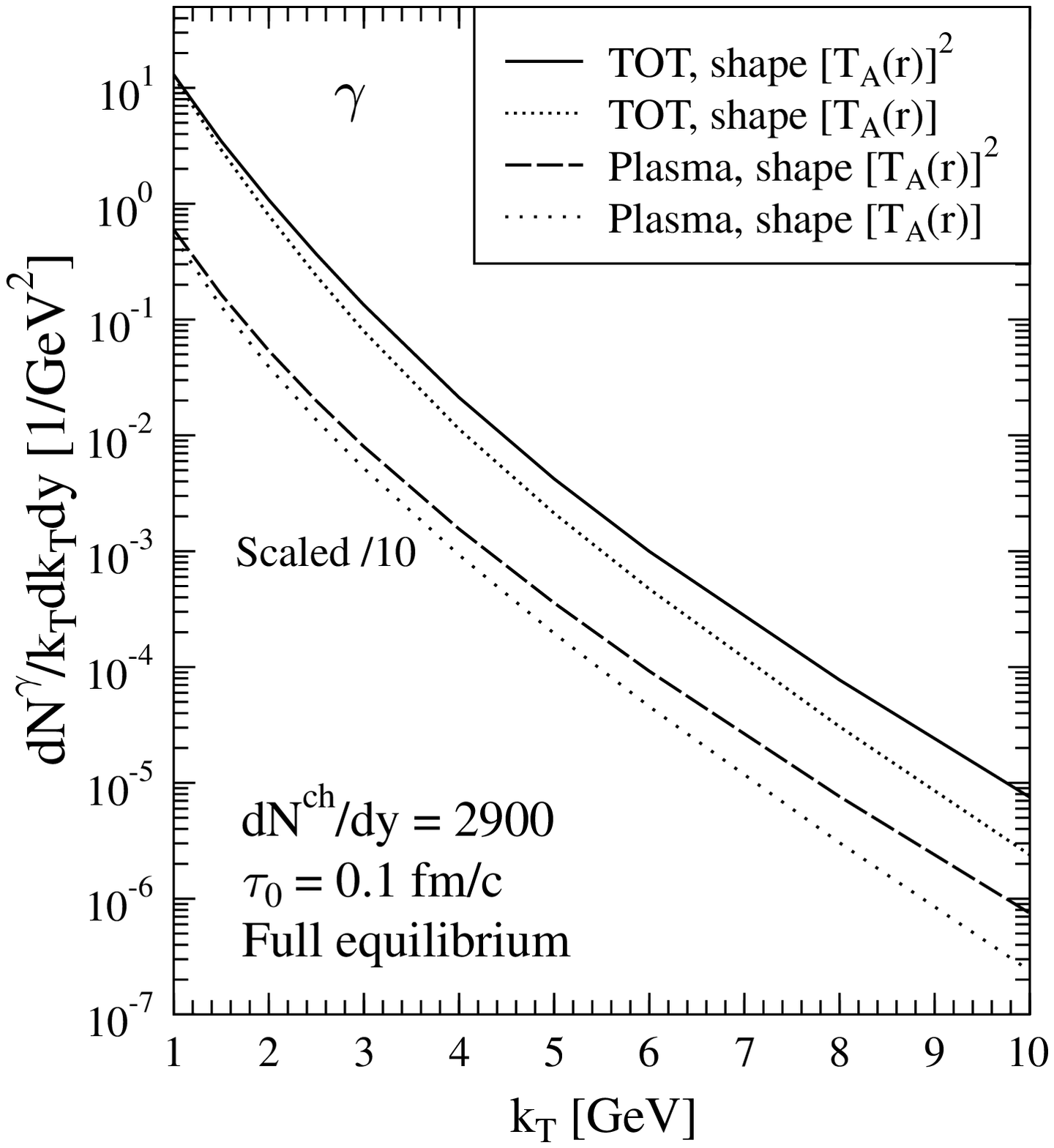}}
    \caption{
      Photon spectra when the shape of initial condition is is taken
      to be $[T_A(r)]^\alpha$ with $\alpha=1$ or 2.
      $dN_{\rm ch}/dy\simeq2900$ and $\tau_0=0.1$ fm/c in both cases.}
\label{fig:phot_shape}
\end{minipage}
\hfill
\begin{minipage}[c]{7.6cm}
\centerline{\includegraphics[width=9.0cm]{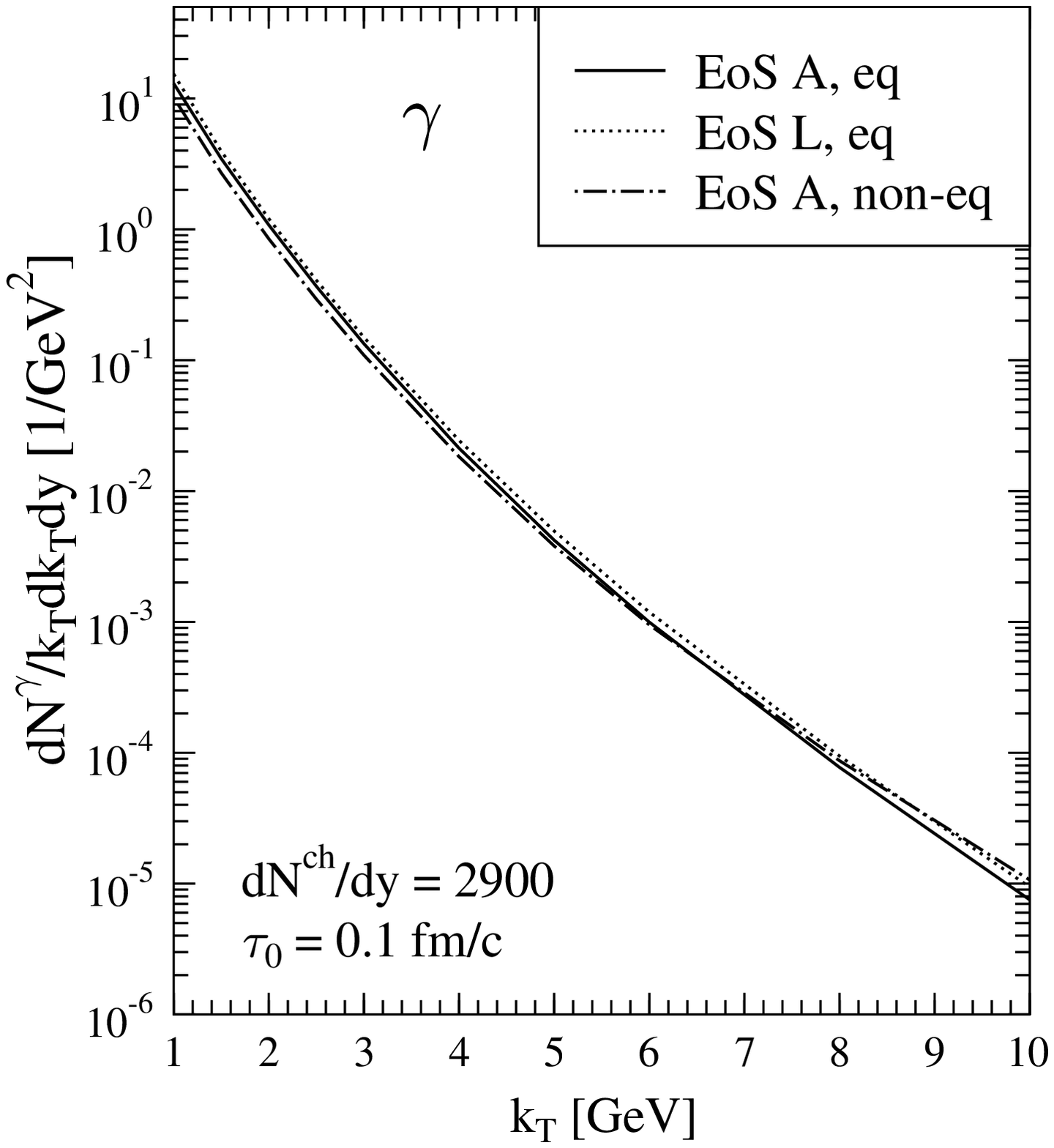}}
\vspace{-2.cm}
   \caption{
   Our baseline result (solid line) is shown with the
   the photon spectrum for an out of
   chemical equilibrium calculation (dashed-dotted line) and for a
   calculation using the lattice EoS (dotted line). In all cases
   $dN_{\rm ch}/dy\simeq2900$ and $\tau_0=0.1$ fm/c.
   }
\label{fig:phot_fug_lat}
\end{minipage}
\end{figure}

In calculating the spectra in Figs.~\ref{fig:gamma_dNchdy} and
\ref{fig:tau_0} we have assumed that the radial dependence of initial
distributions is proportional to $[T_A(r)]^\alpha,\ \alpha=2$.
Fig.~\ref{fig:phot_shape} shows the difference between the cases
$\alpha=$ 1 and 2. With $\alpha=2$ the distribution is steeper and more
peaked in the center than in the case $\alpha=1$.  Consequently, for
fixed multiplicity the maximum temperature is higher in the former case
and the strong temperature dependence of emission rate leads to stronger
overall emission and more shallow dependence on $p_{_T}$ of the spectrum.

We next discuss how the deviation from chemical equilibrium affects the
photon emission.  The initial values and the evolution of fugacities are
determined as described in Sec.~\ref{sec:fugacities}.  When the values
of fugacities are less than one, $\lambda_i<1$, the numbers of
degrees of freedom are effectively less than in plasma in full
equilibrium.  Comparing with the equilibrium case at a given energy
density, parton densities are smaller but temperatures higher.  
For the photon emission spectrum, it turns out that the increase in
temperature approximately  compensates the decrease of parton densities, over
the $p_{_T}$  range considered.

In Fig.~\ref{fig:phot_fug_lat} we compare our baseline case in full equilibrium
with the non-equilibrium case specified in Sec.~\ref{sec:fugacities}.  In both
cases the charged particle multiplicity is 2900 and the initial time
$\tau_{~0}=0.1$ fm/c.  
The spectra approximately coincide, the off equilibrium case giving slightly
smaller results at small $p_{_T}$ and about 20\% higher at $ p_{_T} \gsim 10$
GeV.

We also show in Fig.~\ref{fig:phot_fug_lat} the result of a calculation
using an equation of state with the high temperature phase fitted to the
lattice data. The calculation starts from the same initial conditions as
in the baseline case. As is seen in Fig.~\ref{fig:EoS}, the lattice
parametrization corresponds to higher temperature at given energy
density than our EoS$\,$A. In the photon spectrum this shows up as a
slight increase in the photon emission. There are indications that if
lattice results are interpreted in terms of quasiparticle excitations,
the quasiparticle masses do not exactly correspond to the thermal masses
of partons. The photon emission is calculated with the standard
perturbative rates but it might be more consistent to replace the
thermal masses in the rates with the (temperature dependent)
quasiparticle masses obtained from the lattice information.

Recently an interesting new photon source was proposed
\cite{Fries:2002kt}. It is an outcome of the
interactions of energetic jets with the thermalized plasma.  The idea is similar
to that in jet quenching:  the bulk of the plasma comes from
partons close to the saturation momentum but the high energy partons do not
thermalize.  They will traverse the matter interacting with its quanta.  In
some of these interactions photons will be emitted and if the properties of
matter and the cross section for photon production are known, the emission rate
can be calculated.

\begin{figure}[thb]
\begin{minipage}[c]{7.6cm}
\centerline{\includegraphics[width=7.0cm]{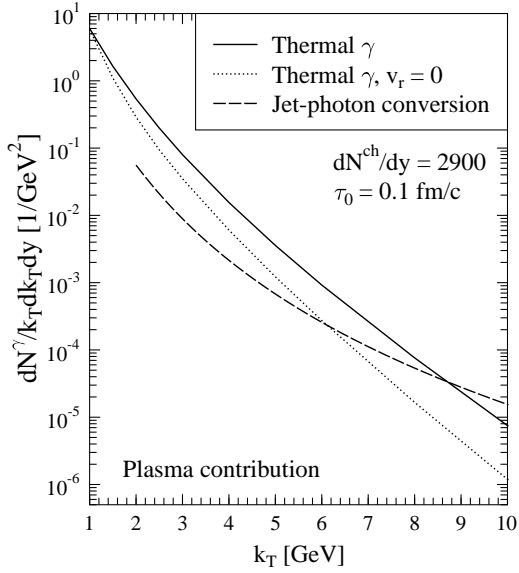} }
\end{minipage}
\hfill
\begin{minipage}[c]{7.6cm}
   \caption{
    Photons from jets interacting with plasma. Thermal photons dominate
    at small $p_{_T}$ but have a steeper slope than those from jets. Since
    the jet + plasma calculation is performed with $v_r=0$, we show also
    the thermal spectrum in that case.
     }
\label{fig:jetonthermal}
\end{minipage}
\end{figure}

In Fig.~\ref{fig:jetonthermal} the spectrum of photons emitted from jets
traversing the plasma is compared with our baseline thermal spectrum.
The calculations of jets interacting with plasma has been performed in the case
of no transverse expansion and this complicates the comparison with the results
from thermal emission. In any case, the emission from jet + plasma
interactions is related to the existence of plasma; it is not a
competing signal with thermal emission but they add together, enhancing
the possibility to observe a photon excess due to the formation of
plasma.

\subsection{Comparing thermal and decay photons in a hydrodynamical model}
\label{sec:pion_photon}
{\it H.~Niemi, P.V.~Ruuskanen, S.S.~R\"as\"anen }

Hadron decays, in particular those of $\pi^0$'s, form the most copious
source of photons.  In this section we show the photon spectra from the
hadron decays when the hadron spectra are obtained from the
hydrodynamical calculation and compare them with the spectra of thermal
photons.  The decay photons from all hadrons in the EoS are included.
In a hydrodynamical model the thermalized part of produced matter is
assumed to dominate the multiplicity and the transverse momentum
distributions up to few GeV.  A hard parton, say $p_{_T}\gsim 10$ GeV will
not thermalize but in traversing the thermal part of matter it may loose
energy before fragmenting into the jet of final hadrons.  The loss of
energy of these hard partons leads to a shift to smaller transverse
momenta of the large-$p_{_T}$ hadrons. The jets still dominate at large
transverse momentum part of the hadron spectra.  In this picture decay
photons with $p_{_T}\lsim 2-4$ GeV come from the hydrodynamic part of the
hadron spectra and the large-$p_{_T}$ photons from hadrons produced in
jets.  There is no obvious way to join the two regions smoothly since
there is no well-defined way to cut off the jet calculation at low
transverse momenta.

\begin{figure} [bht]

\begin{minipage}[c]{7.6cm}
\centerline{\includegraphics[width=10cm]{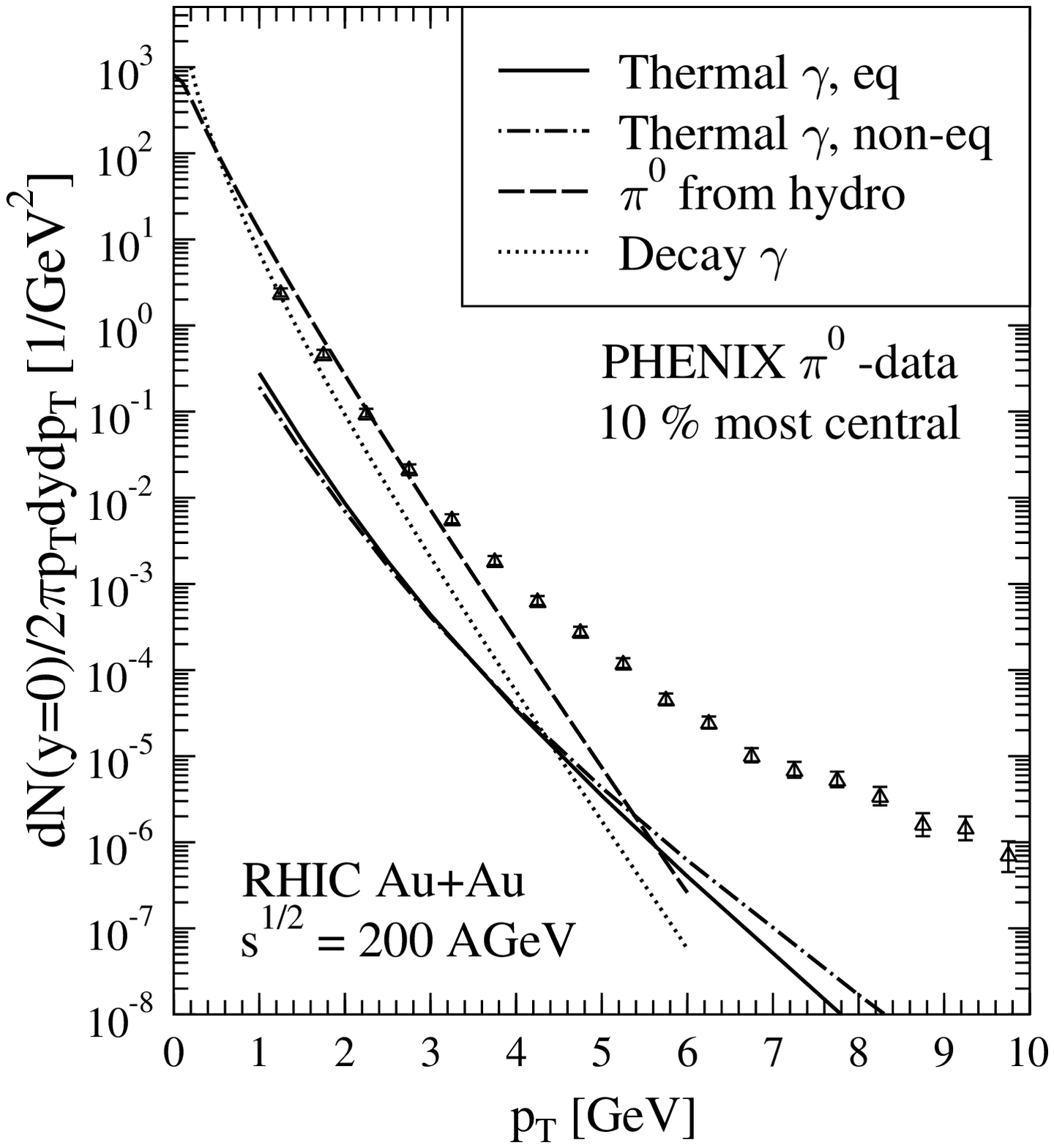}}
\vspace{-2.cm}
    \caption{
     Spectra of (i) $\pi^0$'s (dashed), (ii) decay photons from these $\pi^0$'s
   (dotted) and (iii) thermal photons (solid line: full equilibrium; 
   dash-dotted line: off chemical equilibrium) from a
   hydrodynamical calculation with initial conditions from a minijet
   calculation at $\sqrt{s}=200$ GeV. 
   Also the $\pi^0$ spectrum, measured by the PHENIX collaboration 
   is shown.
     }
\label{fig:thpi0gamma200}
\end{minipage}
\hfill
\begin{minipage}[c]{7.6cm}
\centerline{\includegraphics[width=10cm]{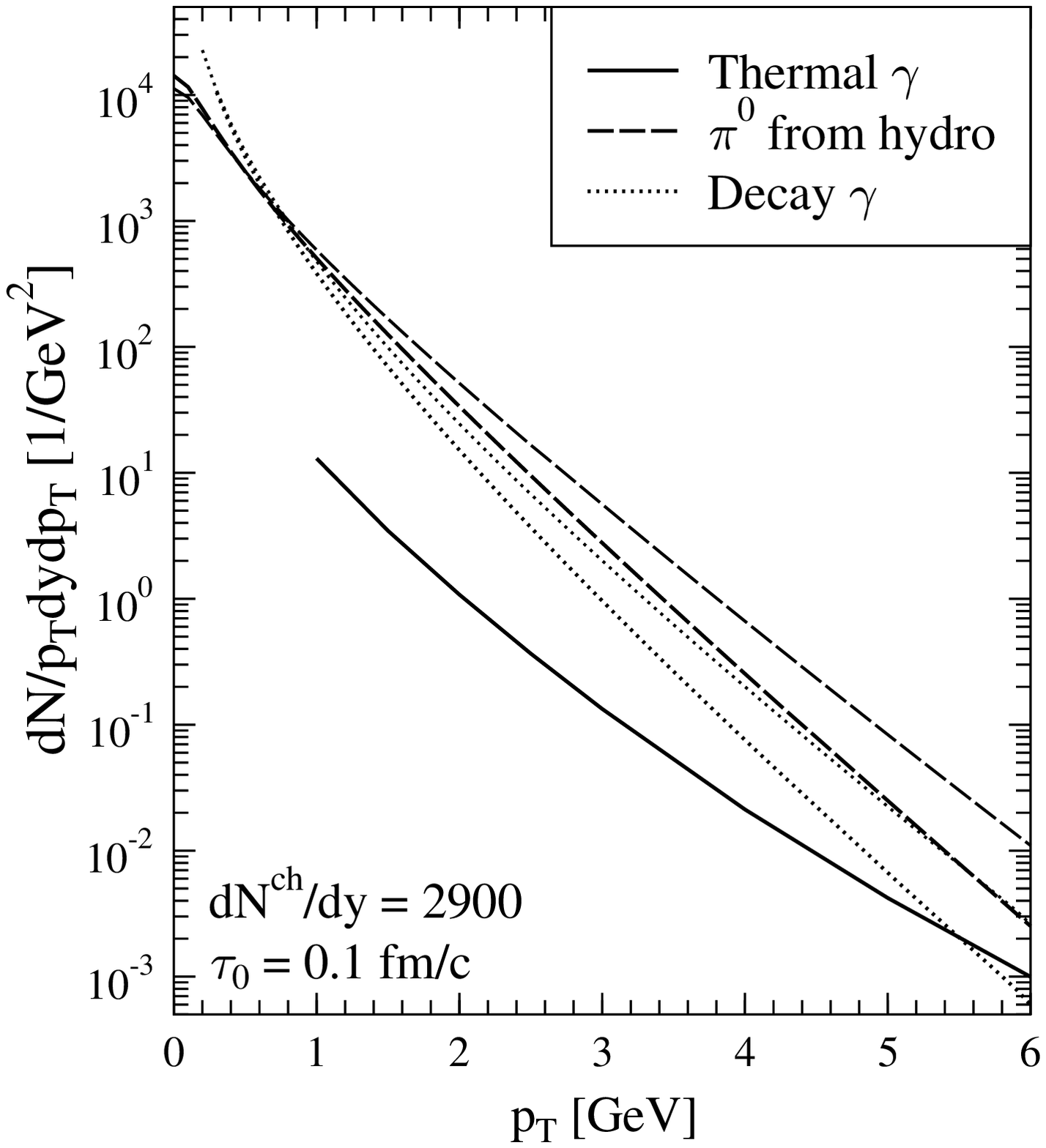}}
\vspace{-2.cm}
   \caption{
   As in Fig.~\ref{fig:thpi0gamma200} but at $\sqrt{s}=5.5$ TeV.
   The bands for $\pi^0$'s and decay photons are shown for $\Tdec=120$
   (upper) and 150 (lower) MeV. 
   The predictions for thermal photons with full chemical equilibrium and off
   chemical equilibrium cannot be distinguished and are shown by the solid 
   line.
   }
\label{fig:thpi0gamma5500}
\end{minipage}
\end{figure}

We would like to remind that in the hydrodynamical model thermal photons
are emitted throughout the expansion of the produced fireball whereas
the final hadrons emerge only from rarefied low-temperature matter on
the decoupling hyper surface.  Consequently the large transverse
momentum photons and hadrons come from different regions of the
fireball:  high-$p_{_T}$ photons are emitted at the earliest times from the
initial quark-gluon matter at highest temperature reached in the
interior of the fireball.  On the other hand, the strong flow develops
at the surface of expanding matter and the large-$p_{_T}$ hadrons originate
from the fastest matter at the edge of the fireball.  Expressed
differently, the slope of the tail of thermal photon spectrum is
determined by the highest values of the initial temperature in the
interior of the matter but those of hadron spectra by the fastest flow
on the surface of the matter.

We show the calculation from the minijet initial conditions only.  The
initial time is $\tau_{~0}=1/p_{\rm sat}$, equal to $0.1$ fm/c at the LHC
energy, $\sqrt{s}=5.5$ TeV, and $0.18$ fm/c at the RHIC energy,
$\sqrt{s}=200$ GeV.  As is seen from
Fig.~\ref{fig:thpi0gamma200}, the magnitude
of the calculated $\pi^0$ spectrum agrees with that of the measured one
for $p_{_T}\lsim 3$ GeV.  Above this region the shallow tail of $\pi^0$'s
comes from the jet fragmentation.  The spectra of decay and thermal
photons cross at $p_{_T}\simeq~4.5$~GeV at RHIC. 

Results at the LHC energy are shown in Fig.~\ref{fig:thpi0gamma5500}.
The spectra of $\pi^0$'s and the decay photons are shown for two
decoupling temperatures.  The lower decoupling temperature, $\Tdec=120$
MeV gives the upper boundaries and the higher $\Tdec=150$ MeV the lower
ones.  
The solid line shows the thermal photon spectrum for both the full chemical
equilibrium and the off equilibrium cases: they cannot be distinguished. 		
This curve is, of course, not the
lower limit for the thermal photon emission. 
For example, as is seen from
Figs.~\ref{fig:gamma_dNchdy} and \ref{fig:tau_0}, thermal production
depends very strongly on the assumptions made of the initial conditions.
In the case of Fig.~\ref{fig:thpi0gamma5500} the thermal and decay
photons cross at $\sim 5$ GeV and for $p_{_T}\lsim~2$~GeV the amount of decay
photons is more than 10 times the number of thermal photons. Depending on the
amount of $\pi^0$'s from jet fragmentation  a window with thermal photons  of
the order of 10 \% of the background could exist in the region of $p_{_T}$
above 2-3~GeV.  All sources of photons, thermal, prompt pQCD from primary
interactions between incoming partons, and decays of both the thermal and jet
hadrons, are compared in the next section.

\subsection{Thermal small mass lepton pairs}
{\it P. Aurenche, F. Gelis, H.~Niemi, P.V.~Ruuskanen and
S.S.~R\"as\"anen}

Like real photons, also virtual photons, decaying to lepton pairs, can be
emitted during the thermal stage of collisions.  As mentioned in
sec.~\ref{thermaldileptonrates} the dynamics of production of a small mass pair
at large momentum is similar to that of a real photon and experimentally this
channel provides a complementary signature to real photon production.

\begin{figure}[thb]
\begin{minipage}[c]{7.6cm}
\centerline{\includegraphics[width=7.0cm]
            {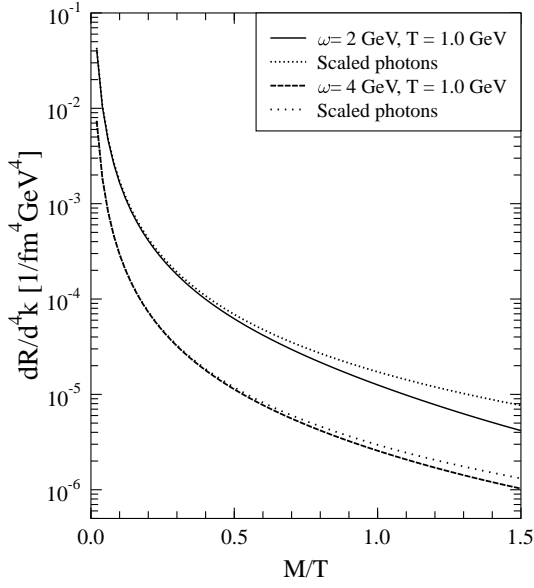}}
\end{minipage}
\hfill
\begin{minipage}[c]{7.6cm}
   \caption{
     Emission rates for lepton pairs as function of $M/T$ are compared
     with the photon emission rate scaled with $\alpha/3\pi M^2$
     (dotted lines) for $E=\omega=2.0$ (solid line) and 4.0 GeV (dashed
     line).
     }
\label{fig:pair_gamma_scale}
\end{minipage}
\end{figure}

From Eq.~(\ref{eq:realphot}) and Eq.~(\ref{eq:virtphot}) one can derive the
approximate scaling relation which should hold true when the virtual photon
mass becomes small compared to all other scales in the problem: 
\bea
 {{dN_{_{l^+l^-}}}\over{dt d^3{\imb x}\ dM^2 d^3{\imb p}}}&=&
 {{dN_{_{l^+l^-}}}\over{dt d^3{\imb x}\ 2 E dE d^3{\imb p}}} \\
 &\simeq& {\alpha \over \pi M^2}\;
 {{dN_{_\gamma}}\over{dt d^3{\imb x}\ d^3{\imb p}}}
\label{eq:pair_gamma_scale}
\eea
where $P=(E,{\imb p})$ and $P^2=M^2$ for the lepton pair. 

In Fig.~\ref{fig:pair_gamma_scale} the dilepton rate for $E=2$ and 4 GeV at
$T=0.5$ GeV is shown as a function of $M/T$ and compared to the scaled photon
rate at the same energies $E$. It is seen that for the pair mass up to 0.5
GeV/c$^2$ the scaling works very well.
The scaling holds true also for the emission from the plasma integrated over
its whole expansion time in the nucleus-nucleus collision. This is shown in
Fig.~\ref{fig:pair_gamma_QGP} for the transverse distribution at different
values of the pair mass. Even for $M=0.5$ GeV/c$^2$ the scaling works well except for
the small-$p_{_T}$ region where the ratio $M/E$ becomes large and, as expected,
the scaling formula overestimates the rate.

\begin{figure} [bht]
\begin{minipage}[c]{7.0cm}
\centerline{\includegraphics[width=7cm]{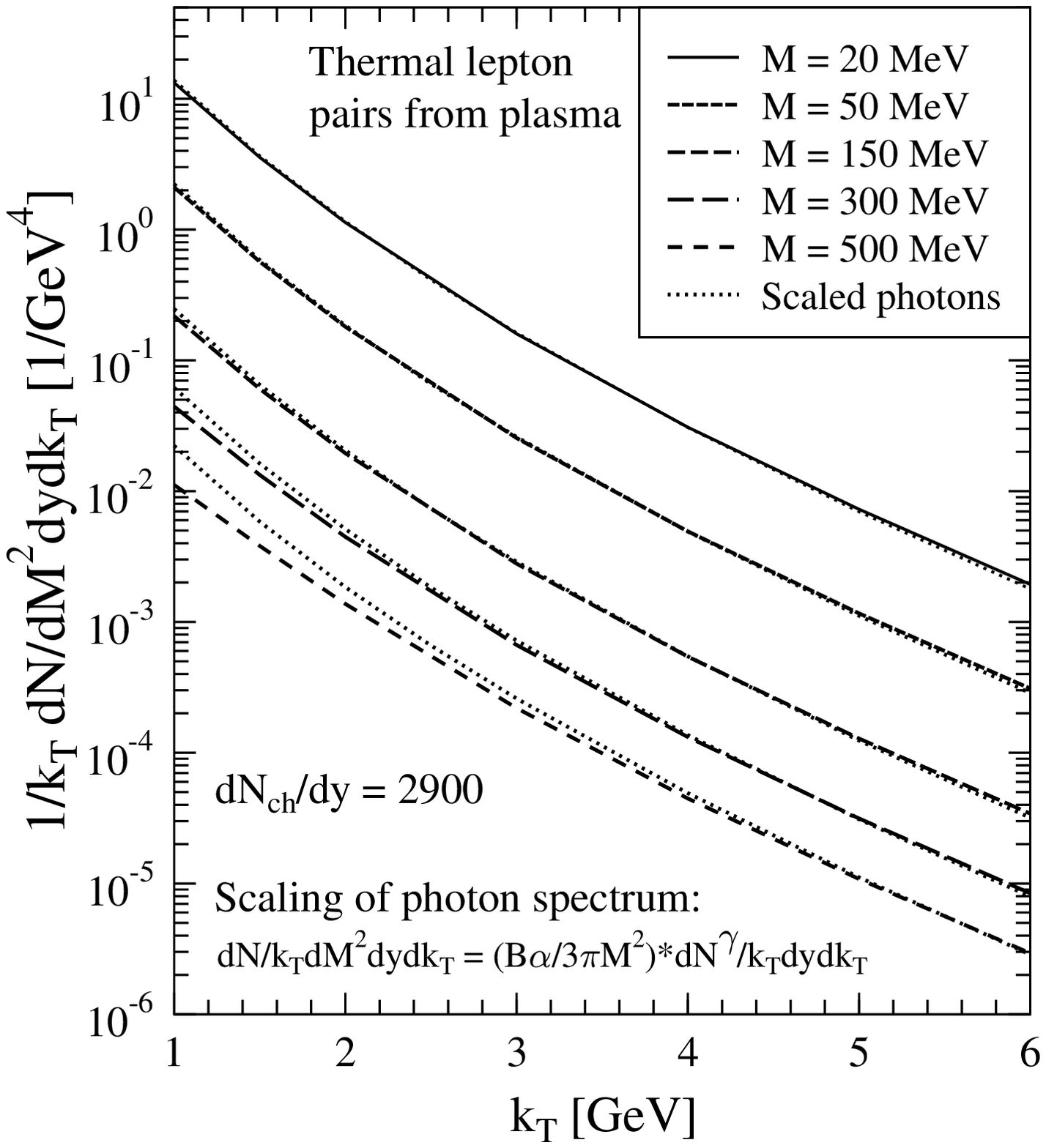}}
\vspace{-0.2in}
    \caption{
     Transverse spectra for thermally emitted pairs from the plasma
     phase in a Pb+Pb collision at LHC for charged multiplicity
     $dN_{\rm ch}/dy=2900$. Photon spectrum from the same hydrodynamical
     calculation with emission rate scaled according to
     Eq.~\ref{eq:pair_gamma_scale}.
     }
\label{fig:pair_gamma_QGP}
\end{minipage}
\hfill
\begin{minipage}[c]{7.0cm}
\centerline{\includegraphics[width=7cm]{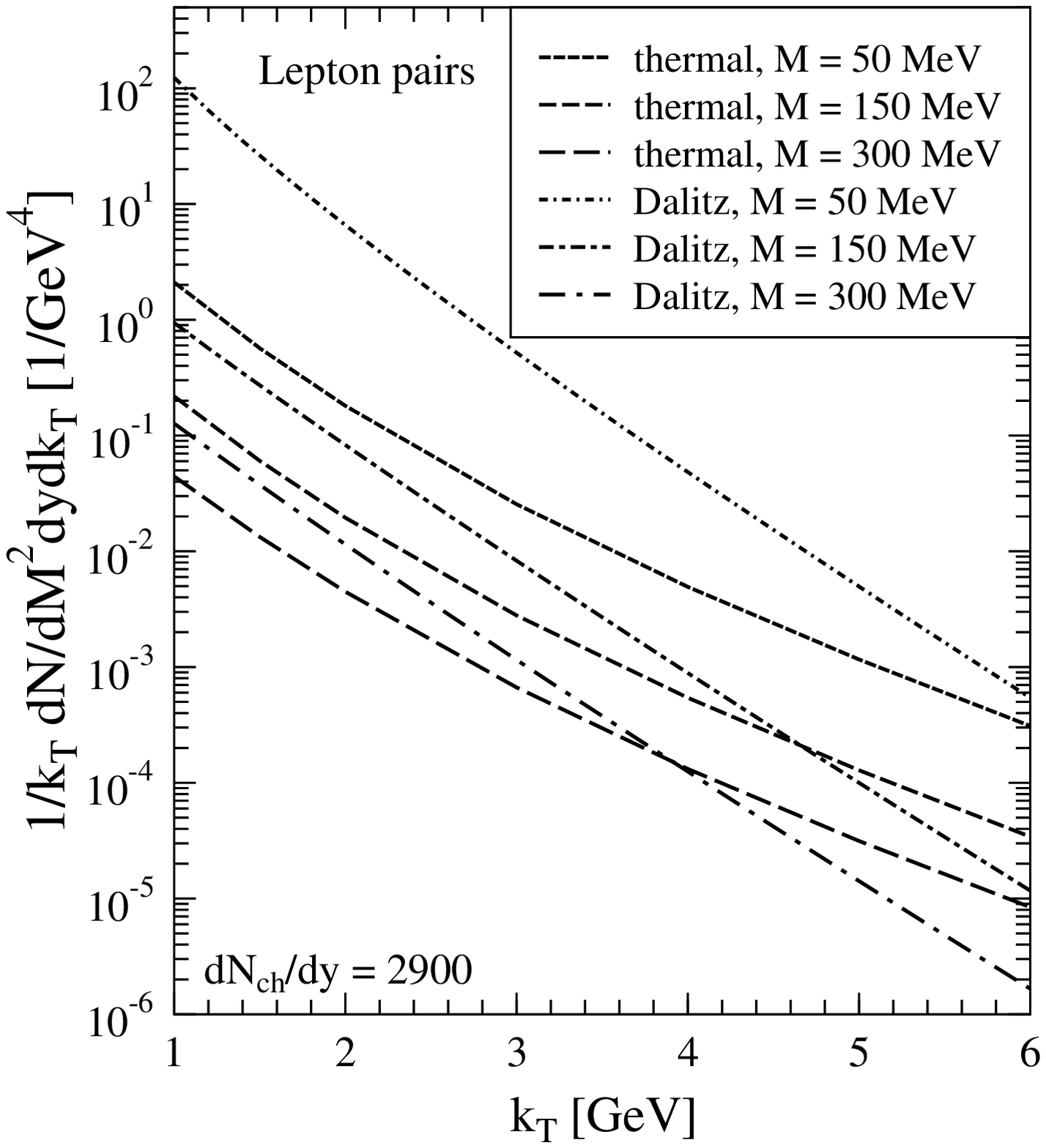}}
\vspace{-0.2in}
   \caption{
    Dalitz pairs from $\eta$ and $\pi^0$ compared with thermal pairs in
    a Pb+Pb collision at LHC with $dN_{\rm ch}/dy=2900$. The spectra of
    $\eta$'s and $\pi^0$'s are obtained from the same hydrodynamical
    calculation as the thermally emitted pairs.
   }
\label{fig:Dalitz_ther_comp}
\end{minipage}
\end{figure}

For small-mass pairs between 0.2 and 0.4 GeV/c$^2$ the dominant background
comes from the Dalitz decays of $\eta$'s.  For $M\sim 0.4-0.5$ GeV/c$^2$ pairs
from $\eta'\to ee\gamma$, $\omega\to ee\pi^0$ and direct vector meson decays
become comparable and for larger masses (up to 1~GeV/c$^2$) dominant.  To get an idea
of the relative size of the thermal and background contributions we
calculate the Dalitz pairs using for $\pi^0$'s and $\eta$'s the spectra
obtained from the hydrodynamic calculation.  For the Dalitz decay distribution
we use \cite{Samios:1961, Cleymans:1991ps} the form
\be
\frac{d\Gamma^a}{dM}=\frac{4\alpha}{3\pi}\frac{\Gamma(a\to\gamma\gamma)}{M}
\left(1-\frac{M^2}{m_a^2}\right)^2\left(1+\frac{2m_e^2}{M^2}\right)
\left(1-\frac{4m_e^2}{M^2}\right)^{1/2}
\ee
which, when integrated over the pair mass, reproduces well the measured
branching ratios \cite{Hagiwara:fs}.

In Fig.~\ref{fig:Dalitz_ther_comp} thermally emitted lepton pairs are
compared with Dalitz pairs for three different mass values.  At the
smallest mass value, $M=50$ MeV/c$^2$, the Dalitz pairs are dominated by those
from $\pi^0$ decays and they are well above the thermal rate.  The
situation with the relative size of thermal and decay contributions is
qualitatively the same as in the case of real photons, see 
Fig.~\ref{fig:thpi0gamma5500}.  For the mass values above the $\pi^0$ mass it
becomes more favourable.  Even at small transverse momentum, the ratio
of Dalitz to thermal pairs is at most 5 and the contributions cross at
$p_{_T}\sim 4-5$ GeV/c.

It should be noticed that the discussion here concerns the correlated pairs
only.  Since the multiplicity of neutral pions is expected to be more than
1000, the number of Dalitz pairs from $\pi^0$ decays with the 0.012 branching
ratio will be $\sim 15$ per unit rapidity.  In the experiment it is not
possible to identify unambiguously if both particles of an observed $e^+e^-$
pair came from the same decay or from a single thermal emission (correlated
pair), or if they came from a different interaction, either decay or thermal,
and would constitute an uncorrelated pair.  The number of observed uncorrelated
pairs goes like $\sim N_{\rm pair}(N_{\rm pair}-1)$ and therefore should
rapidly decrease with $\pt$. An estimate of the uncorrelated background was
already discussed in sec.~\ref{dilepton-non-thermal} in the context of the 
\textsc{Dpmjet} model. In principle, the subtraction of the uncorrelated pairs
can be done on statistical basis (using like-sign pairs) but the available
overall intensities may limit the accuracy and consequently the minimum
signal-to-background ratio for resolving the signal.  Since the decays from the
same hadrons produce the background both for photons and small-mass lepton
pairs, the simultaneous measurement of photons and leptons, or any independent
measurement on $\pi^0$ and $\eta$ spectra, can be used to improve the
background subtraction.


%



%

\section{COMPARING $\pi^0$'s AND PHOTONS FROM DIFFERENT SOURCES}
\label{comparing}

{\it P. Aurenche, H. Delagrange and P.V. Ruuskanen}

In this Chapter we present a compilation of the results of
Chapters~\ref{sec:pertQCD} and \ref{thermal} and we
compare the strength of different sources of $\pi^{~0}$'s and photons. For the
reader's convenience, we briefly summarize the features of each of the three
models we have considered:\\
-- \textsc{Dpmjet} which is able to deal with the whole $p_{_T}$
spectrum, down to  $p_{_T} = 0$, as it contains both soft physics (string
formation) and pQCD physics in the LO approximation. In A+A collisions a high
density of soft strings is produced and final state interactions are taken into
account by fusion and percolation of strings. The parameter determining this
feature has been fitted to RHIC data. No reference to final state energy loss is needed in
this approach;\\
-- Standard NLO QCD calculations which are valid at ``high'' $p_{_T}$ only,
where ``high'' is arbitrarily chosen to be $p_{_T} > 3$ GeV/c: this rather low
value for LHC is justified by the fact that DGLAP physics seems to hold true
down to low $x$ values, and more practically, because down to 3 GeV/c
\textsc{Dpmjet} and NLO predictions are still in good agreement. The NLO
predictions describe the ``primary" collisions modified, in the case of A+A 
scattering, by shadowing and energy loss;\\
-- Hydrodynamic evolution of the quark-gluon plasma and hot hadronic matter
during which hadrons and photons are produced essentially at low transverse
momentum. This is refered to as ``secondary" particle production.

We naively expect to obtain a full description of a nucleus-nucleus
collision by adding the NLO contribution to the QGP and hot hadronic matter
contributions which describe secondary interactions.
We therefore have at our disposal two alternative models: \textsc{Dpmjet} and
``NLO QCD + Hydro'' (the ``standard approach") to make predictions which we are
now going to discuss.

Since the photons from the decays of hadrons, in particular from $\pi^{~0}$'s,
form a background which is larger than the total emission from all other photon
sources we begin by comparing the spectra of $\pi^{~0}$'s calculated assuming
different production mechanisms.
In the second part of the Chapter we present the compilation and comparison of
photon spectra from different sources and discuss the implications on the
possibility to observe different dynamical features, such as the  production
mechanism and the equilibration, which are believed to be important in the
formation and evolution of matter in lead--lead collisions at LHC.
A third part briefly deals with lepton pair production.

\subsection{Features of neutral pion spectra}

Hadrons at large transverse momenta originate from hard
collisions of incoming energetic partons followed by the formation of hadron
jets through the fragmentation of scattered partons. In proton-proton
collisions this is firmly established and it should also take place in
nucleus-nucleus collisions. However, as discussed in Chapter~\ref{sec:pertQCD}, nuclear effects
like shadowing, energy loss in traversing the system of other produced quanta
and possible collective effects are expected to be important and to modify the
spectra from those in hadron-hadron collisions. Most extreme of these
modifications would be the thermalization of all produced quanta. However, this
is unlikely on theoretical grounds from estimates of energy loss and the
experimental results at RHIC~\cite{ppg014,starbacktoback,star2003} clearly show the persistence
of the high transverse momentum tails in hadron spectra. These measurements
also show a clear change in the shape of spectra indicating the energy loss of
jet-generating partons in the final state.

On the other hand the very fact of energy loss of high energy partons in
the final state points to the equilibration of lower energy partons. An
interesting question is the characterization of the energy region where the
transition from partons able to traverse the whole final state to those
stopping and participating to the collective behaviour of the rest of produced
matter takes place. It should be remembered that the high energy partons, say
$\pt \gsim 10$ GeV/c, leading to the jet formation, carry only a small fraction of
the energy of all final state particles. This facilitates the hydrodynamic
modelling since it should be a good approximation to assume that all the energy
is equilibrated.

In the following we summarize the results of the previous chapters by showing
those from each calculation which are estimated as the most likely.   The
results for $\pi^0$ production at RHIC are shown in
Fig.~69.
This figure is the same as Fig.~\ref{fig:auau200etrap} to which we have added
the $\pt$ spectrum originating from thermal production. At around $\pt =
3$~GeV/c the rapidly falling thermal spectrum crosses the NLO QCD one which is
somewhat below the experimental points. The thermal spectrum agrees with the
data at $\pt < 3$~GeV/c while the NLO QCD predictions agree with the data above
$\pt \sim 4$~GeV/c where thermal production becomes irrelevant. It thus appears
that the
``NLO QCD + Hydro" picture gives a very reasonable agreement with data at
200~GeV/c as does \textsc{Dpmjet}. These two models are therefore good starting
points for extrapolation to LHC energies.
%
\begin{figure}[tb]   
\begin{center}
\includegraphics[height=11cm]{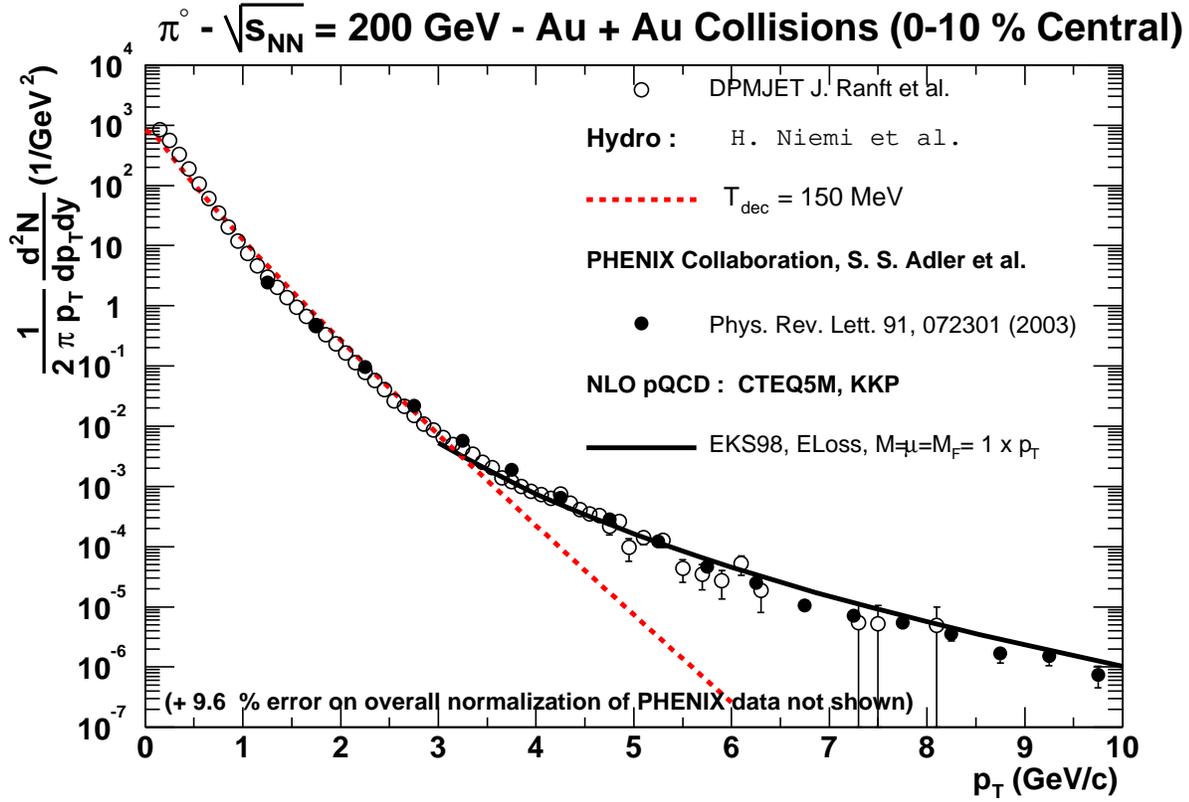}
\vspace{-.5cm}
\caption{
  Spectra of $\pi^0$'s in Au+Au collisions at $\sqrt{s} = 200$~GeV shown
for the following calculations: (i) pQCD jet production and fragmentation
(solid line), (ii) \textsc{Dpmjet} model (open circles), and
(iii) pQCD+saturation for production followed by hydrodynamics with decoupling
at $T_{\rm dec}=150$ (dashed line).
 }
\end{center}
\label{fig:AuAu-neutralpions}
\end{figure}
%

In Fig.~70
we show  the results at $\sqrt{s} = 5.5$~TeV.
We display: the \textsc{Dpmjet} model $\pi^0$
predictions (open circles), the  spectra from the NLO perturbative QCD
calculation (solid line) and from the hydrodynamical calculation with full
thermalization (freeze-out at $T_{\rm ~dec}=120$ MeV, dotted line and 150 MeV,
dashed line).   The basic input parameters for pQCD calculation (including
shadowing and energy loss) are summarized in the figure (see
Chapter~\ref{sec:pertQCD} for more details and in particular
Fig.~\ref{fig:pbpb55etrap}). The effect of the choice of renormalization and
factorization scales is shown as a shaded band.
%
\begin{figure}[htb]   
\begin{center}
\includegraphics[height=11cm]{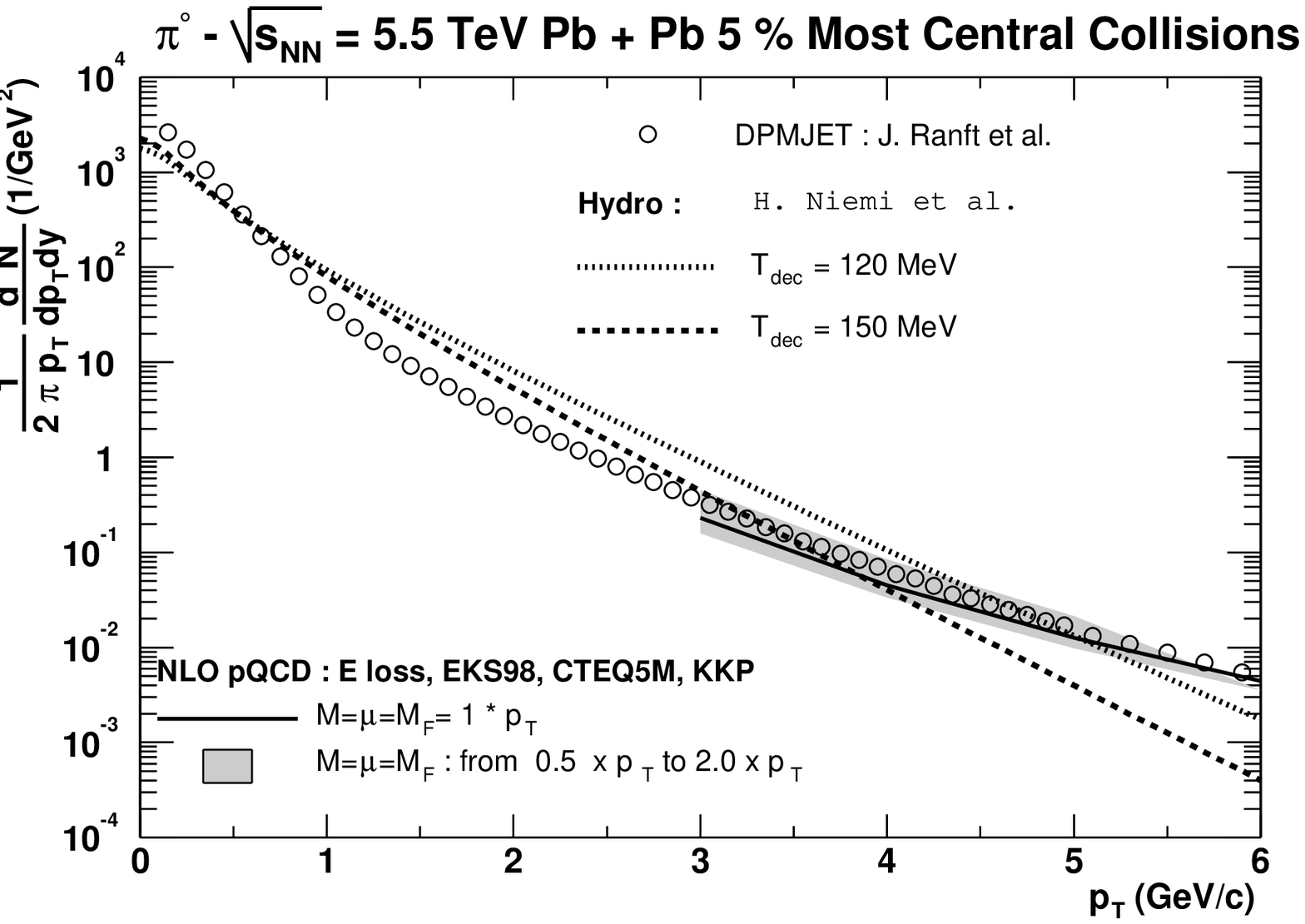}
\vspace{-.5cm}
 \caption{
Spectra of $\pi^0$'s in Pb+Pb collisions at $\sqrt{s} = 5.5$~TeV.
The following calculations are shown: (i) pQCD jet
production and fragmentation (solid line with the grey band), (ii) \textsc{Dpmjet} model
(open circles), and (iii) pQCD+saturation for production followed by
hydrodynamics with decoupling at $T_{\rm dec}=120$ (dotted line) and 150 MeV
(dashed line).
 }
\end{center}
\label{fig:neutralpions}
\end{figure}

The results from hydrodynamics dominate the spectrum at low $p_{_T}$ possibly
up to intermediate momenta 4 \ldots5 GeV/c where they cross below the NLO QCD
predictions. So the transverse momentum regions where pQCD calculations or the
hydrodynamic description apply are complementary.  The full prediction in
the  ``NLO QCD + Hydro'' is the sum of the two curves but unfortunately, there
is no clear way to join the two smoothly. It seems that the difference in
slopes is so large that the region of crossing is quite stable against the
uncertainties in the calculations.
At large transverse momenta the \textsc{Dpmjet} results agree quite well with
the pQCD calculation up to the highest values where the models have been
compared, as can be seen in Fig.~\ref{fig:pbpb55etrap}: this agreement occurs
despite the fact that the two models have very different mechanisms to describe
the interaction of jets with hot matter.

One can note the rather different shape of the $p_{_T}$ distributions, in the
range  0 to 5 Gev/c, between the two approaches. Interestingly, the \textsc{Dpmjet}
results show an even stronger change of slope than that from the high--$p_{_T}$
pQCD region and the low--$\pt$ hydrodynamic region. The change also occurs at
lower momenta, around 1\ldots2 GeV/c, and the small-$p_{_T}$ part is much steeper
than that obtained from the hydrodynamic calculation. If the final multiplicity
(the total particle multiplicity from the hydrodynamic calculation $\sim4500$)
is as high as or close to what follows from the pQCD + saturation model initial
conditions, it seems difficult to obtain such a steep slope from a
hydrodynamical calculation. The reason for this is that it takes a long time to
dilute the dense initial energy of a high-multiplicity collision; a strong
transverse collective flow has time to build up and boost the particles to high
transverse momenta even in case of relatively low thermal momenta at
freeze-out. These differences show that the shape of the transverse momentum
spectra of hadrons, even at relatively low transverse momenta, contains
important information on the dynamics in the nucleus--nucleus collision. These
differences in shape are much more marked at LHC than at RHIC.


\subsection{Comparing photon spectra from main sources}
\label{sec:compare-photon}

Photon spectra from different sources are shown in
Figs.~71 and 72
for RHIC and LHC, respectively.  The figures are more complicated than
in the case of pions since photons can be produced either
directly or through the decay of hadrons.

\begin{figure}[thbp]   
\begin{center}
\includegraphics[height=12cm]{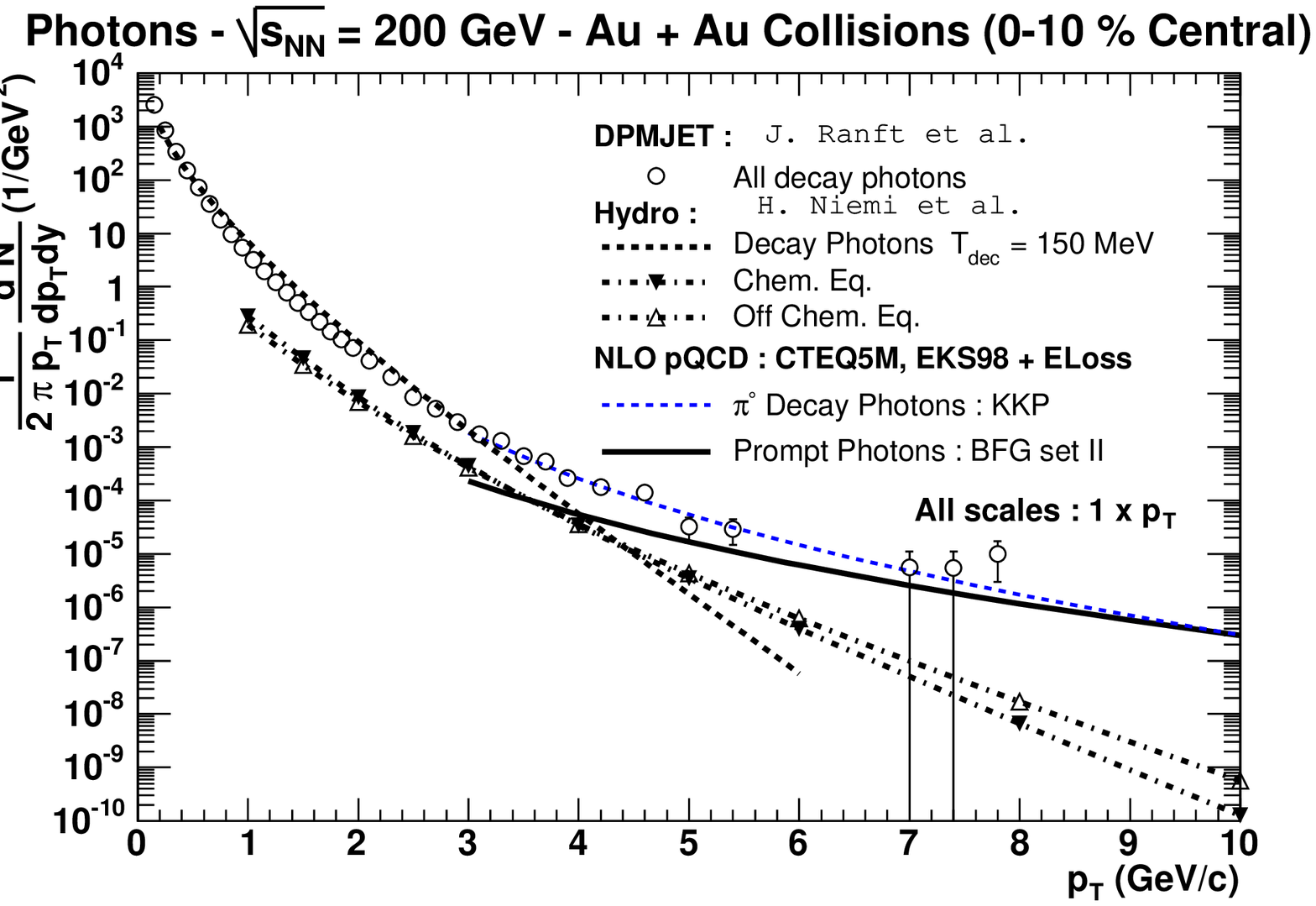}
\vspace{-.5cm}
\caption{
Photon spectra in Au+Au collisions at $\sqrt{s} = 200$~GeV.  Thermal
emission is calculated assuming either full thermal equilibrium in the
plasma phase (dashed-dotted line with dark triangle) or a gluon-rich
plasma (dashed-dotted lines with open triangle).  Decay photons from
thermal hadrons (thick dashed line) and decay photons from NLO QCD pions
(thin dashed line) are also shown together with the prompt photons
(solid line) and decay photons from the \textsc{Dpmjet} calculation
(open circles).
 }
 \end{center}
\label{fig:AuAu-allphotons}
\end{figure}

At RHIC we see that around $\pt\sim$ 3.5 to 4 GeV/c the NLO QCD production of
prompt photons (solid line) is of the same order of magnitude as the
production of thermal photons (dashed-dotted lines) and the decay
photons from hadrons as obtained in the hydrodynamic calculation with
$\tau_{~0}=0.19$ fm/c (thick dashed line).  Thermally emitted photons
decrease rapidly with $\pt$ and compared to the prompt QCD photons and
photons from hadronic decays they become negligible already around 6
GeV/c.  There is no sensitivity to whether or not the plasma is in chemical
equilibrium except at the higher values of $\pt$ where the thermal rate is
negligibly small. One can also observe an overall agreement between DPMJET and
the NLO QCD + hydrodynamic estimates over the whole $\pt$ spectrum.

We next discuss the LHC predictions shown in 
Fig.~72.
The prompt NLO QCD
photon spectrum is given by the lower solid line and the spectrum of
decay photons from $\pi^0$'s, produced through NLO QCD jet
fragmentation, by the upper solid line.  The dotted and dashed lines
show the decay photons from all hadrons in the hydrodynamic calculation
with $T_{\rm dec}$ either 120 MeV (dotted) or 150 MeV (dashed).
Thermal photons are shown both from a calculation when the quark-gluon
plasma is assumed to be in full thermal and chemical equilibrium
(dashed-dotted line with filled triangles) and when the plasma is
dominated by gluons and is not in chemical equilibrium (dashed-dotted
line with open triangles).  As mentioned before kinetic equilibrium is assumed
also in the latter case and in both cases the initial time in the hydrodynamic
calculation is $\tau_0=0.1$~fm/c. 
The two sets of predictions cannot be distinguished. Given the same initial
energy density from the primary production, the deficit of quarks and
antiquarks, which reduces the photon emission, leads to higher values of the
temperature which enhance the emission. The slope of thermal production is
steeper than for the prompt QCD photons from the interactions of incoming
partons but the crossing of thermal and prompt QCD photon spectra takes place
at rather high value of $\pt$ around 6$\ldots$7~GeV/c.  However, if the
effective thermalization time is longer the crossing region could be as low
as~$3\ldots 4$~GeV/c.

%
\begin{figure}[thbp]
\hspace{5mm}
\begin{center}
\includegraphics[height=12cm]{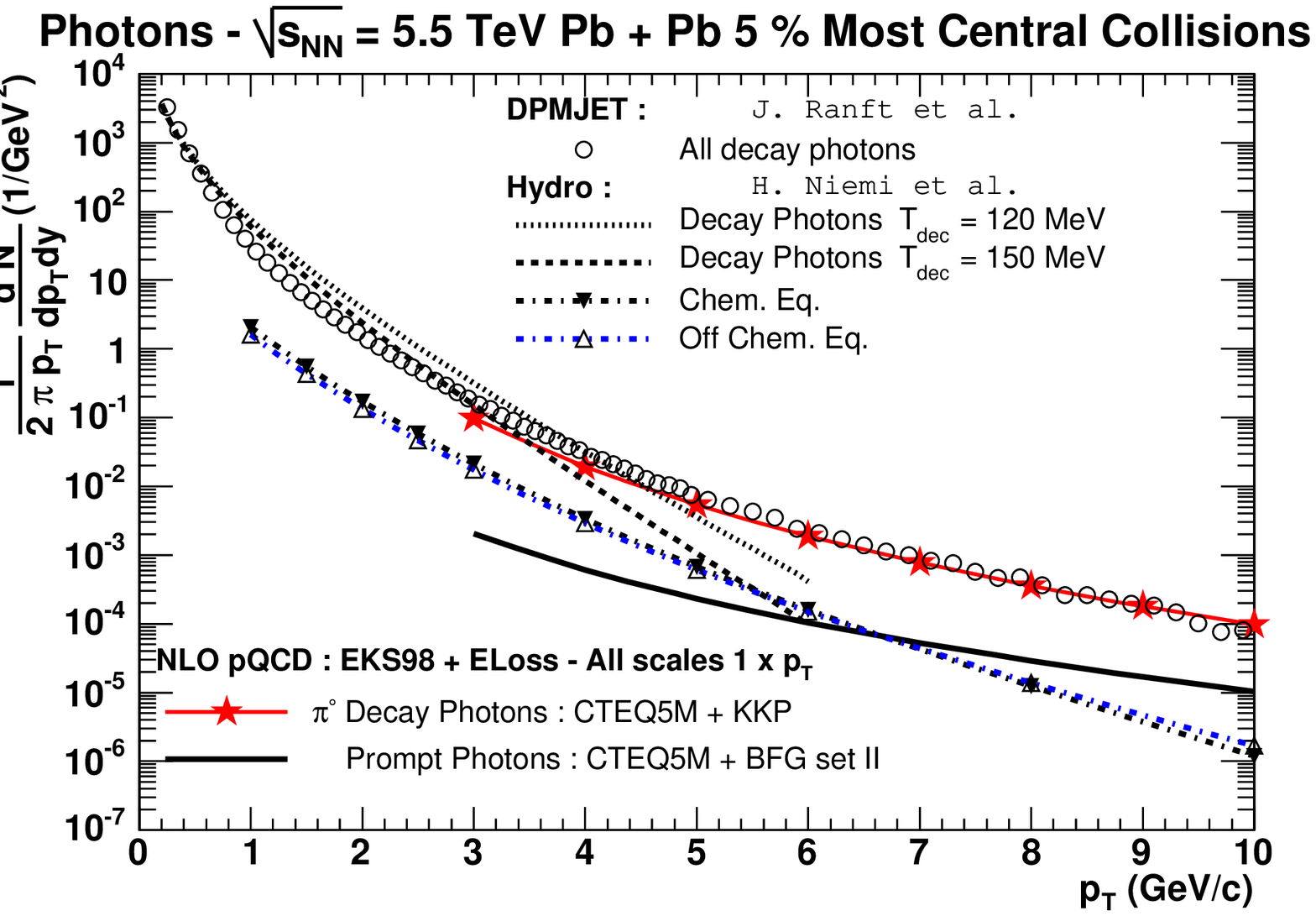}
\vspace{-.5cm}
 \caption{
Photon spectra in Pb+Pb collisions at $\sqrt{s} = 5.5$~TeV. Thermal emission is
calculated assuming either full thermal equilibrium in the plasma phase
(dashed-dotted line with dark triangle) or a gluon-rich plasma (dashed-dotted
lines with open triangle). Decay photons from thermal hadrons are indicated by
dotted or dashed lines depending of the decoupling temperature. The spectrum of
NLO prompt and decay photons are shown by the lower and upper solid lines
respectively.
}
 \end{center}
\label{fig:allphotonsa}
\end{figure}

\begin{figure}[thp]   
\begin{center}
\vspace{-2.cm}
\includegraphics[height=10.5cm]{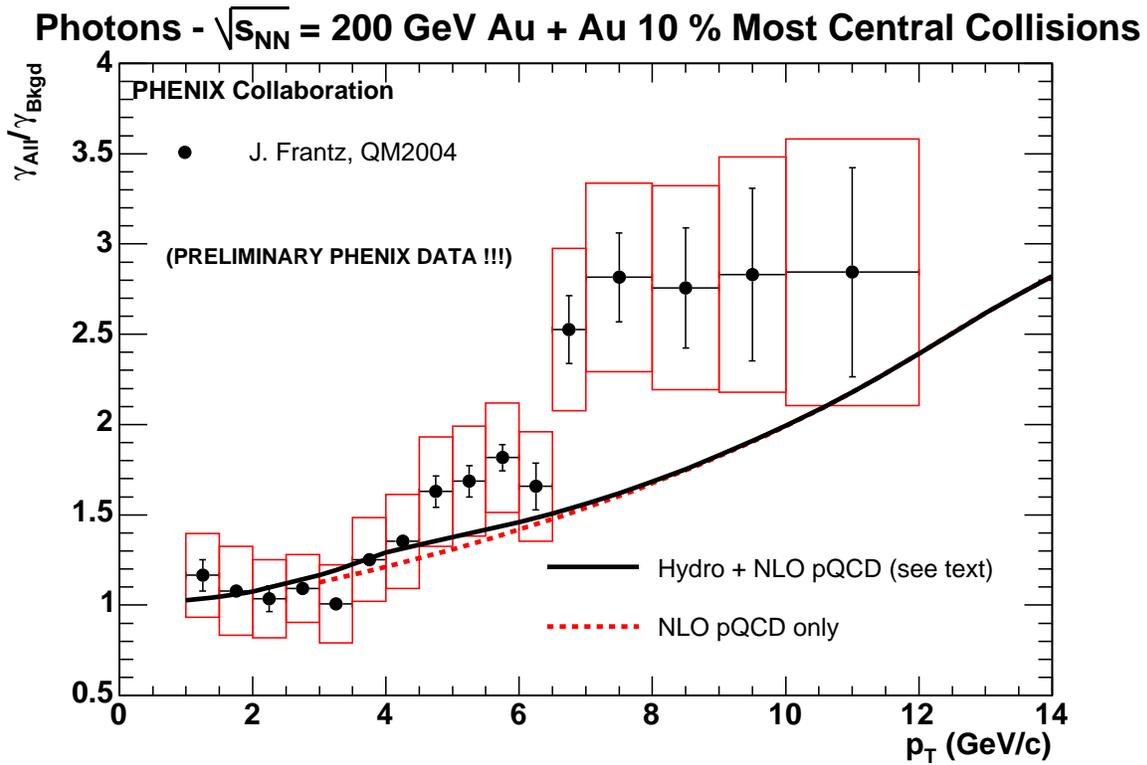}
\caption{
Ratio of all (NLO+Hydro) photons over all decay (NLO QCD + Hydro) 
photons as a function of $p_{_T}$, in Au+Au collisions at $\sqrt{s} =$~200~GeV.
Predictions for primary production of prompt photons (NLO QCD)
are shown by the dotted line. Preliminary data from PHENIX are also displayed.
 }
\end{center}
\label{fig:direct_over_all_rhic}
\end{figure}

\begin{figure}[thp]   
\begin{center}
\vspace{-.5cm}
\includegraphics[height=10.5cm]{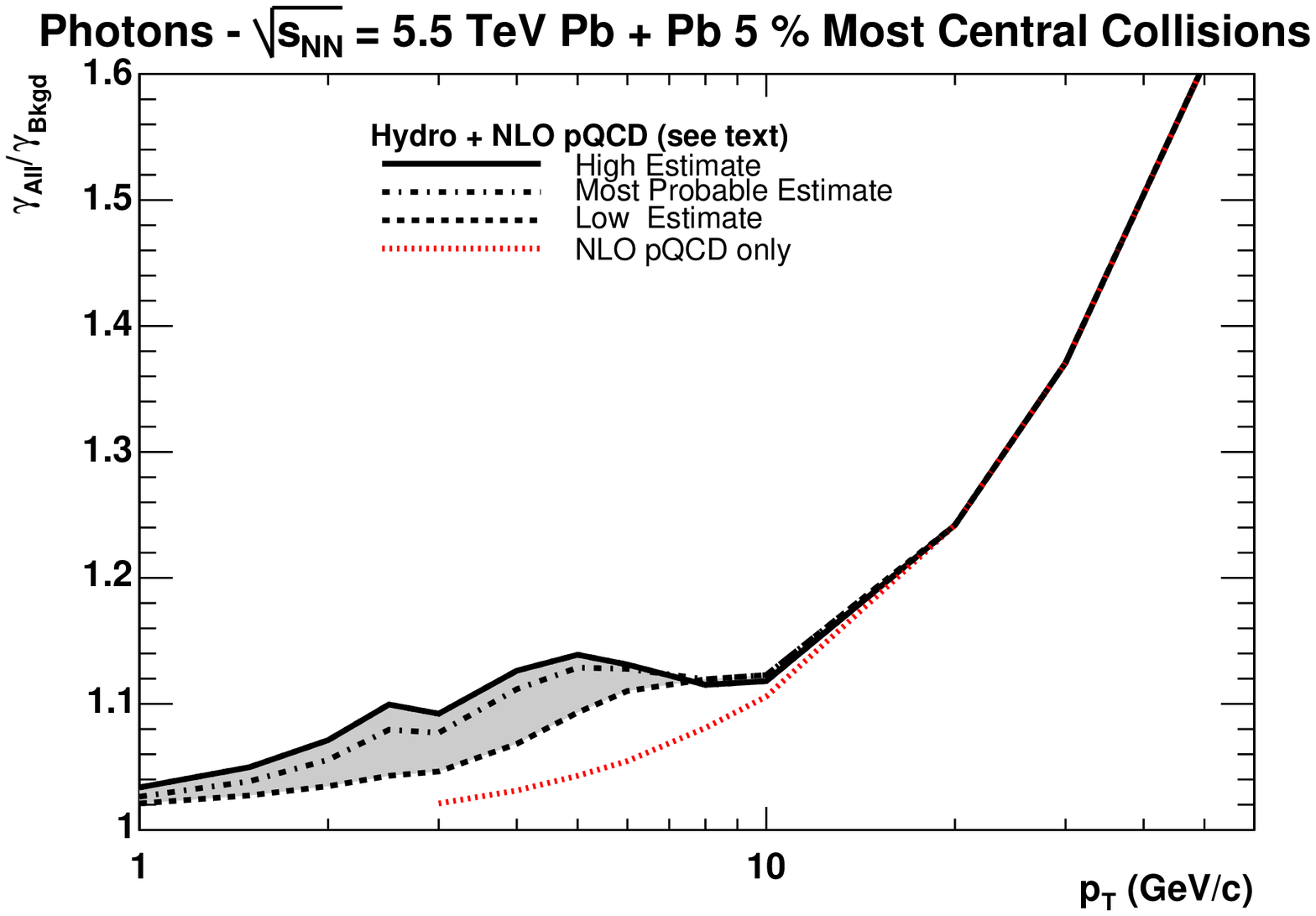}
 \caption{
Ratio of all (NLO+Hydro) photons over all decay (NLO QCD + Hydro)
photons as a function of $p_{_T}$, in Pb+Pb collisions at $\sqrt{s} =$~5.5~TeV.
Predictions for primary production of prompt photons (NLO QCD)
are shown by the dotted line.
 }
\end{center}
\label{fig:direct_over_all}
\end{figure}
%

In order to compare more clearly the different sources, we have plotted in
Figs.~73 and 74,
for RHIC and LHC, respectively, the ratio of the spectrum of ``all" photons 
to the background (from hadronic decays) photon spectrum.

In Au+Au collisions at $\sqrt{s} =$~200~GeV, the thermal photons from the
hydrodynamic calculation (solid line) provide a small increase 
in the ratio  $\gamma_{\rm All} / \gamma_{\rm Bkgd}$ in the range 3~GeV/c to
6~GeV/c compared to a pure pQCD estimate (dotted line). However this increase
appears too small to be observed experimentally. The thermal results are
insensitive to whether the plasma is in chemical equilibrium or not. 
With increasing $\pt$, due to prompt QCD photons, the fraction of direct
photons grows quite rapidly reaching 50~\% already at 10 GeV/c. Comparison with
PHENIX preliminary data \cite{Frantz:04} shows a good agreement for  $\pt
<$~4~GeV/c (where thermal effects play a major role) while the theoretical
predictions tend to underestimate the ratio at larger $\pt$ values where the NLO
QCD production mechanism is dominant. This disagreement can be related to an
overestimate, at large $\pt$, of $\pi^0$ rates, and hence of decay photons,
compared to data (see Fig. 69). 

In Fig.~74
the ratio $\gamma_{\rm All} / \gamma_{\rm Bkgd}$ is shown for Pb+Pb
collisions at LHC.  The upper curve is obtained by using the prediction
from the plasma in equilibrium for the thermal photons and 150 MeV for
$T_{\rm dec}$ which, for the calculation of the decay photons, gives the
steeper slope for the hadron spectra.  In the case of the lower curve
off-chemical equilibrium is assumed in calculating the thermal photons and
$T_{\rm dec}=120$~MeV in obtaining the hadron spectra for the calculation of
the decay photons. The  dash-dotted line indicates the prediction obtained
under the hypothesis of a gluon rich plasma  and a decoupling temperature of
150~MeV consistent with the RHIC data (see Fig. 68). Note also the rather
flat dependence, at around the 10 to 15\% level, of the ratio for $\pt$
between  3 and 10~GeV/c where thermal effects play a role.  Above 10 GeV/c the
ratio increases rapidly with $\pt$ so that direct photon production represents
25\% at $\pt=20$~GeV/c and 60\% at $\pt=60$~GeV/c.

In conclusion, the overall behaviour of $\gamma_{\rm All} / \gamma_{\rm Bkgd}$
is similar with that in the RHIC results but the ratios differ in details, in
particular the enhancement due to thermal production is shifted up to the
interval from 3 to 10~GeV/c and is larger.  


We should add to the rate of photons produced in A+A collisions also the
possibility that before a hard final state parton fragments into a jet
of hadrons it may convert to a photon in the parton medium via Compton
scattering or annihilation as proposed in ~\cite{Fries:2002kt}.  The
resulting spectrum should have a power behavior and could dominate over
the thermal photons already for $\pt\sim6\ldots9$ GeV/c (see
Fig.~\ref{fig:jetonthermal}).  This mechanism is hard to quantify,
however, and the presently available calculation has been done assuming
plasma evolution without transverse expansion which is not compatible
with the one used in our discussion.  Therefore we do not attempt a
quantitative comparison.  Nevertheless, it would be important to further
develop the modelling of this process.

There are other uncertainties in these results for $\gamma_{\rm
All}/\gamma_{\rm Bkgd}$, the most serious for the thermal contribution
being the uncertainty in $\tau_0$, the initial time in the
hydrodynamical calculation when the matter is assumed to reach an
(approximate) kinetic equilibrium.  If the time for the beginning of the
production of thermal photons is more than 1 fm/c, possibilities for
observing them are not very encouraging.  However, one should not be too
discouraged by those estimates which predict kinetic thermalization
times over 1 fm/c.  It may take time for the initially produced parton
system to get equilibrated, but even then we can expect it to be a dense
system of {\it interacting} partons right after their production.  When
interacting (and evolving towards equilibrium), these partons
would emit photons.  The uncertainty then is if the hydrodynamical
calculation with the assumption of kinetic equilibration gives a
reasonable estimate of the emission of photons from a nonequilibrium but
dense parton matter.

If the production of photons from secondary collisions is large enough
to be observed, then even if the produced parton system never reaches
equilibrium, these photons carry information on the typical scale of
relative momenta in the parton matter and on its density.  In an
equilibrated system the momentum scale is given by the temperature, and
the density and the momentum scale are related.  If early equilibration
does not take place the observation of photons from secondary collisions
would still be a direct window into the properties of the densest phase
of partonic matter formed in nuclear collisions.

As the discussion above shows, the phenomenology of photon production is
far from being on a quantitative basis yet.  However, our studies seem
to indicate that, at LHC, the ratio of direct photons over decay
photons, whatever their origin, should be 10\% or more for $p_{_T} >$
3 to 4~GeV/c.  At the lowest values of $p_{_T}$ the direct photons
would be mostly produced in secondary collisions while around 10 Gev/c
they would be of mixed origin with the perturbatively produced ({\em i.
e.} in primary collisions) photons becoming dominant with increasing
$\pt$.  Many of the uncertainties concerning thermal production would be
reduced if the model parameters could be fixed from hadron data leaving
then the photon spectrum as a crucial consistency check of the whole
approach.

\subsection{Comparing lepton pair spectra from main sources}
\label{sec:comp-leptons}


In this section we briefly compare lepton pairs from NLO QCD
interactions, from hadron decays, and from thermal interactions in the
plasma phase.  We present the transverse momentum distributions of pairs
in the mass interval [0.2,0.6]~GeV/c$^2$ and consider correlated pairs
only.  The results, discussed separately for each source in previous
sections, are summarized in
Fig.~75.
In all models and calculations the same input is used as for the photon
summary in the previous section.

\begin{figure}[thp]
\hspace{5mm}
\begin{center}
\includegraphics[height=10cm]{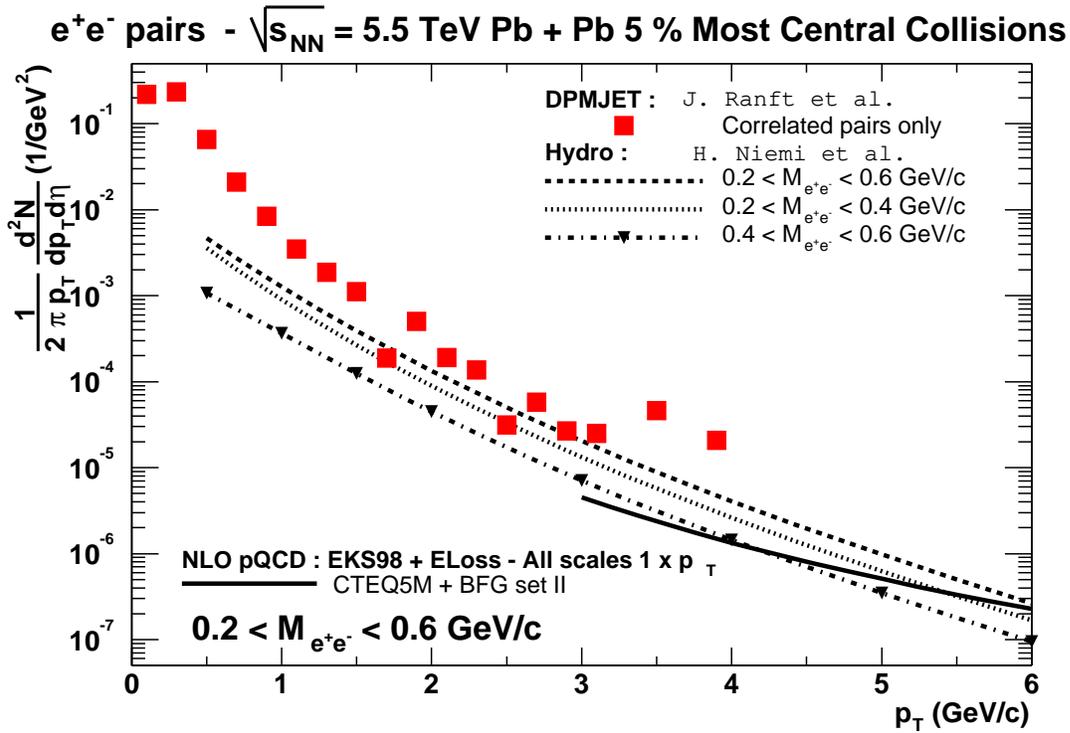}
\caption{
Lepton pairs from thermal emission are compared with those from prompt
emission in the NLO QCD approach.
}
\end{center}
\label{fig:all-leptons}
\end{figure}

The main background coming from hadron decays is illustrated with DPMJET
results (filled squares) since the DMPJET calculation agrees with the
NLO QCD results for hadrons at large and with the hydrodynamic
calculation at low transverse momentum.  Results from the NLO QCD
calculation on pair production are shown as the solid line and thermal
pairs from the hydrodynamic calculation as a dashed line.  Contributions
of thermal pairs from each half of the mass interval are shown also
separately to draw attention to the rather strong mass dependence in the
thermal production.

Statistics of the DPMJET calculation is not large enough for obtaining
results beyond $\pt \sim 3\ldots4$~GeV/c but the indications of the
low--$\pt$ behaviour seem clear:  pairs from hadron decays dominate at
low $\pt$ but they have a steeper $\pt$ dependence and seem to join the
distribution of thermal pairs around $\pt\sim3$~GeV/c.  Pairs from hard
interactions of incoming partons, given by the NLO QCD result, overtake
the thermal production around $\pt\sim6$~GeV/c.  If the $\pt$ dependence
of different contributions is similar to that in the photon production,
we would conclude that for $\pt~\gsim~3$~GeV/c the background from the
hadron decay pairs has the same size as the sum of thermal and NLO QCD
pairs and might go below it as $\pt$ increases.

In summary, considering the signal-to-background ratio the $e^+e^-$
channel appears at first sight quite promising for the study of hot
matter in nucleus-nucleus collisions.  However, our discussion is
preliminary and lots of work remains to be done in order to confirm if
the spectrum of direct lepton pairs can be extracted from the large
background of uncorrelated pairs within the statistics which could be
expected to be available at LHC.


%
\section{PRODUCTION OF PHOTONS AT ``LARGE" TRANSVERSE MOMENTUM AND CORRELATION
STUDIES}
\label{correlations}
{\em F.~Arleo, P.~Aurenche, J.-Ph.~Guillet,O.L.~Kodolova, I.P.~Lokhtin,
A.~Nikitenko, I.N.~Vardanyan, M.~Werlen}



It was shown in previous chapters that the signal for direct, thermal or
non-thermal, photon production in A+A collisions is not too large compared to
various backgrounds, making it a challenging but not easy task to use inclusive
photon production as a signal for quark-gluon plasma formation in heavy ion
collisions. In this chapter we consider correlation observables in a
kinematical regime where the photon is produced ``promptly" in a hard QCD
process and we study the hadrons recoiling from the photon. These hadrons are
decay products of a jet, most probably  a quark. Since the fragmentation
properties of jets are supposed to be modified by medium effects, differences
in the shape of the correlations should be seen when comparing $pp$ and A+A
collisions. In the next section one looks at the global jet and study, from an
experimental (CMS) point of view, the effect of smearing and efficiency  on an
assumed energy loss. Then a discussion is given on photon-photon and
photon-hadron correlations where the hadron or the second photon are decay
products of the jet. One takes a more phenomenological point of view and
discuss various functions in the hope of determining the correlation most
sensitive to medium effects. All studies are preliminary in the sense that they
are conducted at the leading order and therefore will be modified by NLO
corrections which may be very important in some corners of phase space.

%
%
\subsection{Photon-jet correlation at CMS}
{\em O.L.~Kodolova, I.P.~Lokhtin, A.~Nikitenko, I.~Vardanian}

Among other proposed signals of jet quenching (see Chapter "Jets" of this
Report~\cite{wiedemannYR}), the  $\pt$-imbalance between a produced jet with a
gauge boson in $\gamma+$jet~\cite{hsw,Wang:1996} and
$Z+$jet~\cite{Kartvelishvili:1996} production has been identified as being
observable in  heavy ion collisions using the CMS detector~\cite{Baur:2000}. 
The dominant leading order  diagrams for high transverse momentum $\gamma$ +
jet production are shown in  figure~\ref{gamjet:fig1}. Contrary to the
gluon-dominated jet pair production where one could  investigate jet quenching
due to mostly gluon energy loss in dense matter,  $\gamma+$jet  channel gives a
possibility to study quark energy loss.  The main background here is hard jet
pair  production when one of the jet in an event is misidentified as a photon.
The leading $\pi ^{0}$ in 
the jet is a main source of the misidentification. Table~\ref{gamjet:tab1} presents the event rates 
for signal and background processes in one month of Pb+Pb beams (a half of the time is 
supposed to be devoted to data taking), $R= 1.2 \times 10^6$ s, assuming luminosity 
$L = 5 \times 10^{26}~$cm$^{-2}$s$^{-1}$ to that $$N(events)= R \sigma^{h}_{AA} L ,$$ where 
production cross sections in minimum bias nucleus-nucleus collisions is obtained from those in 
$pp$ interactions at the same energy ($\sqrt{s} = 5.5$ TeV) using simple parameterization 
$\sigma^h_{AA}=A^2 \sigma^h_{pp}$. The cross section in $pp$ collisions were evaluated
using the PYTHIA\_$6.1$ Monte-Carlo generator~\cite{pythia6-a} with the CTEQ5L parton 
distribution function. Note that the influence of nuclear shadowing is practically negligible for the 
region of  sufficiently hard $\gamma+$jet production, $x_{1,2} \sim
\sqrt{\widehat{s}/s} \ga  0.2$. However, large theoretical uncertainties in
absolute rates in $pp$ collisions come  from choice of the parton distribution
functions, next-to-leading corrections, etc.  It means that measurements in
$pp$ or D+D collisions at  the same or similar energies per nucleon as in the
heavy ion runs are strongly desirable to  determine the baseline rate
precisely. 

\begin{table}[htb]
\begin{center}
\label{gamjet:tab1} 
\caption{\small Expected rates for $\gamma$+jet and  $\pi_0 (\rightarrow 2\gamma)$+jet channels 
in one month of Pb$-$Pb beams.} 

\medskip 

\begin{tabular}{|l|c|c|} \hline  
Channel & Barrel, $|\eta|<1.5$ & Barrel+Endcap  $|\eta|<3$ \\ \hline 
$\gamma$+jet, $E_T^{jet,\gamma}>100$ GeV & 1.6$\times$10$^3$ & 3.0$\times$10$^3$  
\\ \hline 
$\pi_0 (\rightarrow 2\gamma)$+jet, $E_T^{jet,\gamma}>100$ GeV & 8.4$\times $10$^3$ 
& 2.2$\times$10$^4$  \\ \hline  
\end{tabular}
\end{center}
\end{table}

One can see that even for events with $E_T>100$ GeV the background is still
dominant. Signal-to-background ratio becames close to $1$ only above $200$ GeV.
The identification of the influence of the dense medium formation on signal
spectra requires the reduction of the background. One of the possibilities is
to apply some kind of the photon isolation, so called "zero suppression
criteria", which requires no energy above a given threshold around photon (see
section "Photon detection at CMS" for details).  In this case the
signal-to-background ratio at $E_T>100$ GeV can be improved by a factor about 2.3
at the cost of a $14\%$ reduction of the signal. Another possibility is to apply
some kinematical cuts, which do not have influence on the $\pt$-imbalance of
the process (i.e. not result in shift of maximal value of 
$E_T^{\gamma}-E_T^{jet}$ distribution).  

The possibility to observe the medium-induced energy loss of quark-initiated
jet  using photon-jet correlation in heavy ion collisions with CMS detector has
been  investigated in~\cite{Kodolova2:1998,Baur:2000}. It has been found that
initial state gluon radiation and finite jet energy resolution (which is much
larger than photon energy resolution)  result in significant smearing of
distribution of differences in transverse energies between the  photon and jet.
But it is still symmetric: $E_T^{\gamma}=E_T^{jet}$ only in average  (not for
each given event). The non-symmetric shape of the distribution appears if  a
jet loses energy: the maximal value of the distribution is equal to the
average  energy loss of the quark-initiated jet at the given energy detection
threshold, $E_T^{jet} \sim 100$ GeV in CMS case. Note  that we are not
measuring energy loss of a leading quark by such a   way, but getting total
loss of quark-initiated jet outside the given jet cone. 

\begin{figure}[htbp]
\begin{center} 
\resizebox{75mm}{75mm} 
{\includegraphics{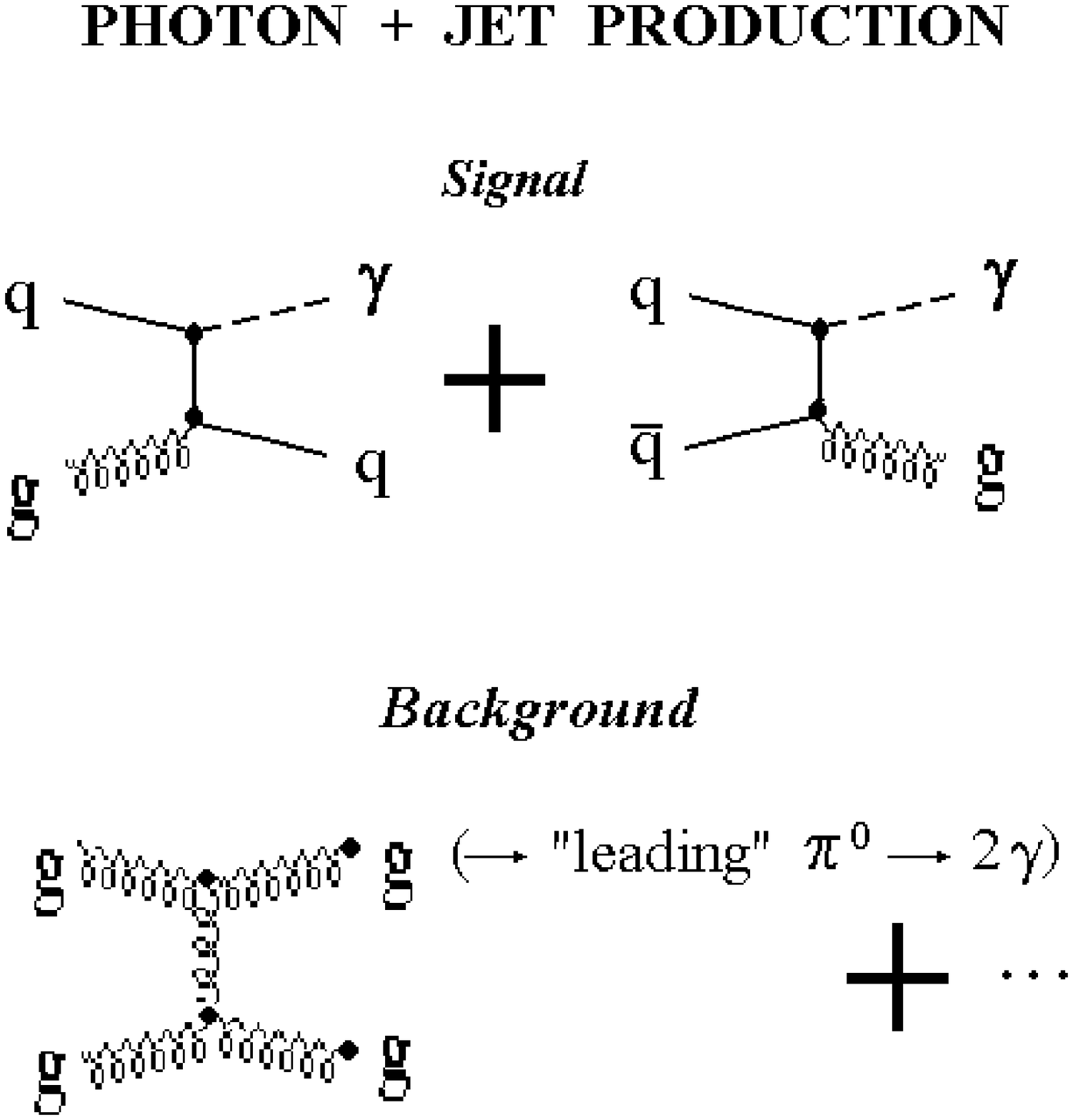}} 
\caption{\small Leading order diagrams for  $\gamma$+jet (signal) and  $\pi_0 (\rightarrow 
2\gamma)$+jet (background) production. }
\label{gamjet:fig1}
\end{center} 
\end{figure}

\begin{figure}[htbp]
\begin{center} 
\resizebox{110mm}{120mm} 
{\includegraphics{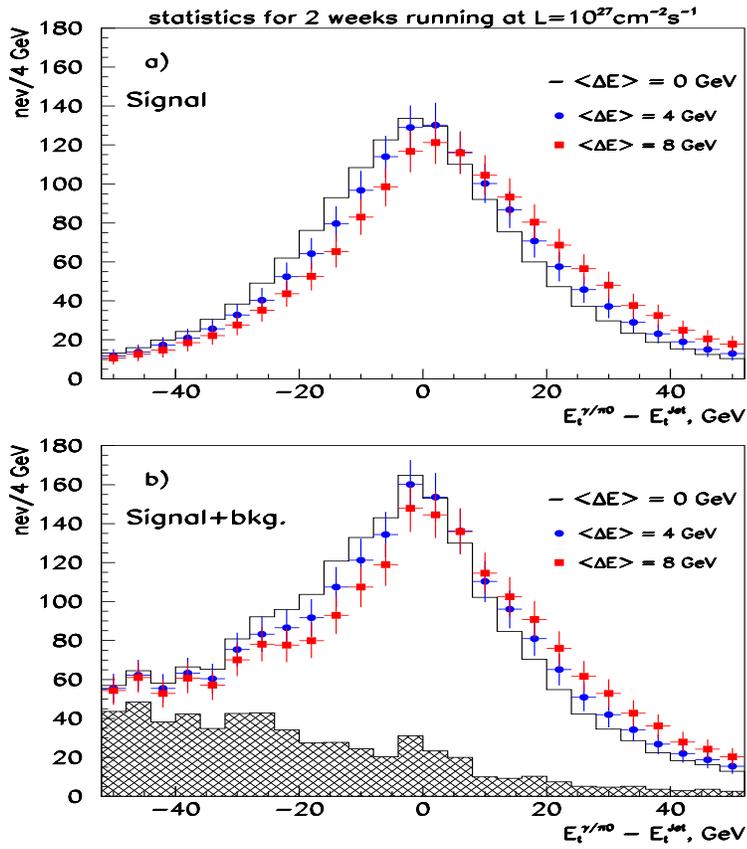}} 
\caption{\small The distributions of differences in transverse energy between the $\gamma$ and 
jet with $E_t^{\gamma,~jet} > 120$ GeV: a) without ($\pi^0$ + jet) background,  b) with 
($\pi^0$ + jet) background. The pseudorapidity coverage is  $\mid \eta_{\gamma,~jet} \mid < 1.5$. 
Different values of jet in-medium energy loss, initial state gluon radiation and finite jet energy
resolution are taken into account.}
\label{gamjet:fig2}
\end{center} 
\end{figure}

In order to test the sensitivity of $\gamma$+jet production to jet quenching,
it was   considered three scenarios with average collisional energy loss of a
jet:  $<\Delta E_{q}> \simeq 0$, $4$ and $8$ GeV respectively ($<\Delta E_{g}>
= 9/4 \cdot   <\Delta E_{q}>$). The jet energy resolution at mid-rapidity
obtained for Pb+Pb  collisions has also been used to smear the energy of the
recoiling parton, as well as  the jet rejection factor and signal
efficiency~\cite{Kodolova2:1998}. Figure~\ref{gamjet:fig2}  shows the
distributions of differences in transverse energy between the photon and  jet
with $E_T^{\gamma,~jet} > 120$ GeV  in one month of Pb+Pb beams without and  
with ($\pi^0$ + jet) background in the pseudorapidity region $\mid
\eta^{\gamma,~jet} \mid < 1.5$  for different values of jet energy loss. In
this case a luminosity $L =   10^{27}~$cm$^{-2}$s$^{-1}$ was assumed and
PYTHIA\_$5.7$ version with the default CTEQ2L pdf choice was used. The jet
energy resolution leads to a difference between the input values  $<\Delta
E_{q}>$ and the ones obtained from the spectra. The background of 
$\pi^0$-contamination results in non-zero negative values of the final
distributions  (figure ~\ref{gamjet:fig2}b) already in the case without jet
energy loss. However one can see that the shape of the distribution is well
distinguished for the scenarios considered. For the region of 
$(E_T^{\gamma}-E_T^{jet}) > 0$ there is a difference for almost every bin
greater than  $1$ standard deviation for the rather small jet energy loss $8$
GeV and even for the loss $4$ GeV.  In the real experiment it would be possible
to estimate the number of background events using  the region without the
signal ($E_T^{\gamma}-E_T^{jet}) < - 100$ GeV/c and background shape  from
Monte-Carlo simulation and/or from $pp$ data. A significant difference in the
shape of  the $E_T^{\gamma}-E_T^{jet}$ distribution can allow to optimize
extraction of the signal from the experimental spectra.

%
%

\subsection{Photon-hadron and photon-photon correlations in $pp$ and A+A
collisions}
{\em F.~Arleo, P.~Aurenche, J.-Ph.~Guillet}

In the previous chapters it was suggested that thermal production of photons
could be seen for rather low $\pt$ values. In the following we consider, on the
contrary, the case of a photon with a high enough $\pt$ so that it is produced
promptly in a hard QCD process and therefore not affected by thermal effects
and we study the decay products of the jet recoiling from this photon. The
analogue of the inclusive spectrum Eq.~(\ref{eq:inclu-rate}), at the leading
logarithmic order, can be written~\cite{cfg1996}
\begin{eqnarray}
{d \sigma^{^{AB \rightarrow CD}} \over d p_{_T{_3}} dy_3 d p_{_T{_4}} dy_4} =
{1 \over 8 \pi s^2} & \sum_{a,b,c,d} & 
 \int_{z_{3_{min}}}^1 {dz_3 \over z_3}\  D_{C/c}(z_3,M_{_F})
\ \int_{z_{4_{min}}}^1 {dz_4 \over z_4}\  D_{D/d}(z_4,M_{_F})\ k_{_T{_3}}\nonumber\\ 
&&  \delta(k_{_T{_3}}-k_{_T{_4}}) \ 
{F_{_{a/A}}(x_1,M) \over x_1} \ { F_{_{b/B}}(x_2,M) \over x_2}  \ \ 
|{\overline M}|^2_{ab \rightarrow cd}
\label{eq:correl-rate}
\end{eqnarray}
where the $p_{_T{_i}}$ (resp. $k_{_T{_i}}$) are the final state particle (resp.
partonic) transverse momenta and the scaling variables $z_i$ are defined by
$z_i = p_{_T{_i}} / k_{_T{_i}}$. The quantity  $|{\overline M}|^2_{ab
\rightarrow cd}$ is the matrix element squared, averaged over spin and color,
of the partonic sub-process $ab \rightarrow cd$.   Higher order corrections to
Eq.~(\ref{eq:correl-rate}) have been calculated~\cite{{cfg1996}} and, based on
existing data~\cite{wa70-gam-gam,e706-gam-gam,d0-gam-gam},
extensive NLO phenomenological studies of correlations have been carried from
fixed target energies to the LHC~\cite{bgpw2002,bgpw2001}. If one looks at a photon at
large enough $p_{_T{_3}}$, it will be directly produced and the corresponding
fragmentation will reduce to $\delta(1-z_{~3})$. Studying various correlation
variables will then allow the mapping of the fragmentation function 
$D_{D/d}(z_4,M_{_F})$ since all other ingredients (structure functions, matrix
elements) are known. One could then observe the effect of partonic energy loss
in the medium by comparing correlation functions in A+A collisions with  the
equivalent ones in $p p$ collisions. In the following we restrict ourselves to
the case of promptly produced photon and do not impose isolation cuts. We do
not consider the ``background" contribution generated by photons decaying from
hadronic resonances, a problem presently under investigation~\cite{arleo2004}. 

\begin{figure}[htbp]
\begin{center}
\vskip-3.cm
\includegraphics[height=14.0cm]{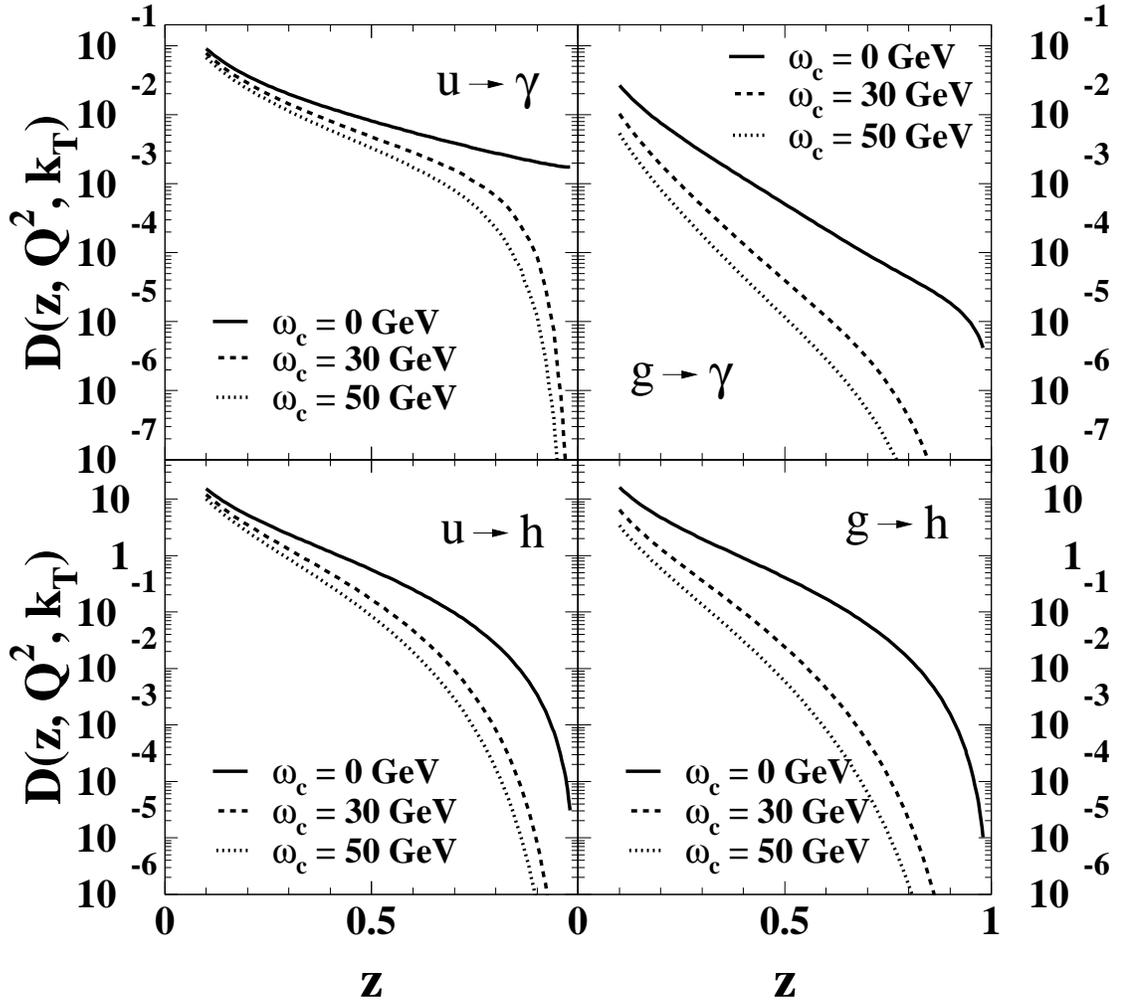}
\end{center}
\vskip-0.5cm
\caption{  
Comparison of partonic fragmentation functions in the vacuum and in the medium
for typical values of the energy loss parameter $\omega_c$. {\em Top two 
figures:} $k_{_{Td}} =$~20~GeV/c up quark and gluon into a photon; {\em bottom
two figures:} $k_{_{Td}} =$~20~GeV/c up quark and gluon into a charged hadron.
}
\label{fig:nuc-frag-func}
\end{figure} 
How to express final state interactions through medium-modified fragmentation
functions  remains however far from being clear. To illustrate the effect of
parton energy loss in dense media on correlation functions, we shall adopt here
the effective model suggested in  Ref.~\cite{Wang:1996}.  Within this
approach, the parton energy shift leads to a rescaling of the momentum
fraction  $z_d$ in  presence of a QCD medium,
\begin{equation}\label{eq:shift}
z_d = \frac{p_{_{Td}}}{k_{_{Td}}} \qquad \to \qquad z_d^* = \frac{p_{_{Td}}}{k_{_{Td}} - \epsilon} = \frac{z_d}{1 -
\epsilon/k_{_{Td}}},
\end{equation}
where $p_{_{Td}}$ (respectively, $k_{_{Td}}$) stands for the transverse
momentum of the photon or hadron (respectively, the parton), and $\epsilon$ the
energy lost by the hard  parton while  going through the medium. Consequently,
the medium-modified fragmentation functions  $D_{D/d}^{med}(z_d, Q^2, k_{_{Td}})$ may
simply be expressed as a function of the standard (vacuum)  fragmentation
functions $D_{D/d}(z_d, Q^2)$ through~\cite{Wang:1996}
\begin{equation}
\label{eq:modelFF}
z_d\,D_{D/d}^{med}(z_d, Q^2, k_{_{Td}}) = \int_0^{k_{_{Td}} - p_{_{Td}}} \, d\epsilon
\,\,{\cal D}(\epsilon, k_{_{Td}})\,\,\, z_d^*\,D_{D/d}(z_d^*, Q^2).
\end{equation}
Here, ${\cal D}(\epsilon, k_{_{Td}})$ denotes the probability for a parton with
transverse energy  $E=k_{_{Td}}$ to lose  an energy $\epsilon$~\cite{bdms2001}.
Note that medium-induced fragmentation functions now depend explicitely on the
parton transverse energy $k_{_T}$. Assuming the soft gluons radiated by the
leading hard parton to be emitted independently, this probability distribution
(or quenching weight) can easily be related to the medium-induced gluon
spectrum determined perturbatively, $dI/d\omega$, characterized by the energy
scale~\cite{eloss}
\begin{equation}
\omega_c = \frac{1}{2}\,\hat{q}\, L^2
\end{equation}
The gluon transport coefficient $\hat{q}$ reflects the medium gluon density
while $L$ is the  length of matter covered by the hard parton. Note that
$\hat{q}$ can be very large in practice in a hot pion gas or quark gluon
plasma~\cite{Baier:2002tc}. Based on the estimate from the pion inclusive
spectrum in A+A collisions at RHIC ($\sqrt{s} = 200$~GeV), we shall 
illustrate our results using the following values $\omega_{~c}=30$ and $50$~GeV
for the  characteristic scale of the dense medium produced in nuclear
collisions at the LHC.  The distribution ${\cal D}(\epsilon, k_{_{Td}})$ has
been given a simple analytic parameterization in~\cite{arleo2002} which we
shall use in the present calculations.

Since fragmentation functions fall steeply with $z$, even a small shift 
$\Delta z_d=z_d^*-z_d\approx z_d\,\epsilon/k_{_{Td}}$ in (\ref{eq:shift}) may
substantially affect  fragmentation processes in presence of a hot and dense
QCD medium. This can be seen for  instance in Figure~\ref{fig:nuc-frag-func}
where fragmentation functions into a photon ({\it top}) or a charged hadron
({\it bottom}) are computed for $k_{_T} = 20$~GeV/c up quark ({\it left}) and
gluon ({\it right}) traversing the medium. In particular, we expect the effects
of parton  energy loss to be more pronounced as $z$ gets larger. This will be
further discussed below when correlation functions in nuclear collisions are
computed perturbatively.

As an example we consider the following kinematical constraints. Both observed
particles (photon or charged hadron) are required to be in a rapidity interval
[$-0.5,+0.5$] and furthermore one photon should have a transverse
momentum greater than 10 GeV/c while the other photon or the charged hadron is
required to have $\pt > 3$~GeV/c. Integrated over all kinematical variables,
with the above constraints, we display in Fig.~\ref{fig:pp-cross-section} the
photon spectrum $d \sigma / d {\pt}$ for the case of two photon events
({\em left}) and the charged hadron spectrum for the photon + hadron
events ({\em right}). The various dynamical components are shown. In the two
photon case  they are the ``direct" (both photons produced in the hard
sub-process), the ``one-f" (one photon produced directly and the other by
bremsstrahlung from a fragmenting parton) and the ``two-f" (both photons produced by bremsstrahlung of
the final state partons)~\footnote{
	We follow for the labeling of the various terms the conventions used in
	the DIPHOX code~\cite{diphox-program}. In the definition of Chap.~2 a
	photon produced ``directly in the hard sub-process" or a photon
	produced by bremsstrahlung of a parton produced in a hard QCD
	sub-process are both ``prompt" photons since they emerge from a primary
	collision and are not decay products of a resonance.
}.
\begin{figure}[htbp]
\begin{center}
\includegraphics[height=7.0cm]{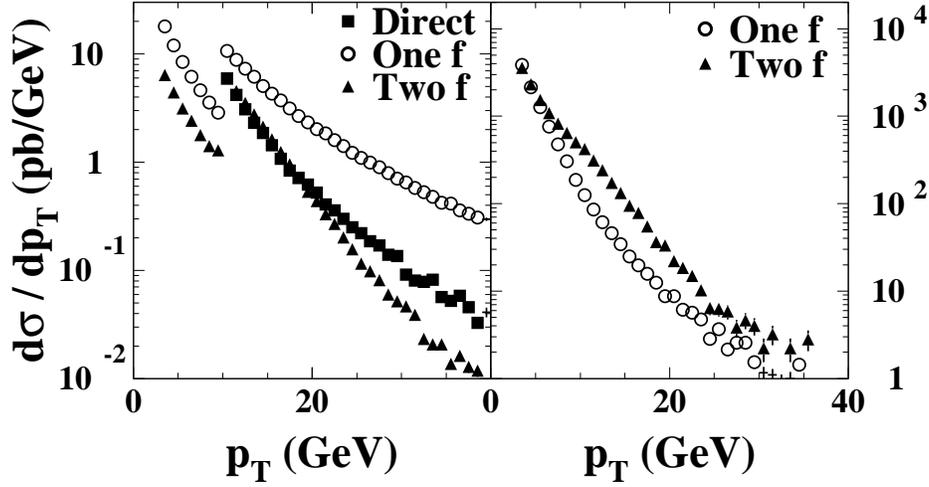}
\end{center}
\vskip-0.5cm
\caption{  
Details of the various dynamical components in $pp$ collisions at $\sqrt s =
5.5$~TeV. {\em Left}: the photon spectrum in the two photon events; {\em right}: 
the charged hadron spectrum in the photon-hadron events.
}
\label{fig:pp-cross-section}
\end{figure}  
The interpretation of the photon-photon figure is easy. For $\pt < 10$~GeV/c,
the observed photon is the less energetic of the two photons because of the 10
GeV/c cut imposed on the other photon: the observed photon can only be produced
by bremsstrahlung  and the figure shows that the ``one-f" process dominates by
far. When $\pt > 10$~GeV/c, the direct component comes into play but again
the ``one-f" process is the largest. This confirms that, with the cuts used,
the cross section Eq.~(\ref{eq:correl-rate}) for the case of two photon
production is mainly sensitive to one fragmentation function. For the
photon-hadron case the ``one-f" (the photon produced directly and the hadron
a fragment of a jet) and the ``two-f" (both particle fragments of jets)
processes are comparable. 
A word of caution is necessary: as usual in perturbative QCD, the distinction
between direct, one-f and two-f is only indicative as the relative weights of
the components depends on the choice of the (factorization, renormalization)
scales, chosen in this discussion to be all equal to $\pt$.

In the following we define various correlation functions in $\gamma-\gamma$
events~\cite{diphox-program} (see Fig.~\ref{fig:AA-gam-gam-cross-section}) and
study their modifications as a function of the assumed energy loss
characterized by the value of the parameter $\omega_c$. They are :\\
-- the photon $\pt$ distribution;\\
-- the diphoton invariant mass $m_{\gamma \gamma}$ spectrum;\\
-- the spectrum in the transverse momentum of the pair defined as  
${{\imb q}_{_T}} = {{\imb p}_{_{T3}}} + {{\imb p}_{_{T4}}}$;\\
--  the distribution in the scaled momentum fraction
$z_{~34} = - {\imb p}_{_{T3}} \cdot  {\imb p}_{_{T4}} / {\imb p}^2_{_{T3}}$.

\begin{figure}[bhtp]
\includegraphics[height=15.0cm]{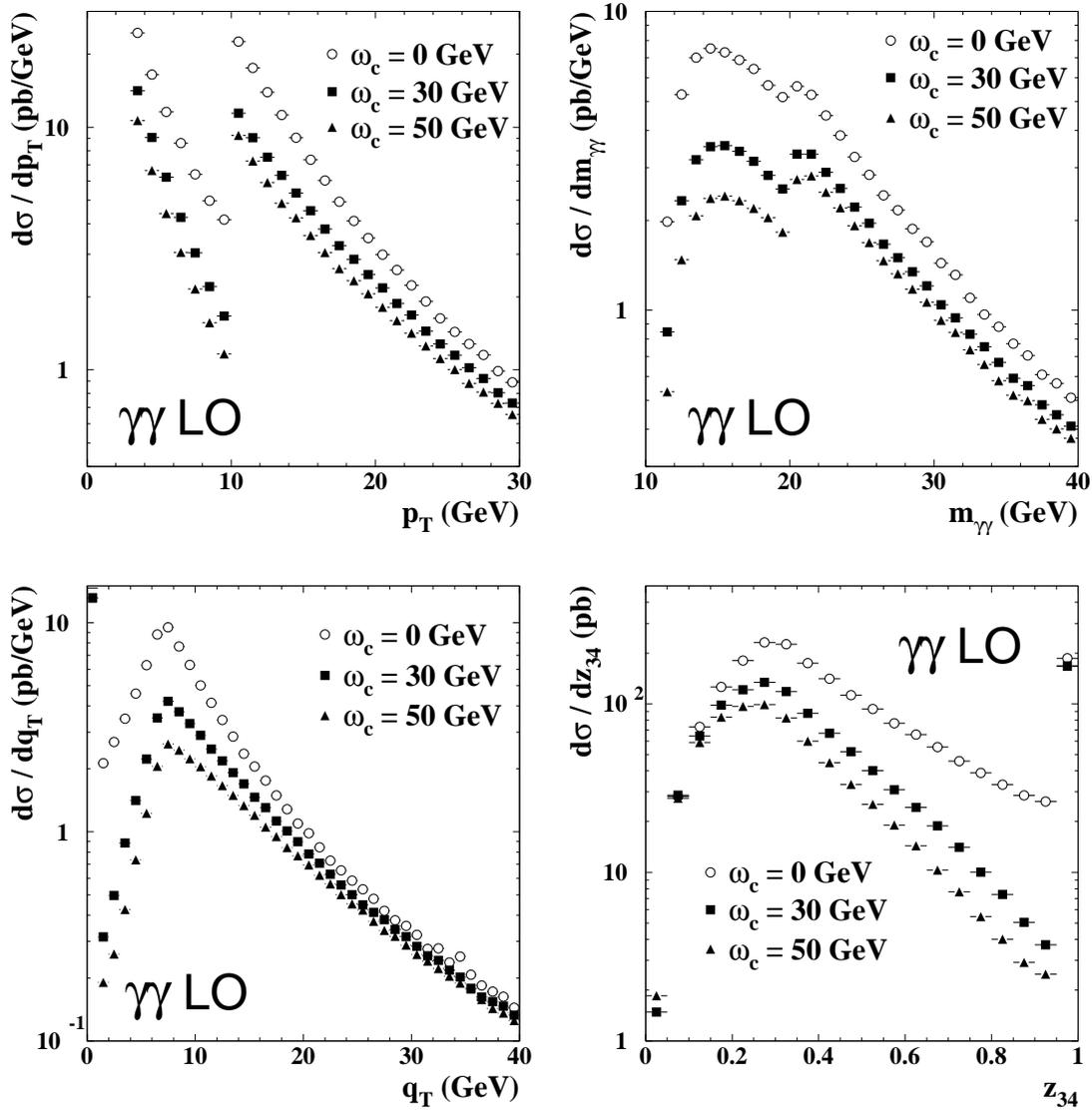}
\caption{  
Photon-photon correlation functions in Pb+Pb collisions at $\sqrt s =$~5.5~TeV.
}
\label{fig:AA-gam-gam-cross-section}
\end{figure} 
Considering the spectrum $d \sigma / d \pt$ in
Fig.~\ref{fig:AA-gam-gam-cross-section} one sees that the curve below $\pt =
10$~GeV/c decreases when $\omega_c$ increases (energy loss increases), very
similarly to the fragmentation functions in Fig.~\ref{fig:nuc-frag-func}. The
relative suppression is larger as $\pt$ increases corresponding to an increase
of the effective fragmentation variable $z$. For $\pt$ above the cut, on the
contrary, the medium and the vacuum spectra come closer together when the
transverse momentum increases. This can be understood as follows: the detected
photon with large momentum is produced directly while the other photon is
produced predominantly near the  3 GeV/c threshold and its effective 
$z \sim z_{~34}$
decreases when the $\pt$ of the directly produced photon increases: since at
small $z$ the energy loss in the medium is small (see
Fig.~\ref{fig:nuc-frag-func}) the in-medium curve and the ``vacuum" curve
approach each other.

The di-photon invariant mass spectrum (upper right plot in
Fig.~\ref{fig:AA-gam-gam-cross-section}) shows, in vacuum as well as in the
medium, a characteristic bump above $m_{\gamma \gamma} =$ 20~GeV/c due to the
onset of the ``direct" production mechanism. Only ``one-f" and ``two-f"
processes contribute below this threshold. When increasing the energy loss,
these last two processes are suppressed, leading to a larger suppression of the
invariant mass distribution at low masses and an increasingly marked shoulder
since the direct channel is not affected by the medium.
\begin{figure}[bhtp]
\includegraphics[height=15.0cm]{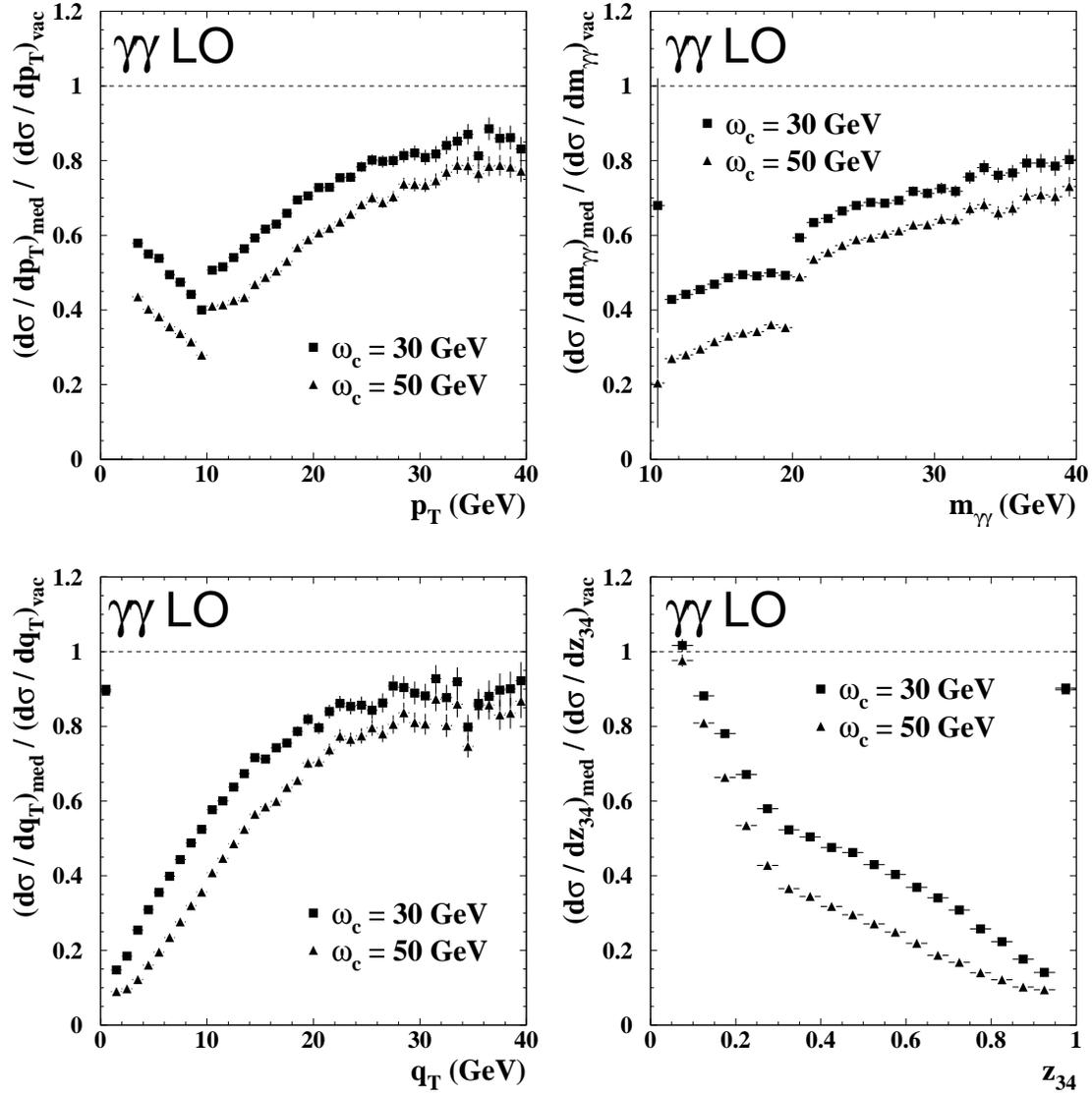}
\caption{  
Ratio of photon-photon correlation functions in Pb+Pb collisions and $p p$
collisions at $\sqrt s =$~5.5~TeV.
}
\label{fig:AA-gam-gam-ratio}
\end{figure} 

The peaked shape of the transverse momentum distribution of the pair (lower
left plot in  Fig.~\ref{fig:AA-gam-gam-cross-section}) at $\pt \sim$ 7~GeV/c is
an effect of the imposed asymmetric lower cuts on the photon momenta and the
decrease of the spectrum at the lower $q_{_T}$ values when the energy loss
increases is easy to understand: to obtain a small $q_{_T}$ value both photons
should have similar $\pt$ which means that the bremsstrahlung photon will be
emitted from a parton with a large $z$ value, a configuration highly unfavored
in the medium. The suppression of the spectrum for small $q_{_T} = 0$ will be
somewhat modified by higher order corrections which will smear the direct
process contribution shown, in the figure,  by a point at $q_{_T} = 0$. On the
other hand at large $q_{_T}$ no dramatic medium effect is expected since the
dominant kinematics will be a photon (directly produced) at large  $\pt$ and a
bremsstrahlung photon at very small $\pt$, equivalently small $z$, both of which
are not affected by medium effects

Finally, the distribution in the scaling variable $z_{~34}$ (bottom right plot)
nicely reflects the behavior of the fragmentation function when medium effects
are increasing: the scaling variable and the fragmentation variable $z$ are
very closely related and it is natural that the $z_{~34}$ be reminiscent  of the
behavior of the fragmentation functions shown in Fig.~\ref{fig:nuc-frag-func}
when the energy loss increases.
\begin{figure}[htbp]
\includegraphics[height=15.0cm]{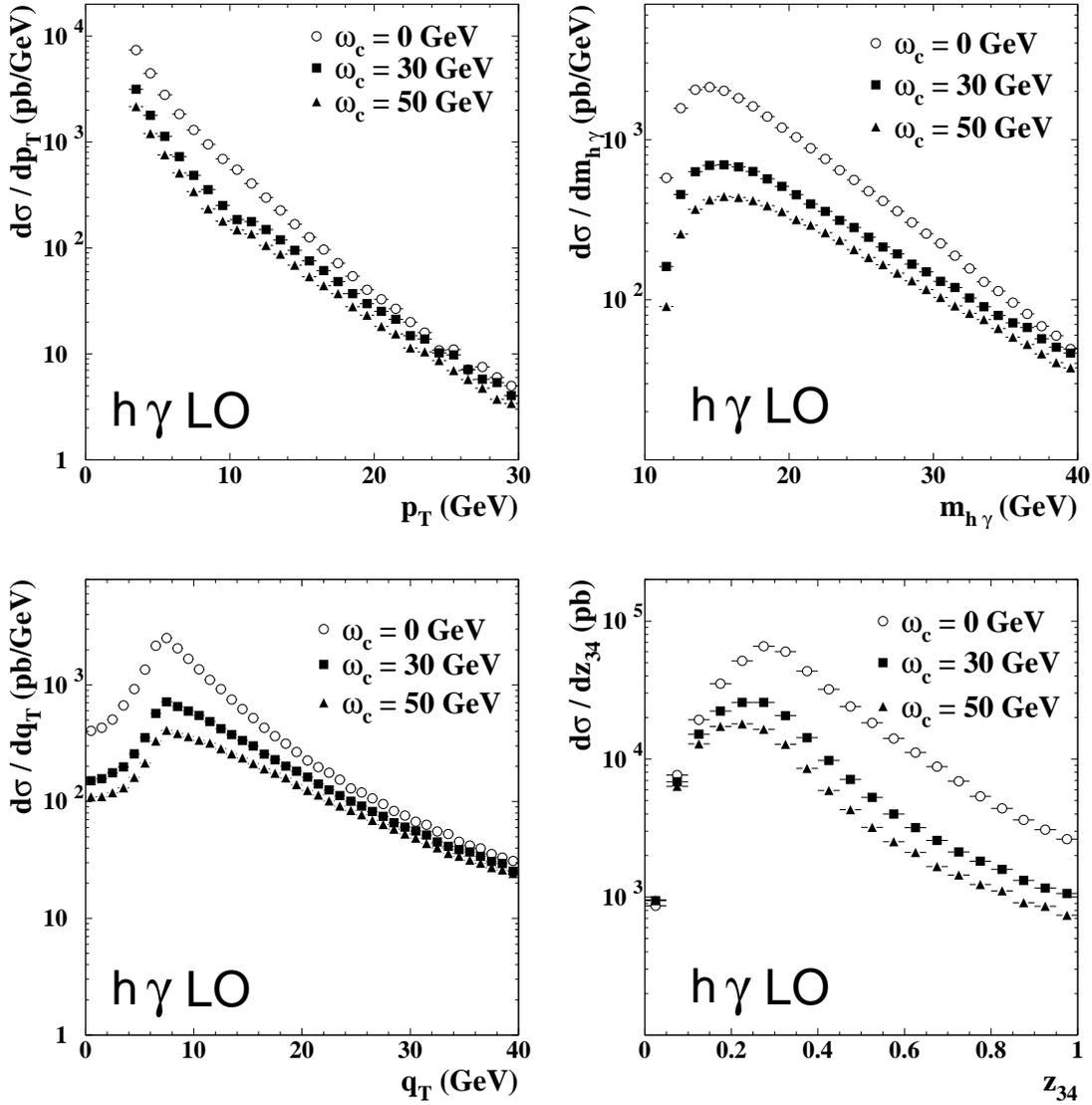}
\caption{  
Photon-charged hadron correlation functions in Pb+Pb collisions at $\sqrt s
=$~5.5~TeV. 
}
\label{fig:AA-chg-gam-cross-section}
\end{figure} 

The energy loss features of the observables we just discussed are emphasized if
we plot the ratio of the correlation functions in A+A collisions over the same
in $p p$ collisions. This is done in Fig.~\ref{fig:AA-gam-gam-ratio}. The effect 
of the energy loss does not reduce simply to a change of normalisation of the
correlations but also to a change of shape. For the normalized $d \sigma / d
\pt$ distribution ({\em top left}) one recognizes the increase in the
suppression as $\pt$ increases below the threshold value of 10~GeV/c,
corresponding to an increasing $z$ value of the detected photon while above $\pt
=$~10~GeV/c the effect of the energy loss is, as expected, becoming less and
less visible. Particularly interesting are also the normalized 
$d \sigma / d q_{_{~T}}$ ({\em bottom left}) and $d \sigma / d z_{34}$ ({\em bottom
right}) correlations which display a marked dependence on the variable 
$q_{_{~T}}$ and $z_{34}$ respectively. 

All these studies have been carried out in the LO approximation. Clearly,
higher order corrections will somewhat smear the results specially near
thresholds or infra-red singular points ($z=1$ or $\q_{_{~T}}=0$ for example).
Very preliminary studies in the NLO approximation, assuming that higher order
corrections can be calculated in the same way in $pp$ and in A+A collisions,
show that the effects of the energy loss can still be clearly seen in the
correlation functions~\cite{arleo2004}.

As a further illustration of correlation studies we present in
Fig.~\ref{fig:AA-chg-gam-cross-section} the results for photon-charged hadron
observables, with the same kinematical constraints as before: $\pt >$~10~GeV/c
for the photon and  $\pt >$~3~GeV/c for the hadron. The interpretation of the
curves is very similar to the case of photon-photon correlations the main
advantage of the photon-hadron channel being a much higher counting rate.

%
%
%




\section{OUT-OF-EQUILIBRIUM PHOTON PRODUCTION}

\noindent {\em I.~Dadi\'c, F.~Gelis, G.D.~Moore, J.~Serreau}

\vspace{.5cm}

In the previous sections, \ref{thermal} and \ref{comparing}, the rate of
thermal photon production in the quark-gluon plasma is calculated ignoring the
finite life time of the plasma. Recently however it has been argued that
finite life-time effects could considerably increase the rate of production.
We discuss this class of models and show that they are based on an unrealistic
modeling of the initial and final state conditions which leads to an infinite
amount of energy being radiated away by the photons. These models are
therefore unphysical. A possible approach to non-equilibrium aspects
is then briefly sketched.

\subsection{Transient effects in the real-time approach}
\noindent {\it I. Dadi\'c, F. Gelis, G.D.~Moore}

In equilibrium, one usually defines the initial density operator at an
initial time $t_i$ which is infinitely remote in the past. However,
one may wonder if any changes are to be expected when the initial
state is defined at some finite time $t_i$. Some recent papers by
Wang, Boyanovsky and Ng \cite{WangB1,WangBN1} suggested that one can
see transient effects in the photon radiation of a thermal system that
exists only for a finite amount of time. A similar model has been
considered by Dadi\'c, based on \cite{Dadic2,DadicE1,Dadic4}, but there
are no published results on photon production at the time of writing
this report.  The main feature of the photon emission spectrum
obtained in \cite{WangB1,WangBN1} is that it decreases as a power of
the photon energy instead of the usual exponential spectrum. This
implies that already at moderate photon energies, the prediction of
\cite{WangB1,WangBN1} dominates by several orders of magnitude over
the usual thermal rates.

We start this section by discussing some general properties of an equilibrated
system prepared at a finite initial time, and discuss what to expect
if the particles in the initial bath are non interacting. We proceed
with a derivation of the results of \cite{WangB1,WangBN1} for the
photon emission rate in such a model, and we discuss also briefly the
approach of \cite{Dadic2,DadicE1,Dadic4}. Finally, we close the
section by discussing the fact that this model leads to infinite
``vacuum'' contributions, precluding any phenomenological application
of these rates.

\subsubsection{Starting the equilibrium evolution at a finite time}
Let us first assume that the density operator at $t_i$ is
$\rho(t_i)=\exp(-\beta H)$, where $H$ is the {\sl full} Hamiltonian,
including all the interactions. In order to calculate perturbatively a
correlator like $\left<{\rm T} \phi(x_1)\cdots \phi(x_n)\right>$ where
$\left<A\right>\equiv {\rm Tr}\,(\rho(t_i)A)/{\rm Tr}\,(\rho(t_i))$,
one has to extract the two sources of dependence on the coupling
constants.  The first one is the usual dependence of the Heisenberg
field $\phi(x)$ on the coupling constants, and it is readily extracted
by going to the interaction picture:
\begin{equation}
\phi(x)=U(t_i,x_0)\phi_{_{I}}(x)U(x_0,t_i)\; ,
\label{eq:int-pict}
\end{equation}
where $U$ is an evolution operator defined as
\begin{equation}
U(t_2,t_1)={\rm P}\,\exp i \int_{t_1}^{t_2} d^4x 
{\cal L}_{\rm int}(\phi_{_{I}}(x))
\label{eq:U}
\end{equation}
with ${\cal L}_{\rm int}$ the part of the Lagrangian density that
contains the interactions and $\phi_{_{I}}$ a free field that
coincides with the Heisenberg field at $x_0=t_i$.

However, there is a second source of dependence on the coupling
constants, which is present in the density operator itself
\cite{Gelis1,Gelis8}. This can be extracted thanks to the following
formula that relates the full density operator to the one defined with
the non-interacting Hamiltonian $H_0$ \cite{RammeS1}:
\begin{equation}
e^{-\beta H}=e^{-\beta H_0} {\rm P}\,\exp i\int_{t_i}^{t_i-i\beta} d^4x
{\cal L}_{\rm int}(\phi_{_{I}}(x))\; .
\label{eq:rho}
\end{equation}
Eqs.~(\ref{eq:int-pict}), (\ref{eq:U}) and (\ref{eq:rho}) explain the
structure of the time integration contour in the real-time formalism
\cite{Schwi1,Keldy1,ChouSHY1,LandsW1,Bella1}: it starts on the real
axis at $t_i$ and goes along the real axis up to some final time $t_f$
(which can be $+\infty$, but needs in fact only to be larger than any
physical time in the problem, since causality forbids any dependence
of physical quantities upon this time), then goes back to $t_i$ along
the real axis, and finally goes down to $t_i-i\beta$ following a
segment parallel to the imaginary axis.

The bare propagators in this formalism obey the so-called
Kubo-Martin-Schwinger (KMS) symmetry \cite{Kubo1,MartiS1}, which reads
\begin{equation}
G_0(t_i,y_0)=\alpha G_0(t_i-i\beta,y_0)\; ,
\end{equation}
where $\alpha=1$ for a real scalar field, $\alpha=\exp(\beta \mu q)$
for a boson that carries the conserved charge $q$ with an associated
chemical potential $\mu$, and $\alpha=-\exp(\beta \mu q)$ for a
fermion carrying the conserved charge $q$. The same symmetry holds for
the second point of the propagator, with $q$ changed into
$-q$. Because the number of fermion lines arriving at a vertex is even
and because the total conserved charge arriving at a vertex is zero,
it is easy to verify that any Feynman diagram made of these bare
propagators and where the time integrations at the vertices run on the
previously defined contour is {\sl independent of the initial time
$t_i$} \cite{Gelis1,BellaM1}. A corollary of this property is that all
the Feynman diagrams depend only on differences of their external
times, even if they have been calculated on a contour that breaks
invariance under time translation because of the finite $t_i$.

Physically, this result means the following: {\it if one prepares a
system in a state of thermal equilibrium at some finite time $t_i$,
all the Green's functions of the theory are independent of $t_i$ and
there is no way to tell what the initial time was.} In other words,
there are no observable transient effects in a system initially
prepared in a state of thermal equilibrium (provided the particles are
{\sl interacting} in the initial statistical ensemble).

\subsubsection{Non interacting initial statistical ensemble}
In the previous argument, the contribution of the vertical branch of
the contour is crucial in order to achieve a result that does not
depend on $t_i$. Indeed, the KMS symmetry that relates the values of
the propagators at both ends of the contour is of no help if we
terminate the contour at $t_i$. If we perform the time integrations at
the vertices using only the two horizontal branches of the contour
(i.e. the so-called Keldysh contour), then we lose the property of
independence on $t_i$, as well as the invariance under time
translations of the Green's functions, and there are now transient
effects.

Given that the vertical branch of the time path arises from
Eq.~(\ref{eq:rho}), not using the vertical part is equivalent to using
an initial density operator $\rho(t_i)=\exp(-\beta H_0)$ which does
not contain the interaction terms of the Hamiltonian. In other words,
this is equivalent to having initially a statistical ensemble of {\sl non
interacting} particles. An identical situation would be the following:
start at $t=-\infty$ with the density operator
$\rho(-\infty)=\exp(-\beta H_0)$ and assume that all the coupling
constants are zero before $t_i$ and jump to their normal value at
$t_i$. Indeed, if we start with $\rho(-\infty)=\exp(-\beta H_0)$ and
if we have a non interacting dynamics between $-\infty$ and $t_i$,
then we will have $\rho(t)=\exp(-\beta H_0)$ as long as the
interactions remain zero, by virtue of
$\dot\rho(t)=-i[H_0,\rho(t)]=0$.

\subsubsection{Photon production}
Let us now describe how one could calculate the production of photons
when the initial ensemble is a non-interacting statistical ensemble. This
system is implemented by taking $\rho(t_i)=\exp(-\beta H_0)$, with a
time path that has only the two horizontal branches.

{\noindent \bf Derivation of the results of \cite{WangB1,WangBN1}}

The phase-space density of photons at time $t$ in the system is
given by:
\begin{eqnarray}
2p_0\frac{dN_\gamma}{d^3\x d^3\p}=\frac{1}{(2\pi)^3} \frac{1}{V} 
\sum_{{\rm pol\ }\lambda}\;{\rm Tr}\left(
\rho(t) a^\dagger_\lambda(\p)a_\lambda(\p)
\right)
=\frac{1}{(2\pi)^3} \frac{1}{V} 
\sum_{{\rm pol\ }\lambda}\;{\rm Tr}\left(
\rho(t_i) a^\dagger_\lambda(t,\p)a_\lambda(t,\p)\right)\; ,
\label{eq:nk}
\end{eqnarray}
where $\rho(t)$ is the time-dependent density operator,
$a^\dagger_\lambda(\p)$ is the photon creation operator in the
Schr\"o\-din\-ger picture and $V$ is the volume of the system. In the
second equality, we use the cyclicity of the trace in order to
transpose the time dependence from the density operator into the
number operator, and we have defined:
\begin{equation}
a_\lambda(t,\p)\equiv e^{iH(t-t_i)}a_\lambda(\p)e^{-iH(t-t_i)}\; .
\end{equation}
This time-dependent creation operator is related to the Heisenberg
field via:
\begin{equation}
a^\dagger_\lambda(t,\p)=-i\int d^3\x\; e^{-ip\cdot x}
\stackrel{\leftrightarrow}{\partial}_{x_0} A^\mu(x)\epsilon_\mu^\lambda(\p)\; ,
\label{eq:ak}
\end{equation}
where $\epsilon_\mu^\lambda(\p)$ is the appropriate polarization
vector.  Inserting Eq.~(\ref{eq:ak}) in Eq.~(\ref{eq:nk}), taking the
time derivative, and using the equation of motion of the Heisenberg
fields, we obtain\footnote{The spatial coordinates have been Fourier
transformed.}:
\begin{eqnarray}
2p_0\frac{dN_\gamma}{d^4x d^3\p}&=\frac{1}{(2\pi)^3} \frac{1}{V}
\sum_{{\rm pol\ }\lambda} 
\epsilon_\mu^{\lambda *}(\p) \epsilon_\nu^{\lambda}(\p)
\lim_{y_0\to x_0}&e\,{\rm Tr}\Big(\rho(t_i)
\big[J^\mu(x_0,\p) (\partial_{y_0}-ip_0)A^\nu(y_0,\p)\nonumber\\
&&+(\partial_{y_0}+ip_0)A^\nu(y_0,\p) J^\mu(x_0,\p)
\big]
\Big)\; ,
\end{eqnarray}
where $J^\mu(x_0,\p)$ is the spatial Fourier transform of the current
$\overline{\psi}(x)\gamma^\mu \psi(x)$ ($\psi$ is the Heisenberg
fer\-mio\-nic field). One can also write $J^\mu(x_0,\p)
A^\nu(y_0,\p) ={\rm P}\, J^\mu_-(x_0,\p) A^\nu_+(y_0,\p)$, where ${\rm
P}$ denotes the path ordering along the Keldysh contour, and the
indices $\pm$ indicate on what branch of the contour a field is kept.
At this point, one can expand the Heisenberg fields in terms of the
free fields of the interaction picture. Since ${\rm Tr}\,(\rho(t_i)
A^\nu_{\rm in})=0$, one must at least expand one order further in the
electromagnetic coupling $e$. Truncating the expansion at the first
non-zero order in $e$, but keeping all orders in the strong
interactions, we obtain the following formula
\begin{eqnarray}
&&2p_0\frac{dN_\gamma}{d^4x d^3\p}=\frac{1}{(2\pi)^3} \frac{1}{V}
\sum_{{\rm pol\ }\lambda} 
\epsilon_\mu^{\lambda *}(\p) \epsilon_\nu^{\lambda}(\p)
\lim_{y_0\to x_0}e^2 \int_{\cal C}dz_0 \int\frac{d^3\k}{(2\pi)^3}\nonumber\\
&&\!\!\!\!\!\!\!\!\!\!\!\!\times
\,\Big[
(\partial_{y_0}-ip_0)
\left<{\rm P}\;A^\nu_{\rm in,+}(y_0,\p)A^\alpha_{\rm in}(z_0,\k)\right>
\left<{\rm P}\;J^\mu_{\rm in,-}(x_0,\p) J_\alpha{}_{\rm in}(z_0,\k)
e^{i\int_{\cal C}{\cal L}_{\rm int}^{QCD}}\right>
\nonumber\\
&&\!\!\!\!\!\!\!\!\!\!+(\partial_{y_0}+ip_0)
\left<{\rm P}\;A^\nu_{\rm in,-}(y_0,\p)A^\alpha_{\rm in}(z_0,\k)\right>
\left<{\rm P}\;J^\mu_{\rm in,+}(x_0,\p) J_\alpha{}_{\rm in}(z_0,\k)
e^{i\int_{\cal C}{\cal L}_{\rm int}^{QCD}}\right>
\Big]\; ,
\label{eq:rate1}
\end{eqnarray}
where ${\cal L}_{\rm int}^{QCD}$ is the QCD part of the interactions,
and where ${\cal C}$ is the Keldysh contour that runs from $t_i$ to
$+\infty$ and then back to $t_i$. $\left<\cdots\right>$ denotes the
average over the initial ensemble, and we have used the fact that we
can factorize the photon fields and the quark fields because the
initial density operator does not couple photons and quarks. The first
of these correlators contains a factor $(2\pi)^3\delta(\p+\k)$ which
is used to integrate out the vector $\k$. Then the second correlator
brings a factor $(2\pi)^3\delta(0)$ which should be interpreted as the
volume of the system, and cancels the factor $1/V$. At this point, it
is a simple matter of algebra to work out the derivatives of the
photon-photon correlator, and we get:
\begin{eqnarray}
&&2p_0\frac{dN_\gamma}{d^4x d^3\p}=\frac{e^2}{(2\pi)^3}
\sum_{{\rm pol\ }\lambda} 
\epsilon_\mu^{\lambda *}(\p) \epsilon_\nu^{\lambda}(\p)
\int_{-\infty}^{+\infty} \frac{d\omega}{\pi}
\frac{\sin((p_0-\omega)(x_0-t_i))}{p_0-\omega}
\nonumber\\
&&\qquad\qquad\qquad\qquad\times
\Big[
(1+n_\gamma^0(p_0))\Pi_{+-}^{\mu\nu}(\omega,\p)
-n_\gamma^0(p_0)\Pi_{-+}^{\mu\nu}(\omega,\p)
\Big]\; ,
\label{eq:ph-rate}
\end{eqnarray}
where $n_\gamma^0(p_0)$ is the photon distribution in the initial
ensemble, and where $\Pi_{\mp\pm}^{\mu\nu}(\omega,\p)$ is the Fourier
transform of the current-current correlator, possibly including QCD
corrections to all orders. The standard assumption is to assume that
the system is small enough so that photons do not accumulate in the
system. Therefore, in the ``photon production rate'' one drops all the
terms proportional to $n_\gamma^0(p_0)$, i.e. the photon absorption
and the blocking effects. Note that there are memory effects in this
formula since the rate at time $x_0$ depends on the initial time
$t_i$. Moreover, the self-energy under the integral over $\omega$ is
evaluated off-shell as long as the time difference $x_0-t_i$ remains
finite. The limit of infinite initial time $t_i\to -\infty$ is then
obtained by using $\lim_{a\to +\infty}\sin(a k)/k=\pi\delta(k)$, and
one recovers trivially the usual formula \cite{Weldo3,GaleK1} for the
photon production rate, with an on-shell self-energy.

Note also that the result derived here is the same as the one obtained
in \cite{WangB1,WangBN1}, even if the starting point is enunciated
differently. {\it This implies that the model of \cite{WangB1,WangBN1}
is equivalent to a description in which a non interacting statistical
ensemble would be prepared at $t=-\infty$, followed by a
non-interacting evolution of the system until the finite time $t_i$,
at which point one switches on all the interactions.} Let us also
emphasize that the same calculation can be performed while keeping
explicitly the vertical branch ($[t_i,t_i-i\beta]$) of the contour, in
which case the residual $t_i$ dependence drops out and one obtains the
standard equilibrium result without having to take a limit. This is
consistent with the fact that keeping the vertical branch of the time
path just amounts to have $\rho(t_i)=\exp(-\beta H)$ (i.e. an
interacting initial statistical ensemble) instead of
$\rho(t_i)=\exp(-\beta H_0)$ (a non-interacting initial statistical
ensemble).

{\noindent \bf On the approach of \cite{Dadic2,DadicE1,Dadic4}}

The papers \cite{Dadic2,DadicE1,Dadic4} do not discuss photon
production, but develop a formalism (called ``projected functions
formalism'') that can be used to calculate Green's functions of an
out-of-equilibrium system. This formalism also assumes that the
initial conditions are set at some finite time $t_i$, which is taken
to be $t_i=0$ for simplicity. Assuming also the initial ensemble to be
a bath of non interacting particles, the time path is a Keldysh
contour that starts at $t_i=0$, goes to $t=+\infty$ and then back to
$t_i$. For this reason, all the Green's functions are always evaluated
with time arguments in the range $0\le x_0 < +\infty$. For a 2-point
function with time variables $x_0$ and $y_0$, one can introduce the
usual combinations $s_0\equiv x_0-y_0$ and $X_0\equiv (x_0+y_0)/2$,
and one has the constraint: $-2X_0\le s_0\le 2X_0$. This can be
enforced in the Wigner transform (Fourier transform with respect to
the relative variable) as follows:
\begin{equation}
G_{X_0}(p_0,\p)=\int_{-\infty}^{+\infty}dp'_0 P_{X_0}(p_0,p'_0) G_\infty(p_0,\p)\; ,
\end{equation}
where $G_{X_0}$ is the Wigner transform restricted to the range
$s_0\in[-2X_0,2X_0]$, $G_\infty$ is the unrestricted Wigner transform
and $P_{X_0}$ is a projection operator given by:
\begin{equation}
P_{X_0}(p_0,p'_0)=\frac{\sin(2X_0(p_0-p'_0))}{\pi(p_0-p'_0)}\; .
\end{equation}
Since the time evolution starts at $t_i=0$, all the diagrams must be
evaluated with time integrations at the vertices restricted by factors
$\theta(x_0)$. In Fourier space, this brings factors like:
\begin{equation}
\frac{i}{\sum_i p_{0,i}+i\varepsilon}\; ,
\end{equation}
where the $p_{0,i}$ are the energies carried by the propagators
attached to a particular vertex (all defined to be outgoing), instead
of the usual delta functions that would ensure energy conservation at
the vertex.

The phase space density of photons at the time $x_0$ is
defined\footnote{This formula is written by analogy with a similar
relation known to be valid in thermal equilibrium. However,
\cite{Dadic4} does not give a first principles justification of this
formula in the case of a system out-of-equilibrium.}  in \cite{Dadic4}
as:
\begin{equation}
1+2 (2\pi)^3\frac{dN_\gamma}{d^3\x d^3\p}=p_0\int_{-\infty}^{+\infty}
\frac{d\omega}{2\pi} G_{S,x_0}(\omega,\p)\; ,
\end{equation}
where $G_S\equiv G_{+-}+G_{-+}$ is the third non-zero component of the
matrix propagator (the other two components being respectively the
retarded and advanced propagators) in the Keldysh basis of the
real-time formalism (we use here the standard notations of
\cite{EijckKW1}, and we refer the reader to this reference for more
details). One can then use the formalism of
\cite{Dadic2,DadicE1,Dadic4} in order to evaluate the photon
production rate:
\begin{eqnarray}
&&2p_0\frac{dN_\gamma}{d^4x d^3\p}=\frac{e^2}{(2\pi)^3}
\sum_{{\rm pol\ }\lambda} 
\epsilon_\mu^{\lambda *}(\p) \epsilon_\nu^{\lambda}(\p)
\int_{-\infty}^{+\infty}\frac{d\omega}{\pi}\;{\rm Im}\,\left[
\widetilde{\Pi}_{S,\infty}^{\mu\nu}(\omega,\p) 
\frac{p_0 e^{-i\omega x_0} \sin(p_0 x_0)}{\omega^2-p_0^2+2i\varepsilon\omega}
\right]\; ,
\label{eq:ph-rate-dadic}
\end{eqnarray}
where $\widetilde{\Pi}_{S}(\omega,\p)\equiv \Pi_S(\omega,\p)- {\rm
  sign}(\omega)\big[\Pi_R(\omega,\p)-\Pi_A(\omega,\p)\big]$. At the
  time of writing this report, it is not clear whether
  Eqs.~(\ref{eq:ph-rate}) and (\ref{eq:ph-rate-dadic}) are equivalent
  or not.

\subsubsection{Infinite vacuum contributions and other problems...}
We now discuss some of the consequences of Eq.~(\ref{eq:ph-rate}).
Because the self-energy $\Pi_{+-}^{\mu\nu}(\omega,\p)$ can be
evaluated off-shell\footnote{Note that Eqs.~(\ref{eq:ph-rate}) and
(\ref{eq:ph-rate-dadic}) contain the usual on-shell
contributions. They start when the photon polarization tensor is
evaluated at the 2-loop order.} in Eq.~(\ref{eq:ph-rate}), we can have
a non zero photon production rate already from the lowest order
self-energy, i.e. the diagram with only one quark loop. This
contribution was evaluated in \cite{WangB1,WangBN1}, where one can
read:
\begin{eqnarray}
&&p_0\frac{dN_\gamma}{d^4x d^3\p}=\frac{6\pi e^2}{(2\pi)^3}
\int_{-\infty}^{+\infty} \frac{d\omega}{\pi}
\frac{\sin((p_0-\omega)(x_0-t_i))}{p_0-\omega}\times\nonumber\\
&&\qquad\qquad\times
\int \frac{d^3\q}{(2\pi)^3}
\Big\{
2[1-(\hat\p\cdot\hat\k)(\hat\p\cdot\hat\q)]n_{\rm q}(q)(1-n_{\rm q}(k))
\delta(\omega+k-q)\nonumber\\
&&\qquad\qquad\qquad\qquad
+[1+(\hat\p\cdot\hat\k)(\hat\p\cdot\hat\q)]n_{\rm q}(q)n_{\rm q}(k)
\delta(\omega-k-q)
\Big\}\; ,
\label{eq:rate-1L}
\end{eqnarray}
where we denote $\k\equiv\p+\q$, $\hat\p\equiv \p/p$, $\hat\q\equiv
\q/q$ and $\hat\k\equiv \k/k$, and where $n_q(k)\equiv 1/(\exp(\beta
k)+1)$ is the quark distribution in the system.  

The first shortcoming of this formula is that it is incomplete: its
authors have dropped a term whose integrand is proportional to
$(1-n_{\rm q}(q))(1-n_{\rm q}(k))\delta(\omega+q+k)$. This term can be
non zero only for negative values of $\omega$, but since the
integration variable $\omega$ and the energy $p_0>0$ of the produced
photon do not need to have the same sign, the contribution of this
term to the photon rate is different from zero. Moreover, this
contribution is divergent at any photon momentum $\p$. Indeed, the
result of the integration over $\q$ for this term behaves like
$\omega^2$ at large $\omega$, so that the integral over $\omega$ is
not defined.  Physically, it corresponds to the decay of the vacuum
into a $q\bar{q}$ pair and a photon, and is due to the mismatch
between the vacuum of the free theory and the vacuum of the
interacting theory. In other words, the initial (non interacting)
vacuum is an excited state in the interacting theory, and tends to
decay spontaneously into particles. Naturally, this problem disappears
if $t_i\to-\infty$ since
${\sin((p_0-\omega)(x_0-t_i))}/{(p_0-\omega)}\to
\pi\delta(p_0-\omega)$ in this limit, and has no support at negative
$\omega$. Note also that the same problem arises with
Eq.~(\ref{eq:ph-rate-dadic}): here also, there are infinite vacuum
contributions.

Even if, following \cite{WangB1,WangBN1}, we decide to ignore the
vacuum contributions, Eq.~(\ref{eq:rate-1L}) still has another very
serious problem: the asymptotic behavior (large $p_0=|\p|$) of the
power spectrum is too hard for being integrable.  More precisely, one
has:
\begin{equation}
p_0\frac{dN_\gamma}{d^4x d^3\p} \empile{\sim}\over{p_0\to +\infty} p_0^{-2}\; ,
\end{equation}
which implies that the integrated energy yield per unit time and per
unit volume $\int {d^3\p} p_0({dN_\gamma}/{d^4x d^3\p})$ is
infinite because the integral does not converge in the ultraviolet.

It has been suggested that the divergence of the total energy might be due
to the fact that the coupling constant in this model is switched on
instantaneously. Indeed, even if the self-energy
$\Pi_{+-}^{\mu\nu}(\omega,\p)$ vanishes exponentially when
$|\omega|\to +\infty$, the power law decrease of the power spectrum is
due to the fact that the function
${\sin((p_0-\omega)(x_0-t_i))}/{(p_0-\omega)}$ has itself a power law
decrease with respect to $p_0$. However, this function arises as a
result of the integration over the intermediate time $z_{~0}$:
\begin{eqnarray}
\frac{\sin((p_0-\omega)(x_0-t_i))}{p_0-\omega}&=&\int_{t_i}^{x^0} dz_0
\cos((p_0-\omega)(z_0-x_0))\nonumber\\
&=&
\int_{-\infty}^{x^0} dz_0 \theta(z_0-t_i)
\cos(p_0-\omega)(z_0-x_0)\; ,
\label{eq:time-int}
\end{eqnarray}
where the cosine comes from factors like $\lim_{y_0\to
x_0}(\partial_{y_0}-ip_0)\big<{\rm P}\;A^\nu_{\rm
in,+}(y_0,\p)A^\alpha_{\rm in}(z_{~0},\k)\big>$ in Eq.~(\ref{eq:rate1}),
and where the $\theta(z_0-t_i)$ reflects the time-dependence of the
coupling constant ($z_0$ is the time attached to a vertex coupling the
photon to a quark line). As a toy model, one can smoothen the behavior
of the coupling constant by replacing $\theta(z_{~0}-t_i)$ by
$1/(\exp((t_i-z_{0})/\tau)+1)$ where $\tau$ is the typical time during
which the coupling constant evolves from $0$ to its normal
value. Doing this substitution in Eq.~(\ref{eq:time-int}) gives:
\begin{eqnarray}
\int\limits_{-\infty}^{x^0}\!\!\! dz_0 
\frac{\cos((p_0-\omega)(z_0-x_0))}{e^{(t_i-z_0)/\tau}+1} 
=
{\frac{2\pi\tau \sin((p_0-\omega)(x_0-t_i))}{e^{\pi(p_0-\omega)\tau}-e^{-\pi(p_0-\omega)\tau}}}&+&
\frac{\tau e^{-(x_0-t_i)/\tau}}{1+(p_0-\omega)^2\tau^2}\nonumber\\
&&+{\cal O}(\tau e^{-2(x_0-t_i)/\tau})\; ,
\label{eq:time-int1}
\end{eqnarray}
which is a good approximation for $x_0-t_i\gg\tau$. In this
expression, the first term vanishes exponentially when
$|p_0-\omega|\to +\infty$. The power law behavior has been restricted
to the second term, which has an exponentially suppressed prefactor
when $x_0-t_i\gg \tau$, i.e. when one looks at the rate at a time
which is far from the region where the coupling constant is changing.
In particular, one can see that the power law behavior becomes
dominant over the exponential behavior only for $|p_0-\omega|\gtrsim
(x_0-t_i)/\pi\tau^2$. In other words, there is a considerable range of
$p_0-\omega$ in which one does not see effects of the power law
behavior. However, this does not prevent the ultraviolet divergence
when one tries to integrate Eq.~(\ref{eq:rate-1L}) over the photon
energy. 

This residual ultraviolet divergent power spectrum seems to be due to
the fact that Eq.~(\ref{eq:rate-1L}) evaluates the photon number of an
interacting system with a non-interacting definition of the photon. As
a consequence, this definition of the photon number ``measures'' also
the virtual photon clouds that surround all charged particles after
the electromagnetic interaction has been turned on. This
interpretation is also supported by the fact that the photon
self-energy in Eq.~(\ref{eq:rate-1L}) is always evaluated off-shell.

\subsubsection{What have we learned?}
It appears that the models considered in \cite{WangB1,WangBN1} and
\cite{Dadic2,DadicE1,Dadic4} suffer from very serious problems which
forbid any quantitative prediction at this point. In particular, a
common problem in these models is that they lead to infinite vacuum
contributions, which correspond to the spontaneous decay into
particles of the vacuum of the non-interacting theory.
Note that in a more recent paper \cite{BoyanV1}, an ad-hoc
subtraction is proposed in order to dispose of all these
divergences\footnote{
	Although this is not said explicitly in \cite{BoyanV1}, these new
	results completely invalidate the results and the phenomenological
	discussion of \cite{WangB1,WangBN1}.
}. 
What remains to be seen is whether this subtraction can be justified from first
principles, and whether one can get rid of the inherent arbitrariness that
comes with subtracting infinities.

From a more formal point of view, the important lesson taught to us by
the study of this model is that in generic non-equilibrium problems,
the initial ensemble should be specified with respect to the spectrum of the
interacting theory. An approach exploring this issue has been proposed in
\cite{Serreau2003}.

\subsection{The 2PI effective action approach}
\noindent {\it J. Serreau}

To follow the space-time evolution of an out-of-equilibrium system 
of quantum fields one may solve the Schwinger-Dyson equations of the 
theory -- written in the form of an initial value problem -- for given 
initial conditions. This cannot be done exactly and, in practice, one has 
to rely on approximation schemes. 
A powerful way of deriving systematic approximations to the dynamical 
equations is to work at the level of effective actions, from which
one obtains the equations of motion by functional differentiation with 
respect to the fields. An example is given by the well-known 
one-particle-irreducible (1PI) effective action, which is the 
generating functional of 1PI Green's functions. In particular, the use 
of such functional methods ensures that the global symmetries 
of the underlying theory are preserved by the approximate equations of 
motion (for instance, this guarantees energy conservation).

It has however been observed that standard approximations of the 1PI 
effective action can be secular in time \cite{BetteW1,MihaiACDH1}: the 
validity of a given expansion in a small parameter -- such as a coupling
or a $1/N$--expansion -- can be spoiled by the occurrence of terms which 
grow indefinitely with time.  This has been a major obstacle for
practical studies of the real time dynamics of far from equilibrium 
quantum fields beyond so-called mean-field approximations (leading 
order in large--$N$, Hartree). The latter, which neglect direct scattering 
between (quasi-)particles, are free of secular terms and have been
extensively used over the last decades in various physical 
situations \cite{CoopeHKMP1,BoyanVHS1}. However, these ``collisionless''
approximations are known to fail to describe important physical effects
such as late time thermalization, or early time damping of unequal-time 
correlations (see e.g. \cite{Berge2}). Another important case where one 
can perform explicit calculations of the non-equilibrium dynamics corresponds 
to the classical statistical field theory limit 
\cite{AartsBW1,Serre1,ProkoR1}. This can provide a good description 
of the quantum dynamics when the typical occupation numbers of the field 
modes are large. A great advantage is that one can solve exactly the full 
nonlinear classical dynamics by means of standard Monte Carlo methods 
together with numerical integration techniques. However, there are important 
physical questions, such as the description of quantum thermalization at 
late time, which cannot be addressed with these methods. 

Important progress have been made in recent years with the use of 
approximation schemes based on the two-particle-irreducible (2PI) 
effective action \cite{CornwJT1,Baym1}.
The latter is a generating functional for the correlation functions 
of the theory, parameterized in terms of the connected one and two-point 
functions, i.e.\ the average value of the field and the propagator (for 
comparison, the 1PI effective action is parameterized in terms of the 
one-point function only). This approach allows for practicable and systematic
calculations of the non-equilibrium dynamics beyond mean-field and classical 
field approximations.\footnote{This approach, supplemented by a gradient 
expansion, also provides a very efficient way to derive Boltzmann 
equations in quantum field theory \cite{CalzeH1}.}
In particular, it has been possible to demonstrate for the first time
the late time quantum thermalization of a scalar field in $1+1$ 
dimensions from a first principle calculation, by using a three-loop 
approximation of the 2PI effective action \cite{BergeC1}.

\subsubsection{Current status}

In recent years, the 2PI effective action approach to 
out-of-equilibrium phenomena has been extensively investigated 
and applied to various situations of physical 
interest in the context of scalar and fermionic field theories
(for a recent review see \cite{BergeS1}). 
The most studied case is that of scalar field theories, where the 
question of reliable approximations has received a lot of attention.
Various schemes have been worked out and used e.g.\ to study thermalization. 
These include a coupling-expansion up to three-loop order, which is the 
simplest approximation including direct scattering and describing
thermalization \cite{BergeC1}, as well as a $1/N$--expansion 
up to next-to-leading order (NLO) \cite{Berge2,AartsABBS1}.
In a slightly different context, namely the Schwinger-Dyson approach, 
motivated by a direct truncation of Schwinger-Dyson equations, a 
similar approximation has been studied: the so-called bare vertex 
approximation (BVA) \cite{BlagoCDM1}. In terms of the 2PI $1/N$--expansion, 
the latter partially include field-dependent\footnote{In particular, the 
BVA and the NLO approximation are equivalent in the symmetric regime (see 
e.g. \cite{AartsABBS1})} NNLO contributions \cite{AartsABBS1}. 
These approximations have been compared to exact Monte Carlo 
simulations in the framework of classical statistical field 
theory in $1+1$ dimensions \cite{BlagoCDM1,AartsB1,CoopeDM1}.
It has been shown that the 2PI $1/N$--expansion at NLO provides a 
quantitative description of the dynamics already for moderate values 
of $N$. For instance, the damping rate of unequal-time two-point functions 
is accurately described at NLO for $N\ge4$ \cite{AartsB1}. 
Similar studies of the time evolution of the one and two-point function in 
the limit $N=1$ \cite{BlagoCDM1,CoopeDM1} show that, although quantitatively 
different, both the NLO approximation and the BVA are in qualitative 
agreement with the exact result near the symmetric regime. Studies of the 
quantum dynamics demonstrate that the 2PI $1/N$--expansion at NLO does not 
exhibit any spontaneous symmetry breaking for $N=1$ in $1+1$ dimensions 
\cite{CoopeDM2}, in agreement with general results. Although it is not a 
controlled limit for a $1/N$--expansion, such a qualitative agreement provides 
a very sensitive test of this systematic expansion scheme. For instance, the 
inclusion of part but not all NNLO contributions, as in the BVA, leads to 
(spurious) spontaneous symmetry breaking in one spatial dimension \cite{CoopeDM2}. 
This may, however, be particular to the limit $N=1$ and a direct comparison between
these two approximations, both in classical and quantum field theory, shows that 
they agree rather well with each other for $N\ge 4$ \cite{Mihai1}. First 
results in $3+1$ dimensions exhibiting the phenomenon of spontaneous symmetry 
breaking have been presented in \cite{BergeS1}, using the 2PI 
$1/N$--expansion at NLO.

The 2PI technique has been demonstrated to be a powerful tool to study 
realistic particle physics applications. In particular, it has been used to
perform the first quantitative study of the phenomenon of parametric resonance 
in quantum field theory beyond mean field approximations \cite{BergeS2}. 
This phenomenon provides a paradigm for situations with nonperturbatively 
large particle densities, where neither gradient nor coupling expansions are 
applicable. In contrast, the 2PI $1/N$--expansion remains valid in such 
situations. In particular, the use of the NLO approximation have been shown 
to solve the problem of an analytic description at large densities.
A full numerical solution of the corresponding equations of motion as well 
as approximate analytic results concerning the nonlinear dynamics have been 
obtained \cite{BergeS2}.
Another important point which has been studied in details in the context of 
scalar field theories concerns the renormalization of 2PI approximation schemes 
\cite{HeesK1,BraatP1,BlaizIR1}. It has been shown that approximations based on 
a systematic loop-expansion of the 2PI effective action (the so-called 
$\Phi$--derivable approximations) can be renormalized at any truncation order 
if the theory under consideration is perturbatively renormalizable \cite{HeesK1}.
Finally, the 2PI approach has been successfully applied to the study of fermionic 
field theories. In particular, the late time thermalization of fermionic quantum 
fields has been demonstrated for the first time from a first principle calculation 
\cite{BergeBS1}. This has been done for a realistic $3+1$ dimensional theory of 
Dirac fermions (``quarks'') coupled to scalars (``pions'') in a chirally 
invariant way, employing a perturbative expansion of the 2PI effective action 
at lowest non-trivial (two-loop) order. Bose-Einstein and Fermi-Dirac 
distributions have been shown to emerge from the non-equilibrium dynamics 
without further approximation.

One of the main open questions of the 2PI effective action approach
concerns the description of the dynamics of gauge fields. Gauge 
invariance makes it a non trivial issue: the 2PI effective action 
being a functional of the one and two-point functions only, it 
is difficult to find truncations which are consistent with Ward 
identities. In order to deal with this problem, different directions 
have been followed. One of them is to try to enforce Ward identities
at a given truncation order by modifying the 2PI scheme (see e.g. 
\cite{Motto1}). A similar strategy has proven useful for the calculation 
of thermodynamic properties of a QCD medium in equilibrium \cite{BlaizIR2},
where a hard thermal loop approximation has been applied on the top of
a 2PI two-loop truncation. The price to pay is that one may loose the 
description in terms of an effective action, which, as emphasized above, 
is very useful for non-equilibrium systems. Another interesting possibility 
is to analyze the magnitude of the gauge-fixing dependence of physical 
results within a given approximation scheme. This has been investigated
within the 2PI loop-expansion in Ref. \cite{ArrizS1,CarringKZ} and further 
developments are to be expected.

\subsubsection{Application to photon production?}

From very general considerations, the problem of photon emission can be 
reduced, at lowest order in the electromagnetic coupling constant 
$\alpha_{\rm em}$, to the calculation of the following connected 
current-current correlator
\begin{equation}
\label{correlator}
 \left< J^\mu (t,\x) \, J^\nu (t',\x') \right>  \, ,
\end{equation} 
where $J^\mu (t,\x)$ is the current which couples to the photon 
field and where the brackets denote the average with respect to the 
(initial) density matrix of the emitting system. The 2PI approach 
provides in principle a powerful tool to compute the time dependence
of the above two-point function. The current is generically a bilinear in the
fields describing the emitting system and the correlator Eq.~\ref{correlator}
can be expressed in terms of the corresponding four-point function. The latter
can be obtained from the 2PI effective action by taking appropriate functional 
derivatives. More details concerning the possible use of these methods in the
context of out-of-equilibrium electromagnetic  radiation can be found in Ref.
\cite{Serreau2003}.

As emphasized above, the 2PI effective action approach allows one to
compute the two-time dependence of the current--current correlator
without making {\it a priori} assumptions about this dependence. For 
example, contrarily to the gradient expansion, it does
not require any separation of scales between the $(t+t')$ and the
$(t-t')$--dependence of two-point functions. In the context of heavy
ion collisions, there are various situations where one does not expect
any such separation of scales. This is for example the case when the
system rapidly cools down through a phase transition. Other typical
non-equilibrium effects which can play an important role are initial
time and finite time effects. First, the presence of the initial
condition alone breaks time-translation invariance and it is known
that the early time behavior cannot be described by a gradient
expansion.  Of course, because of interactions, the system forgets
about the initial condition on a time scale governed by the inverse
damping rate. Only in the case where the total photon production rate
(integrated over the whole time history of the system) is dominated by
times larger than the latter can one neglect the presence of the
initial time. Moreover, because invariance under time-translation is
broken, off-shell processes may contribute to the production rate, as
discussed in the previous sections. The 2PI approach may be a way
to address these issues, taking these effects into account in a
consistent manner. 

Other effects which can be of importance, are those 
related to ``internal'' aspects of the time evolution of the emitter.
An immediate example is provided by situations where the lifetime of some
physical excitations of the system cannot be neglected on the timescales 
under consideration. This could be the case for a gas of resonances and 
may also be of importance when chiral symmetry gets restored. Just as the 
off-shell contributions mentioned above, these effects would automatically 
be taken into account in the 2PI effective action approach, applied to the 
appropriate microscopic theory. On the contrary, they are known to be very 
difficult to include in a gradient expansion. Other interesting topics 
include for example the question of photon emission from a disoriented 
chiral condensate \cite{BoyanVHK2}, which may be formed during the 
out-of-equilibrium chiral phase transition. 

In conclusion, the 2PI effective action approach has been much developed  in
recent years in the context of non-equilibrium quantum field theory  and has
been demonstrated to provide a powerful tool to study realistic  physical
situations. These methods might prove useful for the study of photon and 
dilepton production in general non-equilibrium situations.


  


%
\section{\bf LATTICE CALCULATION OF THE VECTOR SPECTRAL FUNCTION}

\noindent
{\em  P.~Petrezcky, F.~Gelis, G.D.~Moore}

\vspace{.5cm}

In this report we are interested mostly in the production of energetic real or
quasi-real photons and the calculations of rates is done in the HTL framework.
The same technics allows the computation of the production rate of massive
static photons~\cite{BraatPY1} which have recently also been considered using
non-perturbative lattice technics. Based on existing litterature there appears
a contradiction between the two methods: in the lattice approach a strong
threshold effect is observed and the production rate tends to 0 when the photon
mass decreases, in contrast to the HTL approach where the rate is increasing. In
this section we discuss these matters further.

Lattice calculations provide information on the imaginary time current-current
correlator
\begin{equation}
G_{\mu \nu}(\tau,\p)=\langle J_{\mu}(0,\p)
J_{\mu}^{\dagger}(\tau,-\p) \rangle, 
\label{corr}
\end{equation}
where $J_{\mu}(\tau,\p)=\sum_{\x} \, e^{i \x \cdot
\p} \, J_{\mu}(\tau,\x)\equiv
\sum_{\x} \, e^{i \x \cdot \p} \, 
\bar q(\tau, \x) \gamma_{\mu} q(\tau,\x)$. 
This is related to the spectral function $\sigma_{\mu \nu}$ by an
integral relation,
\begin{equation}
G_{\mu \nu}(\tau,\p)=\int_0^{\infty} d \omega \> \sigma_{\mu
\nu}(\omega, \p)\, K(\tau, \omega),  
\label{integral_rep}
\end{equation}
where the kernel
$K(\tau,\omega)=\cosh(\omega(\tau-1/2T))/\sinh(\omega/2T))$ is the
imaginary time free boson propagator \footnote{On the lattice, this should
  be replaced by its lattice version} .  The spectral function
$\sigma_{\mu \nu}(\omega,\p)$ can be related to the emission rate of
real or virtual photons \cite{Weldo3,GaleK1}.

From the lattice data on $G_{\mu \nu}(\tau,\p)$ the spectral
function can be reconstructed using 
the Maximum Entropy Method (MEM) \cite{NakahAH1,AsakaHN1}. The first calculations
of $\sigma_{\mu \nu}$ at finite 
temperature using MEM were done in \cite{KarscLPSW1}. There it was found that
$\sigma_{\mu \mu}(\omega) \equiv \sum_{\mu=\nu} \sigma_{\mu 
\nu}(\omega,\p=0)$ is slightly enhanced over the free correlator in
the region $4T<\omega<7T$ and strongly 
suppressed for $\omega<3T$.  
These features of the spectral function are clearly illustrated
by Fig.~\ref{m076}, where results at $T=3T_c$ are shown. Also
shown there is the one-loop order HTL resummed result for
the spectral function showing strong enhancement for small $\omega$.~\footnote{
        The two-loop corrections increase the one-loop result even
        more~\cite{AurenGKZ2}.}
\begin{figure}[htbp]
\begin{center}
  \resizebox*{8cm}{!}{\includegraphics{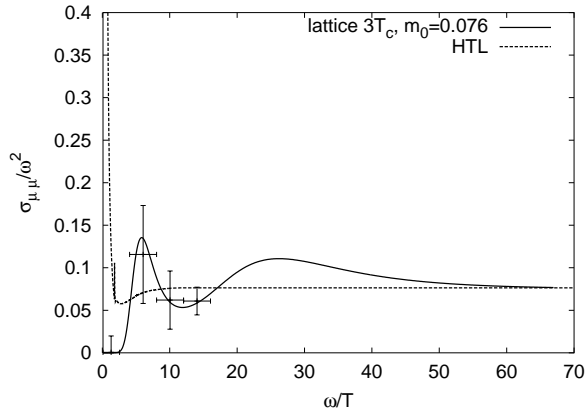}}
\end{center}
\caption{
  The vector spectral function $\sigma_{\mu \mu}(\omega)$ at $T=3T_c$.
  The error bars in Fig. \ref{m076} are statistical and were estimated
  using the method of   \cite{NakahAH1,AsakaHN1}. The systematic errors will
  be discussed latter in the text.  }
\label{m076}
\end{figure}
One concern about this result is that the reconstruction only uses
$N_{\tau}=16$ points in the imaginary time direction.  However, as
noted in \cite{KarscLPSW1}, the correlator $G_{\mu \nu}(\tau,\p)$
alone already provides stringent constraints on $\sigma_{\mu
  \nu}(\omega,\p)$ and thus allows to check the validity of the
perturbative calculations of this quantity. In particular, the value
of the correlator at $\tau=1/(2T)$ is not sensitive to the behavior of
$\sigma_{\mu \nu}(\omega,\p)$ for $\omega>16T=1/a$ (for
$N_{\tau}=16$ the lattice spacing is $1/(16T)$) and thus to the
lattice artifacts \cite{KarscLPSW1}. The lattice data on $G_{\mu
  \mu}(\omega,p)$ provide an integral constraint (sum rule) on
$\sigma_{\mu \mu}(\omega,p)$,
\begin{equation}
\displaystyle
\int_0^{\infty} d \omega \sigma_{\mu \mu}(\omega,p){1\over
\sinh({\omega\over 2 T})}=G_{\mu \mu}(\tau=1/(2T),p). 
\label{sumrule}
\end{equation}
In table~5. the ratio of $G_{\mu \mu}(\tau=1/(2T),p)$ 
calculated on a $64^3\times 16$ lattice to the corresponding free (but
lattice) value is listed for several $p=|\p|$. Note that
$G_{\mu \mu}(\tau=1/(2T),p)$ is always close to the corresponding value
in the free theory. For example, for 
$\p=0$ it is only $9\%$ larger than the free value. This implies
that a strong enhancement of the 
vector spectral function relative to its lowest order perturbative
(free) value is ruled out. 

\begin{table}[htbp]
\begin{center}
\begin{tabular}{|c|c|c|c|c|c|}
\hline
$T/T_c$   &   $p=0$    &  $p=1.57T$   &  $p=2.22T$  &    
$p=3.14T$     &    $p=4.44T$   \\
\hline
1.5       & 1.095(19)  & 1.079(14)    &  1.052(20)  &   
1.001(23)      &   0.898(85)    \\
\hline
3.0       & 1.083(14)  & 1.071(10)    &  1.060(14)  &   
1.039(10)      &   1.007(15)    \\
\hline
\end{tabular}
\end{center}
\label{tab:latt-1}
\caption{The value of $G(\tau=1/(2T),p)/G_{free,latt}(\tau=1/(2T),p)$ at
$T/T_c=$ 1.5 and 3.0} 
\end{table}

Recently, lattice calculations of $G_{i i}(\tau)$
using unimproved Wilson fermions%
\footnote{
        In   \cite{KarscLPSW1} a non-perturbatively 
        improved fermion action was used, i.e. discretization errors of order
        ${\cal O}(a)$ present for Wilson 
        fermions were completely removed.} 
on anisotropic lattices \cite{AsakaHN2} have helped address the
question about the 
small number of temporal points used to reconstruct 
$\sigma_{i i}(\omega)$ (only the spatial part was 
considered in this study).
The use of anisotropic lattices allowed to use 
more points ($N_{\tau}=32-96$) in the imaginary time direction while
keeping the temperature fixed. On the other hand the 
use of the unimproved Wilson action for quarks introduces additional lattice
artifacts. The vector spectral functions 
obtained in \cite{AsakaHN2} qualitatively agree with those obtained
in \cite{KarscLPSW1}  
for $\omega<16T$. They have a broad peak around $\omega/T \sim 5-6$ and
suppression%
\footnote{
        At $T \sim 2T_c$ strong enhancement of the spectral function
        over the free spectral function is observed; however, it is not
        clear how statistically significant this enhancement is.}  
for $\omega<3T$. 
For $T \sim (1.4-1.5)T_c$ ($N_{\tau}=54$) the position of the peak of
the vector spectral function 
roughly coincides with the findings of    \cite{KarscLPSW1}. 

The drop of the spectral function below $\omega=3T$ is in
contradiction with HTL perturbation theory, which predicts a
$1/\omega$ behavior for $\sigma_{\mu \mu}(\omega)$ at small $\omega$
(see Fig.  \ref{m076}), due to the contribution of processes like
bremsstrahlung.  However, it is expected that this lowest order
behavior is modified into $1/\sqrt{\omega}$ by multiple scattering
effects when $\omega$ becomes small. At even smaller $\omega$, this
behavior will be further reduced by strong dissipative effects, so
that $\sigma_{\mu \mu}(\omega)$ vanishes linearly with $\omega$.
Indeed, the small $\omega$ limit of $e^2 \sigma_{\mu \mu}/6\omega$ is
the electrical conductivity in the quark-gluon plasma. The small
$\omega$ behavior of the MEM-extracted spectral function would imply a
vanishing electrical conductivity, but at the moment we lack a natural
explanation that would lead to a very small electrical conductivity in
a quark-gluon plasma.  Related to this, there are also questions about
whether $\sigma_{\mu \mu}(\omega)$, or $\sigma_{ii}(\omega)$ (just the
spatial part) is the appropriate quantity to reconstruct.

There are systematic uncertainties in the reconstructed spectral
function. The most important uncertainties apart from those coming
from the limited number of points in the time direction ($N_{\tau}$) are
those associated with discretization errors and with the choice
of the so-called default model $m(\omega)$, the prior assumption on the form
of $\sigma_{\mu \mu}(\omega)$. For large $\omega$, $\sigma_{\mu \mu}$
can be calculated using perturbation theory (due to asymptotic
freedom) and one has $\sigma_{\mu \mu}(\omega)=m_0 \omega^2$ for
$\omega \gg 1GeV$ with $m_0 \simeq 0.076$ \cite{ReindRY1}.
It is natural to build this assumption into the MEM analysis
by choosing the default model $m(\omega)=m_0 \omega^2$ with
$m_0=0.076$. The above form of $\sigma_{\mu \mu}(\omega)$
(or $m(\omega)$ ) is valid only in the continuum; on the lattice
it will certainly be modified and this fact should be taken
into account when choosing the default model. The simplest
way out, suggested in \cite{NakahAH1,AsakaHN1}, is to use 
$m_0 \omega^2$ for the default model but to vary $m_0$ and study
the dependence of the result on $m_0$. We note that the result
varies little if one varies the default model at small $\omega$.
For instance, one can consider $m(\omega)=m_0 \omega^2 \tanh(\omega/(4T))$
or $m_0 \omega^2 +m_1 \omega$ with $m_1 \sim T$ and find only
small changes in the corresponding spectral function. The
result is sensitive to the choice of $m_0$ (this is not
the case at $T=0$ \cite{NakahAH1,AsakaHN1}). In Fig. \ref{m152}
we show the vector spectral function $\sigma_{\mu \mu}(\omega)$
using $m(\omega)=m_0 \omega^2$ with $m_0=0.152$, i.e. with
twice the perturbative value. The gross structure of 
the spectral function remains unchanged; however, the
statistical significance of the suppression for small $\omega$
is considerably reduced.  One expects that dependence on
the default model should become weaker if larger $N_{\tau}$
are considered, but it is unlikely that it will disappear
completely. The most practical way to solve this problem is
to fix the large $\omega$ behavior of $\sigma_{\mu \mu}$ at
$T=0$ and use this as the default model, as at large $\omega$
$\sigma_{\mu \mu}$ is expected to be temperature independent.
\begin{figure}[htbp]
\begin{center}
  \resizebox*{8cm}{!}{\includegraphics{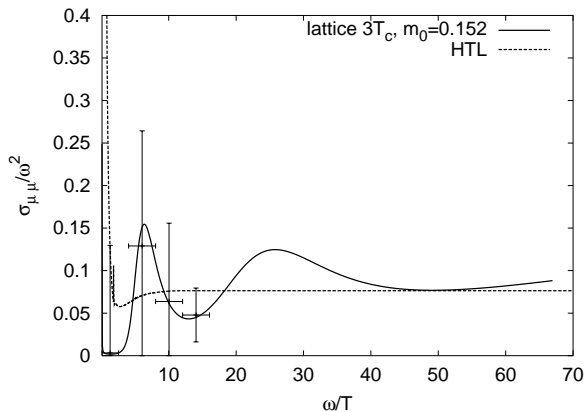}}
\end{center}
\caption{
  The vector spectral function at $T=3T_c$.  The error bars are
  statistical and estimated according to  \cite{NakahAH1,AsakaHN1}. For further
  details see text.}
\label{m152}
\end{figure}

Before closing this section, we should mention an alternative
approach to the problem of extracting the behavior of the vector
spectral function at zero momentum and small energy, proposed recently
by S.~Gupta \cite{Gupta1}. This method tries to extract a Pad\'e
approximation of the energy dependence of the spectral function from
the lattice data for the vector propagator. The extracted slope of the
spectral function at $\omega=0$ gives the following value for the
electrical conductivity:
\begin{equation}
\sigma_{\rm el}\approx 7.0 (\sum_s e_s^2) T\; ,
\end{equation}
where $\sum_s e_s^2$ is the sum of the electric charges squared
of all the species present in the plasma. The discrepancy between this
value of the electrical conductivity and the apparently very small
value one would obtain from the results of \cite{KarscLPSW1} has not
been elucidated at the time of writing this report.

Hopefully, once all these systematic effects are under control, the
above issues can be sorted out and the analysis of the vector spectral
function can be extended to finite momenta to give an estimate on the
emission rate of quasi-real photons.

%
%
%

\section{APPENDIX I -- HARD SCATTERING CROSS SECTIONS: FROM $p p$ TO $p$A AND
A+A }

\noindent
{\em D. G. d'Enterria, J.~Ranft}

\vspace{.5cm}


\subsection{Proton-nucleus ($p$A) collisions}

\subsubsection*{Glauber formalism}

The inelastic cross-section of a $p+$A reaction, $\sigma_{pA}$, 
can be derived in the eikonal limit (straight line trajectories of colliding nucleons)
from the corresponding inelastic nucleon-nucleon $NN$ cross-section, 
$\sigma_{\mbox{\scriptsize{\it{NN}}}}$, and the geometry 
of the $p$A collision simply determined by the impact parameter $b$ of the reaction. 
In the Glauber multiple collision model~\cite{enterria:glau}, such a cross-section reads 

\begin{equation}
\sigma_{pA}=\int d^2b \left[1-e^{-\sigma_{\mbox{\scriptsize{\it{NN}}}}T_A(b)}\right]\; ,
\label{eq:glauber_pA}
\end{equation}
where $T_A(b)$ is the {\it nuclear thickness function} (or {\it nuclear profile function}) 
of the nucleus $A$ at impact parameter $b$:
\begin{equation}
T_A(b)=\int dz \;\rho_A(b,z).
\label{eq:nuc_profile}
\end{equation}
$T_A(b)$ gives the number of nucleons in the nucleus $A$ per unit area along a direction 
$z$ separated from the center of the nucleus by an impact parameter $b$. The nuclear
density, $\rho_A(b,z)$, is usually parametrized by a Woods-Saxon distribution with 
nuclear radius $R_A = 1.19\cdot A^{1/3} - 1.61\cdot A^{-1/3}$ fm and surface thickness $a=0.54$ fm 
as given by the experimental data \cite{enterria:deJae} and normalized so that
\begin{equation}
\int d^2b\; T_A(b) \;= \; A.
\label{eq:norm_T_A}
\end{equation}

\subsubsection*{Hard scattering cross-sections}

Though Eq. (\ref{eq:glauber_pA}) is a general expression for the {\it total} inelastic cross-section, 
it can be applied to an inclusive $p+A\rightarrow h+X$ process of production of particle $h$. 
When one considers hard scattering processes (e.g. direct photon or high-$p_T$ $\pi^0$ production),
the corresponding cross-section $\sigma_{\mbox{\scriptsize{\it{NN}}}}^{hard}$ is small and one can 
expand Eq. (\ref{eq:glauber_pA}) in orders of $\sigma_{\mbox{\scriptsize{\it{NN}}}} T_A(b)$ and then, 
to first approximation 
\begin{equation}
\sigma_{pA}^{hard}\approx \int d^2b\; \sigma_{\mbox{\scriptsize{\it{NN}}}}^{hard}\;T_{A}(b)
\label{eq:glauber_pA_2}
\end{equation}

\subsubsection*{``Minimum bias'' hard scattering cross-sections} 

Integrating Eq. (\ref{eq:glauber_pA_2}) over impact parameter, and using (\ref{eq:norm_T_A}),
one gets the {\it minimum bias} ($MB$) cross-section for a given hard process 
in $p$A collisions
relative to the same cross-section in $pp$ (or $NN$) collisions:
\begin{equation}
(\sigma_{pA}^{hard})_{\mbox{\scriptsize{\it{MB}}}} = \;A\cdot\sigma_{\mbox{\scriptsize{\it{NN}}}}^{hard}
\label{eq:glauber_pA_minbias}
\end{equation}
From this expression it is easy to see that the corresponding 
minimum bias multiplicity (invariant yield per nuclear reaction: 
$N^{hard}_{\mbox{\scriptsize{\it{NN,pA}}}}=\sigma^{hard}_{\mbox{\scriptsize{\it{NN,pA}}}}/\sigma_{\mbox{\scriptsize{\it{NN,pA}}}}^{geo}$) for a given hard-process in a $p$A collision compared to that of a $pp$ collision is
\begin{equation}
\langle N_{\mbox{\scriptsize{\it{pA}}}}^{hard}\rangle_{\mbox{\scriptsize{\it{MB}}}} = \;A\cdot\frac{\sigma_{\mbox{\scriptsize{\it{NN}}}}}{\sigma_{\mbox{\scriptsize{\it{pA}}}}^{geo}}\cdot N_{\mbox{\scriptsize{\it{NN}}}}^{hard} = \frac{A}{\sigma_{\mbox{\scriptsize{\it{pA}}}}^{geo}}\cdot\sigma_{\mbox{\scriptsize{\it{NN}}}}^{hard}\; ,
\label{eq:Nhard_pA_minbias}
\end{equation}
where $\sigma_{\mbox{\scriptsize{\it{pA}}}}^{geo}$ is the geometrical 
$p$A cross-section given, in its most general form, by Eq. (\ref{eq:glauber_pA}). 
The average nuclear thickness function for {\it minimum bias} reactions 
[making use of Eq. (\ref{eq:norm_T_A})] reads:
\begin{equation}
\langle T_{\mbox{\scriptsize{\it{A}}}}\rangle_{\mbox{\scriptsize{\it{MB}}}} \equiv \frac{\int d^2b \; T_{\mbox{\scriptsize{\it{A}}}}}{\int d^2b} \;= \frac{A}{\pi \; R_A^2}=\frac{A}{\sigma_{\mbox{\scriptsize{\it{pA}}}}^{geo}}.
\label{eq:T_pA_minbias}
\end{equation}
Thus, for a $p$Pb ($A(Pb)$ = 208) collision at LHC energies $\sqrt{s_{\mbox{\scriptsize{\it{NN}}}}}$ = 8.8 TeV with
\begin{eqnarray}
\sigma_{\mbox{\scriptsize{\it{NN}}}} & \approx & 77 \; \mbox{ mb \cite{NNcross-section}, and } \\
\sigma_{\mbox{\scriptsize{\it{pPb}}}}^{geo} & \approx & 2162 \; \mbox{ mb \cite{pPbcross-section},} 
\end{eqnarray}
one obtains: 
$\langle N_{\mbox{\scriptsize{\it{pPb}}}}^{hard}\rangle_{\mbox{\scriptsize{\it{MB}}}} \approx 7.4
\cdot N_{\mbox{\scriptsize{\it{NN}}}}^{hard}$, and the average nuclear thickness
function amounts to $\langle T_{\mbox{\scriptsize{\it{Pb}}}}\rangle_{\mbox{\scriptsize{\it{MB}}}}\; 
=$ 0.096 mb$^{-1}$ = 0.96 fm$^{-2}$.

\subsection{Nucleus-nucleus (A+B) collisions}

\subsubsection*{Glauber formalism}

As in the proton-nucleus case, the inclusive inelastic cross-section $\sigma_{\mbox{\scriptsize{\it{AB}}}}$ 
for a collision of nuclei $A$ and $B$ is given in the multiple-scattering Glauber approximation by:
\begin{equation}
\sigma_{\mbox{\scriptsize{\it{AB}}}}=\int d^2b \left[1-e^{-\sigma_{\mbox{\scriptsize{\it{NN}}}}T_{\mbox{\scriptsize{\it{AB}}}}(b)}\right]\; ,
\label{eq:glauber_AB}
\end{equation}
where now $T_{\mbox{\scriptsize{\it{AB}}}}(b)$ is the {\it nuclear overlap function} of the nuclei $A$ and $B$ separated
by impact parameter $b$. $T_{\mbox{\scriptsize{\it{AB}}}}(b)$ can be written as a convolution of the corresponding
thickness functions of $A$ and $B$ over the element of overlapping area $d^2{\vec s}$
 ($\vec{s}=(x,y)$ is a 2-D vector in the transverse plane, and $\vec{b}$ is the impact parameter 
between the centers of the nuclei):
\begin{equation}
T_{\mbox{\scriptsize{\it{AB}}}}(b)\; = \;\int d^2{\vec s} \;T_A({\vec s})\;T_B(|{\vec b}-{\vec s}|).
\label{eq:nuc_overlap}
\end{equation}
$T_{\mbox{\scriptsize{\it{AB}}}}(b)$ is normalized so that integrating over all impact parameters one gets:
\begin{equation}
\int d^2b\; T_{\mbox{\scriptsize{\it{AB}}}}(b)= A\,B.
\label{eq:norm_T_AB}
\end{equation}

\subsubsection*{Hard scattering cross-sections}

As in the $p$A case, for hard processes of the type $A+B\rightarrow h+X$, Eq. (\ref{eq:glauber_AB}), 
can be approximated by:
\begin{equation}
\sigma_{\mbox{\scriptsize{\it{AB}}}}^{hard}\approx \int d^2b\; \sigma_{\mbox{\scriptsize{\it{NN}}}}^{hard}\;T_{\mbox{\scriptsize{\it{AB}}}}(b).
\label{eq:glauber_AB_2}
\end{equation}

\subsubsection*{``Minimum bias'' hard scattering cross-sections and yields}

Integrating Eq. (\ref{eq:glauber_AB_2}) over impact parameter and using (\ref{eq:norm_T_AB}),
one gets the {\it minimum bias} ($MB$) cross-section for a given hard process in A+B collisions
relative to the corresponding $pp$ cross-section:
\begin{equation}
(\sigma_{\mbox{\scriptsize{\it{AB}}}}^{hard})_{\mbox{\scriptsize{\it{MB}}}} = \;A\cdot B\cdot\sigma_{\mbox{\scriptsize{\it{NN}}}}^{hard}
\label{eq:sigma_glauber_AB_minbias}
\end{equation}
Again the corresponding {\it minimum bias} multiplicity (invariant yield per nuclear reaction: 
$N^{hard}_{\mbox{\scriptsize{\it{NN,AB}}}}=\sigma^{hard}_{\mbox{\scriptsize{\it{NN,AB}}}}/\sigma_{\mbox{\scriptsize{\it{NN,AB}}}}^{geo}$)
for a given hard-process in a A+B collision compared to that of a $pp$ collision is
\begin{equation}
\langle N_{\mbox{\scriptsize{\it{AB}}}}^{hard}\rangle_{\mbox{\scriptsize{\it{MB}}}} = \;A\cdot B\cdot\frac{\sigma_{\mbox{\scriptsize{\it{NN}}}}}{\sigma_{\mbox{\scriptsize{\it{AB}}}}^{geo}}\cdot N_{\mbox{\scriptsize{\it{NN}}}}^{hard} = \frac{A\cdot B}{\sigma_{\mbox{\scriptsize{\it{AB}}}}^{geo}}\cdot\sigma_{\mbox{\scriptsize{\it{NN}}}}^{hard}\; ,
\label{eq:Nhard_AB_minbias}
\end{equation}
where $\sigma_{\mbox{\scriptsize{\it{AB}}}}^{geo}$ is the geometrical 
A+B cross-section given, in its most general form, by Eq. (\ref{eq:glauber_AB}). 
The average nuclear overlap function for {\it minimum bias} reactions 
[making use of Eq. (\ref{eq:norm_T_AB})] reads now:
\begin{equation}
\langle T_{\mbox{\scriptsize{\it{AB}}}}\rangle_{\mbox{\scriptsize{\it{MB}}}} \equiv \frac{\int d^2b \; T_{\mbox{\scriptsize{\it{AB}}}}}{\int d^2b} \;= \frac{A\cdot B}{\pi (R_A+R_B)^2}=\frac{AB}{\sigma_{\mbox{\scriptsize{\it{AB}}}}^{geo}},
\label{eq:T_AB_minbias}
\end{equation}
Thus, for a Pb+Pb ($A^2(Pb)$ = 43264) collision at LHC energies
$\sqrt{s_{\mbox{\scriptsize{\it{NN}}}}}$ = 5.5 TeV with
\begin{eqnarray}
\sigma_{\mbox{\scriptsize{\it{NN}}}} & \approx & 72 \;\; \mbox{ mb \cite{NNcross-section}, and} \\
\sigma_{\mbox{\scriptsize{\it{PbPb}}}}^{geo} & \approx & 7745 \;\; \mbox{ mb \cite{PbPbcross-section},} 
\end{eqnarray}
one gets: 
$\langle N_{\mbox{\scriptsize{\it{PbPb}}}}^{hard}\rangle_{\mbox{\scriptsize{\it{MB}}}}
\approx 400 \cdot N_{\mbox{\scriptsize{\it{NN}}}}^{hard}$, and the average
nuclear overlap function amounts to $\langle
T_{\mbox{\scriptsize{\it{PbPb}}}}\rangle_{\mbox{\scriptsize{\it{MB}}}}\;  =$
5.59 mb$^{-1}$ = 55.9 fm$^{-2}$.

\noindent
For comparison for RHIC, at 200 GeV, one has
$\sigma_{\mbox{\scriptsize{\it{NN}}}} \approx 
40.83 \; \mbox{mb}$ and 
$\sigma_{\mbox{\scriptsize{\it{AuAu}}}}^{geo}  \approx  7085 \pm 33\;\mbox{mb}$. 


\subsubsection*{Binary collision scaling}

For a given impact parameter $b$ the {\it average} hard scattering yield can be obtained by 
multiplying each nucleon in nucleus $A$ against the density it sees along the 
$z$ direction in nucleus $B$, then integrated over all of nucleus $A$, i.e.
\begin{equation}
\langle N_{\mbox{\scriptsize{\it{AB}}}}^{hard}\rangle (b) \;=\; 
\sigma_{\mbox{\scriptsize{\it{NN}}}}^{hard}\;\int d^2{\vec s} \;\int \rho_A({\vec s,z'})\;\int \rho_B(|{\vec b}-{\vec s}|,z'')\;dz''dz'\;
\equiv \; \sigma_{\mbox{\scriptsize{\it{NN}}}}^{hard}\cdot T_{\mbox{\scriptsize{\it{AB}}}}(b)\;,
\label{eq:N_AB}
\end{equation}
where we have made use of expressions (\ref{eq:nuc_profile}) and (\ref{eq:nuc_overlap}).
In the same way, one can obtain a useful expression for the probability
of an inelastic $NN$ collision or, equivalently, for the {\it average} number of 
binary inelastic collisions, $\langle N_{coll} \rangle$, in a nucleus-nucleus reaction
with impact parameter $b$:
\begin{equation}
\langle N_{coll}\rangle (b) \; = \; \sigma_{\mbox{\scriptsize{\it{NN}}}}\cdot T_{\mbox{\scriptsize{\it{AB}}}}(b)
\label{eq:N_coll}
\end{equation}
From this last expression one can see that the nuclear overlap function, 
$T_{\mbox{\scriptsize{\it{AB}}}}(b)=N_{coll}(b)/\sigma_{\mbox{\scriptsize{\it{NN}}}}$ 
[mb$^{-1}$], can be thought as the luminosity (reaction rate per unit of cross-section) per $AB$ collision 
at a given impact parameter. 
\comment{
As an example, the average number of binary collisions in minimum bias  Pb+Pb
reactions at LHC ($\sigma_{\mbox{\scriptsize{\it{NN}}}}$ = 72 mb = 7.2
fm$^{2}$) is:
\begin{equation}
\langle N_{coll}\rangle_{\mbox{\scriptsize{\it{MB}}}} = 7.2\;fm^{2}\;\cdot\;55.9\;fm^{-2}\;=\;400.
\end{equation}
}
From (\ref{eq:N_AB}) and (\ref{eq:N_coll}), we get so-called ``binary (or point-like) scaling'' 
formula for the hard scattering yields in heavy-ion reactions:
\begin{equation}
\langle N_{\mbox{\scriptsize{\it{AB}}}}^{hard}\rangle (b) \approx 
\langle N_{coll}\rangle (b) \cdot N_{\mbox{\scriptsize{\it{NN}}}}^{hard}
\label{eq:binary_scaling}
\end{equation}

\subsubsection*{Hard scattering yields and cross-sections in a given centrality class}

Equation (\ref{eq:glauber_AB_2}) gives the reaction cross-section for a given hard process 
in A+B collisions {\it at a given impact parameter} $b$ as a function of the corresponding
reaction cross-section in $pp$ collisions. Very usually, however, in nucleus-nucleus collisions we 
are interested in calculating such a reaction cross-section for a given {\it centrality class}, 
$(\sigma_{AB}^{hard})_{C_1-C_2}$, where the centrality selection 
$C_1-C_2$ corresponds to integrating Eq. (\ref{eq:glauber_AB_2}) between impact parameters $b_1$ and $b_2$.
It is useful, in this case, to define two parameters \cite{enterria:vogt,enterria:star}:
\begin{itemize}
\item The fraction of the total cross-section for hard processes occurring at impact parameters
$b_{1}<b<b_{2}$  ($d^2b = 2\pi bdb$):
\begin{equation}
f_{hard}(b_1<b<b_2)\;=\;\frac{2\pi}{AB}\int_{b_1}^{b_2} bdb\; T_{\mbox{\scriptsize{\it{AB}}}}(b).
\label{eq:f_AB}
\end{equation}
\item The fraction of the geometric cross-section with impact parameter $b_{1}<b<b_{2}$:
\begin{equation}
f_{geo}(b_1<b<b_2)\;=\;\left [2\pi\; \int_{b_1}^{b_2} bdb \left(1-e^{-\sigma_{\mbox{\scriptsize{\it{NN}}}}
T_{\mbox{\scriptsize{\it{AB}}}}(b)}\right)\right]/\sigma_{\mbox{\scriptsize{\it{AB}}}}^{geo}\; ,
\label{eq:f_geo}
\end{equation}
[$f_{geo}$ simply corresponds to a 0.$X$ (e.g. 0.1) factor for the $X$\%(10\%) centrality.]
\end{itemize}
Hard scattering production is more enhanced for increasingly central reactions 
(with larger number on $N_{coll}$) as compared to the total reaction cross-section 
(which includes ``soft'', - scaling with the number of participant nucleons $N_{part}$ -, 
as well as ``hard'' contributions). The growth with $b$ of the geometric cross-section is 
slower than that of the hard component. For this reason, the behaviour of $f_{\mbox{\scriptsize{\it{AB}}}}$ and $f_{geo}$
as a function of $b$, although similar in shape is not the same (see \cite{enterria:vogt}): 
$f_{hard}\approx$ 1 for $b$ = 2$R_A$, but $f_{geo}\approx$ 0.75 for $b$ = 2$R_A$.

Similarly to (\ref{eq:T_AB_minbias}), we can obtain now the nuclear overlap function for 
any given centrality class $C_1-C_2$:
\begin{equation}
\langle T_{\mbox{\scriptsize{\it{AB}}}}\rangle_{C_1-C_2} \;\equiv\; \frac{\int^{b_2}_{b_1} d^2b \; T_{\mbox{\scriptsize{\it{AB}}}}}
{\int^{b_2}_{b_1} d^2b} = \frac{A\cdot B}{\sigma_{\mbox{\scriptsize{\it{AB}}}}^{geo}} \cdot \frac{f_{hard}}{f_{geo}}\;
\label{eq:T_AB_centrality}
\end{equation}
The number of hard processes per nuclear collision for reactions with
impact parameter $b_1<b<b_2$ is given by
\begin{equation}
\langle N_{\mbox{\scriptsize{\it{AB}}}}^{\mbox{\scriptsize{\it{hard}}}}\rangle_{C_1-C_2} = 
\frac{\sigma_{\mbox{\scriptsize{\it{AB}}}}^{hard}\mbox{\small{$(b_1<b<b_2$)}}}
{\sigma_{\mbox{\scriptsize{\it{AB}}}}^{geo}\mbox{\small{$(b_1<b<b_2)$}}} = A\cdot B\cdot \frac{\sigma_{\mbox{\scriptsize{\it{NN}}}}
^{\mbox{\scriptsize{\it{hard}}}}}{\sigma_{\mbox{\scriptsize{\it{AB}}}}^{geo}} \cdot \frac{f_{hard}}{f_{geo}} \; ,
\label{eq:N_hard_centrality}
\end{equation}
which we could have just obtained directly from (\ref{eq:N_AB}) and (\ref{eq:T_AB_centrality}). From 
(\ref{eq:Nhard_AB_minbias}) and (\ref{eq:N_hard_centrality}) it is also easy to see that:
\begin{equation}
\langle N_{\mbox{\scriptsize{\it{AB}}}}^{\mbox{\scriptsize{\it{hard}}}}\rangle_{C_1-C_2} = 
\langle N_{\mbox{\scriptsize{\it{AB}}}}^{\mbox{\scriptsize{\it{hard}}}}\rangle_{\mbox{\scriptsize{\it{MB}}}} \cdot \frac{f_{hard}}{f_{geo}}
\label{eq:N_hard_centrality2}
\end{equation}

Finally, the cross-section for hard processes produced in the centrality class $C_1-C_2$ (corresponding
to a fraction $f_{geo}$ of the reaction cross-section) is:
\begin{equation}
(\sigma_{\mbox{\scriptsize{\it{AB}}}}^{\mbox{\scriptsize{\it{hard}}}})_{C_1-C_2} = 
A\cdot B \cdot \frac{f_{hard}}{f_{geo}} \cdot \sigma_{\mbox{\scriptsize{\it{NN}}}}^{\mbox{\scriptsize{\it{hard}}}}
\label{eq:sigma_hard_centrality}
\end{equation}
Figure \ref{fig:fgeo_vs_fhard}, extracted from \cite{enterria:vogt}, plots the (top) fraction of the hard 
cross-section, $f_{hard}(0<b<b_2)$ (labeled in the plot as $f_{\mbox{\scriptsize{\it{AB}}}}$), as a function of 
the top fraction of the total geometrical cross-section, $f_{geo}(0<b<b_2)$,
for several nucleus-nucleus reactions.
\begin{figure}[htbp]
\begin{center}
\includegraphics[height=7.0cm]{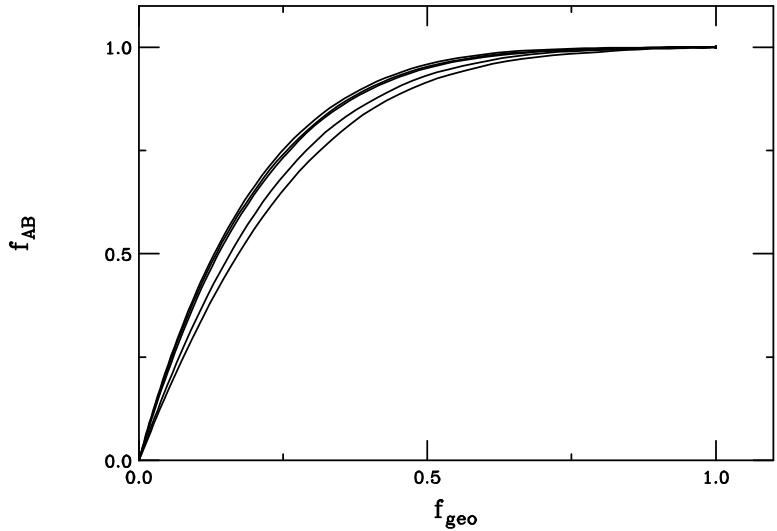}
\end{center}
\caption{Figure 4 of ref. \protect\cite{enterria:vogt}. Fraction of the hard cross-section, $f_{hard}(0<b<b_2)$, vs. 
the fraction of the total geometrical cross-section, $f_{geo}(0<b<b_1)$, for several heavy-ion collisions 
(from left to right): 197+197, 110+197, 63+193, 27+197, and 16+197.}
\label{fig:fgeo_vs_fhard}
\end{figure}
As a practical application of Eq. (\ref{eq:sigma_hard_centrality}) and the results of Fig. \ref{fig:fgeo_vs_fhard}, 
the hard-scattering cross-sections in Pb+Pb for the top 0-10\% ($f_{hard}$ = 0.41 for $f_{geo}$ = 0.1
from the practically equivalent Au+Au system of figure \ref{fig:fgeo_vs_fhard}) 
and 0-20\% ($f_{hard}$ = 0.664 for $f_{geo}$ = 0.2) central collisions relate to the $pp$ cross-section, 
in the absence of nuclear effects, respectively as:
\begin{eqnarray}
(\sigma_{\mbox{\scriptsize{\it{AB}}}}^{\mbox{\scriptsize{\it{hard}}}})_{0-10\%} & = & (208)^2\cdot \frac{0.41 }{0.1}\cdot \sigma_{\mbox{\scriptsize{\it{NN}}}}^{\mbox{\scriptsize{\it{hard}}}}\;\approx\;1.7\cdot 10^{5}\cdot\sigma_{\mbox{\scriptsize{\it{NN}}}}^{\mbox{\scriptsize{\it{hard}}}}\\
(\sigma_{\mbox{\scriptsize{\it{AB}}}}^{\mbox{\scriptsize{\it{hard}}}})_{0-20\%} & = & (208)^2\cdot \frac{0.664}{0.2}\cdot \sigma_{\mbox{\scriptsize{\it{NN}}}}^{\mbox{\scriptsize{\it{hard}}}}\;\approx\;1.4\cdot 10^{5}\cdot\sigma_{\mbox{\scriptsize{\it{NN}}}}^{\mbox{\scriptsize{\it{hard}}}}
\label{eq:N_hard_values}
\end{eqnarray}


A straightforward way to compute the invariant yield for a given hard 
process in a given centrality class of a nucleus-nucleus collision from the corresponding yield 
in $pp$ collisions consists in determining, via a Glauber MC calculation, the number of inelastic 
$NN$ collisions corresponding to that centrality class via 
\begin{equation}
\langle N_{coll}\rangle_{C_1-C_2} \; =\; \langle T_{\mbox{\scriptsize{\it{AB}}}}\rangle_{C_1-C_2} \cdot 
\sigma_{\mbox{\scriptsize{\it{NN}}}}\; ,
\end{equation}
and then use this value in the ``binary-scaling'' formula 
\begin{eqnarray}
\langle N_{\mbox{\scriptsize{\it{AB}}}}^{hard}\rangle_{C_1-C_2} & = &
\langle N_{coll}\rangle_{C_1-C_2} \cdot N_{\mbox{\scriptsize{\it{NN}}}}^{hard}\;,\;\;\;\;\;\mbox{ or}
\label{eq:binary_scaling_yields1}
\\
(\sigma_{\mbox{\scriptsize{\it{AB}}}}^{\mbox{\scriptsize{\it{hard}}}})_{C_1-C_2}  & = & \langle N_{coll}\rangle_{C_1-C_2} \cdot 
\frac{\sigma_{\mbox{\scriptsize{\it{AB}}}}^{geo}}{\sigma_{\mbox{\scriptsize{\it{NN}}}}}\cdot\sigma_{\mbox{\scriptsize{\it{NN}}}}^{\mbox{\scriptsize{\it{hard}}}}
\label{eq:binary_scaling_yields2}
\end{eqnarray}
The same two formulae above apply to $p$A collisions (of course substituting A+B by $p$A and computing
$N_{coll}$ from $T_{A}$ instead of from $T_{AB}$).

Finally, to obtain the {\it experimental rates}, 
$({\cal N}_{\mbox{\scriptsize{\it{AB}}}}^{\mbox{\scriptsize{\it{hard}}}})_{C_1-C_2}$, 
actually measured in a given centrality bin one needs to take into account the expected 
{\it integrated luminosity} ${\cal L}_{int}$ [mb$^{-1}$] as follows:
\begin{eqnarray}
({\cal N}_{\mbox{\scriptsize{\it{AB}}}}^{\mbox{\scriptsize{\it{hard}}}})_{C_1-C_2} & = & {\cal L}_{int} \cdot 
(\sigma_{\mbox{\scriptsize{\it{AB}}}}^{\mbox{\scriptsize{\it{hard}}}})_{C_1-C_2} 
\label{eq:experimental_rates}
\end{eqnarray}

\subsection{Hard scattering yields and cross-sections for $p$Pb and Pb+Pb collisions}

As a practical application of the Glauber approach described here, in Table
\ref{tab:glauber_Ncoll}, the values of $\langle N_{coll}\rangle$ are quoted for
different centrality classes obtained from a Monte Carlo calculation \cite{enterria:klaus}
for $p$Pb ($\sqrt{s_{\mbox{\scriptsize{\it{NN}}}}}$ = 8.8 TeV and 
$\sigma_{\mbox{\scriptsize{\it{NN}}}}$ = 77 mb) and Pb+Pb 
($\sqrt{s_{\mbox{\scriptsize{\it{NN}}}}}$ = 5.5 TeV for an inelastic $pp$ cross-section 
of $\sigma_{\mbox{\scriptsize{\it{NN}}}}$ = 72 mb) collisions 
(Woods-Saxon Pb density parametrization with $R_A$ = 6.78 fm 
and $a$ = 0.54 fm).
\begin{table}[htb]
\begin{center}
\caption{Number of inelastic $NN$ collisions, $\langle N_{coll}\rangle$, 
and nuclear thickness $\langle T_{pPb} \rangle$ or overlap $\langle 
T_{PbPb} \rangle$ function
per centrality class, in $p$Pb ($\sqrt{s_{\mbox{\scriptsize{\it{NN}}}}}$ 
= 8.8 TeV,
$\sigma_{\mbox{\scriptsize{\it{NN}}}}$ = 77 mb) and Pb+Pb collisions at LHC
($\sqrt{s_{\mbox{\scriptsize{\it{NN}}}}}$ = 5.5 TeV, 
$\sigma_{\mbox{\scriptsize{\it{NN}}}}$ = 72 mb)
obtained with the Glauber Monte Carlo code of ref. 
\protect\cite{enterria:klaus}. The errors,
not shown, are of the same order as the current uncertainty in the value 
of the nucleon-nucleon
inelastic cross section, $\sigma_{\mbox{\scriptsize{\it{NN}}}}$, at
LHC energies ($\sim$ 10\%).}
\vskip0.4cm
\begin{tabular}{c|c|c|c|c}
\hline\hline
\hspace{1mm}
Centrality ($C_1-C_2$)\hspace{1mm} & \multicolumn{2}{c|}{p+Pb}  & 
\multicolumn{2}{c}{Pb+Pb} \\
 & \hspace{6mm} $\langle N_{coll} \rangle$ \hspace{6mm} & \hspace{1mm} 
$\langle T_{pPb} \rangle$ (mb$^{-1}$) \hspace{1mm} &
   \hspace{6mm} $\langle N_{coll} \rangle$ \hspace{7mm} & \hspace{1mm} 
$\langle T_{PbPb} \rangle$ (mb$^{-1}$) \hspace{1mm} \\\hline
\hline
  0- 5\% & 15.7 & 0.203                & 1876.0  & 26.0\\
  0-10\% & 15.3 & 0.198                & 1670.2  & 23.2 \\
 10-20\% & 13.8 & 0.179                & 1019.5  & 14.2 \\
 20-30\% & 12.0 & 0.155                &  612.4  &  8.50 \\
 30-40\% &  9.9 & 0.128                &  351.8  &  4.89 \\
 40-50\% &  7.8 & 0.101                &  188.0  &  2.61 \\
 50-60\% &  5.6 & 7.27$\cdot$10$^{-2}$ &   92.9  &  1.29 \\
 60-70\% &  3.8 & 4.93$\cdot$10$^{-2}$ &   41.4  &  5.75$\cdot$10$^{-1}$ 
\\
 70-80\% &  2.6 & 3.37$\cdot$10$^{-2}$ &   16.8  &  2.33$\cdot$10$^{-1}$ 
\\
 80-90\% &  1.7 & 2.20$\cdot$10$^{-2}$ &    6.7  &  9.31$\cdot$10$^{-2}$ 
\\
90-100\% &  1.2 & 1.55$\cdot$10$^{-2}$ &    2.7  &  3.75$\cdot$10$^{-2}$ 
\\
min. bias&  7.4 & 9.61$\cdot$10$^{-2}$ &  400.0  &  5.58\\\hline\hline
\\\hline\hline
\end{tabular}
\label{tab:glauber_Ncoll}
\end{center}
\end{table}
Using (\ref{eq:binary_scaling_yields1}), (\ref{eq:binary_scaling_yields2}) and Table~\ref{tab:glauber_Ncoll}, 
we can now easily get the scaling factors of the 
cross-sections and yields from $pp$ to, e.g., central (0-10\%), minimum bias, 
and semi-peripheral (60-80\%, from the combined average 60-70\% and 70-80\%) $p$Pb (8.8 TeV) 
and Pb+Pb (5.5 TeV) collisions :

For $p$Pb collisions ($\sigma_{\mbox{\scriptsize{\it{pPb}}}}^{geo}$ = 2162 mb):

\begin{eqnarray}
\langle N_{\mbox{\scriptsize{\it{pPb}}}}^{\mbox{\scriptsize{\it{hard}}}}\rangle_{0-10\%}  = 15.3 \cdot N_{\mbox{\scriptsize{\it{NN}}}}^{hard} 
& \; \Longrightarrow \;  & 
(\sigma_{\mbox{\scriptsize{\it{pPb}}}}^{\mbox{\scriptsize{\it{hard}}}})_{0-10\%} = 15.3 \cdot \frac{2162}{77}\cdot\sigma_{\mbox{\scriptsize{\it{NN}}}}^{\mbox{\scriptsize{\it{hard}}}} \;\approx\; 4.5\cdot 10^{2}\cdot\sigma_{\mbox{\scriptsize{\it{NN}}}}^{\mbox{\scriptsize{\it{hard}}}}\\
\langle N_{\mbox{\scriptsize{\it{pPb}}}}^{\mbox{\scriptsize{\it{hard}}}}\rangle_{60-80\%}  = 3.2 \cdot N_{\mbox{\scriptsize{\it{NN}}}}^{hard} 
& \; \Longrightarrow \;  & 
(\sigma_{\mbox{\scriptsize{\it{pPb}}}}^{\mbox{\scriptsize{\it{hard}}}})_{60-80\%} = 3.2 \cdot \frac{2162}{77}\cdot\sigma_{\mbox{\scriptsize{\it{NN}}}}^{\mbox{\scriptsize{\it{hard}}}} \;\approx\; 10^{2}\cdot\sigma_{\mbox{\scriptsize{\it{NN}}}}^{\mbox{\scriptsize{\it{hard}}}}\\
\langle N_{\mbox{\scriptsize{\it{pPb}}}}^{\mbox{\scriptsize{\it{hard}}}}\rangle_{MB} = 7.4 \cdot N_{\mbox{\scriptsize{\it{NN}}}}^{hard} 
& \; \Longrightarrow \;  & 
(\sigma_{\mbox{\scriptsize{\it{pPb}}}}^{\mbox{\scriptsize{\it{hard}}}})_{MB} = 7.4 \cdot \frac{2162}{77}\cdot\sigma_{\mbox{\scriptsize{\it{NN}}}}^{\mbox{\scriptsize{\it{hard}}}} \;\approx\; 2\cdot 10^{2}\cdot\sigma_{\mbox{\scriptsize{\it{NN}}}}^{\mbox{\scriptsize{\it{hard}}}}
\label{eq:N_hard_pPb}
\end{eqnarray}

For Pb+Pb collisions ($\sigma_{\mbox{\scriptsize{\it{PbPb}}}}^{geo}$ = 7745 mb):

\begin{eqnarray}
\langle N_{\mbox{\scriptsize{\it{PbPb}}}}^{\mbox{\scriptsize{\it{hard}}}}\rangle_{0-10\%}  = 1670 \cdot N_{\mbox{\scriptsize{\it{NN}}}}^{hard} & \; \Longrightarrow \;  & 
(\sigma_{\mbox{\scriptsize{\it{PbPb}}}}^{\mbox{\scriptsize{\it{hard}}}})_{0-10\%} = 1670 \cdot \frac{7745}{72}\cdot\sigma_{\mbox{\scriptsize{\it{NN}}}}^{\mbox{\scriptsize{\it{hard}}}} \;\approx\; 1.6\cdot 10^{5}\cdot\sigma_{\mbox{\scriptsize{\it{NN}}}}^{\mbox{\scriptsize{\it{hard}}}}\\
\langle N_{\mbox{\scriptsize{\it{PbPb}}}}^{\mbox{\scriptsize{\it{hard}}}}\rangle_{60-80\%} = 29.1 \cdot N_{\mbox{\scriptsize{\it{NN}}}}^{hard} & \; \Longrightarrow \; & 
(\sigma_{\mbox{\scriptsize{\it{PbPb}}}}^{\mbox{\scriptsize{\it{hard}}}})_{60-80\%} = 29.1 \cdot \frac{7745}{72}\cdot\sigma_{\mbox{\scriptsize{\it{NN}}}}^{\mbox{\scriptsize{\it{hard}}}} \;\approx\; 3.1\cdot 10^{3}\cdot\sigma_{\mbox{\scriptsize{\it{NN}}}}^{\mbox{\scriptsize{\it{hard}}}}\\
\langle N_{\mbox{\scriptsize{\it{PbPb}}}}^{\mbox{\scriptsize{\it{hard}}}}\rangle_{MB} = 400 \cdot N_{\mbox{\scriptsize{\it{NN}}}}^{hard} & \; \Longrightarrow \; & 
(\sigma_{\mbox{\scriptsize{\it{PbPb}}}}^{\mbox{\scriptsize{\it{hard}}}})_{MB} = 400 \cdot \frac{7745}{72}\cdot\sigma_{\mbox{\scriptsize{\it{NN}}}}^{\mbox{\scriptsize{\it{hard}}}} \;\approx\; 4.3\cdot 10^{4}\cdot\sigma_{\mbox{\scriptsize{\it{NN}}}}^{\mbox{\scriptsize{\it{hard}}}}
\label{eq:N_hard_valuesPbPb}
\end{eqnarray}

\subsection{Nuclear effects in $p$A and A+B collisions}

Eqs. (\ref{eq:glauber_pA_2}) and (\ref{eq:glauber_AB_2}) for the hard scattering cross-sections in $p$A and A+B 
collisions have been derived within an eikonal framework which only takes into account the geometric aspects 
of the reactions. Any differences of the experimentally measured $\sigma_{pA,AB}^{hard}$ with respect to these 
expressions 
indicate ``de facto'' the existence of ``nuclear effects'' (such as e.g. ``shadowing'', ``Cronin enhancement'',
or ``parton energy loss'') not accounted for by the Glauber formalism. Indeed, in the multiple-scattering Glauber 
model each nucleon-nucleon collision is treated incoherently and thus, unaffected by any other scattering 
taking place before (initial-state) or after (final-state effects) it.\\

If the Glauber approximation holds, from (\ref{eq:glauber_pA_2}) and (\ref{eq:glauber_AB_2}) 
one would expect a $\propto A^1$, and $\propto A^2$ growth of the 
hard processes cross-section with system size respectively. Equivalently, since 
$N^{hard}_{\mbox{\scriptsize{\it{NN,AB}}}}=\sigma^{hard}_{\mbox{\scriptsize{\it{NN,AB}}}}/\sigma_{\mbox{\scriptsize{\it{NN,AA}}}}^{geo}$ and $\sigma_{\mbox{\scriptsize{\it{NN,AB}}}}^{geo}\propto R_A^2$ with $R_A \propto A^{1/3}$, one would expect a
growth of the {\it number} of hard process as $\propto A^{1/3}, \propto A^{4/3}$ for $pA, AA$ collisions respectively.
Experimentally, in minimum bias $pA$ and $AB$ collisions, it has been found that the 
production cross-sections for hard processes actually grow as:
\begin{equation}
(\sigma_{pA}^{hard})_{\mbox{\scriptsize{\it{MB}}}} = \;A^\alpha\cdot\sigma_{\mbox{\scriptsize{\it{NN}}}}^{hard} \;,\;\;\;\;
\mbox{ and }\;\;\; 
(\sigma_{\mbox{\scriptsize{\it{AB}}}}^{hard})_{\mbox{\scriptsize{\it{MB}}}} = \;(AB)^\alpha\cdot\sigma_{\mbox{\scriptsize{\it{NN}}}}^{hard} \;,\;\;\;\; \mbox{ with } \alpha\neq 1
\label{eq:minbias_exp}
\end{equation}
More precisely, in high-$p_T$ processes in $pA$ and heavy-ion collisions at SPS energies one founds $\alpha>$ 1 
(due to initial-state p$_T$ broadening or ``Cronin enhancement''); whereas $\alpha<$ 1 at RHIC energies 
(``high-$p_T$ suppression''). Theoretically, one can still make predictions on the hard probe yields 
in $pA,AB$ collisions using the pQCD 
factorization machinery for the $pp$ cross-section complemented with the Glauber formalism while
modifiying effectively the nuclear PDFs and parton fragmentation functions 
to take into account any initial- and/or final- state nuclear medium effects.

\subsection{Summary of useful formulae}

Finally, let us summarize a few useful formulae derived here to determine the hard-scattering 
invariant yields, cross-sections, or experimental rates, from $pp$ to $p$A and A+B 
collisions for centrality bin $C_1-C_2$ (corresponding to a nuclear thickness $T_{A}$ or nuclear 
overlap function $T_{AB}$ and to an average number of $NN$ inelastic collisions $\langle N_{coll}\rangle$):
\begin{eqnarray}
\frac{(d^2 N_{\mbox{\tiny{\it{pA,AB}}}}^{hard})_{C_1-C_2}}{dp_Tdy} & = &
\langle T_{A,AB}\rangle_{C_1-C_2} \cdot \frac{d^2
\sigma_{\mbox{\scriptsize{\it{pp}}}}^{hard}}{dp_Tdy} 
\\
\frac{(d^2
\sigma_{\mbox{\tiny{\it{pA,AB}}}}^{\mbox{\scriptsize{\it{hard}}}})_{C_1-C_2}}{dp_Tdy} 
& = & \langle T_{A,AB}\rangle_{C_1-C_2} \cdot
\sigma_{\mbox{\tiny{\it{pA,AB}}}}^{geo} \cdot 
\frac{d^2\sigma_{\mbox{\scriptsize{\it{pp}}}}^{\mbox{\scriptsize{\it{hard}}}}}{dp_Tdy}
\\
\frac{(d^2 {\cal
N}_{\mbox{\tiny{\it{pA,AB}}}}^{\mbox{\scriptsize{\it{hard}}}})_{C_1-C_2}}{dp_Tdy} 
& = &  {\cal L}_{int} \cdot \langle T_{A,AB}\rangle_{C_1-C_2} \cdot
\sigma_{\mbox{\tiny{\it{pA,AB}}}}^{geo} \cdot  \frac{d^2
\sigma_{\mbox{\scriptsize{\it{pp}}}}^{\mbox{\scriptsize{\it{hard}}}}}{dp_Tdy}
\\
\frac{(d^2 N_{\mbox{\tiny{\it{pA,AB}}}}^{hard})_{C_1-C_2}}{dp_Tdy} & = &
\langle N_{coll}\rangle_{C_1-C_2} \cdot \frac{d^2
N_{\mbox{\scriptsize{\it{pp}}}}^{hard}}{dp_Tdy} 
\label{eq:binary_scaling_yiel1} \\
\frac{(d^2
\sigma_{\mbox{\tiny{\it{pA,AB}}}}^{\mbox{\scriptsize{\it{hard}}}})_{C_1-C_2}}{dp_Tdy} 
& = &  \langle N_{coll}\rangle_{C_1-C_2} \cdot
\frac{\sigma_{\mbox{\tiny{\it{pA,AB}}}}^{geo}}{\sigma_{\mbox{\scriptsize{\it{NN}}}}}
\cdot \frac{d^2
\sigma_{\mbox{\scriptsize{\it{pp}}}}^{\mbox{\scriptsize{\it{hard}}}}}{dp_Tdy}
\label{eq:binary_scaling_yiel11} \\
\frac{(d^2 {\cal
N}_{\mbox{\tiny{\it{pA,AB}}}}^{\mbox{\scriptsize{\it{hard}}}})_{C_1-C_2}}{dp_Tdy} 
& = &  {\cal L}_{int}\cdot \langle N_{coll}\rangle_{C_1-C_2} \cdot 
\frac{\sigma_{\mbox{\tiny{\it{pA,AB}}}}^{geo}}{\sigma_{\mbox{\scriptsize{\it{NN}}}}}
\cdot \frac{d^2
\sigma_{\mbox{\scriptsize{\it{pp}}}}^{\mbox{\scriptsize{\it{hard}}}}}{dp_Tdy}
\label{eq:binary_scaling_yiel111}
\end{eqnarray}

\newpage

\section{APPENDIX II -- STANDARDS FOR LUMINOSITIES AND ACCEPTANCES}
\label{app:acceptance}

The list of considered collision systems is:
Pb+Pb, D+Pb or $p$Pb, Ar+Ar and $pp$.

\vspace{.4cm}

\noindent
{\bf Luminosities: in (cm$^{-2}$ s${^-1}$)}

   Pb+Pb :      $5*10^{26}$
   
   D+Pb :       $5*10^{28}$
   
   $p$Pb  :     $10^{29}$
   
   Ar+Ar :      $10^{29}$  but $3*10^{27}$ for ALICE central part
   
   $pp$   :      $10^{34}$  but $3*10^{30}$ for ALICE

\vspace{.4cm}

\noindent
{\bf Center of mass energy}

\noindent
The center of mass energy of a collision of nucleus $(A1,\ Z1)$ with nucleus
$(A2,\ Z2)$ is given by $\sqrt s =$ 14 TeV * $\sqrt{Z1*Z2/(A1*A2)}$

\vspace{.4cm}

\noindent
{\bf CMS Acceptance}

\noindent
For electron, muon, photon and jet reconstruction, essentially no hadronic
particle identification (PID) is available.
In heavy ion collisions, CMS understands tracking for
$|\eta| < 1.5$,  $p_T > 4 $ GeV.
The tracking studies for charged hadrons at $p_T > 1$ GeV is in progress.
This should be a conservative standard.  On the other hand, the principal CMS
rapidity acceptance is:

$|\eta| < 2.4$ for muons and charged hadrons,

$|\eta| < 3.$ for photons and electrons, 

$|\eta| < 5$ for jets. 

\noindent
 Depending on event multiplicity and further studies, a larger acceptance
 [$|\eta| <  2.5$  and $p{}_T > 1.5$ GeV in the forward directions] may be
 available for heavy ion collisions. These numbers may be used as more
 realistic standard.

\vspace{.4cm}

\noindent
{\bf ALICE Acceptance}

 For hadrons and electrons: $|\eta| <  0.9$,
 
 PID   pion          $p_T > 100$ MeV
 
 PID   kaon, proton  $p_T > 200$ MeV
 
 HMPID small acceptance RICH allows PID of $p,\ K$
 for $p_T < 5$ GeV, $|\eta| <  0.5$ and 57 degrees azimuth

 PHOS photon spectrometer covers
 $|\eta| <  0.13$ and 100 degrees azimuth

 For muons:    $ 2.5 < \eta < 4$,     $p_T > 1.0$ GeV

 Acceptance for charmonium and bottonium reconstruction from
 electron pairs:  $|\eta| <  0.9$, $p_T > 0$. For more details
 about this acceptance and trigger conditions, contact
 experimentalists.

\newpage
%
\noindent
{\bf Acknowledgments} 

\vspace{.2cm}
\noindent  
The following sources of funding are acknowledged:

Academy of Finland, Project 50338: K.J. Eskola, H. Niemi, P.V. Ruuskanen, S.S.
R\"as\"anen;

Alexander von Humboldt Foundation: K. Redlich;

Department of Energy (U.S.A.): G.~David;

Polish State Committee for Scientific Research (KBN), grant 2P03 (06925):
K. Redlich.

\noindent  
F.~Arleo, F.~Gelis, J.~Ranft and V.~Ruuskanen thank P.~Sorba for hospitality at
LAPTH  where some of the work presented in this report was initiated. Support
from the Institute for Nuclear Theory, University of Washington, Seattle is
also gratefully acknowledged by F. ~Arleo and P.~Aurenche.

\vspace{.5cm}
%
%
%

\end{document}